\documentstyle[12pt,newlfont,
macros,epsf,newcomnd]{article}

\newlength{\pcm}
\newlength{\pmm}
\newcommand {\GB} {\,\epsfxsize=1.2\pcm \parbox{1.2\pcm}{\epsfbox{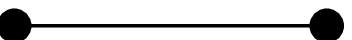}}\,}
\newcommand {\GBdotted} {\,\epsfxsize=1.2\pcm \parbox{1.2\pcm}{\epsfbox{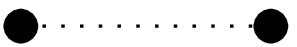}}\,}

\newcommand {\GH} {\,\epsfxsize=0.8\pcm \parbox{0.8\pcm}{\epsfbox{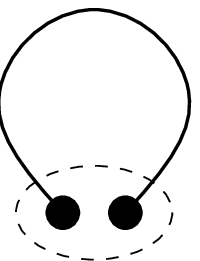}}\,}
\newcommand {\GI} {\,\epsfxsize=2\pcm \parbox{2\pcm}{\epsfbox{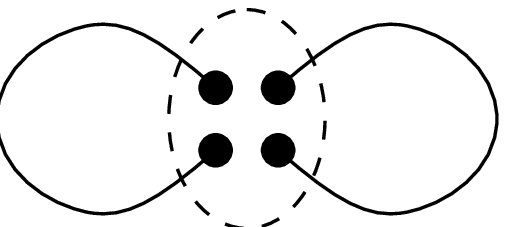}}\,}
\newcommand {\GIsubeins} {\,\epsfxsize=2\pcm \parbox{2\pcm}{\epsfbox{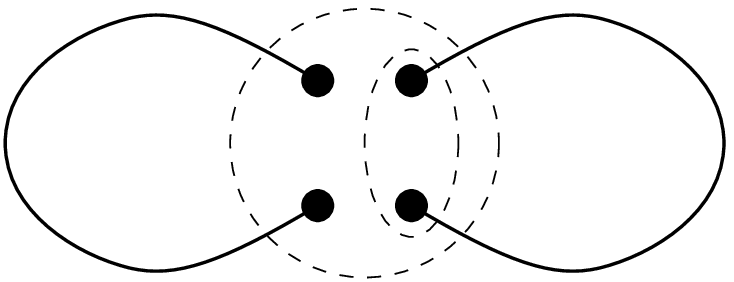}}\,}
\newcommand {\GIsubzwei} {\,\epsfxsize=2\pcm \parbox{2\pcm}{\epsfbox{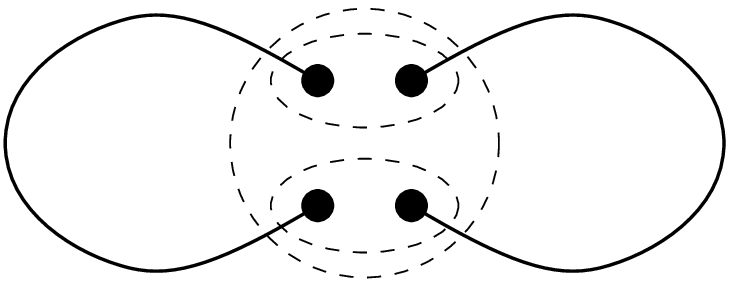}}\,}
\newcommand {\GJ} {\,\epsfxsize=1.3\pcm \parbox{1.3\pcm}{\epsfbox{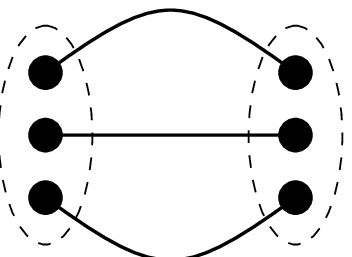}}\,}
\newcommand {\GJsub} {\,\epsfxsize=1.3\pcm \parbox{1.3\pcm}{\epsfbox{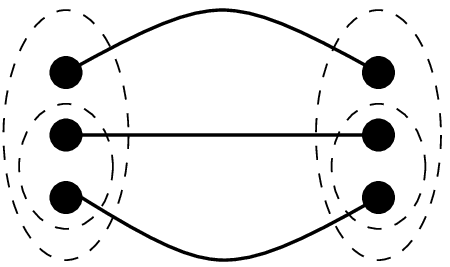}}\,}

\newcommand {\GM} {\,\epsfxsize=1.5\pcm \parbox{1.5\pcm}{\epsfbox{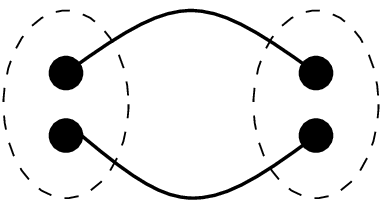}}\,}
\newcommand {\GN} {\,\epsfxsize=0.13\pcm \parbox{0.13\pcm}{\epsfbox{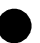}}\,}
\newcommand {\GO} {\,\epsfxsize=0.4\pcm \parbox{0.4\pcm}{\epsfbox{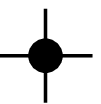}}\,}
\newcommand {\GOalphabeta} {\,\epsfxsize=0.4\pcm 
\parbox{0.4\pcm}{\epsfbox{./eps/go.eps}}
\hspace{-4.5\pmm}\raisebox{-1\pmm}[0\pmm][0\pmm]{$_{\alpha\,\,\beta}$}\,}
\newcommand {\GP} {\,\epsfxsize=2.5\pcm \parbox{2.5\pcm}{\epsfbox{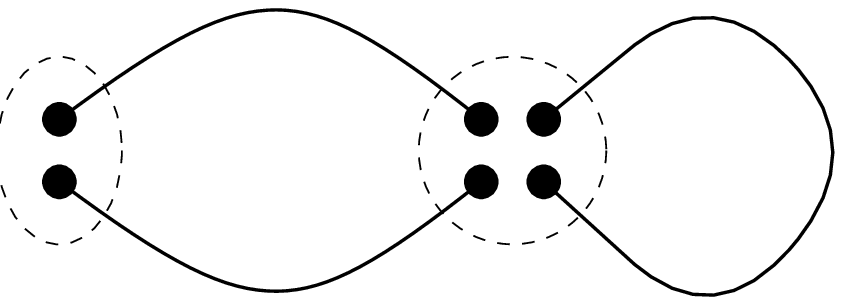}}\,}
\newcommand {\GPpsub} {\,\epsfxsize=2.5\pcm \parbox{2.5\pcm}{\epsfbox{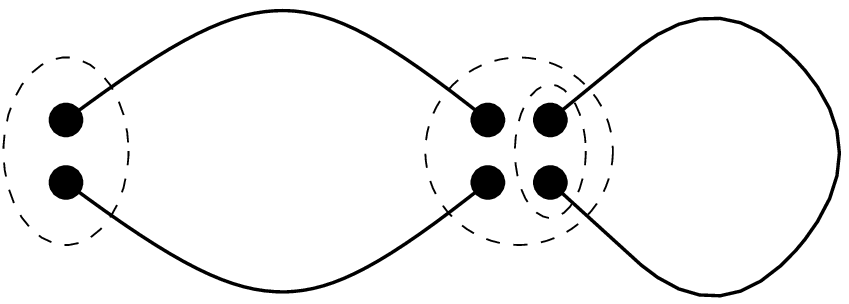}}\,}
\newcommand {\GPrel} {\,\epsfxsize=2.5\pcm \parbox{2.5\pcm}{\epsfbox{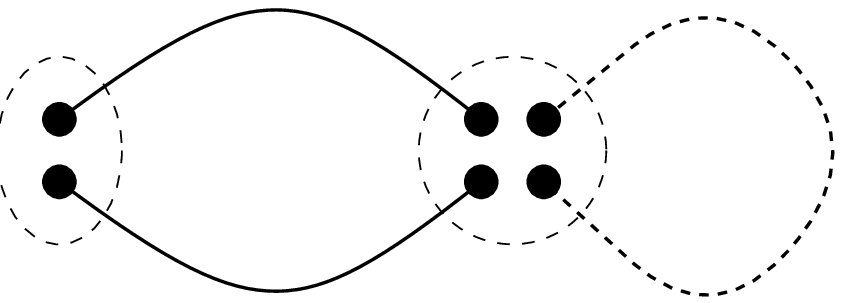}}\,}

\newcommand {\GW} {\,\epsfxsize=1.2\pcm
\parbox{1.2\pcm}{\epsfbox{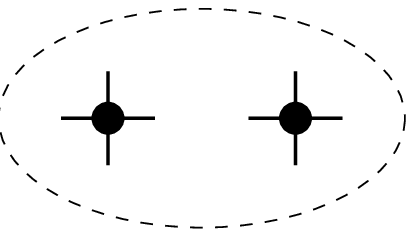}}\,}
\newcommand {\GX} {\,\epsfxsize=2.4\pcm
\parbox{2.4\pcm}{\epsfbox{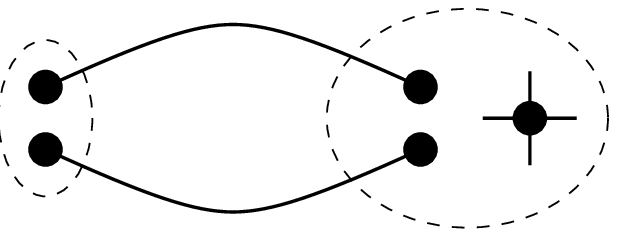}}\,}
\newcommand {\GXalphabeta} {\,\epsfxsize=2.4\pcm
\parbox{2.4\pcm}{\epsfbox{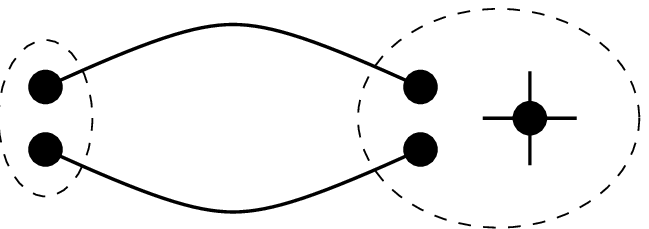}}
\hspace{-6.75\pmm}\raisebox{-0.75\pmm}[0\pmm][0\pmm]{$_{\alpha\,\,\beta}$}\hspace{4.5\pmm}}

\newcommand {\GZ} {\,\epsfxsize=1.1\pcm
\parbox{1.1\pcm}{\epsfbox{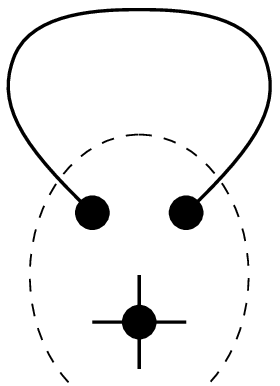}}\,}
\newcommand {\GZp} {\,\epsfxsize=1.1\pcm
\parbox{1.1\pcm}{\epsfbox{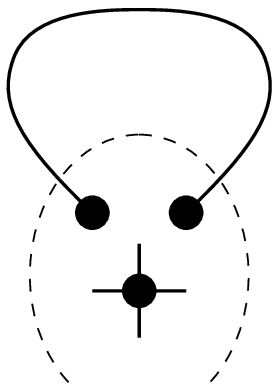}}\,}
\newcommand {\GZalphabeta}{\GZp\hspace{-8.4\pmm}\raisebox{-5\pmm}[0\pmm][0\pmm]{$_{\alpha\,\beta}$}\hspace{4.5\pmm} }

\newcommand {\FD} {\,\epsfxsize=1.7\pcm \parbox{1.7\pcm}{\epsfbox{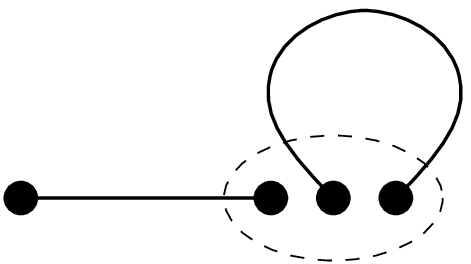}}\,}
\newcommand {\FE} {\,\epsfxsize=1.6\pcm \parbox{1.6\pcm}{\epsfbox{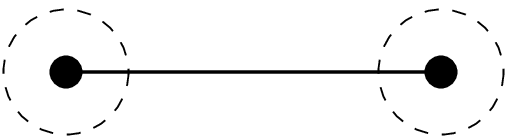}}\,}
\newcommand {\FF} {\,\epsfxsize=1.9\pcm \parbox{1.9\pcm}{\epsfbox{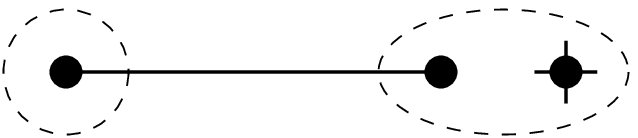}}\,}
\newcommand {\FG} {\,\epsfxsize=0.5\pcm \parbox{0.5\pcm}{\epsfbox{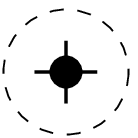}}\,}
\newcommand {\FH} {\,\epsfxsize=3.0\pcm \parbox{3\pcm}{\epsfbox{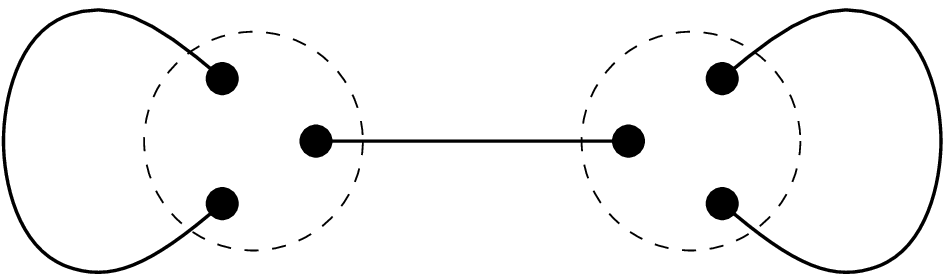}}\,}
\newcommand {\FI} {\,\epsfxsize=1.5\pcm \parbox{1.5\pcm}{\epsfbox{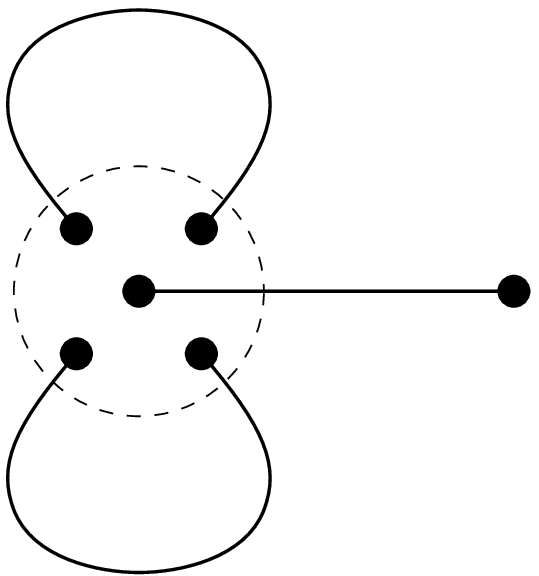}}\,}

\begin{document}

\begin{titlepage}

\renewcommand{\thefootnote}{\fnsymbol{footnote}}
\rightline{\today}
\rightline{T96/089}
\vskip -6.0cm\leftline{\epsfxsize=5.5cm \epsfbox{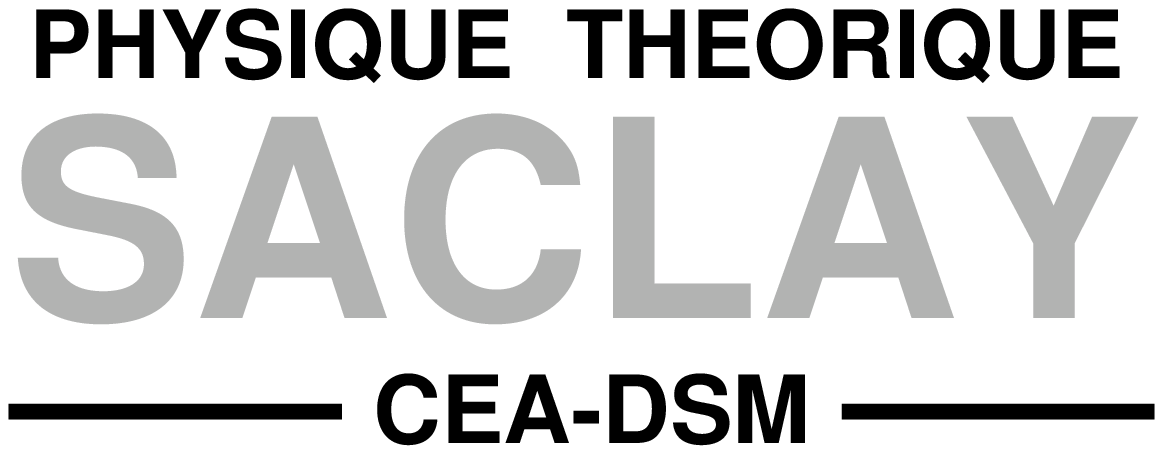}}
\vskip 2.5cm
\centerline{\bf\sf\Large New Renormalization Group Results for Scaling of}
\smallskip
\centerline{\bf\sf\Large Self-Avoiding Tethered Membranes}
\vskip2.5cm
\centerline{\bf\large Kay J\"org Wiese%
\footnote{Email: wiese@amoco.saclay.cea.fr}
and Fran\c cois David%
\footnote{Email: david@amoco.saclay.cea.fr}%
\footnote{Physique Th\'eorique CNRS}}
\smallskip
\centerline{Service de Physique Th\'eorique, C.E.A.\  Saclay}
\centerline{F-91191 Gif-sur-Yvette Cedex, FRANCE}
\vskip 1.5cm

\begin{abstract}
The scaling properties of self-avoiding polymerized 2-dimensional
membranes are studied via renormalization group methods based
on a multilocal operator product expansion.
The renormalization group functions are calculated to second order.
This yields the scaling exponent $\nu$  to order $\E^2$.
Our extrapolations for $\nu$ agree with the Gaussian variational
estimate for large space dimension $d$ and are close to the Flory
estimate for $d=3$.
The interplay between self-avoidance and rigidity at small $d$ is
briefly discussed.
\end{abstract}

\vfill

\centerline{\it Submitted for publication in Nuclear Physics B}

\vspace{1cm}

\end{titlepage}
\addtocounter{page}{1}



\tableofcontents

\pagebreak

\section{Introduction}

The statistical properties of polymerized flexible membranes are an interesting
subject \cite{r:Jerus}, which is still only partially understood.
These objects, also called tethered membranes, are 
two-dimensional generalizations of polymers (one-dimensional flexible chains).
It is expected that such two-dimensional membranes
exhibit a larger variety of behavior than polymers, when the
temperature and the elastic properties of the membranes are varied.
One reason is the following: simple dimensional analysis shows that for
two-dimensional films, the bending rigidity modulus has the dimension of a pure
energy and is therefore marginally relevant (in the sense of the renormalization
group) \cite{r:PelLei85}.
Moreover, two-dimensional membranes may have internal shear elasticity and
this separates two very different classes of flexible membranes: fluid
membranes with zero shear modulus and crystalline (or tethered) membranes
with non-zero shear modulus
\cite{r:NelPel87}.

In this paper, we consider tethered membranes. As long as one
takes into account only the local elastic forces (i.e. bending rigidity, 
compression and shear moduli), numerical simulations and analytical calculations
point towards a consistent and relatively well understood picture
\cite{r:KaKN86,
r:PaKN88,DavidGuitter,r:ArLu88,r:PaKa89,
r:KaNe87}.
For high rigidity or equivalently low temperature, the membrane is in a flat
phase with an average orientation and a classical ``fractal" dimension
$d_F=2$.
The roughness properties of this flat phase are nevertheless non-trivial,
due to the non-linear coupling between undulation modes and phonons.
For low rigidity or equivalently high temperature, the membrane is in a
crumpled phase, without any global orientation. 
In this phase the statistics of the surface is Gaussian
(at large length scales) and its fractal dimension is infinite
($d_F=\infty$).
The two phases are separated by a crumpling transition, which occurs for
a finite bending rigidity (or equivalently a non-zero temperature).
Numerical simulations and some of the analytical calculations indicate 
that this
transition is continuous and characterized by non-trivial critical exponents.

The above results do not take into account steric interactions, that is
local self-avoidance and thus concern ``phantom surfaces".
For membranes ($D=2$) these interactions must be relevant for the crumpled
phase (at large length scales), whatever the bulk dimension $d$ of space is,
since the fractal dimension of phantom surfaces is then infinite.
They are expected to be strongly relevant for physical membranes in
three-dimensional
space ($d=3$). 
This is in contrast with polymers, where steric intercations are relevant only
for $d\le 4$ and lead to a swollen phase in dimensions $1<d<4$ with $1<d_F<2$. 

The study of the effect of self-avoidance for tethered membranes is in fact much
more difficult than for polymers.
Most of the studies rely on numerical simulations, which find evidence for
a flat phase in 3-dimensions
\cite{r:Abr&al89,r:KrollGompper1993}. 
Some experiments have been performed using thin sheets of graphite oxide, but the
results are contradictory.
In \cite{r:graphitefrac} but not in
\cite{r:graphiteflat} a swollen crumpled phase is found.   

In order to study theoretically self-avoiding membranes, one can use
standard approximations, such as the Flory approach or variational methods.
A more systematic renormalization group approach has been initiated in
\cite{r:AroLub87,r:KarNel87}.
This approach is inspired from the direct renormalization method
used for polymers
\cite{r:CloJan90}.
It is a perturbative method and it has been used to calculate the
fractal exponent $\nu^*=2/d_F$ at first order in an $\E$-expansion
\cite{r:AroLub87,r:KarNel87,Hwa}.
An important problem has been to check the internal consistency of this
renormalization method and to extend it properly to all orders in perturbation
theory.
This has been completed by B.~Duplantier, E.~Guitter and one of the authors
in \cite{r:DDG3} (a detailed proof might be released soon
\cite{r:DDG4}), who have shown
that the model of self-avoiding membranes of \cite{r:AroLub87,r:KarNel87},
although corresponding to a {\it non-local} field theory, is renormalizable
in perturbation theory.
This result establishes the validity of the ${\cal O}(\E)$ calculations and
gives a systematic formalism to extend these calculations to
higher orders.
\medskip

In this paper we present for the first time in full details the renormalization
group calculation at second order for the model of self-avoiding tethered
membranes and discuss the results obtained by this approach.
A short presentation of the main results has already appeared in
\cite{r:DaWi96b}.

This study is interesting for several reasons:
\par\noindent
(1) It provides an explicit realization of the renormalization formalism of
\cite{r:DDG3}.
In particular these calculations illustrate nicely how subdominant divergences
are organized.
\par\noindent
(2)
The calculation is a non-trivial task.
As usual in renormalization theory, there is a long way from existence and
convergence theorems to actual calculations, but additional difficulties
are present here.
In particular, beyond first order, the amplitudes (which are called
manifold integrals and generalize in a non-trivial way Feynman integrals
in the Schwinger representation to non-integer dimensions $0<D<2$) cannot
be calculated analytically and in fact describe distributions.
One must rely on numerical estimates for these integrals.
\par\noindent
(3)
It leads to the first estimates at order ${\cal O}(\E^2)$ for the fractal
exponent $\nu^*=2/d_F$.
This allows to check the validity of the ${\cal O}(\E)$ estimates,
as well as the consistency of the extrapolation methods used to extract these
estimates.
Indeed, direct calculations for two-dimensional membranes are impossible,
but the model is extended to ``membranes" with internal dimension
$0<D<2$, thus interpolating  between polymers ($D=1$) and membranes ($D=2$).
Perturbative calculations lead to an $\E$-expansion, where the $\E$-parameter
is given by
\be
\label{e:EpsFirst}
\E\ =\ \E(D,d)\ =\ 2D-d{(2-D)\over 2} \ .
\ee
$D$ is the internal dimension of the membrane and $d$ the dimension
of bulk space (in which the membrane fluctuates).
Using the $\E$-expansion one may start a-priori from any point
$(D_0,d_0)$ such that $\E(D_0,d_0)=0$ to extrapolate to (for instance) the
physical point $(D=2,d=3)$.
In \cite{Hwa} an extrapolation scheme was used for ${\cal O}(\E)$
calculations.
We shall need and will develop more systematic extrapolation schemes for
${\cal O}(\E^2)$ calculations and we shall discuss their respective advantages.
\par\noindent
(4)
Finally, the calculations can be performed for membranes $(D=2)$ in a space with
arbitrary  dimension $d$ and we can compare explicitly our ${\cal O}(\E^2)$
results for $\nu^*$ with other predictions.
It turns out that our estimates are quite reliable for large $d$ and in
remarkable agreement with the result of a Gaussian variational estimate.
We shall explain this fact and argue that this feature persists at
higher order in perturbation theory.
For smaller $d$ the results are less stable, but still good and in 
reasonable
agreement with Flory estimates (we have no good explanation for this fact).
\medskip

The paper is organized as follows:

In section~2 the continuous model of self-avoiding tethered membranes of
\cite{r:AroLub87,r:KarNel87} is introduced. 
It is a generalization of the Edwards-model for polymers.
We recall the basic results of \cite{r:DDG3,r:DDG4} concerning the structure of the
perturbative expansion for the model, the nature of the short distance, or
ultra-violet (UV), divergences and their relation with the so-called
multilocal operator product expansion (MOPE).

In section~3 we recall the renormalization of the model at first order in
perturbation theory (one loop), give the explicit expressions for
the counterterms in the ``minimal subtraction scheme" used in this
paper and derive the renormalization group functions $\beta(b)$ and $\nu(b)$
at one loop. 
The results are of course not new, but this fixes the method and the notations
used throughout the paper.

In section 4 we analyze the UV-divergences at second order (two loops),
and obtain the expressions for the two-loop counterterms, in terms of singular
parts of integrals of MOPE coefficients.
First the UV-singular configurations (divergent and subdivergent diagrams) 
are identified (subsection~4.1), then the leading UV-divergences (poles
in $1/\E^2$) are obtained and their exponentiation
(predicted by renormalization group) checked (subsection~4.2).
This allows to obtain the subleading divergences (poles in $1/\E$), whose
residues give the renormalization group functions at two loops (subsection~4.3).
We then express these residues as combinations of convergent integrals
involving MOPE coefficients (subsection~4.4).

The next seven sections are devoted to the explicit calculation of
these integrals.
In section 5 we briefly introduce some basic analytical methods used in the 
calculations, which have been developed in \cite{r:WieseDavid95}.
These are: (1) the definition of the ``distance measure'' which allows to
define properly the integration over non-integer dimensional space $\R^D$
($0<D<2$); 
(2) the expression of the residue of the UV-poles in ($1/\E$) as a boundary 
term in the distance integrals;
(3) the ``conformal mapping" technique, which allows to map different
domains of integration and leads to crucial simplifications.

Sections 6, 7 and 8 are devoted to the numerical calculation of the
three diagrams which contribute to the coupling constant renormalization.
In all cases, we start from the corresponding MOPE coefficients and determine
explicitly the integrals which have to be computed (this is in general
not straightforward).
Then we evaluate numerically these integrals for values of $1<D<2$ (the internal
dimension of the membrane).
We have in general to decompose the domain of distance integration   into several
pieces, called sectors, and to find ad hoc changes of variables in each sector.
In addition, this requires an adaptive Monte Carlo integration routine, 
first developed in \cite{r:WieseDavid95}, in order to master the rapid
variations of the integrand.
For the diagram  of section~7 there are additional subtleties,
arizing from the fact that
the measure of integration is then a distribution, with non-integrable
singularities on some boundary of the integration domain, which have to
be treated by a finite part prescription.
For each diagram we compute analytically its $D\to 1$ limit.
This provides a check of the numerics. Secondly for $D=1$ our 
model reduces to the Edwards-model for polymers, for which 
2-loop calculations have already been  performed by several authors
and which give an additional check of our calculations.

Sections 9, 10 and 11 are devoted to the numerical calculation of
the three diagrams which contribute to the field renormalization and
are organized in a similar way.

In section 12 we use these two loop results to calculate critical exponents
for self-avoiding membranes.
First we recall how the renormalization group functions at two loops
are related to the counterterms that we have calculated (subsection~12.1) and
thus obtain the $\E^2$ term for the fractal exponent $\nu^*$.
Then as explained above, we have to set up extrapolation methods to
extrapolate from the $\E$-expansion to the physical case $D=2$.
We generalize and systematize the extrapolation method proposed by Hwa \cite{Hwa},
and use our new schemes to evaluate $\nu^*$ for membrane for various
bulk space dimensions \mbox{$2\le d <\infty$} (subsection~12.2).
In subsection~12.3 our results are compared with that of the Gaussian variational
method.
We argue that $\nu^*$, as obtained from the (properly resummed) $\E$-expansion,
must coincide with the variational estimate $\nu_{\mbox{\scriptsize var}}$
for large $d$ and we propose a new $\E$-expansion for $\nu^*$ which
coincides with $\nu_{\mbox{\scriptsize var}}$ at order $\E^0$ and which is shown
to give very good results for large $d$.
In subsection~12.4 we compare our results for $\nu^*$ with that of the Flory
method and consider a similar new $\E$-expansion around
$\nu_{\mbox{\scriptsize Flory}}$.
The results of all the two loop extrapolations  for $\nu^*$ are summarized
in subsection~12.5.
Finally we briefly discuss the case of other scaling exponents for
self-avoiding membranes, namely the correction to scaling exponent $\omega$
and the contact exponent $\theta_2$ (subsection~12.6), as well as
the fractal exponent $\nu^\theta$ for membranes at the tri-critical
$\theta$-point, which has been already calculated at order $\E$.
In subsection 12.8 we briefly summarize the main results from
numerical simulations as well as experimental data. 
We also present a heuristic argument which  explains why
both in numerical simulations and experiments self-avoiding 
tethered membranes are  found in a flat phase (subsection 12.9).

Conclusions and future prospects are given in section~13.
Several technical points or examples are gathered in the appendices.

\section{Definition of the Model}

We start from the continuous model for a $D$-dimensional flexible polymerized
membrane introduced in \cite{r:AroLub87,r:KarNel87}.
This model is a simple extension of the well known Edwards' model for
continuous chains. 
The membrane fluctuates in $d$-dimensional space.
Points in the membrane are labeled by coordinates $x\in\R^D$ and the 
configuration of the membrane in physical space is described by
the field $\vec r: x \in \R^D \longrightarrow \vec r(x) \in \R^d$.
The free energy for a configuration is given by the bare Hamiltonian
\begin{equation}
{\cal H}[\vec r]
= \frac1{2-D} \int_x \half \big(\nabla \vec r(x)\big)^2 +
        b \int_x \int_y \tilde\delta^d\big(\vec r(x)-\vec r(y)\big)
\ .
\label{e:2ptact}
\end{equation}
The integral $\int_x$ runs over $D$-dimensional space and $\nabla$ is the usual
gradient operator.
The normalizations 
(hidden in $1\over 2-D$, $\int_x$ and $\tilde \delta^d(\vec r-\vec r')$ )
are chosen in order to simplify the calculations, but are unimportant
for the general understanding (see appendix A).
The first term is a Gaussian elastic  energy which is known to describe the
free ``phantom" surface.
The interaction term corresponds to a weak repulsive contact interaction
(for $b>0$).
The expectation value of physical observables are obtained by performing the
average over all field-configurations $\vec r(x)$ with the Bolzmann
weight $e^{-{\cal H}[\vec r]}$.

Perturbation theory is constructed by performing the series expansion in
powers of the coupling constant $b$.
This expansion suffers from ultraviolet  (UV) divergencies which have to
be removed by renormalization and which are treated by dimensional 
regularization, i.e.\ analytical continuation in $D$ and $d$.
A physical UV-cutoff could be introduced instead but would render the
calculations more complicated.
Long-range infrared (IR) divergencies also appear.
They can be cured by using a finite membrane, or by studying translationally
invariant observables, whose perturbative expansion is also IR-finite in the
thermodynamic limit (infinite membrane).
Examples of such observables are ``neutral" products of vertex operators 
\begin{equation}
{\cal O}\ =\ \prod_{a=1}^N :{\rme}^{i \vec k_a \vec r(x_a)}:\qquad
\sum_{a=1}^N \vec k_a\,=\,0
\ .
\label{e:prodvo}
\end{equation}
In the following we discuss the renormalization of the model, i.e.\  we only deal with the UV-divergencies.

Let us first analyze the theory by power-counting. In internal 
momentum units, such that $\left[ x \right] =-1$, the dimension 
of the field and of the coupling-constant are:
\begin{equation}
\left[ r\right]= - \nu =-{2-D\over 2}\ ,\quad \left[ b \right] =\varepsilon=2D-  \nu  d
\label{e:dimrg}
\end{equation}
In the sense of Wilson the interaction is relevant for $\E>0$. 
Perturbation theory is then expected to  be UV-finite except for
subtractions associated to relevant operators. We will come back to
this point later. 
For clarity we shall represent graphically the different interaction terms
which have to be considered.
The local operators are
\begin{eqnarray}
1\ &=&\ \GN \\
\half\big(\nabla r(x)\big)^2\ &=&\ \GO 
\ .
\label{e:1ptops}
\end{eqnarray}
The bi-local operator, the dipole, is 
\begin{eqnarray}
\tilde \delta^d\big( r(x)-r(y)\big)\ &=&\ \GB
\ .
\label{e:2ptops}
\end{eqnarray}

The expectation-value of an observable is:
\begin{equation}
	\langle{\cal O}[r]\rangle_b = 
\frac{\int {\cal D}[r]\, {\cal O}[r]\,\mbox{\rm e}^{\tx-{\cal H} \left[ r\right] }}
{\int {\cal D}[r] \,\mbox{\rm e}^{\tx-{\cal H} \left[ r\right] }}
\label{e:expectation}
\end{equation}
Perturbatively, all expectation-values are taken with respect
to the free theory:
\begin{equation}
	\langle{\cal O}[r]\rangle_0\ =\ 
\frac{\int {\cal D}[r]\, {\cal O}[r]\,\mbox{\rm e}^{\tx-\frac{1}{2-D} \int_x \half(\nabla r)^2}}
{\int {\cal D}[r] \,\mbox{\rm e}^{\tx-\frac{1}{2-D} \int_x \half(\nabla r)^2}}
\label{e:free expectation}
\end{equation}
A typical term in the expansion of (\ref{e:expectation}) is 
\be
b^n \int\!\!\!\int \ldots \int\!\!\!\int \langle {\cal O} \GB \ldots \GB \rangle_0
\ ,
\ee
where the integral runs over the positions of all   dipole-endpoints.
The analysis in \cite{r:DDG2}
shows that UV-divergencies appear when some dipole-endpoints approach each
other. 
The divergencies are analyzed via a multilocal operator product expansion
(MOPE).
\begin{figure}[htb]
\centerline{\epsfxsize=15cm \parbox{15cm}{\epsfbox{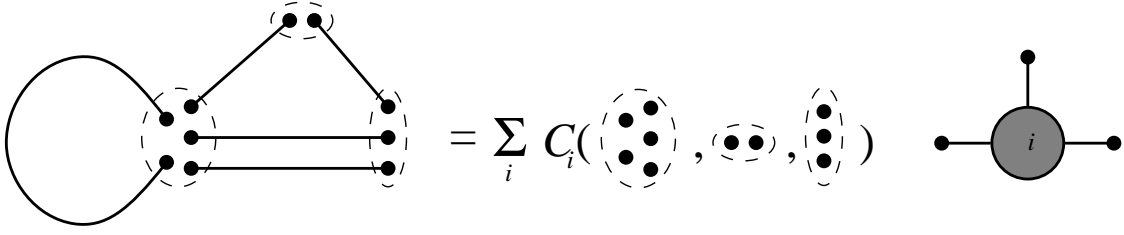}} }
\caption{Example of a contraction}
\label{f:contraction}
\end{figure}
The principle is examplified in figure \ref{f:contraction}.
One considers $n$ dipoles (here $n=5$) and one separates the $2n$ end-points 
into $m$ separate subsets (here $m=3$) delimited by the dashed lines.
The MOPE describes how the product of these $n$ dipoles behaves when the points
inside each of the $m$ subsets are contracted towards a single point $z$.
The result is a sum over multilocal operators $\Phi\{z_1,\ldots ,z_m\}$,
depending on the $m$ points $z_1,\cdots ,z_m$, of the form
\begin{equation}
\label{etheMope}
\sum_{\Phi}C_\Phi\,\Phi\left\{  z_1, z_2, \ldots, z_m  \right\}
\end{equation}
where the MOPE-coefficients $C_\Phi$ depend only on the relative distances
$x_i-x_j$ between the dipole end-point positions $x_i$ {\it
inside each subset}.
This expansion is valid as an operator-identity, i.e. inserted 
in any expectation value and in the limit of small distances between contracted points. 
As the Hamiltonian (\ref{e:2ptact}) does not contain any mass-scale, the
MOPE-coefficients are homogenous function of the relative positions between the
contracted points.
The degree of homogeneity is given by simple dimensional analysis.
In the case considered here, where $n$ dipoles are contracted to an operator
$\Phi$, this degree is simply (for the definition of $\nu$ see \eq{e:dimrg})
\begin{equation}
\mbox{degree}[C_\Phi^n]\,=\ -n\,\nu\,d\,+\,[\Phi]
\end{equation}
where $[\Phi]$ is the canonical dimension of the operator $\Phi$ and 
$d(2-D)/2$ is simply the canonical dimension of the dipole.

In order to evaluate the associated singularity, one has finally to integrate
over all relative distances inside each subsets.
This gives an additional scale factor with degree $D(n-m)$.
A singular configuration, such as depicted on figure~\ref{f:contraction} ,
will be UV-divergent if its degree of divergence, defined as
$D(n-m)+\mbox{deg}[C_\Phi^m]$, is negative.
It is superficially divergent if the degree is zero and convergent otherwise.

The power counting analysis of \cite{r:DDG3,r:DDG4} shows that at the critical
dimension 
$\varepsilon=0$ the identity operator $\GN$ is relevant, while the local
operator $\GO$ and the bi-local dipole operator $\GB$ are
marginally relevant. 
Contractions of $n$ dipoles to $\GN$ give relevant divergences (negative
powers of the short distance cut-off, or poles for some $\varepsilon>0$),
while contractions towards $\GO$ and $\GB$ give superficial divergences
(logarithms of the cut-off, or poles at $\varepsilon=0$).

The analysis of the renormalization in \cite{r:DDG3,r:DDG4} shows that
in order to make perturbation theory UV-finite at $\E=0$, one has to
add to the Hamiltonian counterterms proportional to all relevant and marginal
operators. 
These counterterms are obtained from the MOPE-coefficients, integrated with
some IR-cutoff procedure.

\section{Renormalization at 1-loop order}
\label{s:Renormalization at 1-loop order}
Let us continue on the concrete example of the one-loop divergences,
where we shall also fixe our notations for the MOPE-coefficients.
When the end-points $(x,y)$ of a single dipole are contracted to a point
(taken here to be the center-off-mass $z=(x+y)/2$), the MOPE is
\begin{equation}
\label{e:mopecoef}
 \vphantom{\hbox{\GH}}_x\GH_y 
=
\bigg( \vphantom{\hbox{\GH}}_x\GH_y \bigg| \GN\bigg) \GN \ +\ 
\bigg( \vphantom{\hbox{\GH}}_x\GH_y \bigg| \GO_{\alpha\beta}\bigg)
\GO_{\alpha\beta}  \ + \ldots \nn 
\end{equation}
with the first MOPE-coefficients given explicitly by
\begin{equation}
\bigg( \vphantom{\hbox{\GH}}_x\GH_y \bigg| \GN\bigg) = |x-y|^{-\nu d}
\ ,\qquad
\bigg( \vphantom{\hbox{\GH}}_x\GH_y \bigg| \GO_{\alpha\beta}\bigg)=
 -\half |x-y|^{-\nu(d+2)} (x-y)_{\alpha} (x-y)_{\beta}
\end{equation}
and where $\GO_{\alpha\beta}$ denotes the local tensor  operator
\begin{equation}
\GO_{\alpha\beta}=\half \partial_{\alpha} \vec r \partial_{\beta} \vec r
\end{equation}
The integral over the relative distance $x-y$ for
 $\bigg( \vphantom{\hbox{\GH}}_x\GH_y \bigg| \GO_{\alpha\beta}\bigg) $ is
at $\E=0$ logarithmically divergent. 

The simplest contraction to a dipole is when two dipoles collapse.
The corresponding MOPE-coefficient is
\be
\Bigg( \GM \Bigg| \GB \Bigg)
\Bigg( \GM \Bigg| \GB \Bigg) 
= \left( |x|^{2\nu} + |y|^{2\nu} \right)^{-d/2}
\ee
where $x$ and $y$ are now the relative distances inside the two subsets.
Another possibility is to consider the contraction $\FD$.
But as 
\be
	\Bigg( \FD
\raisebox{-5.5mm}[0mm][0mm]{\makebox[0mm][l]{$\hspace{-9mm}\scriptstyle x \,\, y \,\, z$} }
 \Bigg| \GB \Bigg) = \Bigg( \GB \raisebox{-2mm}[0mm][0mm]{\makebox[0mm][l]{$\hspace{-2mm}\scriptstyle x $} }
\GH\raisebox{-5.5mm}[0mm][0mm]{\makebox[0mm][l]{$\hspace{-6.8mm}\scriptstyle  y \,\, z$} }
 \Bigg| \GB \GN \Bigg) 
 = \Bigg( \GH\raisebox{-5.5mm}[0mm][0mm]{\makebox[0mm][l]{$\hspace{-6.8mm}\scriptstyle  y \,\, z$} }
 \Bigg| \GN \Bigg) 
\label{e:nondiv}
\ee
this does not give a pole at $\E=0$, but a term proportional to the relevant
divergence (that we discuss now), times a regular contribution.

In the next step counterterms have to be introduced in order to subtract
these divergencies.
We have to distinguish between the counterterm for the 
relevant operator and those for marginal operators.
The first one can be defined by analytic continuation, the latter require
a subtraction scale. 
Indeed, the divergence for $\GN$ is given by the integral
\be
	\int_{\Lambda^{-1}< |x-y| < L} \bigg( \vphantom{\hbox{\GH}}_x\GH_y \bigg| \GN\bigg)
	=\int_{\Lambda^{-1}}^{L} \frac{dx}{x} x^{D-\nu d}
	=\frac1{D-\E} \left( \Lambda^{D-\E} -L^{\E-D}\right)
\ee
where $\Lambda$ is a high-momentum UV-regulator and $L$ a large distance
regulator.
For $\epsilon\simeq 0$ this is UV-divergent but IR-convergent.
The simplest way to subtract this divergence is therefore to replace the
dipole operator by 
\be
\strut_x\GB\strut_y\  \longrightarrow\  \strut_x\GB\strut_y\,-\,
\strut_x\GBdotted\strut_y \ ,
\ee
where $\strut_x\GBdotted\strut_y = |x-y|^{-\nu d} $. 
This amounts to add to the bare Hamiltonian (\ref{e:2ptact}) the UV-divergent
counterterm

\be
\Delta{\cal H}[r]\ =\ -{b\over 2}\,	\int_x \int_y |x-y|^{-\nu d} \ , 
\ee
which is a pure number and thus does not change the expectation-value of any
physical observable.
This prescription is sufficient to subtract all relevant UV-divergences
in the calculation of the renormalization-group functions  at 2-loop order, that
we present in the next sections. 

We now treat marginal operators.
Let us come back to the MOPE (\ref{e:mopecoef}).
The integral over the relative distance of 
${\displaystyle\int_{x-y}}\bigg( \vphantom{\hbox{\GH}}_x\GH_y \bigg|
\GO_{\alpha\beta}\bigg) \GO_{\alpha\beta} $
is logarithmically divergent at $\E=0$.
In order to find the appropriate counterterm, we use dimensional
regularization, i.e.\ set $\E>0$.
An IR-cutoff $L$, or equivalently a subtraction momentum scale $\mu=L^{-1}$,
has to be introduced in order to define the subtraction operation.
As a general rule, let us integrate over all distances appearing in the
MOPE-coefficient, bounded by the subtraction scale $L=\mu^{-1}$.
This projects the tensor operator $\GO_{\alpha \beta}$ onto the scalar
$\GO$, times the integral
\be \label{e:MOPEL}
\int_{|x-y|<L} \bigg( {}_x\GH_y \bigg| \GO \bigg) 
\ =\ 
\bigg< \GH \bigg| \GO \bigg>_L 
\ =\ L^\E\,f(\E,D)
\ee
Following \cite{r:DDG3,r:DDG4,r:WieseDavid95} we use a minimal subtraction scheme (MS).
The internal dimension of the membrane $D$ is kept fixed and 
(\ref{e:MOPEL}) is expanded as a Laurent series in $\varepsilon$, which here
starts at $\varepsilon^{-1}$.
Denoting by $\big<\ \big|\ \big>_{\varepsilon^p}$ the term of order
$\varepsilon^p$ of the Laurent expansion of $\big<\ \big|\ \big>_L$ for
$L=1$, the pure pole
part of (\ref{e:MOPEL}) is found to be
\be
\bigg< \GH \bigg| \GO \bigg>_{\E^{-1}} = -\frac{1}{2D}  \frac{1}{\E}
\ee
It is this pole-term that we are going to subtract in the MS-scheme.
This is done by adding to the Hamiltonian a counterterm 
\begin{equation}
\Delta{\cal H}[r]\ =\ -\,b\,\bigg< \GH \bigg| \GO \bigg>_{\E^{-1}}\,
\int_x \GO_x
\end{equation}

Similarly, the divergence arising from the contraction of two dipoles into 
a single dipole is subtracted by a counterterm proportional to the single pole
of
\be
\bigg< \GM \bigg| \GB \bigg>_L =
	\int_{|x|<L}\int_{|y|<L} \Bigg( \GM \Bigg| \GB \Bigg) \nn\\
=\int_{|x|<L}\int_{|y|<L}\left( |x|^{2\nu} + |y|^{2\nu} \right)^{-d/2}
\ee

As a result  the model is UV-finite if we use the
renormalized Hamiltonian ${\cal H}_R$
\begin{equation} \label{e:Ham1}
\!\!\!\!{\cal H}_{R}[\vec r]= \frac{Z}{2-D}\int_x \half 
\big(\nabla\vec r(x)\big)^2
+ b Z_b \mu^\varepsilon \int_x\!\int_y\!
\tilde \delta^d\big (\vec r(x)- \vec r(y)\big ) \ ,
\end{equation}
instead of the bare Hamiltonian ${\cal H}[\vec r]$.
Now $r$ and $b$ are the renormalized field and coupling constant,
and $\mu=L^{-1}$ is the renormalization momentum scale.
The renormalization factors are at one loop 
\begin{eqnarray} 
Z&=& 1-(2-D)\bigg< \GH \bigg| \GO \bigg>_{\E^{-1}} {b}\,+\,{\cal O}(b^2)
\label{e:Z1}
\\
Z_b&=&1+\bigg< \GM \bigg| \GB \bigg>_{\E^{-1}} b \,+\,{\cal O}(b^2)
\label{e:Zb1}
\end{eqnarray}

The renormalized theory can be reexpressed in terms of the bare (unrenormalized)
theory through \begin{equation}
\vec r_0(x)\ =\ Z^{1/2}\,\vec r(x)
\ ,\qquad
b_0\ =\ b\,Z_b\,Z^{d/2}\,\mu^\E
\end{equation}

Following the analysis of \cite{r:DDG3},
the renormalization group $\beta$-function and the anomalous scaling
dimension $\nu$ of $\vec r$ are obtained from the variation of the
coupling constant and the field with respect to the renormalization scale $\mu$,
keeping the bare couplings fixed.
They are written in terms of $Z$ and $Z_b$ as
\begin{eqnarray}
\label{e:beta}
\beta(b) &=&
\mu \frac{\partial}{\partial \mu }\lts_{b_0} b
=\frac{-\E b} {1+ b\frac{\partial}{\partial b } \ln Z_b +
\frac{d}2 b \frac{\partial}{\partial b} \ln Z}\\
\label{e:nu}
\nu (b) &=&
\frac{2-D}{2}-\half \mu \frac{\partial}{\partial \mu }\lts_{b_0} \ln Z =
\frac{2-D}2 -\half \beta(b) \frac{\partial}{\partial b} \ln Z
\end{eqnarray}

\section{Derivation of the Counterterms at 2-Loop Order}
\label{s:Derivation of the Counterterms at 2-Loop Order}
\subsection{The 2-loop counterterms in the MS scheme}
\label{s:The 2-loop counterterms in the MS scheme}
In this section we apply the formalism explained above 
to determine the counterterms which
renormalize the theory at second order.
If we consider only the bare theory, given by the bare Hamiltonian
(\ref{e:2ptact}), power counting gives the three UV-divergent diagrams
(together with their symmetry factors)
\begin{equation}
{1\over 2}\ \GI \ (a)
\ ,\qquad
{2\over 3}\ \GJ\ (b)
\ ,\qquad
{2}\ \GP\ (c)
\label{e:2divdiag}
\end{equation}
which give short distance singularities when the points inside
the subsets are contracted to a single point.
These singularities give double and single poles at $\E=0$.
There are two other potentially dangerous diagrams
\begin{equation}
\FH \ , \qquad \FI
\label{e:2nodiv}
\end{equation}
These diagrams do not give poles at $\E=0$ for reasons similar to what
happens with \eq{e:nondiv}.
Now one has to remember that the model is already renormalized at 1-loop,
i.e. that we use the renormalized Hamiltonian (\ref{e:Ham1}), with the
counterterms (\ref{e:Z1}) and (\ref{e:Zb1}).
As a consequence there are five additional divergent diagrams, which come
from the insertion of the 1-loop counterterms 
\begin{eqnarray}
&&
-2\,\mu^{-\varepsilon}\ \GM\ \bigg<\GM\bigg|\GB\bigg>_{\varepsilon^{-1}}\ 
(d)
\ ,\quad
-\,\mu^{-\varepsilon}\ \GH\ \bigg<\GM\bigg|\GB\bigg>_{\varepsilon^{-1}}\ 
(e)
\nonumber\\
&&\strut\nonumber\\
&&
-\,\mu^{-\varepsilon}\ \GZ\ \bigg<\GH\bigg|\GO\bigg>_{\varepsilon^{-1}}\ 
(f)
\ ,\quad
-2\,\mu^{-\varepsilon}\ \GX\ \bigg<\GH\bigg|\GO\bigg>_{\varepsilon^{-1}}\ 
(g)
\nonumber\\
&&\strut\nonumber\\
&&
{1\over 2}\,\mu^{-2\varepsilon}\ \GW\ 
\left(\bigg<\GH\bigg|\GO\bigg>_{\varepsilon^{-1}}\right)^2\ 
(h)
\label{e:2ctdiag}
\end{eqnarray}
There are other potentially divergent diagrams, analogous to those depicted in (\ref{e:2nodiv}), which factorize into convergent diagrams. 

The first four terms in \eq{e:2ctdiag} are a combination of a diagram divergent at 1-loop order
(giving a single pole) times a divergent 1-loop counterterm (which gives
another single pole).
The fifth term is more peculiar: it is the combination of a convergent diagram
(which corresponds to a contact term)
times two 1-loop counterterms (thus giving also a double pole).

Owing to the MOPE, diagrams $(a)$, $(e)$, $(f)$ and $(h)$ give a divergence
proportional to the insertion of the local operator $\GO$.
With the notations introduced in the previous section, they can be subtracted
by adding a counterterm proportional to the divergent part of the integral
of the corresponding MOPE coefficients
\begin{eqnarray}
&&
{1\over 2}\bigg<\GI\bigg|\GO\bigg>_L
-\mu^{-\varepsilon}
\bigg<\GH\bigg|\GO\bigg>_L
\bigg<\GM\bigg|\GB\bigg>_{\varepsilon^{-1}}
\nonumber\\
&&\strut\nonumber\\
&&
-\mu^{-\varepsilon}
\bigg<\GZ\bigg|\GO\bigg>_L
\bigg<\GH\bigg|\GO\bigg>_{\varepsilon^{-1}}
+{1\over 2}
\mu^{-2\varepsilon}
\bigg<\GW\bigg|\GO\bigg>_L
\left(\bigg<\GH\bigg|\GO\bigg>_{\varepsilon^{-1}}\right)^2
\label{e:Z2div}
\end{eqnarray}
Since we use the minimal subtraction scheme, we want to subtract only the
double and single poles in $\E$ at $\varepsilon=0$.
To isolate these poles, we have to perform a Laurent expansion of the 
various terms in (\ref{e:Z2div})\ and to keep
the terms of order $\E^{-2}$ and $\E^{-1}$ but to drop the analytic part.
Setting the renormalization momentum scale $\mu=L^{-1}$, we
obtain the final expression for the renormalization factor
$Z$ at 2-loop order
\begin{eqnarray}
Z=&&
1\,-\,b\,(2-D)\,\bigg< \GH \bigg| \GO \bigg>_{\E^{-1}} \nonumber\\
&&
+\,b^2\,(2-D)\,\left[
{1\over 2}\bigg<\GI\bigg|\GO\bigg>_{\varepsilon^{-2},\varepsilon^{-1}}
-
\bigg<\GH\bigg|\GO\bigg>_{\varepsilon^{-1},\varepsilon^{0}}
\bigg<\GM\bigg|\GB\bigg>_{\varepsilon^{-1}}
\right.
\nonumber\\
&&\strut\nonumber\\
&&
\left.
-\bigg<\GZ\bigg|\GO\bigg>_{\varepsilon^{-1},\varepsilon^{0}}
\bigg<\GH\bigg|\GO\bigg>_{\varepsilon^{-1}}
+{1\over 2}
\bigg<\GW\bigg|\GO\bigg>_{\varepsilon^{0},\varepsilon^{1}}
\left(\bigg<\GH\bigg|\GO\bigg>_{\varepsilon^{-1}}\right)^2
\right]
\nonumber\\
&&\strut\nonumber\\
&&\strut\qquad+\,{\cal O}(b^3)
\label{e:Z2}
\end{eqnarray}
Here $\big<\ \big|\ \big>_{\varepsilon^{n_1},\cdots ,\varepsilon^{n_p}}$
denotes the sum of the terms of order
$\varepsilon^{n_1},\cdots ,\varepsilon^{n_p}$ 
in the Laurent expansion of $\big<\ \big|\ \big>_L$, taken at $L=1$.

Similarly, the diagrams $(b)$, $(c)$, $(e)$ and $(f)$ give a divergence
proportional to the bilocal operator $\GB$.
An analogous analysis leads to the following expression for the coupling-constant renormalization-factor $Z_b$ at 2 loops
\begin{eqnarray}
&&\hspace{-3mm}Z_b\,=1\,+\,b\ \bigg< \GM \bigg| \GB \bigg>_{\E^{-1}}
\nonumber\\
&&\strut\nonumber\\
&&+\ b^2\,\left[
-{2\over 3}\bigg<\GJ\bigg|\GB\bigg>_{\varepsilon^{-2},\varepsilon^{-1}}
+2\bigg<\GM\bigg|\GB\bigg>_{\varepsilon^{-1},\varepsilon^{0}}
\bigg<\GM\bigg|\GB\bigg>_{\varepsilon^{-1}}\right.
\nonumber\\
&&\strut\nonumber\\
&&
\left.
-2\bigg<\GP\bigg|\GB \bigg>_{\varepsilon^{-2},\varepsilon^{-1}}
+2\bigg<\GX\bigg|\GB\bigg>_{\varepsilon^{-1},\varepsilon^{0}}
\bigg<\GH\bigg|\GO\bigg>_{\varepsilon^{-1}}
\right]
\nonumber\\
&&\strut\nonumber\\
&&\strut\qquad+\,{\cal O}(b^3)
\label{e:Zb2}
\end{eqnarray}

\subsection{The leading divergences (double poles)}
In fact we are interested only in the residues of the single poles, that is
into the residues $c_1$ and $f_1$ in 
the Laurent expansion of the counterterms
\begin{eqnarray}
Z&=&1+{e_1\over\varepsilon}b+\left({f_1\over\varepsilon}+
{f_2\over\varepsilon^2}\right) b^2+{\cal O}(b^3)
\nonumber\\
Z_b&=&1+{a_1\over\varepsilon}b+\left({c_1\over\varepsilon}+
{c_2\over\varepsilon^2}\right) b^2+{\cal O}(b^3) \ .
\label{e:ZZb2}
\end{eqnarray}
Indeed, the finiteness of the renormalization group functions (\ref{e:beta})\ and (\ref{e:nu})\ implies that the residues of the double poles, $f_2$ and $c_2$,
can be expressed in terms of the 1-loop residues $e_1$ and $a_1$.

Let us show explicitly how this happens, since this will be useful in order
to subtract efficiently the double poles in the counterterms.
From the analysis at 1-loop order, we know that
\begin{equation}
{e_1\over\varepsilon}=-(2-D)\bigg< \GH \bigg| \GO \bigg>_{\E^{-1}}
\ ,\qquad
{a_1\over\varepsilon}=\bigg< \GM \bigg| \GB \bigg>_{\E^{-1}}
\label{e:e1f1}
\end{equation}
and we obtain the 1-loop RG functions
\begin{equation}
\beta(b)=-\E b+b^2\left({d\,e_1\over 2}+a_1\right)
\ ,\qquad
\nu(b)={(2-D)\over 2}+b{e_1\over 2} \ .
\label{e:rgfct1}
\end{equation}
Inserting these 1-loop RG functions into (\ref{e:beta})\ and (\ref{e:nu}),
yields differential equations for the counterterms $Z$ and $Z_b$, whose
solution is explicitly
\begin{equation}
Z^{(1)}=\left[1-{b\over\E}\left({d\,e_1\over 2}+a_1\right)\right]
^{-{2\,e_1\over d\,e_1+2\,a_1}}
\ ,\qquad
Z_b^{(1)}=\left[1-{b\over\E}\left({d\,e_1\over 2}+a_1\right)\right]
^{-{2\,a_1\over d\,e_1+2\,a_1}}
\label{e:ZZBlead}
\end{equation}
This shows the exponentiation of the leading divergences, i.e.\ that the
residue of the leading pole in $\E^{-n}$ at order $b^n$ is determined
by the residue
of the simple pole in $\E^{-1}$ at order $b$.
Expanding (\ref{e:ZZBlead})\ to order $b^2$, we obtain
\begin{equation}
Z^{(1)}\ =\ 1+{e_1\over\E}b+{\tilde f_2(\E)\over\E^2}b^2+{\cal O}(b^3)
\ ,\qquad
Z_b^{(1)}\ =\ 1+{a_1\over\E}b+{\tilde c_2(\E)\over\E^2}b^2+{\cal O}(b^3)
\end{equation}
with, in terms of diagrams,
\begin{eqnarray}
{\tilde f_2(\E)\over\E^2}\,=&&\,-\,\nu \,\bigg<\GH\bigg|\GO\bigg>_{\E^{-1}}\ \times
\nonumber\\
&&\quad \times\ 
\left[\bigg<\GM\bigg|\GB\bigg>_{\E^{-1}} - {\nu (d+2)}
\bigg<\GH\bigg|\GO\bigg>_{\E^{-1}}
\right]
\label{e:Z2lead}\\
&&\strut\nonumber\\
{\tilde c_2(\E)\over\E^2}\,=&&
\,{1\over 2}\,\bigg<\GM\bigg|\GB\bigg>_{\E^{-1}}\ \times
\nonumber\\
&&\quad \times\ 
\left[ 2\,\bigg<\GM\bigg|\GB\bigg>_{\E^{-1}}\,-\,{\nu d}\,
\bigg<\GH\bigg|\GO\bigg>_{\E^{-1}}
\right]
\label{e:Zb2lead}
\end{eqnarray}
Both $\tilde f_2(\E)$ and $\tilde c_2(\E)$ contain not only a constant
term but also a term linear in $\E$. The latter is not fixed
by the RG-functions at 1-loop order. 

The leading poles of $Z^{(1)}$ (resp.\ $Z_b^{(1)}$) must be equal to those
of $Z$ (resp. $Z_b$). This implies that
\begin{equation}
f_2\ =\ \tilde f_2(\E)+{\cal O}(\E)\ ;\qquad
c_2\ =\ \tilde c_2(\E)+{\cal O}(\E)
\label{e:f2c2tilde}
\end{equation}
This is compatible with the explicit expressions
(\ref{e:Z2})\ and (\ref{e:Zb2})\ for $Z$ and $Z_b$.
Indeed, one can directly calculate the double poles in (\ref{e:Z2})\ and
(\ref{e:Zb2}).
The simplest case is the diagram $(b)$ in (\ref{e:2divdiag}).
A subdivergence occurs when two dipoles are contracted to a single dipole.
When this contraction is performed first, the MOPE
coefficient factorizes as 
\begin{equation}
\bigg(\GJsub\bigg|\GB\bigg)\ \approx\ \bigg(\GM\bigg|\GB\bigg)
\bigg(\GM\bigg|\GB\bigg)
\label{e:factGB}
\end{equation}
There are three different subdivergences and one finally obtains
that the double pole associated with this diagram is given by
\begin{equation}
\bigg<\GJ\bigg|\GB\bigg>_{\E^{-2}}\ =\ 3\ \times\ {1\over 2}\,\bigg<\GM\bigg|\GB\bigg>_{\E^{-1}}
\bigg<\GM\bigg|\GB\bigg>_{\E^{-1}}
\label{e:dbplGB}
\end{equation}
The factor $1/2$ comes from the nested integration \cite{r:DuHwKa}:
the double pole results from
the integration over a ``sector" where the distances inside the 
subdiagram are smaller than all the other distances.
This will become clear in the explicit calculations of the next
sections.

Similarly, let us consider the diagram $(c)$ in (\ref{e:2divdiag}).
A subdivergence occurs when the single dipole to the right of the diagram
is contracted to a point.
The MOPE coefficient factorizes as
\begin{equation}
\bigg( \GPpsub \bigg| \GB \bigg) \approx \bigg( \GH \bigg| \GO \bigg) 
\bigg( \GX \bigg| \GB \bigg)
\label{e:factGP}
\end{equation}
Consequently, the double pole for this diagram is 
\begin{equation}
\bigg< \GP \bigg| \GB \bigg>_{\E^{-2}}
=  \half\, \bigg< \GH \bigg| \GO \bigg>_{\E^{-1}}
\bigg< \GX \bigg| \GB \bigg>_{\E^{-1}}
\label{e:dbplGP} 
\end{equation}
where the factor $1/2$ again comes from the nested integration.

Finally, let us consider the diagram $(a)$ in (\ref{e:2divdiag}).
Four sectors contribute to the double pole, which correspond to the 
subcontractions depicted here
\begin{eqnarray}
&&
\bigg( \GIsubeins \bigg| \GO \bigg) \approx  \bigg( \GH \bigg| \GO \bigg) 
\bigg( \GZ \bigg| \GO \bigg)
\nonumber\\
&&\strut\nonumber\\
&&\bigg( \GIsubzwei \bigg| \GO \bigg) \approx  
\bigg( \GM \bigg| \GB \bigg)  \bigg( \GH \bigg| \GO \bigg)
\label{e:factGI}
\end{eqnarray}
Each of the contractions appears with a combinatorial factor two.
The double pole for this diagram is therefore
\begin{eqnarray}
&&\bigg< \GI \bigg| \GO \bigg>_{\E^{-2}}\  = \ 2\,\times\,\half\, 
\bigg< \GH \bigg| \GO \bigg>_{\E^{-1}} 
\bigg< \GZ \bigg| \GO \bigg>_{\E^{-1}}
\nonumber\\
&&\strut\nonumber\\
&&
\qquad\qquad \,+\, 2 \times \half\,
\bigg< \GM \bigg| \GB \bigg>_{\E^{-1}}  \bigg< \GH \bigg| \GO \bigg>_{\E^{-1}}
\label{e:dbplGI}
\end{eqnarray}

Inserting (\ref{e:dbplGB}), (\ref{e:dbplGP})\ and (\ref{e:dbplGI})\ in
the explicit expressions for the two-loop counterterms
(\ref{e:Z2})\ and (\ref{e:Zb2})\ we apparently do not obtain for the double poles
the results (\ref{e:Z2lead})\ and (\ref{e:Zb2lead})\ predicted by the
renormalization group.
However we can make use of the equation of motion to
compute the effect of the insertion of the operator $\GO$ in the counterterms.
In  appendix \ref{s:Equation of Motion} we show that
\begin{eqnarray}
&&
\bigg< \GX \bigg| \GB \bigg>_{\E^{-1}}\ =\ -\,{\nu d\over 2}\,
\bigg< \GM \bigg| \GB \bigg>_{\E^{-1}}\!+\ {\cal O}(\E^{0})
\nonumber\\
&&\strut\\
&&
\bigg<\GZ\bigg|\GO\bigg>_{\E^{-1}}\ =\ -\,{\nu (d+2) }\,
\bigg< \GH \bigg| \GO \bigg>_{\E^{-1}}
\nonumber\\
&&
\mbox{\hskip 5.cm}\,+\,
\bigg< \GH \bigg| \GO \bigg>_{\E^{-1}}\bigg< \GW \bigg| \GO \bigg>_{\E^{0}}
\,+\ {\cal O}(\E^{0})
\end{eqnarray}
Using these identities one recovers (\ref{e:Z2lead})\ and (\ref{e:Zb2lead}).

\subsection{The subleading divergences (single poles)}

We can now give the expressions for the residues of the single poles.
For the single pole of $Z$ we find
\begin{eqnarray}
{f_1\over\E} &=& {(2-D)\over 2}\,
\left[
\bigg<\GI\bigg|\GO\bigg>_{\varepsilon^{-2},\varepsilon^{-1}}
+\bigg<\GW\bigg|\GO\bigg>_{\varepsilon^{1}}
\left(\bigg<\GH\bigg|\GO\bigg>_{\varepsilon^{-1}}\right)^2
\right.
\nonumber\\
&&\strut\nonumber\\
&&
-
\bigg<\GH\bigg|\GO\bigg>_{\varepsilon^{-1}}
\bigg<\GM\bigg|\GB\bigg>_{\varepsilon^{-1}}
-2
\bigg<\GH\bigg|\GO\bigg>_{\varepsilon^{0}}
\bigg<\GM\bigg|\GB\bigg>_{\varepsilon^{-1}}
\nonumber\\
&&\strut\nonumber\\
&&
-\left.
\bigg<\GZ\bigg|\GO\bigg>_{\varepsilon^{-1}}
\bigg<\GH\bigg|\GO\bigg>_{\varepsilon^{-1}}
-2
\bigg<\GZ\bigg|\GO\bigg>_{\varepsilon^{0}}
\bigg<\GH\bigg|\GO\bigg>_{\varepsilon^{-1}}
\right]
\label{e:f1}
\end{eqnarray}
One can simplify this expression, since the explicit calculation shows
that
\begin{equation}
\bigg<\GW\bigg|\GO\bigg>_{\varepsilon^{1}}=0 \ ,\qquad
\bigg<\GH\bigg|\GO\bigg>_{\varepsilon^{0}}=0
\label{e:nullCT}
\end{equation}
For the single pole of $Z_b$ the result is
\begin{eqnarray}
{c_1\over\E} & = & \left[
-{2\over 3}\bigg<\GJ\bigg|\GB\bigg>_{\varepsilon^{-2},\varepsilon^{-1}}
+\bigg<\GM\bigg|\GB\bigg>_{\varepsilon^{-1}}
\bigg<\GM\bigg|\GB\bigg>_{\varepsilon^{-1}}\right.
\nonumber\\
&&\strut\nonumber\\
&&
+2\bigg<\GM\bigg|\GB\bigg>_{\varepsilon^{0}}
\bigg<\GM\bigg|\GB\bigg>_{\varepsilon^{-1}}
\nonumber\\
&&\strut\nonumber\\
&&
-2\bigg<\GP\bigg|\GB \bigg>_{\varepsilon^{-2},\varepsilon^{-1}}
+\bigg<\GX\bigg|\GB\bigg>_{\varepsilon^{-1}}
\bigg<\GH\bigg|\GO\bigg>_{\varepsilon^{-1}}\nonumber\\
&&\strut\nonumber\\
&&
\left.
+2\bigg<\GX\bigg|\GB\bigg>_{\varepsilon^{0}}
\bigg<\GH\bigg|\GO\bigg>_{\varepsilon^{-1}}
\right]
\label{e:c1}
\end{eqnarray}

\subsection{Expressing the residues as convergent integrals}
The next step is to evaluate the residues $f_1$ and $c_1$, that is to write them
as convergent integrals involving combinations of MOPE coefficients.
It is convenient not to compute directly $f_1$ and $c_1$, but rather
to consider the 
${\cal O}(\E^{-1})$ parts of the counterterms that are not already contained in
the second order  resummation of the 1-loop divergences, as given by
(\ref{e:Z2lead})\ and (\ref{e:Zb2lead}).
We thus denote
\begin{equation}
{f_2\over\E^2}+{f_1\over\E}\ =\ {\tilde f_2(\E)\over\E^2}+{\tilde f_1\over\E}
\ ,\qquad
{c_2\over\E^2}+{c_1\over\E}={\tilde c_2(\E)\over\E^2}+{\tilde c_1\over\E}
\label{e:f1c1tilde}
\end{equation}
Using $d=4D/(2-D)-2\E/(2-D)$ it is easy to extract the ${\cal O}(\E^{-1})$
part of (\ref{e:Z2lead})\ and (\ref{e:Zb2lead}). With (\ref{e:e1f1})\ we get
\begin{equation}
f_1\ =\ \tilde f_1\,-\,{e_1^2\over 2(2-D)}
\ ,\qquad
c_1\ =\ \tilde c_1\,-\,{e_1\,a_1\over 2(2-D)}
\label{e:fc1fc1tilde}
\end{equation}
Now subtracting (\ref{e:Z2lead})\ from (\ref{e:Z2}) 
yields an equivalent explicit expression for 
$\tilde f_1$ 
\begin{eqnarray}
\frac{\tilde f_1}{2-D} \frac1\E &\ =\ & \half 
\bigg< \GI \bigg| \GO \bigg>_{\E^{-2},\E^{-1}} 
-\,\bigg< \GH \bigg| \GO \bigg>_{\E^{-1}}
\bigg< \GZ \bigg| \GO \bigg>_{\E^{-2},\E^{-1}}
 \nonumber \\
&&\strut\nonumber\\
& & -\,{\nu (d+2)\over 2} \bigg< \GH \bigg| \GO \bigg>_{\E^{-1}}^2 
+\,\half \bigg< \GH \bigg| \GO \bigg>_{\E^{-1}}^2 
\bigg< \GW \bigg| \GO \bigg>_{\E^0} \nonumber \\
&&\strut\nonumber\\
&&  
-\half \, \bigg< \GM \bigg| \GB \bigg>_{\E^{-1}} \bigg< \GH \bigg| \GO \bigg>_{\E^{-1}} 
\label{f1_res}
\end{eqnarray}
Finally, one has to remember that $\bigg<\cdots\bigg|\cdot\bigg>_{\E^p}$
are the terms of order $\E^p$ of the Laurent series of the integral over distances of the corresponding MOPE coefficient
\begin{equation}
 \bigg<\cdots\bigg|\cdot\bigg>_L\,=\,\int_{\mbox{\scriptsize distances}\le L}
\bigg(\cdots\bigg|\cdot\bigg)_L
\label{e:remindLau}
\end{equation}
Using this fact one obtains the following decomposition
\be
\frac{\tilde f_1}{2-D} \frac1\E  =\ {\cal F}_1+{\cal F}_2+{\cal F}_3
\,+\,{\cal O}(\E^0)
\label{e:f1dec}
\ee
where each term can be written as a convergent integral. These terms are:
\bea
{\cal F}_1&=&\half \bigg< \GI \bigg| \GO \bigg>_L -\half 
\bigg< \GH \bigg| \GO \bigg>_L	 
\bigg< \GZ \bigg| \GO \bigg>_L \nn \\
& & \qquad - \half \bigg< \GM \bigg| \GB \bigg>_L 
\bigg< \GH \bigg| \GO \bigg>_L 
\label{e:F1}\\
{\cal F}_2&=&  -\half \bigg< \GH \bigg| \GO \bigg>_{\E^{-1}} \times
\weiter \times \left(  \bigg< \GZ \bigg| \GO \bigg>_L +
\frac{(2-D)(d+2)}2  \bigg< \GH \bigg| \GO \bigg>_L
-\bigg< \GH \bigg| \GO \bigg>_L \bigg< \GW \bigg| \GO \bigg>_L \right) 
\nonumber \\
& & \label{e:F2}\\
{\cal F}_3&=&\half \bigg< \GH \bigg| \GO \bigg>_{\E^{-1}} 
\left( \bigg< \GM \bigg| \GB \bigg>_L
-\bigg< \GM \bigg| \GB \bigg>_{\E^{-1}}
\right)
\label{e:F3}
\eea
Even if not written explicitly, we will only calculate the 
residue of ${\cal F}_1, \ldots, {\cal F}_3$ at $L=1$. 
Similarly, subtracting (\ref{e:Zb2lead})\ from (\ref{e:Zb2})\ we obtain for
$\tilde c_1$
\bea
\frac{\tilde c_1}\E &=& - \bigg< \GM \bigg| \GB \bigg>_{\E^{-1}}^2 
+ {(2-D)d\over 4}  \bigg< \GM \bigg| \GB \bigg>_{\E^{-1}} 
\bigg< \GH \bigg| \GO \bigg>_{\E^{-1}}
\weiter +2 \bigg< \GM \bigg| \GB \bigg>_{\E^{-1}} \bigg< \GM \bigg| \GB \bigg>_{\E^{-1},\E^{0}} 
-\frac23 \bigg< \GJ \bigg| \GB \bigg>_{\E^{-2},\E^{-1}} 
\weiter -2 \bigg< \GP \bigg| \GB \bigg>_{\E^{-2},\E^{-1}}
+2 \,\bigg< \GH \bigg| \GO \bigg>_{\E^{-1}} 
\bigg< \GX \bigg| \GB \bigg>_{\E^{-1},\E^{0}}
\weiter   \label{c1_res}
\label{e:c1_res}
\eea
that we decompose as 
\be
\frac{\tilde c_1}\E\  =
\  {\cal C}_1 + {\cal C}_2 + {\cal C}_3 +{\cal O}(\E^0) \ ,
\label{e:c1dec}
\ee
with
\bea 
\hspace{-1cm}
{\cal C}_1 &=& -\frac23 \bigg< \GJ \bigg| \GB \bigg>_L + \bigg< \GM \bigg| \GB \bigg>_L^2 \label{e:C1}\\
\hspace{-1cm}
{\cal C}_2 &=&  -2 \bigg< \GP \bigg| \GB \bigg>_L +
\bigg< \GH \bigg| \GO \bigg>_{L} \bigg< \GX \bigg| \GB \bigg>_L 
\label{e:C2}\\
\hspace{-1cm}
{\cal C}_3 &=& \bigg< \GH \bigg| \GO \bigg>_{\E^{-1}} \left(  \bigg< \GX \bigg| \GB \bigg>_L 
+{(2-D) d \over 4} \bigg< \GM \bigg| \GB \bigg>_{\E^{-1}} \right)\nonumber\\
\label{e:C3}&& 
\eea 
The coefficients  ${\cal C}_1$,
${\cal C}_2$ and ${\cal C}_3$ can like ${\cal F}_1$, ${\cal F}_2$ and 
${\cal F}_3$ be expressed as convergent integrals and
will be calculated in the next sections.

\section{General Strategy and Technical Tools}
In this section we give a brief overview over the analytical 
tools involved in the calculation of the Feynman-diagrams. 
These are the measure (subsection \ref{Analytic continuation of the measure}), the procedure to 
extract the residue and the conformal mapping
(subsection \ref{s:extract residue}).

\subsection{Analytic continuation of the measure}
\label{Analytic continuation of the measure}
We first define the explicit form for the integration measure in non-integer
dimension $D$, that will be used in the calculations.
We use the general formalism of distance geometry \cite{r:DDG1}, which has
been already used to construct other, but equivalent, representation for
such measures.

The general problem is to integrate a function $f(x_1,\cdots,x_N)$, 
which is invariant under Euclidean displacements (and therefore depends
only on the $N(N-1)/2$ relative distances $|x_i-x_j|$ between these points)
over the $N-1$ first points (the last point is fixed, using translational
invariance) in $\R^D$ for non-integer $D$.
 In order to define the integration, let us take $D\ge N-1$
and integer. 
For $i<N$ we denote by $y_i=x_i-x_N$ the $i$'th distance-vector and by
$y_i^a$ its $a$'th component ($a=1,\ldots,D$).

The integral over $y_1$ is simple:
Using rotation invariance, we fix $y_1$ to have only the $a=1$ component
non-zero. The measure becomes
\begin{equation}
\int \rmd^D\!y_1 = S_D \int_0^\infty \rmd y_1^1 \, (y_1^1)^{D-1}
\ ,\qquad y_1=(y_1^1,0,\ldots,0)
\label{e:inty1}
\end{equation}
where $S_D$ is the volume of the unit-sphere in $\R^D$, defined 
by
\be
S_D = 2 \frac{\pi^{D/2}}{\Gamma(D/2)}
\ee
We now fix $y_2$ to have only $a=1$ and $a=2$ as non-zero components.
The integral over
$y_2$ consists of the integration along the direction fixed
by $y_1$ and the integration in the orthogonal space $\R^{D-1}$:
\begin{equation} \label{measure1}
\int \rmd^D\! y_2= S_{D-1} \int_{-\infty}^{\infty}
\rmd y_2^1 \int _0^{\infty} \rmd y_2^2 \, (y_2^2)^{D-2}
\ ,\qquad y_2=(y_2^1,y_2^2,0, \ldots,0)
\end{equation}
For the $j$-th point, 
one proceeds recursively to integrate first over the hyperplane defined by 
$y_1, \ldots , y_{j-1}$ and then the orthogonal complement:
\begin{equation}
\int \rmd^D\! y_j= S_{D-j+1} \prod_{a<j} \int_{-\infty}^{\infty}
\rmd y_j^a \int _0^{\infty} \rmd y_j^j \, (y_j^j)^{D-j}
\ ,\qquad y_j=(y_j^1,\ldots,y_j^j,0,\ldots,0)
\end{equation}
The final result for an integral over all configurations of $N$ points is 
\begin{equation} \label{N points integral}
\int \prod_{j=1}^{N-1} \rmd^D\! y_j= 
S_D S_{D-1} \ldots S_{D-N+2}
\prod_{j=1}^{N-1} \left( \prod_{a=1}^{j-1}
\int_{-\infty}^{\infty}
\rmd y_j^a \int _0^{\infty} \rmd y_j^j \, (y_j^j)^{D-j} \right)
\end{equation}
This expression for the measure, now written in terms of the $N(N-1)$ variables
$y_j^a$, can be analytically continued to non-integer $D$.
For $D\le N-1$ this measure is not integrable when some $y_j^a=0$.
For $D$ not integer, the integration is defined through the standard
finite-part prescription.
This means that the measure \eq{N points integral}\ becomes a distribution.

Let us made this explicit on the example of $N=3$ points, following
\cite{r:DDG1,r:WieseDavid95}.
The measure is then
\be
S_D S_{D-1}\,\int_0^\infty \rmd y_1^1\,(y_1^1)^{D-1}\,
\int_{-\infty}^{+\infty} \rmd y_2^2\int_0^\infty \rmd y_2^1\,(y_2^1)^{D-2}
\label{e:measure 3 points}
\ee
It is well defined and integrable for $D>1$.
For $D=1$ the integral over $y_2^2$ diverges logarithmically at $y^2_2\to 0$,
but this singularity is cancelled by the zero of $S_{D-1}$ and the measure
becomes
\be
2\,\int_0^\infty \rmd y_1^1\,\int_{-\infty}^{+\infty} \rmd y_2^1\,\int_{0}^{\infty}
\rmd y_2^2\,\delta(y_2^2)\ =\ \int_{\scriptsize\R} d y_1\,\int_{\scriptsize\R} d y_2
\label{e:meas3ptD=1}
\ee
thus it reduces to the measure for two points on a line.
For $0<D<1$ the integral over $y_2^2$ diverges at $y_2^2\to 0$, but this
divergence is treated by a finite part prescription.

For integrals over $N>3$ points, a finite part prescription 
has already be used for $D<2$. This will be
shown explicitly later.
The expression \eq{N points integral}\ 
is equivalent to the measures defined in \cite{r:DDG1}.

\subsection{Extraction of the residue}\label{s:extract residue}
We now explain how we extract the residue of the pole at $\E=0$ for
the example of the 1-loop counterterms.
Note from \eq{e:mopecoef} that 
\be
\Bigg< \GH \Bigg| \GO \Bigg>_L = -\frac1{2D} \int_{x<L} x^{D-\nu d}
= -\frac 1 {2D} \frac1\E L^\E \label{e:sss}
\ee
We used the
normalization of the measure
\be
\int_x = \frac1{S_D} \int \rmd^Dx \ , \qquad S_D=2\frac{\pi^{D/2}}{\Gamma\left(D/2\right)}
\ee
which was chosen to simplify the calculations 
(see appendix \ref{s:Normalizations}).
The residue can most easily be extracted by applying 
$L\frac{\partial}{\partial  L}$ to \eq{e:sss}. This yields:
\be 
\E \,\Bigg< \GH \Bigg| \GO \Bigg>_L =-\frac1{2D} L \int_{x=L} x^{D-\nu d}
=-\frac{1}{2D} L^{\E}
\ee
So the residue  of \eq{e:sss}\ is 
\be  \E \,\Bigg<\GH \Bigg| \GO \Bigg>_{\E^{-1}}\  = \ 
\lim_{\E\to 0} L\frac{\partial}{\partial  L}
\Bigg<\GH \Bigg| \GO \Bigg>_L = -\frac 1 {2D}
\ee

We can apply this recipe to the second 1-loop counterterm:
\be 
\Bigg< \GM \Bigg| \GB \Bigg>_L =\int_{x<L} \int_{y<L} 
\Bigg( \GM \Bigg| \GB \Bigg)
\ee
since it is also proportional to $L^\E$.
We thus have to calculate
\be 
L\frac{\partial}{\partial  L} \Bigg< \GM \Bigg| \GB \Bigg>_L 
=
L\left[ \int_{x<y=L} + \int_{y<x=L}\right] \left(x^{2\nu}+y^{2\nu}\right)^{-d/2}
\label{e:ssss}
\ee

We now introduce a general method which is very useful to manipulate
and simplify such integrals.
It relies on (global) conformal transformations in position
space and is called conformal mapping of sectors.
It has first been introduced in \cite{r:WieseDavid95},
where a geometric interpretation can be found. 
We will explain the method on a concrete example and then state the general
result.

Let us consider the second integral on the r.h.s.\ of \eq{e:ssss}
\be
	L\int_{y<x=L} \left(x^{2\nu}+y^{2\nu}\right)^{-d/2} 
= L \int_0^{\infty} \frac{\rmd x}{x}x^D
\int_0^{\infty} \frac{\rmd y}{y}y^D
\left(x^{2\nu}+y^{2\nu}\right)^{-d/2} 
\delta(x-L) \Theta(y<x)
\ee
Now two changes of variables are performed:
The first one
\be
x\ \to\ \tilde x\ ,\qquad
x=\tilde x y L^{-1}
\ee
leads to
\be
L^{1-D+\nu d}  \int_0^{\infty} \frac{\rmd \tilde x}{\tilde x}\tilde x^D
\int_0^{\infty} \frac{\rmd y}{y}y^{2D-\nu d}
\left(\tilde x^{2\nu}+L^{2\nu}\right)^{-d/2} 
\delta(\tilde x y L^{-1} -L) \Theta (L<\tilde x)
\ee
The second one
\be
y\ \to\ \tilde y\ ,\qquad
	y= \tilde y \tilde x^{-1} L
\ee
finally gives
\be
	L^{1+\E} \int_0^{\infty} \frac{\rmd \tilde x}{\tilde x}\tilde x^{D-\E}
	\int_0^{\infty} \frac{\rmd \tilde y}{\tilde y}\tilde y^{D}
\left(\tilde x^{2\nu}+\tilde y^{2\nu}\right)^{-d/2} 
\delta(\tilde y-L) \Theta(\tilde y < \tilde x)
\label{e:sssS}
\ee
Replacing the second integral in the r.h.s. of \eq{e:ssss} by \eq{e:sssS}\ gives
\be
L\frac{\partial}{\partial  L}
\Bigg< \GM \Bigg| \GB \Bigg>_L 
=L^{1+\E} \int_0^{\infty} \frac{\rmd x}{x}x^D
\int_0^{\infty} \frac{\rmd y}{y}y^D
\left(x^{2\nu}+y^{2\nu}\right)^{-d/2} \max(x,y)^{-\E}
\delta(y-L) 
\ee
Now one distance (here $y$) is fixed, whereas the integral over the other 
distance (here $x$) runs from 0 to $\infty$.
The former constraint $\max(x,y)=L$ has been transformed into the factor
$\max(x,y)^{-\E}$ times the constraint $y=L$.

Before generalizing this formula, we shall show how it can be used in practice.
The residue in $\frac 1\E$ (which determines the corresponding 1-loop counterterm)
is given by the simple formula ($d_c(D)=4D/(2-D)$):
\be \label{e:a1}
\E \Bigg< \GM \Bigg| \GB \Bigg>_{\E^{-1}} =  \int_0^{\infty} \frac{\rmd x}{x}x^D
\left(x^{2\nu}+1\right)^{-d_c(D)/2}
=\frac 1 {2-D} \frac{\Gamma\left( \frac D{2-D}\right)^2}
{\Gamma\left( \frac{2D}{2-D} \right)}
\ee
The subleading term can analogously be calculated by expanding
$(x^{2\nu}+1)^{-(d-d_c(D))/2}$ and $\max(x,y)^\E$ in $\E$.
We obtain the convergent integral representation
\be
\Bigg< \GM \Bigg| \GB \Bigg>_{\E^0} = \int_0^{\infty}\frac{\rmd x}{x}x^D
\left( 
\frac1{2-D} \ln(x^{2\nu} +1) - \ln(\max(x,1)) 
\right)
\ee

\bigskip

This method extends to the integrals which appear in the counterterms
associated to the contraction of any number of points.
In general we have to compute integrals over $N(N-1)$ distances
$x,y,\cdots$, of the form
\be
I(\E)\ =\ \int_{\max(x,y,\ldots)\le L} f(x,y,\ldots)
\label{e:I integrale}
\ee
with an homogeneous function $f$ such that the integral has a conformal weight
(dimension in $L$) $\kappa$.
For the integrals which appear in $n$-loop diagrams, this weight is simply
\be
\kappa\,=\ \,n\,\E
\label{e:kappa weight}
\ee
The integral over the distances is defined by the $D$-dimensional measure
\eq{N points integral}.
The residue is extracted from the dimensionless integral
\be
J(\E)\ =\ L^{-\kappa}\,
L{\partial\over\partial L}I(\E)
\ =\ n\E\, L^{-\kappa}\, I(\E)
\ =\ 
L\,\int_{\max(x,y,\ldots)= L} f(x,y,\ldots)\,\max(x,y,\ldots)^{-\kappa}
\label{e:J integrale}
\ee
The domain of integration can be decomposed into ``sectors", for instance
\be
\{\cdots<y<x=L\}\ ,\ \{\cdots<x<y=L\}\ ,
\label{e:secteurs}
\ee
and we can map these different sectors onto each other by global conformal
transformations.
For instance we can rewrite the integral
\eq{e:J integrale}
\be
J(\E)\ =\ 
L\,\int_{x= L} f(x,y,\ldots)\,\max(x,y,\ldots)^\kappa
\ =\ 
L\,\int_{y= L} f(x,y,\ldots)\,\max(x,y,\ldots)^\kappa
\label{e:diff sect}
\ee
The constraint on the maximum of the distances is replaced by the constraint on a
(arbitrarily chosen) distance.

This mapping of sectors  is one of the basic tools used in the following
to explicitly calculate the 2-loop-diagrams.

\section{Coupling constant renormalization, first graph}
\subsection{The counterterm}
\label{s:C1, counter-term}
\begin{figure}[htb]
\centerline{
	\epsfxsize=7.0cm \parbox{7.0cm}{\epsfbox{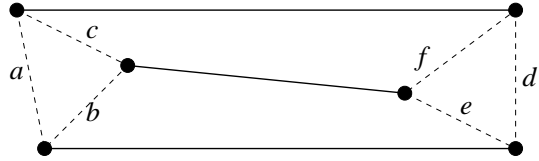}}
}
\caption{The distances in (\protect{\ref{e:ccr1:1}})}
\label{f:ccr1:1}
\end{figure}
We are now going to calculate the first diagram of section \ref{s:Derivation of the Counterterms at 2-Loop Order}.
It contributes to the coupling-constant renormalization 
in 2-loop order and is  
\be \label{e:ccr1:C1}
-\frac32\, {\cal C}_1 = \Bigg< \GJ \Bigg| \GB \Bigg>_L -\frac32 \Bigg< \GM \Bigg| \GB \Bigg>_L^2 \ .
\ee
With the distances labeled as in figure \ref{f:ccr1:1},
the first MOPE-coefficient is:
\begin{eqnarray}
& & \hspace{-0.3cm} \Bigg( \GJ \Bigg| \GB \Bigg)
	=	\nonumber \\
& & \Bigg[ \frac{1}{4} \left( \sqrt{a^{2 \nu}+d^{2 \nu}} +\sqrt{b^{2 \nu}+e^{2 \nu}}
	+\sqrt{c^{2 \nu}+f^{2 \nu}}\right) 
	\left( \sqrt{b^{2 \nu}+e^{2 \nu}} + \sqrt{c^{2 \nu}+f^{2 \nu}} 
          - \sqrt{a^{2 \nu}+d^{2 \nu}}\right) \nonumber \\
& &  \left( \sqrt{a^{2 \nu}+d^{2 \nu}} +\sqrt{c^{2 \nu}+f^{2 \nu}}
          -\sqrt{b^{2 \nu}+e^{2 \nu}}\right) 
   \left( \sqrt{a^{2 \nu}+d^{2 \nu}} +\sqrt{b^{2 \nu}+e^{2 \nu}}
          -\sqrt{c^{2 \nu}+f^{2 \nu}}
          \right) 
   \Bigg]^{-d/2} \nonumber \\
\label{e:ccr1:1}
& & 
\end{eqnarray}
The derivation of this expression can be found in 
appendix~\ref{s:Example of the MOPE}.
Recall that the factorization of the divergences, which are subtracted
in (\ref{e:ccr1:C1}), is:
\be
\Bigg( \GJ \Bigg| \GB \Bigg)
\approx  \Bigg( \GM \Bigg| \GB \Bigg)\Bigg( \GM \Bigg| \GB \Bigg) \nn \\
\ee
We therefore represent the MOPE-coefficients associated to the second term 
in (\ref{e:ccr1:C1}) as:
\begin{eqnarray}
&& \hspace{-1cm}-\frac 3 2  \Bigg( \GM \Bigg| \GB \Bigg)\Bigg( \GM \Bigg| \GB \Bigg) \nn \\
&=& -\half \left(a^{2 \nu}+d^{2 \nu}\right)^{-d/2} \left(b^{2 \nu}+e^{2 \nu}\right)^{-d/2}
-\half \left(a^{2 \nu}+d^{2 \nu}\right)^{-d/2} \left(c^{2 \nu}+f^{2 \nu}\right)^{-d/2} \nonumber \\
& & -\half \left(b^{2 \nu}+e^{2 \nu}\right)^{-d/2} \left(c^{2 \nu}+f^{2 \nu}\right)^{-d/2}
\label{e:ccr1:3}
\end{eqnarray}
There are 3 subdivergencies subtracted due to the 3 possible contractions and a factor $1/2$ due 
to symmetry, i.e.\ due to the nested contraction. 

We can now proceed to calculate the diagram \eq{e:ccr1:C1}, which is of order
$1/\E$ and not of order $1/\E^2$ as the single terms. 
To do so, let us consider the integral $I(L)$, 
which is defined as the integral of the MOPE coefficients,
with all mutual distances appearing in figure~\ref{f:ccr1:1}\ 
restricted to be smaller than $L$:
\be
I(L)=\int\!\!\!\int\!\!\!\int\!\!\!\int\!\!\!\int\!\!\!\int_{a,b,c,d,e,f<L} \nonumber \\
\Bigg( \GJ \Bigg| \GB \Bigg) -\frac 3 2
\Bigg( \GM \Bigg| \GB \Bigg)\Bigg( \GM \Bigg| \GB \Bigg)
\label{e:cct1:3}
\ee
Note that $I(L)$ is {\em not} exactly equal to (\ref{e:ccr1:C1}).
However we shall show below that the leading term, i.e.\ the pole in $1/\E$
in which we are interested, is the same.

$I(L)$ has the following Laurent-expansion:
\be
I(L)=L^{2\E}\left(\frac{a}\E +{\cal O}(\E^0) \right)
\ee
We now apply the operator $L\frac{\partial}{\partial L}$ to $I(L)$ to extract
the residue $a$ in $1/\E$.
We obtain an integral similar to \eq{e:cct1:3} with the constraint
that $\max(a,b,c,d,e,f)=L$.
Using the trick of conformal mapping explained before, we can rewrite this as an
integral with the constraint that one of the distances (for instance $a$)
is equal to $L$ and  that the other can vary freely.
We thus obtain the integral
\begin{eqnarray}
\hspace{-7mm} 
J(L)&=& L {\partial \over \partial L} I(L) \nn \\
&=&L 
\int\!\!\!\int\!\!\!\int\!\!\!\int\!\!\!\int_{b,c,d,e,f}
\max(a,b,c,d,e,f)^{-2\E} \times \nonumber \\
& & \hspace{1cm}\times \left[ \Bigg( \GJ \Bigg| \GB \Bigg) -\frac 3 2
\Bigg( \GM \Bigg| \GB \Bigg)\Bigg( \GM \Bigg| \GB \Bigg) \right]\ \ 
\label{e:ccr1:4}
\end{eqnarray}
We know that the only divergencies, which lead to poles in $1/\E$,
could appear in \eq{e:ccr1:4}\  when a pair of distances,
i.e.\ ($a,d$), ($b,e$) or ($c,f$) simultaneously tends to 0.
But by construction these divergences cancel between the first and the second
terms of the integrand.
So, the integrand in (\ref{e:ccr1:4}) has only integrable singularities at
short distances and we can perform the limit $d\to d_c$, i.e.\ $\E\to 0$
inside the integral in order to compute
\be
\left.J(L)\right|_{\E=0}\ =\ 2\,a
\label{e:Rcc1:4}
\ee
This convergent integral can be calculated numerically by the methods
developed in \cite{r:WieseDavid95}.
We shall explicit this calculation later. 

The next problem we have to treat is that in (\ref{e:cct1:3}) we subtracted
a counterterm that we had chosen by convenience.
It is however not the counterterm in \eq{e:ccr1:C1}, which is fixed by
our renormalization prescription at 1-loop order. 
The latter can be written as 
\begin{equation}
\Bigg< \GM \Bigg| \GB \Bigg>_L^2 = \int\!\!\!\int\!\!\!\int\!\!\!\int\!\!\!\int\!\!\!\int_{a,b,d,e<L,\ c,f}
\left(a^{2 \nu}+d^{2 \nu}\right)^{-d/2} \left(b^{2 \nu}+e^{2 \nu}\right)^{-d/2}
\label{e:ccr1:5}
\end{equation}
The difference between \eq{e:ccr1:5} and the counterterm in (\ref{e:cct1:3})
is that the integration over $c$ and $f$ is {\em not} restricted. 
To evaluate the residue for the true counterterm,
we apply the operator $L\frac{\partial}{\partial L}$ to (\ref{e:ccr1:5}) 
and get an integral with the constraint that $\max(a,b,d,e)=L$.
We now use the trick of conformal mapping to transform this integral into an
integral with the constraint $a=L$ as in (\ref{e:ccr1:4}).
This yields:
\begin{equation}
 L \int\!\!\!\int\!\!\!\int\!\!\!\int\!\!\!\int_{b,c,d,e,f}
\left(a^{2 \nu}+d^{2 \nu}\right)^{-d/2} \left(b^{2 \nu}+e^{2 \nu}\right)^{-d/2}
\max(a,b,d,e)^{-2\E}
\label{e:ccr1:6}
\end{equation}
This expression has to be compared to the corresponding counterterm in
(\ref{e:ccr1:4}).
The difference between these two counterterms is 
\begin{equation}
\int\!\!\!\int\!\!\!\int\!\!\!\int\!\!\!\int_{b,c,d,e,f}
\left(a^{2 \nu}+d^{2 \nu}\right)^{-d/2} \left(b^{2 \nu}+e^{2 \nu}\right)^{-d/2}
\left( \max(a,b,d,e)^{-2\E} -\max(a,b,c,d,e,f)^{-2\E} \right)
\label{e:ccr1:7}
\end{equation}
A priori this term is ${\cal O}(\E^0)$ and might contribute to the residue.
We state that this term is $\cal O(\E)$, thus subdominant and does not
contribute to the counterterm.
To prove this we set $L=1$ and we develop $\max(\ldots)^{-2\E}$
in powers of $\E$.
This expansion is licit over the domain of integration, since
$\max(\ldots )\ge 1$ and since the integral is 
convergent at large distances.
This implies that \eq{e:ccr1:7} is equivalent to
\begin{equation}
2\E \int\!\!\!\int\!\!\!\int\!\!\!\int\!\!\!\int_{b,c,d,e,f}
\left(a^{2 \nu}+d^{2 \nu}\right)^{-d/2} \left(b^{2 \nu}+e^{2 \nu}\right)^{-d/2}
\left[ \ln(\max(a,b,c,d,e,f)) -\ln(\max(a,b,d,e)) \right]
\label{e:ccr1:8}
\end{equation}
when $\E\to 0$.
Poles  of the integral in $1/\E$ might appear if $b,e \to 0$ simultaneously.
In that case however $c\to a$ and $d\to f$ so that 
the difference in the last factor in (\ref{e:ccr1:8}) vanishes.
The last integral in \eq{e:ccr1:8}\ is therefore convergent, implying that
$\protect\eq{e:ccr1:7}={\cal O}(\E)$ and that $I(L)+3/2{\cal C}_1={\cal O}(\E^0)$.
Therefore the use of \eq{e:cct1:3}\ instead of \eq{e:ccr1:C1}\ to compute
the residue of the pole is justified.

Such a phenomenon is not peculiar to this diagram.
In the other diagrams that we shall calculate similar simplifications occur
when dealing with the counterterms, which allow to take the same constraints
over the distances for the 2-loop diagram and for the counterterms associated
with 1-loop subdivergences.
We call this property the ``2-loop miracle". (A generalization to
the calculation of higher order diagrams is possible.)

\subsection{Numerical calculation}
\label{s:C1num}
\begin{figure}[htb]
\centerline{
\epsfxsize=4cm \parbox{
4cm}{\epsfbox{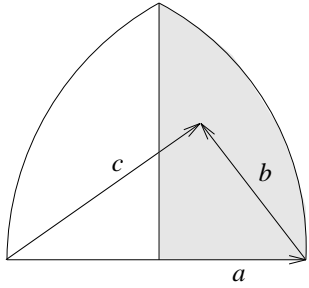}}
}
\caption{The half-sector used for the numerical integration}
\end{figure}

\noindent
We now want to calculate the residue J(L), given by the integral \eq{e:ccr1:4},
numerically.
This calculation will be performed for values of $D$ in the interval
$1<D<2$.
As we already discussed, the integral representation \eq{e:ccr1:4} suffers
from additional divergences for $D<1$, which come from the fact that
the measure over three points becomes a distribution, and which must be
treated by a finite part  prescription.
This will not be done here, since it turns out to be sufficient to
calculate the diagrams for $1<D<2$ in order to have good estimates for the critical
exponent of membranes.

The calculation is considerably simplified by using the symmetries of
the diagram and by integrating over one of the equivalent
sectors only. 
First of all, one can suppose that 
$	\max(a,b,c) > \max(d,e,f)$.
Furthermore an ordering of $a$, $b$ and $c$ is introduced:
$ a>b>c $.
This gives rise to a symmetry-factor $2\times6=12$.
In order to eliminate all divergencies, also $d$, $e$ and $f$ have to be
ordered. We assume that $d>e>f$ but then have to sum over all 
permutations of $d$, $e$ and $f$. 
Denoting by
\be
g(a,b,c,d,e,f) = \Bigg( \GJ \Bigg| \GB \Bigg) -\frac 3 2
\Bigg( \GM \Bigg| \GB \Bigg)\Bigg( \GM \Bigg| \GB \Bigg)
\ee
as in \eq{e:ccr1:1} and \eq{e:ccr1:3}, we get
\bea\hspace{-1cm}
J(L)\ts_{\E=0}&=& \int_{a=1,b,c,d,e,f} g(a,b,c,d,e,f) \nonumber \\
 &=& 12 \int_{a=1>b>c \atop a>d>e>f} g(a,b,c,d,e,f)+g(a,b,c,e,f,d)+g(a,b,c,f,d,e)
\nn \\
& & \qquad\qquad +g(a,b,c,f,e,d)+g(a,b,c,e,d,f)+g(a,b,c,d,f,e)
\eea
We can furthermore divide the remaining integral into 2 sectors where either
$b<e$ or $b>e$.
There will always be an integrable divergence in the smaller distance which
has to be treated by an appropriate variable transformation.  
As in \cite{r:WieseDavid95} we parametrize the integral over $b$ as:
\begin{eqnarray}
 \int_b  &=& \frac{S_{D-1}}{S_{D}} \int_{-\infty}^{\infty}d b_1\, \int_0^{\infty} db_2 \,
    b_2^{D-2}  \\
 a&=&1\ , \quad b=\sqrt{b_1^2+b_2^2}\ , \quad c=\sqrt{(1-b_1)^2+b_2^2}\ ,
 \nonumber
\end{eqnarray} 
restricted to the domain where $b<c<a$: There are singularities for $b_1\to 0$ and for $b_2\to 0$.
They are disentangled by switching to radial and angular coordinates. In these coordinates
the singularities can be eliminated by the following parametrization:
\begin{eqnarray} \label{e:m31A}
\int_b  & = &
\frac{1}{D-\sigma} \frac{1}{D-1} \frac{\pi}{2} \frac{S_{D-1}}{S_D} 
 \int_0^1 d\beta \, \beta^{\frac{2-D}{D-1}} \sin(\alpha)^{D-2} \times \nonumber \\
 & & \hspace{1cm} \times \int_0^1 dt \, b^\sigma  \Theta(1-b) \Theta(c-b) \ ,
\end{eqnarray}
where
\begin{eqnarray}
\alpha & = & \frac{\pi}{2} \beta^{\textstyle \frac{1}{D-1}} \label{e:123454321} \\
b & = & t^{\textstyle \frac{1}{D-\sigma}} \\
\sigma &=&  \left\{  \begin{array}{cl} 2D-2  &,\ e<b \\ 0 &,\ e>b \end{array} \right. \\
c & = & \sqrt{b^2 + 1 - 2b \cos(\alpha)} 
\end{eqnarray}
The change of variables from the angle $\alpha$ to $\beta$, equation (\ref{e:123454321}), 
generates a factor $\beta^{(2-D)/(D-1)}$ which exactly cancels the singularity for $\alpha \to 0$
from the measure, i.e.\ $\sin(\alpha)^{D-2}$.
The other variable-transformations are constructed in a similar way.

Furthermore there is the integral over $d$ and $e$.
\begin{equation}
\int_d \int_e = \int_d d^D \int_{e'} 
\qquad\hbox{with}\qquad
	e'=\frac{e}d
\end{equation}
Therefore we write
\be
\int_{d<a} d^Dd= \frac1{2D}\int_0^1 dr \qquad
d=r^{1/2D}
\ee
The remaining integral over $e'$ is parametrized in analogy to the integral
over $b$ as:
\begin{eqnarray} \label{e:m31B}
\int_{e'} & = &
\frac{1}{D-\tau} \frac{1}{D-1} \frac{\pi}{2} \frac{S_{D-1}}{S_D} 
 \int_0^1 d\delta \, \delta^{\frac{2-D}{D-1}} \sin(\gamma)^{D-2} \nonumber \\
 & & \hspace{3cm} \int_0^1 dt \, (e')^\tau  \Theta(d-f) \Theta(f-e) \ ,
\end{eqnarray}
where
\begin{eqnarray}
\gamma & = & \frac{\pi}{2} \delta^{\textstyle \frac{1}{D-1}}  \\
e & =  & d s^{\textstyle \frac{1}{D-\tau}} \\
\tau &=&  \left\{  \begin{array}{cl} 0  &,\ e<b \\ 2D-2 &,\ e>b \end{array} \right. \\
f & = & \sqrt{d^2 + e^2 - 2ed  \cos(\gamma)} 
\end{eqnarray}
\begin{figure}[t] 
\centerline{
\epsfxsize=12cm \parbox{12cm }{\epsfbox{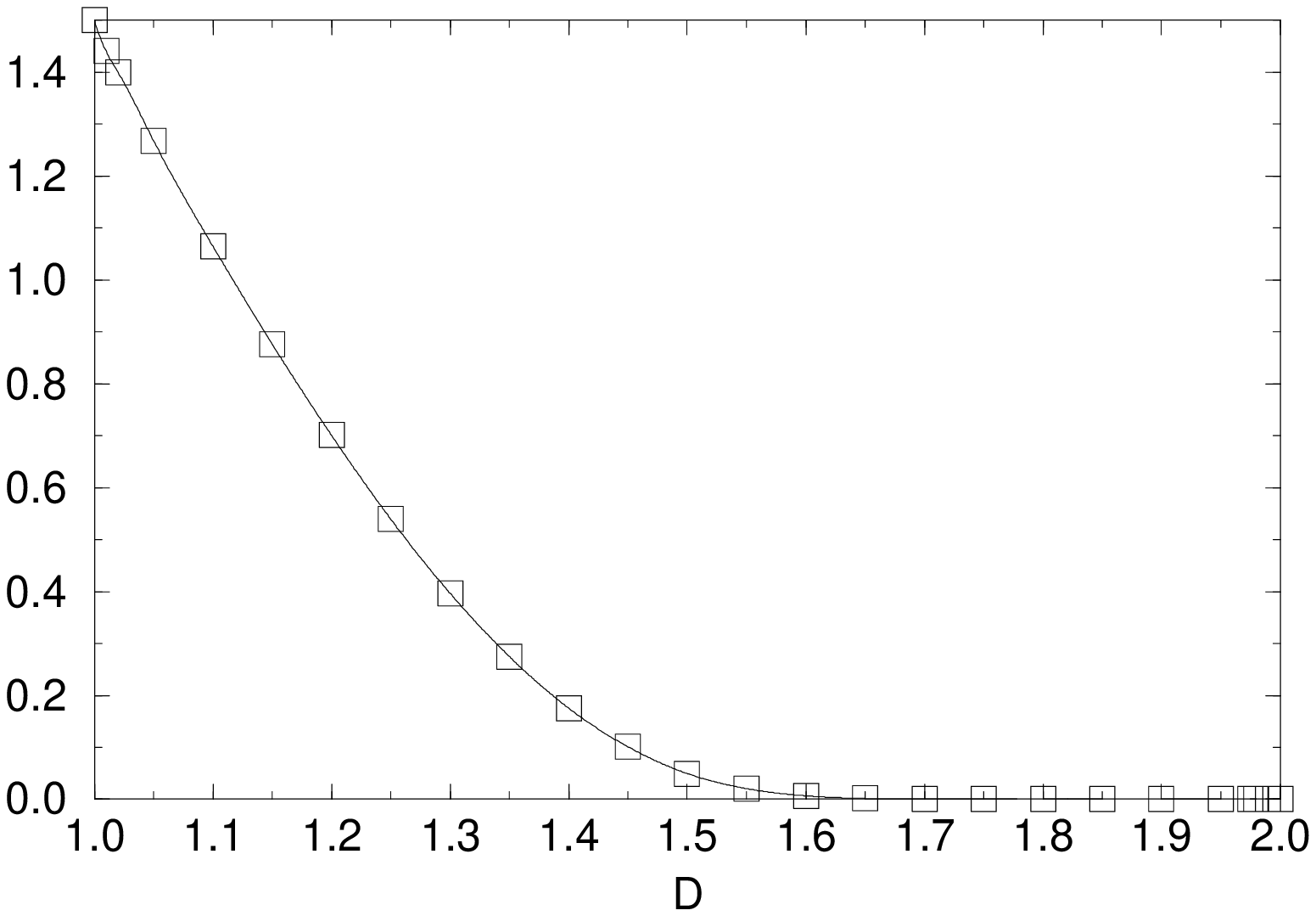}}
}

$$ L \frac{\partial}{\partial L}\lts_{\E=0} \left(\Bigg< \GJ \ \Bigg| \ \GB \Bigg>_L -\frac32
\Bigg< \GM \ \Bigg| \ \GB \Bigg>_L^2  \right) $$
{\tabcolsep5.0mm
\begin{tabular}[t]{|l|l|} \hline
$D$ & \\ \hline \hline
1.00 &	1.5 \\ \hline
1.01 &  $1.441 \pm 5\times 10^{-3}\pm 4\times 10^{-3}$ \\ \hline
1.02 &	$1.399 \pm 5\times 10^{-3} $ \\ \hline
1.05 &	$1.268 \pm 5\times 10^{-3} $ \\ \hline
1.10 &	$1.065 \pm 5\times 10^{-3} $ \\ \hline
1.15 &	$0.876 \pm 4\times 10^{-3} $ \\ \hline
1.20 &	$0.702 \pm 4\times 10^{-3} $ \\ \hline
1.25 &	$0.543 \pm 3\times 10^{-3} $ \\ \hline
1.30 &	$0.399 \pm 2\times 10^{-3}$ \\ \hline
1.35 &	$0.275 \pm 2\times 10^{-3}$ \\ \hline
1.40 &	$0.176 \pm 2\times 10^{-3}$ \\ \hline
1.45 &	$0.1022 \pm 7\times 10^{-4}$ \\ \hline
1.50 & 	$5.16\times 10^{-2} \pm 4\times 10^{-4}$ \\ \hline
\end{tabular}
\hfill
\begin{tabular}[t]{|l|l|} \hline
$D$ & \\ \hline \hline
1.55 &	$2.16\times 10^{-2} \pm 2\times 10^{-3}$ \\ \hline
1.60 &	$7.07\times 10^{-3} \pm 6\times 10^{-5}$ \\ \hline
1.65 &	$1.61\times 10^{-3} \pm 1\times 10^{-5}$ \\ \hline
1.70 &	$2.17\times 10^{-4}  \pm 2\times 10^{-6}$ \\ \hline
1.75 &	$1.23\times 10^{-5}  \pm 2\times 10^{-7}$ \\ \hline
1.80 &	$1.55\times 10^{-7}  \pm 2\times 10^{-9}$ \\ \hline
1.85 &	$1.07\times 10^{-10} \pm 2\times 10^{-12}$ \\ \hline
1.90 &	$4.72\times 10^{-17} \pm 6\times 10^{-19}$ \\ \hline
1.95 &	$3.99 \times 10^{-36} \pm 6\times 10^{-36}$ \\ \hline
1.975 & $2.81\times 10^{-74}\pm 3\times 10^{-76}$ \\ \hline
1.98 & $2.33\times 10^{-93} \pm 2\times 10^{-95}$ \\ \hline
1.99 & $6.89\times 10^{-189} \pm 1 \times 10^{-189}$ \\ \hline
2.00 & $0$  \\ \hline
\end{tabular}
}
\caption{Numerical results for the diagram (\protect{\ref{e:ccr1:4}}). The first error in
the table is the statistical error, the second the systematic error. The latter is only given if it can not be neglected.}
\label{f:graph31}
\end{figure}
With these variable transformations, the integrand is bounded.
This does not mean that the numerical integration is easy.
The main problem is that the integral is localized in some small domain,
in which the integrand (with all the factors of the measure and 
from the variable transformations) is about 100 or 1000 times its mean-value, 
whereas it is much smaller in large domains of integration.
A genuine adaptive Monte-Carlo (AMC) routine has to be used.
It is described in our earlier publication \cite{r:WieseDavid95}.
The idea is to divide the domain of integration into subboxes and to try to
integrate each subbox using standard Monte-Carlo (MC) integration with few
sample-points.
The  MC-routine gives an estimate for the integral and for the standard
deviation from this value.
If the latter is too large, the box is divided into smaller subboxes and
the procedure is repeated.
This algorithm was implemented recursively using the computer-language C.
For details cf.\ \cite{r:WieseDavid95}.
The AMC-routine gives an estimate of the integral and of its statistical error. A systematic error also appears, which is more difficult to estimate.
It comes from the domain of small angular variables, where the numerical
precision of the workstation is no longer sufficient and it appears to be the
most important for $D\to 1$, as can be seen from (\ref{e:m31A}).
It can be estimated by counting exceptions of the floating-point unit and
comparing it to known integrals.
For (\ref{e:ccr1:4}), the systematic error is negligible.
Its treatment will be discussed for the diagrams, where its contribution is
important. 
The results for (\ref{e:ccr1:4}) where obtained within some hours on a
workstation and are listed in figure \ref{f:graph31}.
The numerical results nicely fit with the analytical value for $D=1$,
discussed in the next section. 

We checked the numerical integration routine. This check is provided 
by integrating the function
\be
g(a,b,c,d,e,f)= (a^2+b^2+c^2+d^2+e^2+f^2)^{-2D}
\ee
which has conformal weight $\kappa=0$ and by comparing to the analytical
solution
\be
\int_{a=1,b,c,d,e,f} g(a,b,c,d,e,f)=  \frac{\Gamma(D/2)^4}{8\cdot 3^D\Gamma(2D) } \ .
\ee
This test ensures that the integration can be done for $1<D<2$. 

\subsection{Analytical calculation for $D\to 1$} 
\begin{figure}[htb]
\centerline{\epsfxsize=8cm \parbox{8cm}{\epsfbox{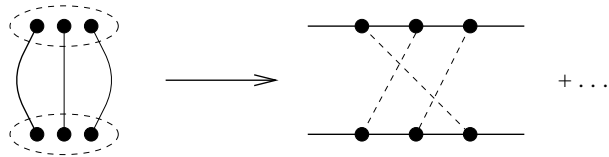}}}
 \label{f:graph31 principle}
\caption{Connection between the  MOPE-coefficient and the diagrams in polymer theory}
\end{figure}To check the numerical calculation, we want to evaluate 
(\ref{e:ccr1:4}) for $D=1$.
This calculation can be performed analytically.

Some subtleties have to be taken care of in order to understand 
the calculation which follows.
We remarked that in (\ref{e:ccr1:4}) the factor
$\max(a,b,c,d,e,f)^{-2\E}$ could be dropped without changing the result as the
contribution to the integral is finite in any subdomain.
For the following calculation we shall drop this factor but shall not take the
limit $d\to d_c$ as we only can calculate the diagram and its counterterms
separately.
Single terms will thus have divergencies in $1/\E$, which are treated by
a finite part integration prescriptions, and which have to cancel at the end.
Through this change only the sum of all terms but not each single term has
a meaning in the limit $\E\to 0$, that we take at the end.

\begin{figure}[htb]
\centerline{\epsfxsize=3.2cm \parbox{3.2cm}{\epsfbox{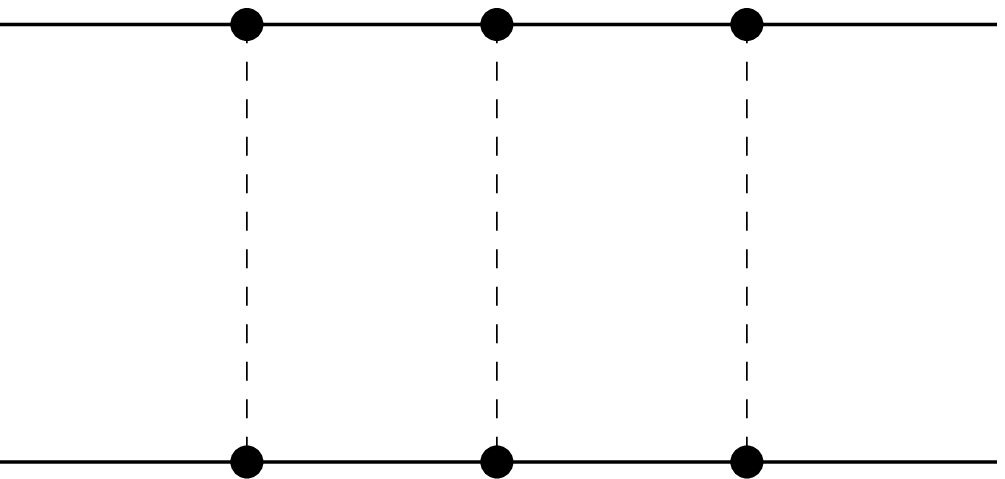}}
\qquad
\epsfxsize=3.2cm \parbox{3.2cm}{\epsfbox{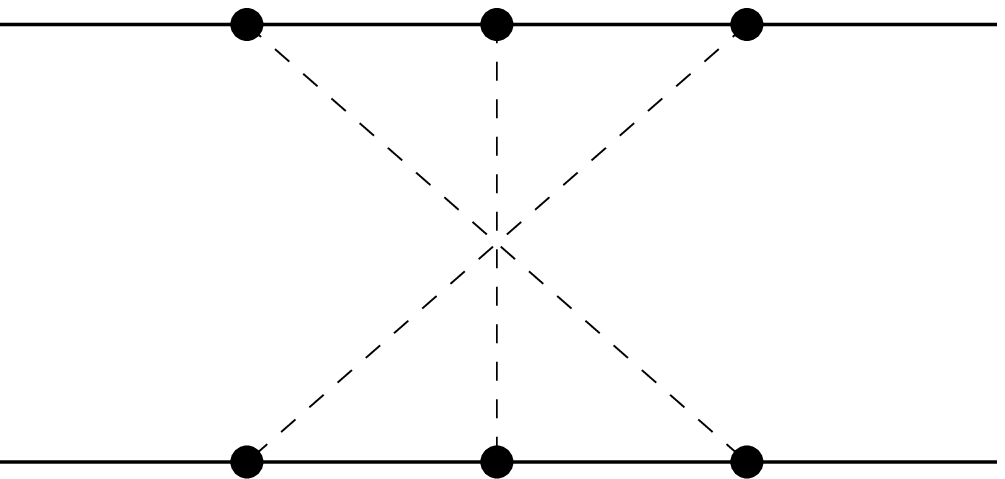}}}
\caption{The 2 untwisted diagrams} \label{f:graph31A}
\end{figure}

\begin{figure}[ht]
\centerline{\epsfxsize=3.2cm \parbox{3.2cm}{\epsfbox{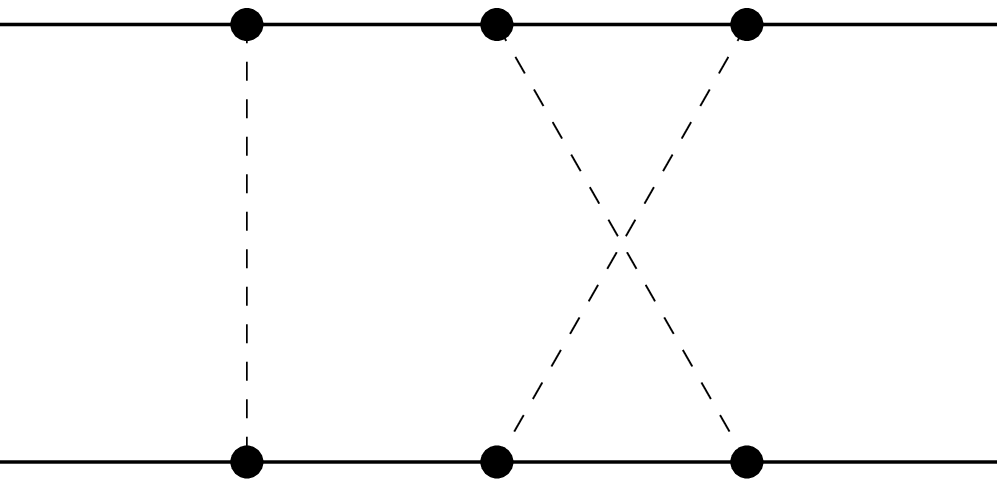}}
\qquad
\epsfxsize=3.2cm \parbox{3.2cm}{\epsfbox{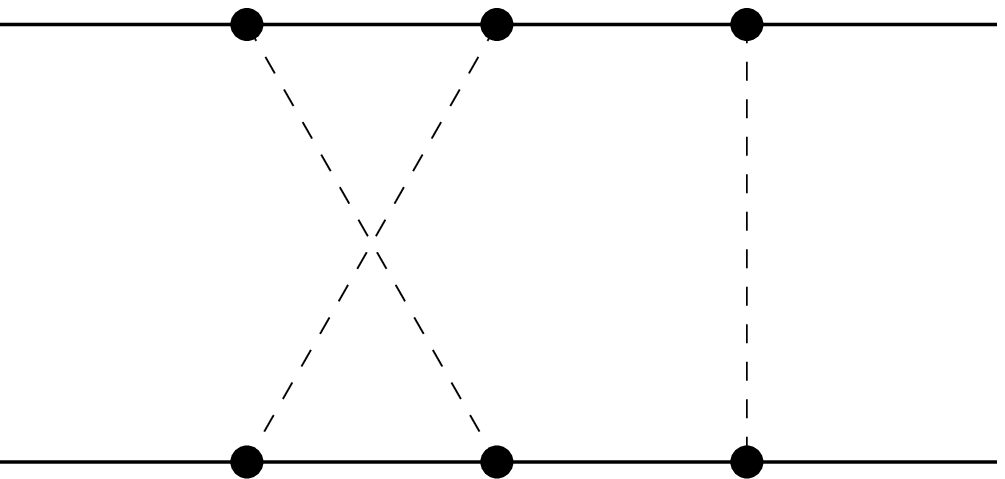}}
\qquad
\epsfxsize=3.2cm \parbox{3.2cm}{\epsfbox{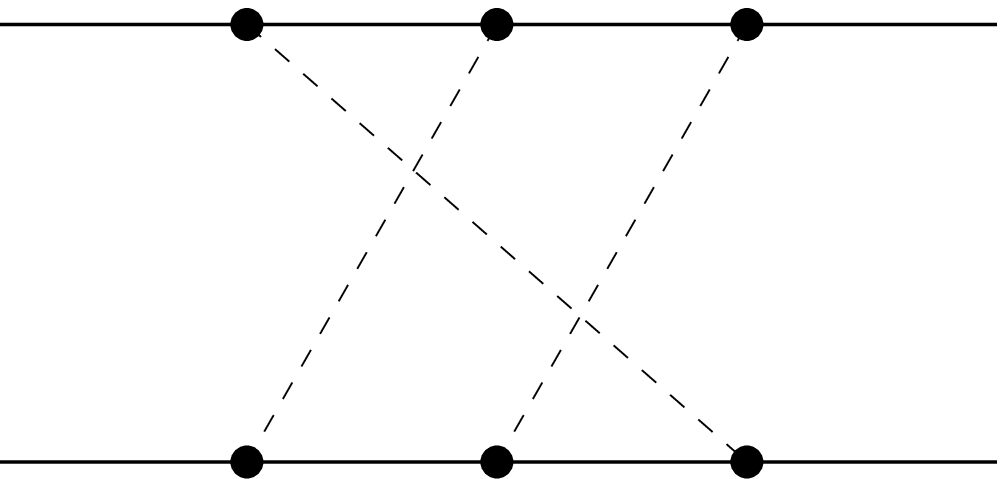}}
\qquad
\epsfxsize=3.2cm \parbox{3.2cm}{\epsfbox{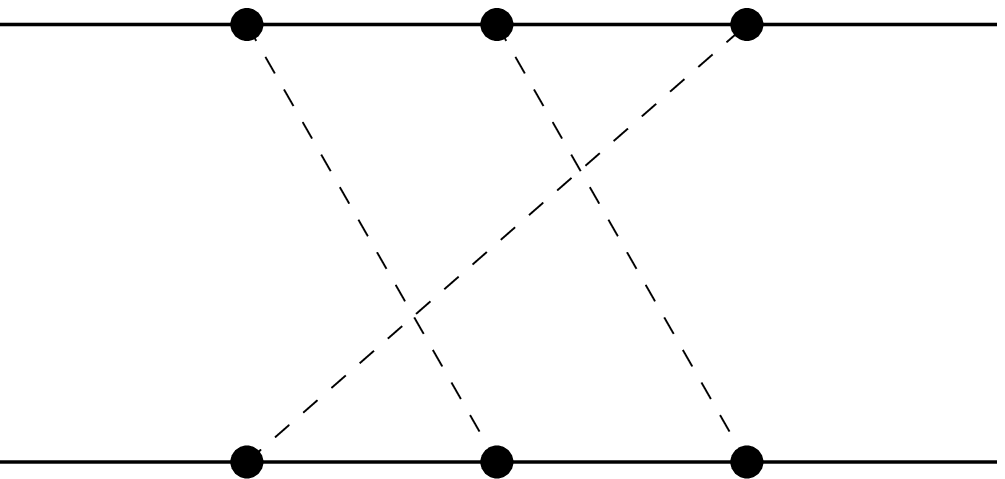}}}
\caption{The 4 twisted diagrams} \label{f:graph31B}
\end{figure}

For $D\to 1$ the measure localizes on a line. This introduces different  
orderings of the distances, which are topologically inequivalent. 
Two types of diagrams appear for $D\to 1$.
There are either the ``untwisted" diagrams (figure \ref{f:graph31A}) or
the ``twisted" diagrams (figure \ref{f:graph31B}).

We remark that the second diagram in figure \ref{f:graph31A}
is untwisted as it can be transformed into the first one by simply exchanging
the orientation of the lower line.
With the same reasoning one deduces that 
the 4 diagrams in figure \ref{f:graph31B} are topologically equivalent.

Let us start to calculate the untwisted diagram (without the counterterm).
We note that $d=4-2\E$.
We find
\begin{eqnarray}
\hspace{-3mm} \epsfxsize=3.2cm \parbox{3.2cm}{\epsfbox{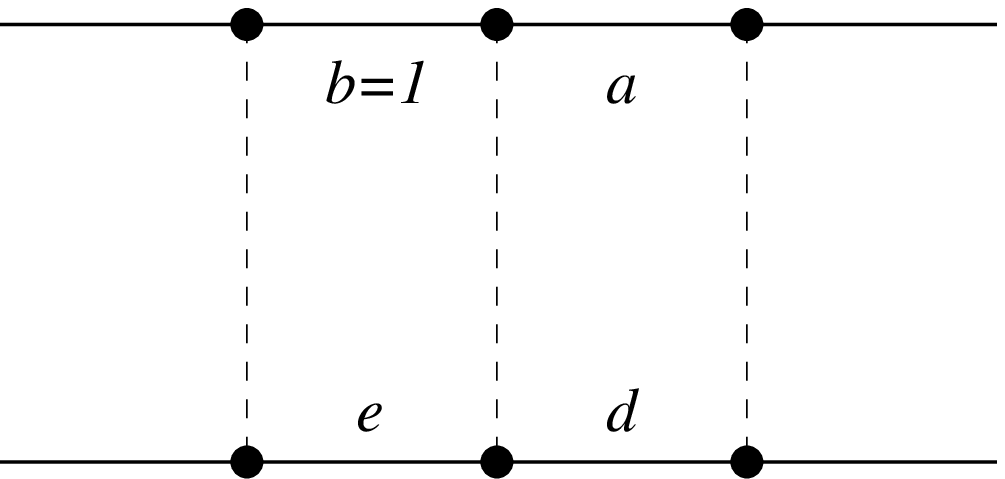}}
\ &=& \int_a\int_d\int_e (ab +ae +bd +ed)^{-d/2} \nn\\
&=& \int_a\int_d\int_e (a+d)^{-d/2}\, (b+e)^{-d/2}=0
\end{eqnarray}
The fact that the last integral is zero is a well known property of
the finite part integration of a homogeneous function
(here $\int_a\int_d(a+d)^{-d/2}$=0).

The counterterms are:
\begin{eqnarray}
& & \int_a \int_d\int_e\, (a+d)^{-d/2}\, (b+e)^{-d/2} = 0  \\
& & \int_a \int_d \int_e\, (a+d)^{-d/2} \,(c+f)^{-d/2} =  
\int_a \int_d \int_e\, (a+d)^{-d/2}\, (1+a+d+e)^{-d/2} \nonumber \\
& & \qquad =\frac{1}{8}\frac1\E + \frac18 + {\cal O}(\E) \\
& &  \int_a \int_d\int_e\, (b+e)^{-d/2}\, (c+f)^{-d/2} = 
\int_a \int_d \int_e\, (1+e)^{-d/2}\, (1+a+d+e)^{-d/2} \nonumber \\
& & \qquad =-\frac{1}{8}\frac{1}{\E} - \frac38  +{\cal O}(\E)
\end{eqnarray}
So together these terms add up to 
\begin{eqnarray}
& & \hspace{-3mm} 6 \left( \ \epsfxsize=2cm \parbox{2cm}{\epsfbox{./eps/fig31a.eps}} \ - \half \sum 
\mbox{counter terms} \right)
= \frac{3}4 +{\cal O}(\E)
\end{eqnarray}
The factor $6$ is the combinatorical factor. 

The untwisted diagram is given by:
\begin{eqnarray}
 \hspace{-10mm} \epsfxsize=3.2cm \parbox{3.2cm}{\epsfbox{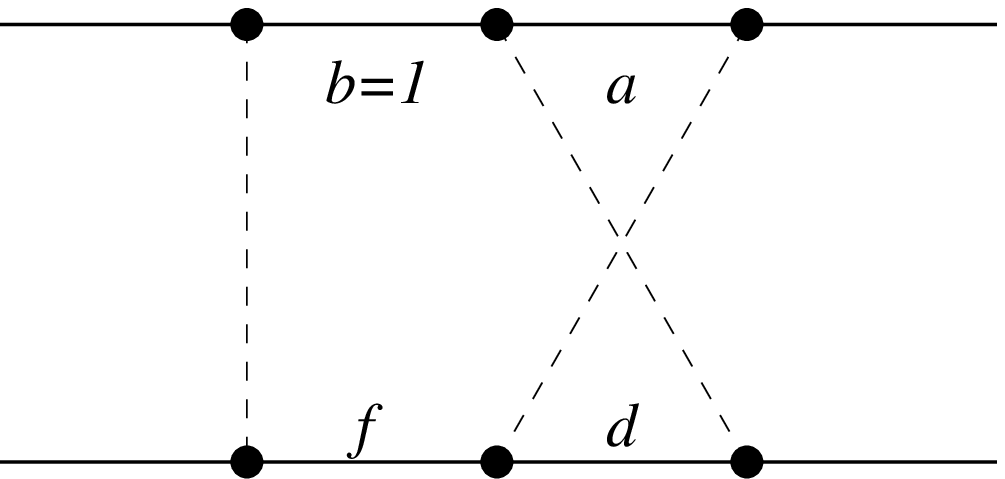}}
&=& \int_a\int_d\int_f \left(ab +af +ad+ bd +df\right)^{-d/2}  \nonumber \\
&=&  \int_a \int_d \int_f \left(a  + af + ad +d + df\right)^{-d/2} =
 \frac18 \frac1\E +\frac3{8} +{\cal O}(\E)
\end{eqnarray}
The counterterms are:
\begin{eqnarray}
& & \int_a \int_d\int_f\, (a+d)^{-d/2}\, (b+e)^{-d/2} = 
\int_a \int_d \int_f \, (a+d)^{-d/2} \, (1+d+f)^{-d/2} \nonumber \\
& & \qquad= \frac18\frac1\E +\frac14+{\cal O}(\E)\\
& & \int_a \int_d \int_f\, (a+d)^{-d/2} \,(c+f)^{-d/2} =
\int_a \int_d \int_f \, (a+d)^{-d/2} \, (1+a+f)^{-d/2} \nonumber \\
& & \qquad =\frac18\frac1\E +\frac14+{\cal O}(\E) \\
& &  \int_a \int_d\int_f\, (b+e)^{-d/2}\, (c+f)^{-d/2} = 
\int_a \int_d \int_f\, (1+f+d)^{-d/2}\, (1+a+f)^{-d/2} \nonumber \\
& & \qquad =\frac18  +{\cal O}(\E)
\end{eqnarray}
These terms add up to 
\begin{eqnarray}
& & \hspace{-3mm} 12 \left( \ \epsfxsize=2cm \parbox{2cm}{\epsfbox{./eps/fig31c.eps}} \ - \half \sum 
\mbox{counter terms} \right)
= \frac{3}4 +{\cal O}(\E)
\end{eqnarray}
The factor $12$ again is the combinatorical factor.

The final result which has to be compared to the numerics thus is (cf.\ (\ref{e:ccr1:4})):
\begin{equation}
J(L=1)\ts_{D=1}=\frac32
\end{equation}

\section{Coupling constant renormalization, second graph}
\label{Coupling constant renormalization, second graph}
\begin{figure}[h]
\centerline{\epsfxsize=10cm \epsfbox{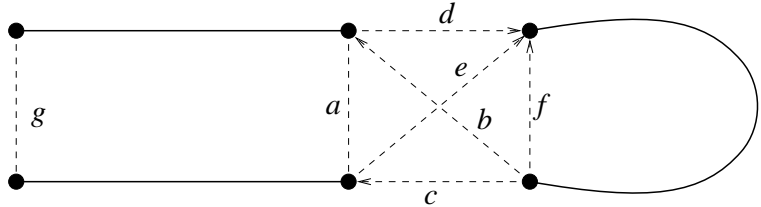}}
\caption{The distances in (\protect{\ref{e:C2L}}). }
\label{f:dist for int211}
\end{figure}

\subsection{Derivation of an analytical expression}
\label{s:C2 der ana exp}
The next diagram which has to be calculated is \eq{e:C2}:
\be \label{e:C2L}
I(L)= -\half {\cal C}_2 = 
\bigg<\GP \bigg| \GB \bigg>_L - \half 
\bigg< \GX \bigg| \GB\bigg>_L \bigg<\GH \bigg|   \GO \bigg>_L
\ee
The second term on the r.h.s.\ subtracts the subdivergence, 
when first the single dipole on the r.h.s.\ of $\GP$ is contracted. In this 
case the MOPE is:
\be \label{e:MOPE of GP}
\bigg(\GP \bigg| \GB \bigg) \approx
\bigg(\GXalphabeta \bigg| \GB \bigg)  
\bigg(\GH \bigg| \GOalphabeta \bigg)
\ee
Note that the MOPE \eq{e:MOPE of GP} has a tensorial 
structure and that a counterterm like 
\be
\bigg(\GX \bigg| \GB \bigg)  
\bigg(\GH \bigg| \GO \bigg)
\ee
would subtract the pole-term from the integral but would
not be sufficient to make  the integral convergent.
With the distances as noted in figure \ref{f:dist for int211}
the MOPE-coefficients are:
\bea
 \bigg(\GP \bigg| \GB \bigg) &=&
\left[ \left( g^{2\nu} + a^{2\nu} \right) f^{2\nu} -\frac14 
\left( b^{2\nu}-c^{2\nu}+e^{2\nu}-d^{2\nu} \right)^2 \right]^{-d/2}
\nn \\ &&
\\
\bigg(\GXalphabeta \bigg| \GB \bigg)  
\bigg(\GH \bigg| \GOalphabeta \bigg) &=& 
\frac d 2 \nu^2 (g^{2\nu} + a^{2\nu})^{-d/2-1} f^{-\nu(d+2)}
 \left( \vec b \vec f b^{-D} -\vec c \vec f c^{-D} \right)^2
\nn \\ 
&=&
\frac d 2 \nu^2 (g^{2\nu} + a^{2\nu})^{-d/2-1} f^{-\nu(d+2)}
\left( \vec e \vec f e^{-D} -\vec d \vec f d^{-D} \right)^2
\nn \\
&& \label{e:CT211dp}
\eea
where two equivalent formulations were given. 

For scalar-products we always use the convention that 
the vectors involved start from the same point, i.e.\
we {\em define} (cf.\ figure \ref{f:dist for int211})
\be
	\vec b \vec f = \half (b^2+f^2-d^2)
\ee
A relevant counter-term, which we did not explicitly
write in ${\cal C}_2$, appears too. It is canceled by
\begin{equation} \label{e:211RCT}
\bigg(\GPrel \bigg| \GB \bigg) = 
(g^{2 \nu} + a^{2 \nu})^{-d/2} f^{-\nu d}
\end{equation}
We state and will show below that
\bea
	I(L)&=&\int_{a,b,c,d,e,f,g<L}  \bigg(\GP \bigg| \GB \bigg)
-\bigg(\GPrel \bigg| \GB \bigg) \nn \\
&&\hspace{2cm}-\bigg(\GXalphabeta \bigg| \GB \bigg)  
\bigg(\GH \bigg| \GOalphabeta \bigg) \Theta(f<\max(a,d,e,g)) 
\nn \\ && \hspace{2cm} + {\cal O}(\E^0) \ ,
\label{e:C2L'}
\eea
where we used the second version of the counter-term in 
\eq{e:CT211dp}.

To compute the residue of the single pole, we
apply $L \partial /\partial L$ on this expression 
and map onto $c=L=1$:
\begin{eqnarray}
&&\hspace{-0.6cm}J(L)=L \frac{\partial }{\partial L} I(L) =  \int\!\!\!\int\!\!\!\int\!\!\!\int\!\!\!\int\!\!\!
\int_{c=1;\,a,b,d,e,f,g} \max(a,b,c,d,e,f,g)^{-2\E} \nonumber \\
&&\left( \left[ \left( g^{2\nu} + a^{2\nu} \right) f^{2\nu} -\frac14 
\left( b^{2\nu}-c^{2\nu}+e^{2\nu}-d^{2\nu} \right)^2 \right]^{-d/2} \right.
-(g^{2 \nu} + a^{2 \nu})^{-d/2} f^{-\nu d} \nonumber \\
&& -\frac d 2 \nu^2
 (g^{2\nu} + a^{2\nu})^{-d/2-1} f^{-\nu(d+2)} 
\left.\left( \vec e \vec f e^{-D} -\vec d \vec f d^{-D} \right)^2 
	\Theta(f<\max(a,d,e,g)) \right)
\label{e:C2 J(L)}
\end{eqnarray}
We  check that this expression is integrable everywhere, as
is indeed the case and can be seen by a (generalized) Taylor-expansion.

We then have to explain that
\bea
&&\hspace{-1cm}\int_{a,d,e,g<L;\, f}
\bigg(\GXalphabeta \bigg| \GB \bigg)\,\bigg(\GH \bigg| \GOalphabeta \bigg)\,\Theta(f<\max(a,d,e,g)) \nn \\
&=&\int_{a,d,e,g<L;\, f}
\bigg(\GX\bigg| \GB \bigg)\,\bigg(\GH \bigg| \GO \bigg)\,\Theta(f<\max(a,d,e,g)) \nn \\
&=& \half
\bigg<\GX\bigg| \GB \bigg>_L\,\bigg<\GH \bigg| \GO \bigg>_L
\eea
The first equality is due to symmetry. The second stems from the 
so-called nested integration. We see this explicitly as follows:
\bea 
&&\hspace{-1cm}\int_{a,d,e,g<L;\, f}
\bigg(\GX\bigg| \GB \bigg)\,\bigg(\GH \bigg| \GO \bigg)\,\Theta(f<\max(a,d,e,g)) \nn \\
&=& \int_{a,d,e,g<L}
\bigg(\GX\bigg| \GB \bigg) \bigg<\GH \bigg| \GO \bigg>_L \left(\frac{\max(a,d,e,g)}{L}\right)^\E
\label{e:C2 1}
\eea
where we used the fact that for any $l$
\be
	\bigg<\GH \bigg| \GO \bigg>_l \sim l^\E
\ee
The whole integral scales like $L^{2\E}$, so that we can apply
$\frac L{2\E} \frac{\partial}{\partial L}$ to \eq{e:C2 1}
without changing it. Doing so we obtain that \eq{e:C2 1}
equals
\bea
  \frac{L}{2\E} \int_{\max(a,d,e,g)=L}
\bigg(\GX\bigg| \GB \bigg) \left(\frac{\max(a,d,e,g)}{L}\right)^\E
\times \bigg<\GH \bigg| \GO \bigg>_L
\eea
The factor $\left(\frac{\max(a,d,e,g)}{L}\right)^\E$ is equal to 
unity and the integral is our usual expression for 
the residue of $\bigg<\GX\bigg| \GB \bigg>_L$.
Taking all this together we obtain
\be
\frac12 \bigg<\GX\bigg| \GB \bigg>_L \bigg<\GH \bigg| \GO \bigg>_L
\ee
This proofs the disered result. The reader can verify that this is
a general feature of these so-called nested contractions. If the 
largest distance in the subdiagram is restricted to be smaller than
the largest distance which rests after complete contraction of the 
subdiagram, then the so restricted integral is $1/2$ times the 
product of the subdiagram and the diagram which rests after contraction.
This demands of course that both of them scale like $L^\E$, where 
$L$ is the IR-cutoff. The reader will be able to generalize this 
rule for a scaling with other exponents, which would be necessary in 
higher order calculations.

The subtle point which we still have to check is
that the changes in the domain of integration
of the marginal counterterm  from \eq{e:C2L} to \eq{e:C2L'} 
do not change the residue, i.e.\ that again the ``2-loop-miracle'' appears.
Analogously to section \ref{s:C1, counter-term} we write down the difference, 
apply $L \partial /\partial L$ on this expression 
and map onto $c=1$. We obtain
\begin{eqnarray}
&&\hspace{-1cm} \frac d 2 \nu^2
\int\!\!\!\int\!\!\!\int\!\!\!\int\!\!\!\int\!\!\!\int_{a,b,d,e,f,g}
 \left(\max(a,b,c,d,e,f,g)^{-2\E}-\max(a,d,e,f,g)^{-2\E}\right) \nonumber\\
& & (g^{2\nu} + a^{2\nu})^{-d/2-1} f^{-\nu(d+2)}
 \left( \vec e \vec f e^{-D} -\vec d \vec f d^{-D} \right)^2 
\Theta(f<\max(a,d,e,g))
\end{eqnarray}
We would like to develop    $\left(\max(a,b,c,d,e,f,g)^{-2\E}-\max(a,b,c,f,g)^{-2\E}\right)$
for $\E$ small
and show that the corrections are of order $\E$. This might be wrong,
if and only if this expression does not vanish at points, where
the integral has a pole. For $f\to 0$ it  vanishes. 
The limit $a,g \to 0$ is a bit more subtle as the
difference does not vanish. However 
in this case $\left( \vec e \vec f e^{-D} -\vec d \vec f d^{-D} \right)^2 $
is of order $a^2$ so that no pole in the integration over $a$ and $g$ appears.

\medskip
Let us now perform in \eq{e:C2 J(L)} the limit $d\to d_c$, i.e. $\E \to 0$. Then we would like
to integrate over $g$ analytically. This is not possible
due to the $\Theta$-function. We therefore modify this constraint from
$\Theta(f<\max(a,d,e,g))$ to $\Theta(f<d)$. Note that 
the modified counterterm still successfully subtracts the marginal
subdivergence. We obtain:
\begin{eqnarray}
&&\hspace{-0.6cm} \tilde J(L)=\int\!\!\!\int\!\!\!\int\!\!\!\int\!\!\!\int\!\!\!
\int_{a,b,d,e,f,g} \nonumber \\
&&\left( \left[ \left( g^{2\nu} + a^{2\nu} \right) f^{2\nu} -\frac14 
\left( b^{2\nu}-c^{2\nu}+e^{2\nu}-d^{2\nu} \right)^2 \right]^{-d_c/2} \right.
-(g^{2 \nu} + a^{2 \nu})^{-d_c/2} f^{-\nu d_c} \nonumber \\
&& -\frac d 2 \nu^2
 (g^{2\nu} + a^{2\nu})^{-d_c/2-1} f^{-\nu(d_c+2)} 
 \left.\left( \vec e \vec f e^{-D} -\vec d \vec f d^{-D} \right)^2 
	\Theta(f<d)\right)
\label{e:tilde J}
\end{eqnarray}
Of course this change affects the result and we shall calculate 
the difference in section \ref{The correction for the unusual marginal counterterm}. 
The modified counterterm has another useful property,
which justifies its choice: The $\Theta$-function is not affected 
by the {\bf R}-operation discussed in the next section.

Now the integration over $g$ can be performed.
The result is:
\be
\tilde J(L) = \int\!\!\!\int\!\!\!\int\!\!\!\int\!\!\!\int_{a,b,d,e,f} 
F(a,b,c,d,e,f)
\label{e:M211}
\ee
with
\begin{eqnarray}
&& \hspace{-1cm} F(a,b,c,d,e,f) = \frac 1{2-D} \frac{\Gamma\left( \frac{D}{2-D}\right)^2}{\Gamma\left( \frac{2D}{2-D}\right)}
 \nonumber \\
&& \bigg\{ f^{-D} \left( a^{2\nu} f^{2\nu} -\frac14 
\left( b^{2\nu}-c^{2\nu}+e^{2\nu}-d^{2\nu}\right)^2 \right)^{-d_c/4}
-a^{-D} f^{-2D}  \nonumber\\
&& \qquad -\frac D 2 \nu a^{-2} f^{-2-D} \Theta(f<d) 
\left( \vec e \vec f e^{-D} -\vec d \vec f d^{-D} \right)^2
\bigg\}\label{e:M211+}
\end{eqnarray}

\subsection{Improvement of the measure}
\label{s:Improvement of the measure}
For $c=1$ the measure, given by (\ref{N points integral}) 
and (\ref{s:Normalizations 2}), simplifies to an integral over the vectors $a$ and $f$:
\begin{equation} \label{all43}
	\frac{S_{D-1} S_{D-2}}{S_D^2} \int_{-\infty}^{\infty}d a_1 \int_{0}^{\infty} \rmd a_2 \,
	a_2^{D-2} \int_{-\infty}^{\infty} \rmd f_1 \int_{-\infty}^{\infty} \rmd f_2 
	\int_0^{\infty} \rmd f_3 \, f_3^{D-3}
	F(a,b,c,d,e,f) 
\end{equation}
For $D<2$ the measure defined in (\ref{all43}) is a distribution and 
suffers from a relevant divergence for 
$f_3 \rightarrow 0$. Geometrically these are 
configurations, where the tetrahedron spanned by $a,b,\ldots,f$ has 
volume 0, i.e.\ is restricted to a plane. A finite part
prescription has to be applied in order to make
the measure finite. This was first discussed in \cite{r:WieseDavid95}.

One may think of implementing this prescription by subtracting the singularity.
This method however imposes at least numerical 
difficulties. 
It is better to eliminate the singularity by a 
partial integration with respect to $f_3$, which is mathematically
equivalent\cite{r:WieseDavid95}. 
As only $d$, $e$ and $f$ depend on $f_3$, the integral 
\begin{equation}
	\mbox{f.p.} \int_0^\infty \rmd f_3 \, f_3^{D-3} F(a,b,c,d,e,f)
\end{equation}
can be converted to
\begin{eqnarray}
& & \frac1{2-D} \int_0^\infty \rmd f_3\, f_3^{D-2} \frac{\partial}{\partial f_3} F(a,b,c,d,e,f) 
	\nonumber \\ 
& & \quad = \frac1{2-D} \int_0^\infty \rmd f_3 \,f_3^{D-1} \mbox{\bf R}
F(a,b,c,d,e,f)
\end{eqnarray}
where {\bf R} is defined via
\begin{equation} \label{mass4pnt1}
\mbox{\bf R}= \frac1d \frac\partial{\partial d} +
	\frac1e \frac\partial{\partial e} + \frac1f \frac\partial{\partial f}
\end{equation}
The strength of the divergences for $a \to 0$ or $f\to 0$ is unchanged.
It is important to remark that this trick cannot be used to 
eliminate the relevant divergences when $a \to 0$ or $f \to 0$.
It works for the divergence in $f_3$, because the integrand does not directly 
depend on $f_3$ but on $d$, $e$ and $f$, which themselves depend on $f_3$.
So the derivation of $F$ with respect to $f_3$ does not produce a factor
$1/f_3$ but factors $1/d$, $1/e$ and $1/f$, which are not singular for 
$f_3 \to 0$.
Explicitly:
\begin{eqnarray}
&& \hspace{-1cm} \mbox{\bf R} F(a,b,c,d,e,f)= 
 \frac 1{2-D} 
\frac{\Gamma\left( \frac{D}{2-D}\right)^2}{\Gamma\left( \frac{2D}{2-D}\right)}
\times
  \nonumber \\
& \times \Bigg\{ & -D f^{-2-D}  \left( a^{2\nu} f^{2\nu} -\frac14 
\left( b^{2\nu}-c^{2\nu}+e^{2\nu}-d^{2\nu}\right)^2 \right)^{-d_c/4}\nonumber \\
 & & -D f^{-D} \left( a^{2\nu} f^{2\nu} -\frac14 
\left( b^{2\nu}-c^{2\nu}+e^{2\nu}-d^{2\nu}\right)^2 \right)^{-d_c/4-1} \times \nonumber \\
&& \qquad \qquad\times \left( a^{2\nu} f^{-D} -\frac12 
\left( b^{2\nu}-c^{2\nu}+e^{2\nu}-d^{2\nu}\right)\left(e^{-D}-d^{-D}\right) \right)
\nonumber \\
&& +2D a^{-D} f^{-2D-2} \nonumber\\
&& +(2+D) \frac{D\nu}{2}a^{-2} f^{-4-D}  \Theta(d>f) 
\left( \vec e \vec f e^{-D} -\vec d \vec f d^{-D} \right)^2
\nonumber \\
&&  -D \nu a^{-2} f^{-2-D} \Theta( d>f) \left( 
\vec e \vec f e^{-D} -\vec d \vec f d^{-D}   \right) \times \nn \\
&&\qquad \qquad \times \left( 2e^{-D} -2d^{-D} -D \vec e \vec f e^{-D-2} +D\vec d \vec f d^{-D-2} \right)
\Bigg\} 
\label{e:ugly}
\end{eqnarray}
Note that $\mbox{\bf R} \Theta( f<d) =0 $ whereas 
$\mbox{\bf R} \Theta( f<\max(a,d,e)) $ gives contributions proportional 
to e.g.\ $\delta(a-f)$ which had to be treated separately. 
This justifies our choice of the modified bound of the counterterm in 
\eq{e:tilde J}.

\subsection{Parametrization of the measure}
\label{all42}
\begin{figure}[htb]
\centerline{
\epsfxsize=6.0cm \parbox{6.0cm}{\epsfbox{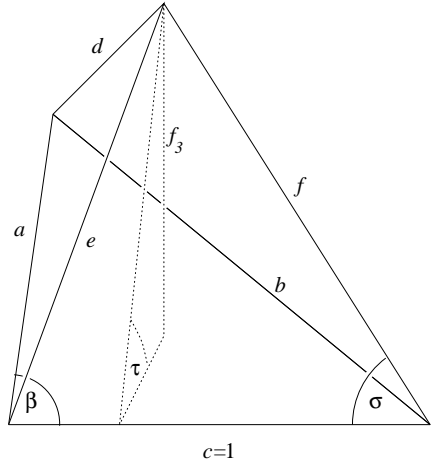}}
}
\caption{Parametrization of the tetrahedron}
\label{f:tetra1}
\end{figure}

The main singularities for small distances appear for $a$ or $f$ small. 
We therefore want to parametrize the
measure with the help of these distances.
 The divergences for small
volume of the tetrahedron spanned by $a, \ldots ,f$,  $a_2 \rightarrow 0$ or $f_3 \rightarrow 0$
shall be treated by a parametrization in angles as by this way small distance and 
small volume singularities are best disentangled. We have chosen the parametrization 
indicated in figure \ref{f:tetra1}. 
One triangle is spanned by $c$ and $a$ with an angle $\beta$ between,
another by $c$ and $f$, where the corresponding angle is $\sigma$.
The angle between the planes spanned by these two triangles is $\tau$.
The distances as functions of $a$, $f$ and $\beta$, $\sigma$, $\tau$
are:
\begin{eqnarray}
	b&=& \sqrt{a^2+1-2a\cos \beta} \nonumber \\
	e&=&\sqrt{f^2 +1 -2f \cos \sigma} \\
	d&=& \sqrt{(a \cos \beta -1 +f \cos \sigma )^2
		+ ( a \sin \beta - f \sin \sigma \cos \tau)^2 +
		(f \sin \tau \sin \sigma)^2 }\nonumber 
\end{eqnarray}
The integrals over $a$ and $f$ run from 0 to $\infty$, the integrals
over $\beta$, $\sigma$ and $\tau$ over the interval $\left[ 0, \pi \right]$. 
As we do not want to map all the points which are far away, we have to find
a reparametrization of the measure which behaves for $a \rightarrow 0$
like $a^\gamma$ and for $a \rightarrow \infty$ like $a^\omega$, by 
this way eliminating the principle divergences.
If $u$ is equally distributed we can use
\begin{eqnarray}
a & = & u^{\textstyle \frac1{D-\gamma}} (1-u)^{\textstyle \frac1{D-\omega}} \ .
\end{eqnarray}
The integral over $\beta$ will  be parametrized as
\begin{eqnarray}
	\beta & = & \left\{ \begin{array}{lcl} \displaystyle
		\frac{\pi}{2} (2\alpha)^{\textstyle \frac1{D-1}} &\qquad& 
			\alpha \le \displaystyle 0.5 \\
		\displaystyle
		\pi-\frac{\pi}{2} (2-2\alpha)^{\textstyle \frac1{D-1}} &\qquad& 
			\displaystyle\alpha > 0.5
		\end{array} \right.
\end{eqnarray}

The integral over $f$ (note that the factor $f_3^2$ came from the partial integration
with respect to $f_3$) 
\begin{equation}
\int \rmd f_1 \, \rmd f_2 \, \rmd f_3 \, f_3^{D-3} (f_3^2)
\end{equation}
can be written as 
\begin{equation}
\int \rmd f\, f^{D-1} \int d\sigma\, (\sin \sigma)^D \int d\tau\, (\sin\tau)^{D-1} (f^2) \ .
\end{equation}
We change variables from $f$ to $v$:
\begin{eqnarray}
f & = & v^{\textstyle \frac1{D-\gamma}} (1-v)^{\textstyle \frac1{D-\omega}}
\end{eqnarray}
Furthermore we choose in the same spirit as for $\beta$
\begin{eqnarray}
\sigma & = & \left\{ \begin{array}{lcl} \displaystyle
	\frac{\pi}{2} (2\eta)^{\textstyle \frac1{D+1}} &\qquad& 
		\eta \le \displaystyle 0.5 \\
	\displaystyle
	\pi-\frac{\pi}{2} (2-2\eta)^{\textstyle \frac1{D+1}} &\qquad& 
		\displaystyle\eta > 0.5
	\end{array} \right. \end{eqnarray}
and 
\begin{eqnarray}
\tau & = & \left\{ \begin{array}{lcl} \displaystyle
	\frac{\pi}{2} (2\zeta)^{\textstyle \frac1{D}} &\qquad& 
		\zeta \le \displaystyle 0.5 \\
	\displaystyle
	\pi-\frac{\pi}{2} (2-2\zeta)^{\textstyle \frac1{D}} &\qquad& 
		\displaystyle\zeta > 0.5
	\end{array} \right.  \end{eqnarray}
So the complete integral over four points is:
\begin{eqnarray} \label{mass4pnt2}
\lefteqn{\frac{S_{D-1} S_{D-2}}{S_D^2} \frac{\pi^3}{(2-D)(D-1)D(D+1)}} & & \nonumber\\
& & \qquad \int_0^1 \rmd u \, a^D \left( \frac{1}{D-\gamma} \frac1u + 
	\frac1{\omega-D} \frac1{1-u} \right)
\int_0^1  \rmd \alpha \, \min(2\alpha,2-2\alpha)^{\textstyle\frac{2-D}{D-1}}(\sin \beta)^{D-2} \nonumber \\
& & \qquad \int_0^1 \rmd v \, f^D \left( \frac{1}{D-\gamma} \frac1v + 
	\frac1{\omega-D} \frac1{1-v} \right)
\int_0^1  \rmd \eta \,  \min(2\eta,2-2\eta)^{\textstyle \frac{-D}{D+1}} 
(\sin \sigma )^D \nonumber \\
& & \qquad \int_0^1  \rmd \zeta \,  \min(2\zeta,2-2\zeta)^{\textstyle \frac{1-D}{D}} 
(\sin \tau )^{D-1}  f^2 \mbox{\bf R} F(a,b,1,d,e,f)
\end{eqnarray}

Another way of parametrizing consists in replacing the integral over the 
vectors $a$ and $f$ by the integral over the vectors $a$ and $d$. This 
parametrization is especially useful to eliminate divergences, when 
$a$ and $f$ simultaneously go to infinity. It will be used for the integration
over one of the sectors in the next section. The new formulas are given here,
a prime indicating new angles and distances as can be deduced from the 
figure.
\begin{figure}[htb]
\centerline{
\epsfxsize=6.0cm \parbox{6.0cm}{\epsfbox{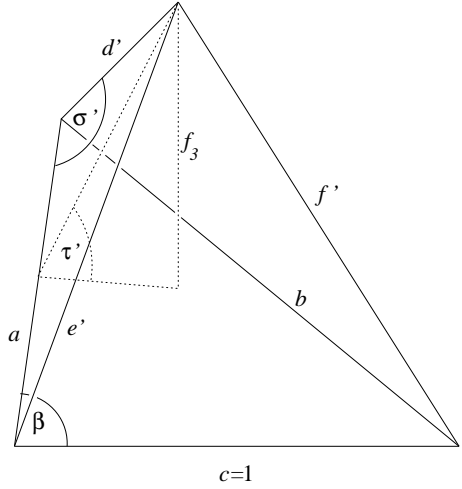}}
}
\caption{Alternative parametrization of the tetrahedron}
\end{figure}
$\sigma'$ and $\tau'$ obey the same relations as $\sigma$ and 
$\tau$. For $d'$ we use the same variable transformation as before 
for $f$:
\begin{equation} \label{secondmeasure}
	d'= v^{\textstyle \frac1{D-\gamma}} (1-v)^{\textstyle \frac1{D-\omega}}
\end{equation}
The other new distances are:
\begin{eqnarray}
e'&=&\sqrt{a^2+d'^2-2 a d' \cos \sigma'} \\
f'&=&\sqrt{(d' \sin \sigma' \sin \tau')^2 + (d' \cos \sigma' -a + \cos \beta)^2 +
	(\sin \beta -d' \sin \sigma' \cos \tau')^2 }
\end{eqnarray}
For the integrand (\ref{e:ugly}), the exponents $\gamma$ and $\omega$
are found by performing a (generalized) Taylor expansion:
\begin{eqnarray}
\gamma &=& 0 \\
\omega &=& 2D 
\end{eqnarray}

\subsection{Decomposition into sectors}
\label{s:decomp}
\begin{figure}[tb] 
\centerline{
\epsfxsize=12cm \parbox{12cm}{\epsfbox{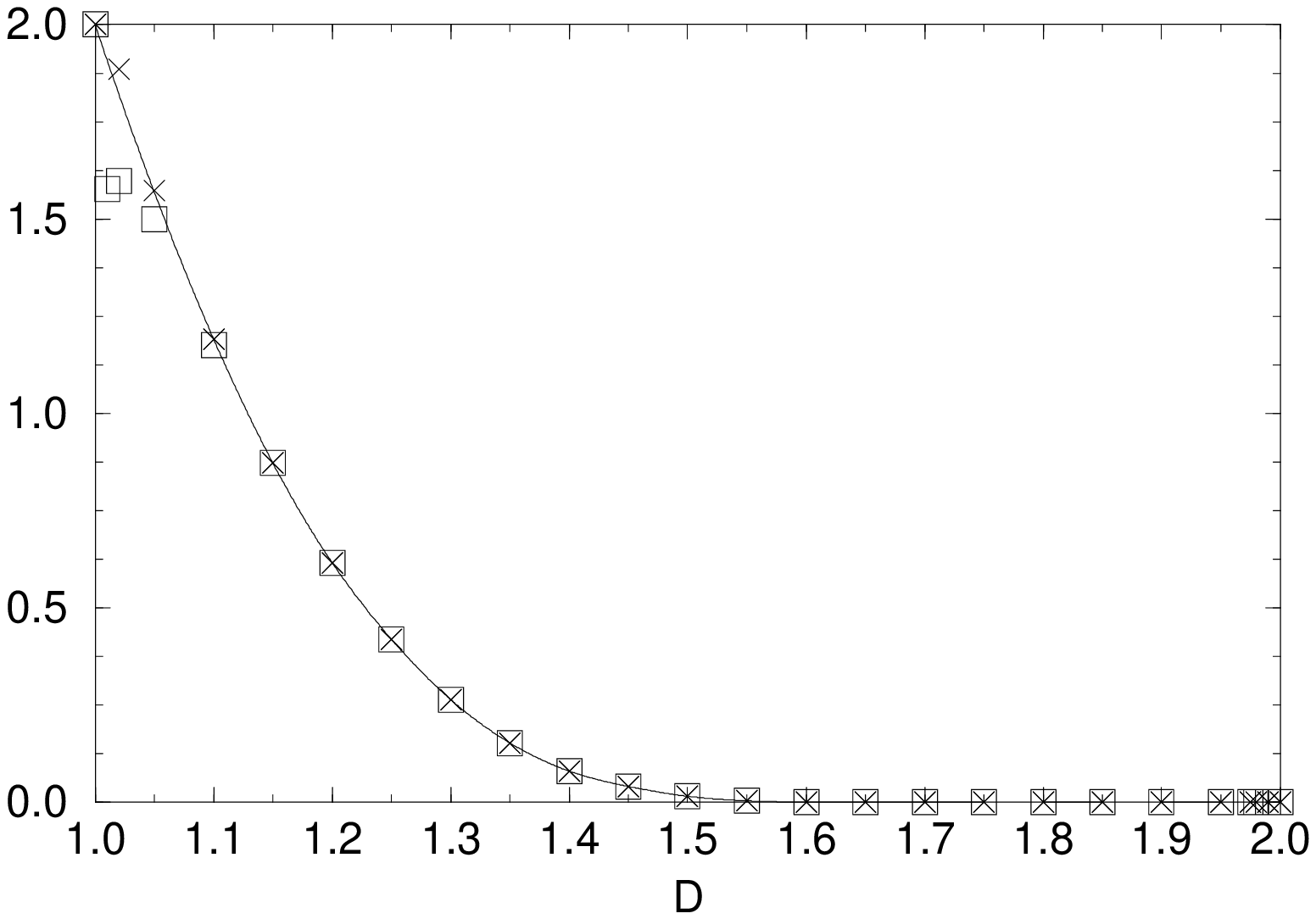}}
}
\vspace{2mm}
{\tabcolsep1.5mm
\begin{tabular}[t]{|l|c|} \hline
$D$ & \\ \hline \hline
1.00 &	$2$ \\ \hline
1.01 &	$1.56 \pm 0.02 + 0.36$\\ \hline
1.02 &	$1.56 \pm 0.02 + 0.25$ \\ \hline
1.05 &	$1.42 \pm 0.02 + 0.07$ \\ \hline
1.10 &	$1.06 \pm 0.02$ \\ \hline
1.15 &	$0.775 \pm 0.009$ \\ \hline
1.20 &	$0.547 \pm0.007$ \\ \hline
1.25 &	$0.379 \pm0.005$ \\ \hline
\end{tabular}\hfill
\begin{tabular}[t]{|l|c|} \hline
$D$ & \\ \hline \hline
1.30 &	$2.52\times 10^{-1} \pm 4\times 10^{-3}$ \\ \hline
1.35 &	$1.62\times 10^{-1} \pm 3\times 10^{-3}$ \\ \hline
1.40 &	$9.88\times 10^{-2} \pm 2\times 10^{-3}$ \\ \hline
1.45 &	$5.62\times 10^{-2} \pm 2\times 10^{-3}$ \\ \hline
1.50 & 	$2.91\times 10^{-2} \pm 6\times 10^{-4}$ \\ \hline
1.55 &	$1.35\times 10^{-2} \pm 3\times 10^{-4}$ \\ \hline
1.60 &	$5.20\times 10^{-3} \pm 2\times 10^{-4}$ \\ \hline
1.65 &	$1.61\times 10^{-3} \pm 5\times 10^{-5}$ \\ \hline
\end{tabular}\hfill
\begin{tabular}[t]{|l|c|} \hline
$D$ & \\ \hline \hline
1.70 &	$3.9\times 10^{-4} \pm 2\times 10^{-4}$ \\ \hline
1.75 &	$4.2\times 10^{-5} \pm3\times 10^{-6}$ \\ \hline
1.80 &	$2.0\times 10^{-6} \pm3\times 10^{-7}$ \\ \hline
1.85 &	$1.3\times 10^{-8}  \pm 3\times 10^{-9}$ \\ \hline
1.90 &	$8.6\times 10^{-13}\pm3\times 10^{-13}$ \\ \hline
1.95 &	$1.7\times 10^{-25} \pm2\times 10^{-25}$ \\ \hline
1.98 &	$ <10^{-48}$ \\ \hline
2.00	 &	0\\ \hline
\end{tabular}}
\caption{Numerical results for (\protect{\ref{e:ugly}}). The first error
is the statistical error, the second an estimate for the
correction due to the systematic error, which becomes important for $D\to1$,
cf.\ the text. In the plot, the boxes are the uncorrected, the crosses
the corrected results.}
\label{f:inte211}
\end{figure}
Although the measure absorbs the principal singularities it cannot
handle all of them. There remains e.g.\  a singularity 
for $b \to 0$ and $e \to 0$.  Two methods may be 
applied to handle the remaining integrable singularities.
The first consists in using the second measure of section
\ref{all42}. The second is to map again some parts
of the domain of integration. 
Thereby, we face the problem
that the measure is no longer symmetric in the 
distances, as we have changed it in order to eliminate the 
relevant singularity for $f_3 \to 0$. In order to restore this symmetry, 
we rewrite the integral (\ref{mass4pnt2}) as
\begin{eqnarray} \label{secondmass}
\lefteqn{\frac{S_{D-1} S_{D-2}}{S_D^2} \frac{\pi^3}{(2-D)(D-1)D(D+1)}} & & \nonumber\\
& & \qquad \int du \, a^{D+2} \left( \frac{1}{D-\gamma} \frac1u + 
	\frac1{\omega-D} \frac1{1-u} \right)
\int  d\alpha \, \min(2\alpha,2-2\alpha)^{\textstyle\frac{2-D}{D-1}}(\sin \beta)^{D} \nonumber \\
& & \qquad \int dv \, f^{D+2} \left( \frac{1}{D-\gamma} \frac1v + 
	\frac1{\omega-D} \frac1{1-v} \right)
\int  d\eta \,  \min(2\eta,2-2\eta)^{\textstyle \frac{-D}{D+1}} 
(\sin \sigma )^D \nonumber \\
& & \qquad \int  d\zeta \,  \min(2\zeta,2-2\zeta)^{\textstyle \frac{1-D}{D}} 
(\sin \tau )^{D-1}   T(a,b,c,d,e,f) \ ,
\end{eqnarray}
with
\begin{equation} \label{int4pnt1}
	T(a,b,c,d,e,f)=\frac{1}{(2\Delta(a,b,c))^2} \mbox{\bf R}F (a,b,c,d,e,f) \ ,
\end{equation}
and 
where $\Delta(a,b,c)=\half ac \sin(\beta )=\frac 14 \sqrt{(a+b+c)(a+b-c)(b+c-a)(c+a-b)}$ is the area of the triangle spanned by $a$, $b$ and
$c$.
This is, except for the geometric prefactor, the invariant measure in 
$D+2$ dimensions.
The integrand now is conformal invariant,
as follows directly from equations \eq{e:I integrale} to \eq{e:diff sect}.

The sectors are decomposed as follows:
\begin{enumerate}
\item $(a<2$ or $ f<2) $ and $ ( b>\half $ or $ e>\half )$ \newline
This sector is convergent: $F(a,b,c,d,e,f)$ is integrated directly,
using the simple measure (\ref{mass4pnt2}).
\item $a>2 $  and $f>2$ \newline
The measure  (\ref{mass4pnt2}) does not eliminate the singularity,
when both $a$ and $f$ simultaneously go to $\infty$. The easiest
way to integrate this sector is to use the second 
measure (\ref{secondmeasure}) ff.\ of section 
\ref{all42}.

The divergences of the integrand could also  be 
eliminated by a mapping. This however 
induces new singularities due to
the measure (the term $1/(2\Delta(a,b,c))^2$ in $T$, equation (\ref{int4pnt1})). This would not be the case, if we had not been 
forced to use the trick of integrating by parts the measure. 
\item $b<\half$ and $e<\half$ \newline
In this sector the mapping can be used successfully:
$a$ has to be exchanged with $b$ and $e$ with $f$. We get 
$T(b,a,c,d,f,e)$ with $ a<\half $ and $ f<\half$. This is integrated
using the 
measure (\ref{secondmass}).
\end{enumerate}

\subsection{Numerical calculations}
The numerical calculations are difficult. We refer the interested 
reader to the discussion in section \ref{s:C1num} and for 
more details to  \cite{r:WieseDavid95}, 
section 6. Here we only give the result, see figure \ref{f:inte211}.
The extrapolation for $D\to1$ is consistent with the analytic result
2, found in subsection \ref{e:211 D to 1}.

We however want to discuss our estimate of the systematic error.
It is  is a general phenomenon that our algorithm fails 
 to correctly integrate for integrals
over 4 points in the limit $D\to 1$. This is
due to problems in the domain $a_2 \to 0$ and $f_2 \to 0$ and
thus occurs even for integrals which are well-behaved at small and
large distances. 
An example of such a function is
\be \label{e:testF}
	F(a,b,c,d,e,f)=(a^2+b^2+c^2+d^2+e^2+f^2)^{-3D/2}
\ee
which is analytically integrated to give
\be
\frac 1{S_D^2 }
\frac{ \Gamma\left(\frac{D}{2}\right)  }{ \Gamma\left(\frac{3D}{2}\right) } 
\left(\frac \pi 4 \right)^D
\ee
The ratio of the numerically and analytically calculated values is 
displayed in figure \ref{f:cor}. As these factors should be independent of the integral which has to 
be performed, we use them to correct the numerical results.\begin{figure}[h] 
{\tabcolsep5mm
\begin{tabular}[t]{|l|c|} \hline
$D$ & \\ \hline \hline
1.01 &	$0.770 \pm 0.002 $\\ \hline
1.02 &	$0.843 \pm0.002 $\\ \hline
\end{tabular}\hfill
\begin{tabular}[t]{|l|c|} \hline
$D$ & \\ \hline \hline
1.05 &	$0.951 \pm0.002 $\\ \hline
1.10 &	$0.993 \pm0.002$\\ \hline
\end{tabular}\hfill
\begin{tabular}[t]{|l|c|} \hline
$D$ & \\ \hline \hline
1.15 &	$0.999 \pm0.002$\\ \hline
1.20 &	$0.999 \pm0.002$\\ \hline
\end{tabular}}
\caption{Numerical results for the integral of \protect{\eq{e:testF}}}
\label{f:cor}
\end{figure}

\subsection{The limit $D \to 1$} \label{e:211 D to 1}
\begin{figure}[thb]
$$
\vbox{\epsfbox{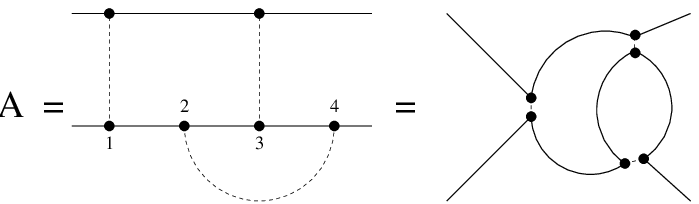}}
$$
$$
\vbox{\epsfbox{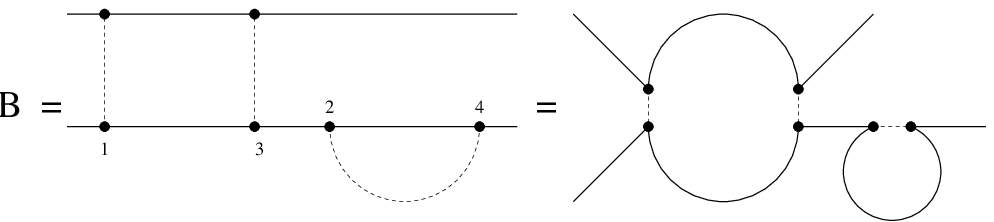}}
$$
$$
\vbox{\epsfbox{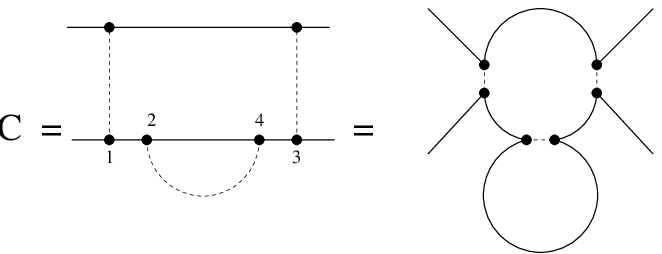}}
$$
$$
\vbox{\epsfbox{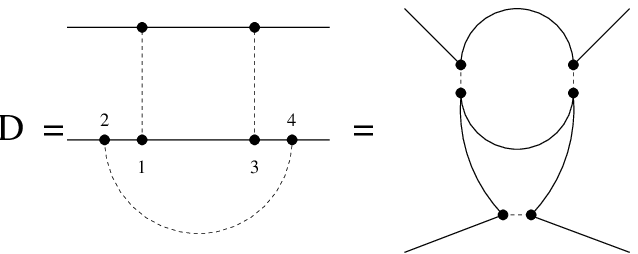}}
$$
\caption{The 4 topological inequivalent classes A, B, C and D and
the equivalent diagrams in the $n\to 0$ limit of scalar $\varphi^4$ theory.}
\label{f:ABCD}
\end{figure}

In the limit $D\to 1$, (\ref{e:M211}) can again
be calculated analytically.
This calculation is interesting as it reveals the connection to 
standard polymer theory and the fact that $\bigg<\GP \bigg| \GB \bigg>_L$
decomposes into three topologically different and non-equiv\-a\-lent 
diagrams.
As in equation (\ref{e:M211}) we keep $c=1$ fixed.
By a direct calculation it can be verified that the measure indeed reduces
to an integral over a line. On this line two points,
the endpoints of $c$, are already fixed. Then there are 12 different 
possibilities to distribute the last two points. They still
can be separated into four topological inequivalent classes A, B, C and D,
cf. figure \ref{f:ABCD}.
These are the
four standard diagrams arising in polymer theory. In each 
of these classes the line with $c=1$ may be chosen to be the line connecting
(12), (14), (23) or (34). Readers more familiar with Feynman
diagrams arising in the framework of a scalar field 
theory may recover the three corresponding diagrams after a 
de Gennes transformation
\cite{r:PGG72}.
They contribute to the renormalization
of the $\varphi^4$-interaction at $d=4$ and are represented on the r.h.s..
Diagrams in one class can be mapped onto each other by the now
well-known mapping of sectors. One subtlety however has to 
be taken into account. The marginal counterterm in equation (\ref{e:M211+})
is not invariant by this mapping. In order to  perform the 
integration only over one sector in every class, the symmetrized version 
will be used:
\bea \hspace{-1cm}
F(a,b,c,d,e,f)&=&f^{-1} (af-{\tx\frac14(b-c+e-d)^2})^{-1} -a^{-1}f^{-2} \nn\\
& & -\frac1{16} a^{-2}f^{-3} \left[
\left( \Theta(f<c)+\Theta(f<b)\right) \left( {\vec b\vec f}{b}^{-1} -
\vec c\vec f c^{-1}\right)^2  \right.\nn \\
& & \qquad\qquad\qquad \left.+\left(\Theta(f<d)+\Theta(f<e)\right) \left( \vec e\vec f{e}^{-1} -\vec d\vec f {d}^{-1}\right)^2
\right] 
\eea
The diagrams give:
\begin{eqnarray} 
{\epsfxsize=2cm \parbox{2cm}{\epsfbox{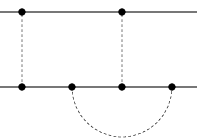}}}
 &=& \frac14 \int_1^\infty \rmd a\, \int_1^\infty \rmd f\, 
 \frac{1}{f(af-1)} -\frac1{af^2} -\frac1{16} \frac1{a^2f} \Theta(a-1-f)  \nonumber \\
  &=& \frac 1 4-\frac1{16} \ln(2) \\
{\epsfxsize=3.05cm \parbox{3.05cm}{\epsfbox{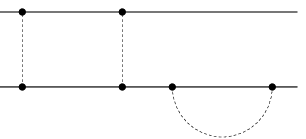}}}
 &=& \frac14 \int_0^\infty \rmd a\, \int_0^\infty \rmd f\, 
 \frac{1}{af^2} -\frac1{af^2} -0  \nonumber \\
  &=& 0 \\
{\epsfxsize=1.78cm \parbox{1.78cm}{\epsfbox{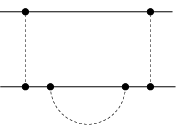}}}
 &=& \frac14 \int_0^\infty \rmd d\, \int_0^\infty \rmd f\, 
\left[ \frac{1}{f^2(1+d)} -\frac1{(1+f+d)f^2} \right. \nn \\
& & \left.\qquad\qquad \qquad \qquad - \frac14 \frac1{(1+f+d)^2f}
	\left( 2+\Theta(f<1)+\Theta(f<d) \right)  \right] \nn \\
&=& \frac14 +\frac18 \ln(2) \\
{\epsfxsize=1.78cm \parbox{1.78cm}{\epsfbox{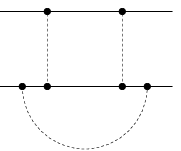}}}
 &=& \frac14 \int_1^\infty \rmd f\, \int_0^{f-1} \rmd a\, 
 \frac{1}{af(f-a)} -\frac1{af^2}  \nonumber \\
  &=& \frac 1 4
\end{eqnarray}
Taking care of the combinatorical factor $4$ for the sectors
A and B and of the factor 2 for C and D, the final result is:
\begin{equation}
4\,\,{\epsfxsize=2cm \parbox{2cm}{\epsfbox{./eps/line4points1ps.eps}}}
\,\,+4\,\,{\epsfxsize=3.05cm \parbox{3.05cm}{\epsfbox{./eps/line4points2ps.eps}}}
\,\,+2\,\,{\epsfxsize=1.78cm \parbox{1.78cm}{\epsfbox{./eps/line4points3ps.eps}}}
\,\,+2\,\,{\epsfxsize=1.78cm \parbox{1.78cm}{\epsfbox{./eps/line4points4ps.eps}}}
\,\,=2
\end{equation}

\subsection{The correction for the unusual marginal counterterm}
\label{The correction for the unusual marginal counterterm}
We recall that $\tilde J(L)$, given in equation \eq{e:tilde J} and
 which we calculated 
numerically, was not exactly the counterterm $J(L)$ but was modified
in order to simplify the calculations. We still have to 
calculate the difference $J(L)-\tilde J(L)$, which also contributes
to the counterterm:
\bea
 \hspace{-0.75cm} J(L)-\tilde J(L) &=& 
\frac{\nu^2 d}2 \int_{c=1; \, a,b,d,e,f,g} \left(a^{2\nu} +g^{2\nu} 
	\right)^{-d/2-1} f^{-\nu(d+2)} \left( \vec b \vec f
	b^{-D} -\vec c \vec f c^{-D} \right)^2 \times
\weiter \qquad\quad \times \left[ 
	\Theta (f<c) -\Theta(f<\max(a,b,c,g)) \right]
	\max(a,b,c,f,g)^{-2\E}
\eea
First of all the integral over $f$ is performed. Since 
\be
\int_{f<l} \left( \vec b \vec f b^{-D} -\vec c \vec f c^{-D} \right)^2
f^{-\nu(d+2)} =\frac 1{D} 
\left( \vec b  b^{-D} -\vec c c^{-D} \right)^2 \frac{l^\E}{\E}
\ee
we get 
\bea
J(L)-\tilde J(L)&=&\frac{\nu^2 d}{2D} \int_{a=1; \, b,c,g} \left(a^{2\nu} +g^{2\nu} 
	\right)^{-d/2-1} \left( \vec b 	b^{-D} -\vec c  c^{-D} \right)^2 \times
\weiter \qquad\qquad \times \frac1{\E}\left[ 
	 c^\E -\max(a,b,c,g)^\E \right]
	\max(a,b,c,f,g)^{-2\E}
\nn \\
&=& \frac{2-D}2 \int_{a=1; \, b,c,g} \left(a^{2\nu} +g^{2\nu} 
	\right)^{-d_c/2-1}  \left( \vec b 
	b^{-D} -\vec c c^{-D} \right)^2 \times
\weiter \qquad \qquad \times \left[\ln(c) -\ln(\max(a,b,c,g)) \right] + {\cal O}(\E) 
\eea
Since $\left( \vec b b^{-D} -\vec c c^{-D} \right)^2$
is symmetric under  the exchange of $\vec b$ and $\vec c$, this can still be 
written as:
\bea
J(L)-\tilde J(L)&=& \frac{2-D}2 \int_{a=1; \, b,c,g} \left(a^{2\nu} +g^{2\nu} 
	\right)^{-d_c/2-1}  \left(  b^{2-2D} +c^{2-2D} +(a^2-b^2-c^2)b^{-D}c^{-D} \right)^2 \times \nn \\
& & \qquad \qquad \times
 \left[\half \ln(bc) -\ln(\max(a,b,c,g)) \right] + {\cal O}(\E) 
\label{e:intecor211}
\eea
\subsubsection*{Numerical integration}
\begin{figure}[tb] 
\centerline{
\epsfxsize=12cm \parbox{12cm}{\epsfbox{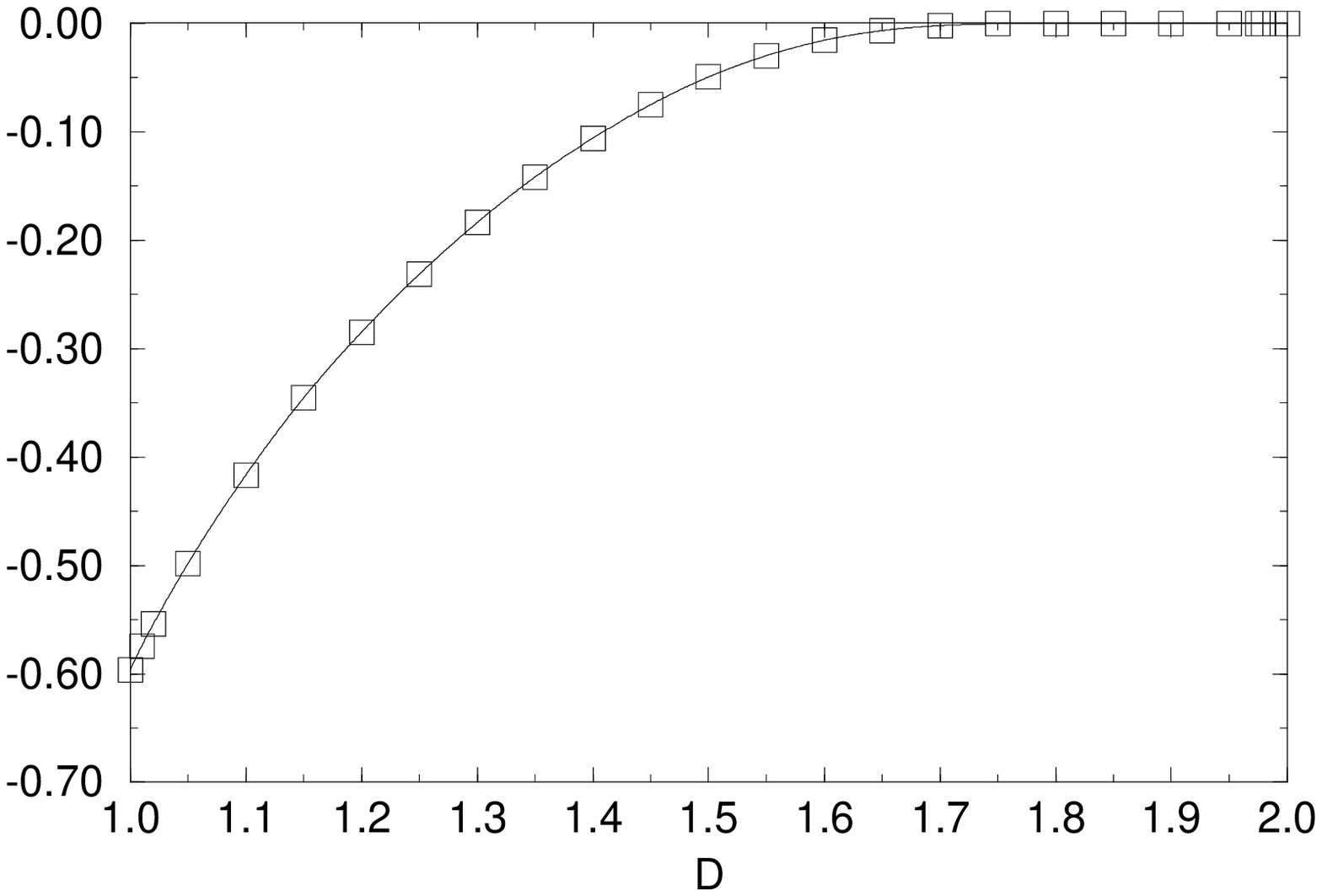}}
}
{\tabcolsep1.2mm
\begin{tabular}[t]{|l|c|} \hline
$D$ & \\ \hline \hline
1.00 &	$-0.596573590$ \\ \hline
1.01 &	$-5.75\times 10^{-1} $\\ \hline
1.02 &	$-5.55\times 10^{-1} $ \\ \hline
1.05 &	$-4.98\times 10^{-1}$ \\ \hline
1.10 &	$-4.16\times 10^{-1} $ \\ \hline
1.15 &	$-3.46\times 10^{-1} $ \\ \hline
\end{tabular}
\hfill
\begin{tabular}[t]{|l|l|} \hline
$D$ & \\ \hline \hline
1.20 &	$-2.85\times 10^{-1} $ \\ \hline
1.25 &	$-2.31\times 10^{-1}$ \\ \hline
1.30 &	$-1.84\times 10^{-1}$ \\ \hline
1.35 &	$-1.42\times 10^{-1}$ \\ \hline
1.40 &	$-1.06\times 10^{-1}$ \\ \hline
1.45 &	$-7.50\times 10^{-2}$ \\ \hline
\end{tabular}
\hfill
\begin{tabular}[t]{|l|r|} \hline
$D$ & \\ \hline \hline
1.50 & 	$-4.96\times 10^{-2}$ \\ \hline
1.55 &	$-2.97\times 10^{-2}$ \\ \hline
1.60 &	$-1.56\times 10^{-2}$ \\ \hline
1.65 &	$-6.64\times 10^{-3}$ \\ \hline
1.70 &	$-2.08\times 10^{-3}$ \\ \hline
1.75 &	$-3.90\times 10^{-4}$ \\ \hline
\end{tabular}
\hfill
\begin{tabular}[t]{|l|c|} \hline
$D$ & \\ \hline \hline
1.80 &	$-2.94\times 10^{-5}$ \\ \hline
1.85 &	$-3.51\times 10^{-7}$ \\ \hline
1.90 &	$-4.05\times 10^{-11}$ \\ \hline
1.95 &	$-4.01\times 10^{-23}$ \\ \hline
1.98 &	$-2.50\times 10^{-59}$ \\ \hline
2.00	 &	0\\ \hline
\end{tabular}}
\caption{Numerical results for equation (\protect{\ref{e:intecor211}}). 
The error is $\pm 10^{-3}$ relative. }
\label{f:intecor211}
\end{figure}The domain of integration has to be split into the two 
sectors where either $b$ or $c$ is the smallest distance. 
These two sectors are equivalent, so the integral will
be performed over $b<c$ only.
The integral is furthermore split into the integral over the radial
and the angular coordinate. We use the following change of variables:
\bea
b&=& u^{1 \over D-\gamma_b} (1-u)^{ { 1\over D-\omega_b }} 
\label{e:m1}
\\
\omega_b &=& \frac32D < 2D \\
\gamma_b &=& 2D-2 \\
\beta & = & \left\{ \begin{array}{lcl} \displaystyle   		\frac{\pi}{2} (2\alpha)^{\textstyle \frac1{D-1}} &\qquad& 
			\alpha \le \displaystyle 0.5 \\
		\displaystyle
		\pi-\frac{\pi}{2} (2-2\alpha)^{\textstyle \frac1{D-1}} &\qquad& 
			\displaystyle\alpha > 0.5
		\end{array} \right. \\
c&=& \sqrt{b^2+1-2b \cos(\beta)}
\label{e:coupcomp}
\eea
This gives 
\be \label{e:m2}
\int_b = \frac{\pi}{D-1} \frac{S_{D-1}}{S_D} \int_0^1 \rmd \alpha \,
	\min(2\alpha,2-2\alpha)^{2-D \over D-1} \sin(\beta)^{D-2} 
\int_0^1 \rmd u\, \left( \frac{1}{D-\gamma_b} \frac1u+
\frac{1}{\omega_b-D} \frac1{1-u}\right) b^D
\ee
The integral over $g$ is independently parametrized as:
\bea
g&=& v^{1 \over D-\gamma_g} (1-v)^{ 1\over D-\omega_g } \\
\omega_g &=& 1+D < 2+D\\
\gamma_g &=& 0
\eea
This implies 
\be
\int_g = \int_0^1 \rmd v \, \left( \frac{1}{D-\gamma_g} \frac1v+
\frac{1}{\omega_g-D} \frac1{1-v}\right)  g^D
\ee
The results of the numerical calculations are given in 
figure \ref{f:intecor211}.
\subsubsection*{The limit $D\to 1$}
For $D=1$ the integral can again be performed analytically. We
get:
\bea
&&\hspace{-1cm}
\int_0^{\infty} \rmd g \, 
\int_0^1 \rmd b \, (1+g)^{-3} \left(\ln(b) -\ln(\max(1,g)) \right)
\nn\\
&&= \int_0^{\infty} \rmd g \, 
 (1+g)^{-3} 
\int_0^1 \rmd b \, \ln(b) -\int_1^{\infty} \rmd g \, 
 (1+g)^{-3} \ln(g)
\nn \\
&& = -\frac14 -\frac12 \ln(2) = -0.5965735903
\eea

\section{Complementary Contribution for the Renormalization of 
the Coupling Constant}
We still have to calculate ${\cal C}_3$, equation \eq{e:C3}: 
\be \label{e:C3 MOPE}
{\cal C}_3 = \Bigg< \GH \Bigg| \GO \Bigg>_{\E^{-1}}  
\left( \Bigg< \GX \Bigg| \GB \Bigg>_L 
	+\half \nu d  \Bigg< \GM \Bigg| \GB \Bigg>_{\E^{-1}} \right)
\ee
The MOPE-coefficient is:
\be
\Bigg( \GX \Bigg| \GB \Bigg)
=-\nu^2 d \left( \vec b b^{-D} -\vec c c^{-D} 
\right)^2 \left( a^{2\nu} +g^{2\nu} \right)^{-d/2-1}
\ee
The non-trivial diagram in \eq{e:C3 MOPE} is:
\be
 \Bigg< \GX \Bigg| \GB \Bigg>_L = \int_{a,b,c,g<L}
\Bigg( \GX \Bigg| \GB \Bigg)
\ee
Apply $L\partial /\partial L$ and map onto $a=L=1$:
\bea
 \Bigg< \GX \Bigg| \GB \Bigg>_L &=& \frac1{\E} \int_{a=L \atop b,c,g}
\Bigg( \GX \Bigg| \GB \Bigg) \max(a,b,c,g)^{-\E} \nn\\
&=& -\frac{\nu d}{\E}\int_{a=L \atop b,c,g} \frac{2-D}{2}
\left( \vec b b^{-D} -\vec c c^{-D} 
\right)^2 \left( a^{2\nu} +g^{2\nu} \right)^{-d/2-1}\times
\weiter \qquad\qquad\qquad
\times \max(a,b,c,g)^{-\E}
\label{e: etoile}
\eea
The second term in \eq{e:C3 MOPE} shall subtract exactly the pole
term of \eq{e: etoile} divided by $d$, {\em not} the pole of 
\eq{e: etoile}, which contains a factor $d=d_c+{\cal O}(\E)$. 
To verify this, remark that
\be
	\int_{a=\mbox{\scriptsize fixed},\, b,c} \frac{2-D}2 
\left( \vec b b^{-D} -\vec c c^{-D} \right)^2 =a^{2-D} \ ,
\ee
which is proven by partial integration. 
This yields:
\bea
&&\hspace{-1cm}\frac1{\E} \int_{a=L \atop b,c,g} \frac{2-D}{2}
\left( \vec b b^{-D} -\vec c c^{-D} 
\right)^2 \left( a^{2\nu} +g^{2\nu} \right)^{-d_c/2-1} 
\nn \\
&&= \frac1{\E} \int_{a=L \atop b,c,g}a^{2\nu} \left( a^{2\nu} +g^{2\nu} \right)^{-d_c/2-1} 
\nn \\
&&= \frac1{\E} \int_{a=L \atop b,c,g}\half\left( a^{2\nu}+ g^{2\nu} \right)\left( a^{2\nu} +g^{2\nu} \right)^{-d_c/2-1}  \ ,
\eea
where in the last step we used the invariance by conformal mapping.
This is equivalent to
\be
	\half \Bigg< \GM \Bigg| \GB \Bigg>_{\E^{-1}} \ ,
\ee
which proves the desired result. We therefore can write:
\bea
&&\hspace{-1cm} \Bigg< \GX \Bigg| \GB \Bigg>_L 
	+\half \nu d  \Bigg< \GM \Bigg| \GB \Bigg>_{\E^{-1}}
\weiter
=-\frac{\nu d}{\E}\int_{a=L \atop b,c,g} \frac{2-D}{2}
\left( \vec b b^{-D} -\vec c c^{-D} 
\right)^2 \times
\weiter
\qquad\qquad\qquad\times
\left\{
\left( a^{2\nu} +g^{2\nu} \right)^{-d/2-1}
 \max(a,b,c,g)^{-\E}
-\left( a^{2\nu} +g^{2\nu} \right)^{-d_c/2-1}
\right\}
\weiter
= (-\nu d) \half \int _{a=1; \, b,c,g} \left( \vec b b^{-D} -\vec c c^{-D} 
\right)^2 \left( a^{2\nu} +g^{2\nu} \right)^{-d_c/2-1} \times
\weiter \qquad \qquad \qquad \times \left[ \ln(a^{2\nu}+g^{2\nu}) -(2-D) \ln(\max(a,b,c,g))  
\right] +{\cal O}(\E) \label{e:coulingfinite1}
\eea

\begin{figure}[tb] 
\centerline{
\epsfxsize=12cm \parbox{12cm}{\epsfbox{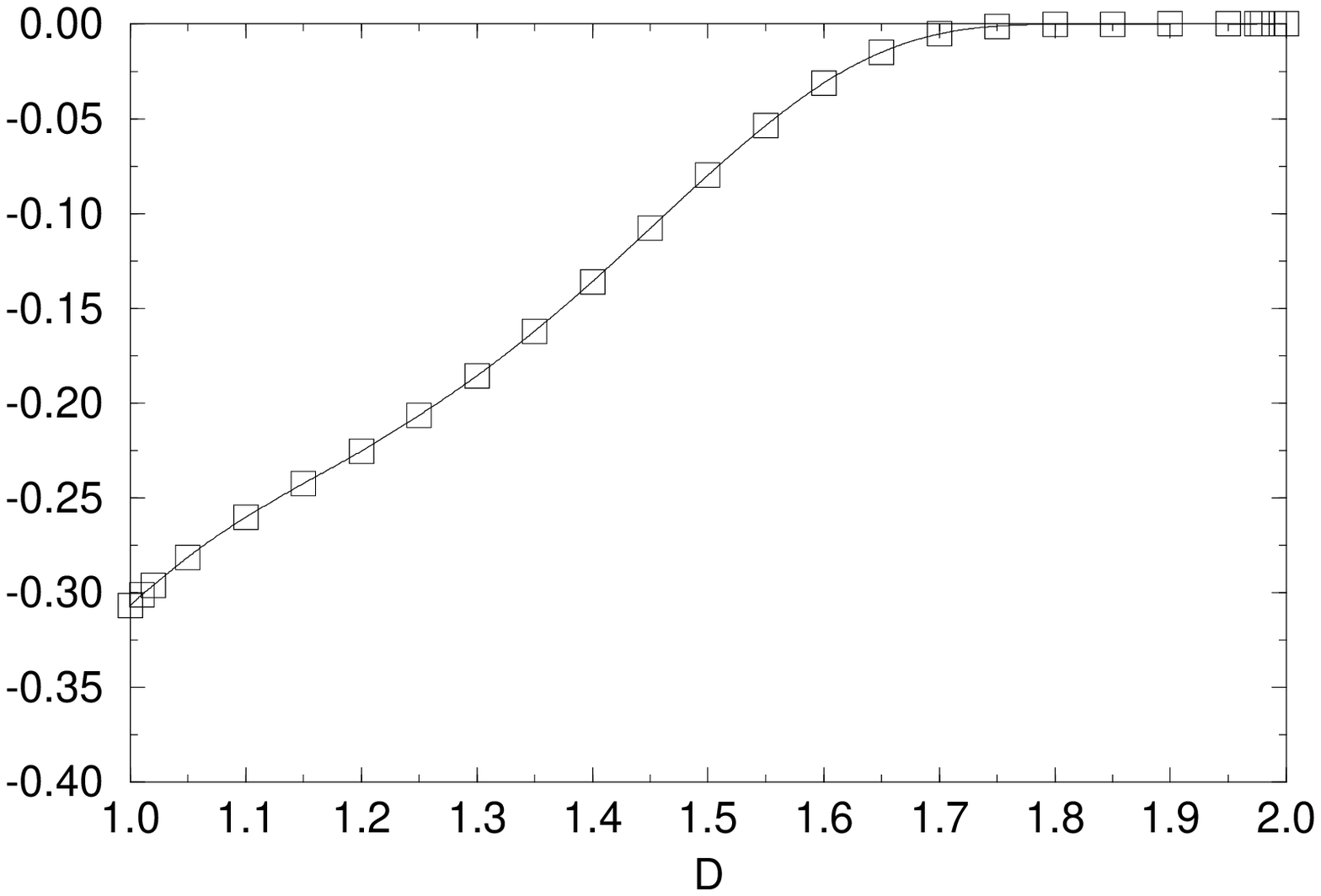}}
}
{\tabcolsep2.3mm
\begin{tabular}[t]{|l|c|} \hline
$D$ & \\ \hline \hline
1.00 &	$-0.306853$ \\ \hline
1.01 &	$-0.301$\\ \hline
1.02 &	$-0.296$ \\ \hline
1.05 &	$-0.282$ \\ \hline
1.10 &	$-0.260$ \\ \hline
1.15 &	$-0.244$ \\ \hline
\end{tabular}
\hfill
\begin{tabular}[t]{|l|l|} \hline
$D$ & \\ \hline \hline
1.20 &	$-0.225$ \\ \hline
1.25 &	$-0.206$ \\ \hline
1.30 &	$-0.185$ \\ \hline
1.35 &	$-0.162$ \\ \hline
1.40 &	$-0.136$ \\ \hline
1.45 &	$-0.108$ \\ \hline
\end{tabular}
\hfill
\begin{tabular}[t]{|l|r|} \hline
$D$ & \\ \hline \hline
1.50 & 	$-7.97 \times 10^{-2}$ \\ \hline
1.55 &	$-5.33 \times 10^{-2}$ \\ \hline
1.60 &	$-3.11 \times 10^{-2}$ \\ \hline
1.65 &	$-1.49 \times 10^{-2}$ \\ \hline
1.70 &	$-5.24 \times 10^{-3}$ \\ \hline
1.75 &	$-1.13 \times 10^{-3}$ \\ \hline
\end{tabular}
\hfill
\begin{tabular}[t]{|l|c|} \hline
$D$ & \\ \hline \hline
1.80 &	$-9.97 \times 10^{-5}$ \\ \hline
1.85 &	$-1.48 \times 10^{-6}$ \\ \hline
1.90 &	$-2.31 \times 10^{-10}$ \\ \hline
1.95 &	$-3.89 \times 10^{-22}$ \\ \hline
1.975 &	$-4.88 \times 10^{-46}$ \\ \hline
2.00	 &	0\\ \hline
\end{tabular}}
\caption{Numerical results for equation (\protect{\ref{e:coulingfinite1}}).
The statistical error is $10^{-2}$.}
\label{f:coupcomp}
\end{figure}
The method to numerically integrate \eq{e:coulingfinite1} is the 
same as in section \ref{The correction for the unusual marginal counterterm}.
We give the results in figure \ref{f:coupcomp}.

In the limit $D\to 1$ 
(\ref{e:coulingfinite1}) reduces to 
\be
 -2\int_0^{\infty}\rmd g\, \frac{\ln(1+g)}{(1+g)^{3}} +2
\int_1^{\infty}  \rmd g\, \frac{\ln(g)}{(1+g)^3} = \ln(2)-1
= -0.306852819
\ee

\section{Renormalization of the Wavefunction, Main Contribution}
\subsection{Derivation of an analytic expression}
\begin{figure}[htb] 
\centerline{
    \epsfxsize=10.0cm \epsfbox{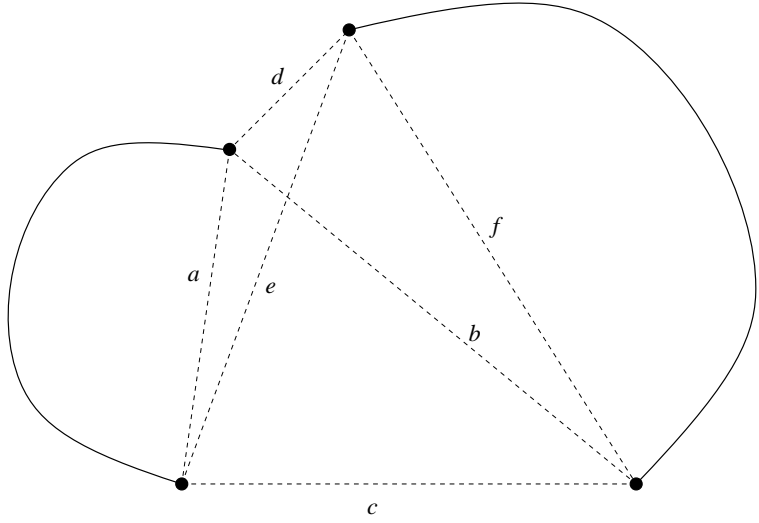}
}
\caption{The distances in (\protect{\ref{e:F3 star}}) }
\label{f:wave1}
\end{figure}
In this section, the calculation of ${\cal F_1}$, defined 
in equation \eq{e:F1}, will be discussed. We write:
\bea
I(L)=2 {\cal F}_1=
 \bigg< \GI \bigg| \GO \bigg>_L &-&
\bigg< \GH \bigg| \GO \bigg>_L	 
\bigg< \GZ \bigg| \GO \bigg>_L \nn \\
& -& \bigg< \GM \bigg| \GB \bigg>_L 
\bigg< \GH \bigg| \GO \bigg>_L 
\label{e:F3 star}
\eea
Let us discuss the MOPE-coefficients involved. Explicit expressions
are given later. The first is
\be	\label{e:F1 first}
	\bigg( \GI \bigg| \GO \bigg)
\ee
\eq{e:F1 first} has two different types of subdivergences, which 
are subtracted by the two counter\-terms in \eq{e:F3 star}. Let us 
symbolically write down the factorization of the MOPE-coefficients. 
If one of the dipoles is contracted first, the MOPE-coefficient 
factorizes as:
\be
	\bigg( \GI \bigg| \GO \bigg)=  \bigg( \GH \bigg| \GN \bigg) 
\bigg( \GH \bigg| \GO \bigg) +\bigg( \GH \bigg| \GOalphabeta \bigg)	 
\bigg( \GZalphabeta \bigg| \GO \bigg) + \ldots  
\ee
The first term is a relevant counterterm, which we did not
mention explicitly in \eq{e:F3 star}. The second subtracts 
the marginal subdivergence. 
Note that we have again to take care of the tensorial structure 
of the factorization. 

If the two dipoles are contracted to a single dipole first, the 
factorization is:
\be
	\bigg( \GI \bigg| \GO \bigg)= 
		\bigg( \GM \bigg| \GB \bigg) \bigg( \GH \bigg| \GO \bigg) 
\ + \ \ldots
\label{e: CT D}
\ee
Note that any of these contractions is obtained with a combinatorial 
factor 2.

We now give a list of the MOPE-coefficients together with the 
$\Theta$-functions, which restrict each counterterm to the 
sector in which the divergence appears and which by integration 
delivers the factor $1/2$ from the nested integration, 
cf.\ section~\ref{s:C2 der ana exp}. First of all
\be
\bigg( \GI \bigg| \GO \bigg)=A
\ee
with 
\bea
A&=&-\frac{1}{2D} 
\bigg[ f^{2 \nu}a^2 +a^{2 \nu}f^2 
+\half(e^{2}+b^{2}-c^2-d^2) (c^{2\nu}-e^{2 \nu} -b^{2 \nu}
	+d^{2 \nu})
\bigg] \times \nonumber \\
 & & \qquad \qquad \times\left[ a^{2 \nu} f^{2 \nu} -\frac{1}{4} \left( c^{2\nu}-e^{2 \nu} -b^{2 \nu}
	+d^{2 \nu} \right)^2 \right]^{-d/2-1}
\eea
The notation for the distances follows  figure \ref{f:wave1}. 
The two relevant counterterms which appear, when  $a$
or  $f$ is contracted first and which can symbolically be
written as 
\be
\bigg( \GH \bigg| \GN \bigg) \bigg( \GH \bigg| \GO \bigg)  + 
\bigg( \GH \bigg| \GO \bigg) \bigg( \GH \bigg| \GN \bigg) = B
\ee
are
\bea 
B&=&-\frac{1}{2D} \left(a^{-\nu d} f^{D-\nu d} + f^{-\nu d} a^{D-\nu d} \right)
\ . 
\eea
The marginal counterterm 
\be
\bigg( \GH \bigg| \GOalphabeta \bigg)	 
\bigg( \GZalphabeta \bigg| \GO \bigg) = C_1,\ C_2,\ \tilde C_1 \ \mbox{or}\  
\tilde C_2
\ee
appears when either $f$ is contracted first
\bea
C_1&=&\left[ \frac{2-D}{8D}\left( b^2+e^2-c^2-d^2\right) 
\left( (f^2+b^2-d^2) b^{-D} -(f^2+c^2-e^2) c^{-D} \right) 
a^{-\nu (d+2)} f^ {-\nu (d+2)} \right.\hspace{-3mm}\nonumber \\
& & \left. -\frac{d+2}{64D} (2-D)^2 \left( (f^2+b^2-d^2) b^{-D} -(f^2+c^2-e^2) c^{-D} \right)^2 a^{D-\nu (d+2)} f^ {-\nu (d+2)} 
\right] \times \nonumber \\
& & \qquad \qquad \times\, \Theta(f<\max(a,b,c)) \\
\tilde C_1&=&\left[ \frac{2-D}{8D}\left( b^2+e^2-c^2-d^2\right) 
\left( (f^2+e^2-c^2) e^{-D} -(f^2+d^2-b^2) d^{-D} \right) 
a^{-\nu (d+2)} f^ {-\nu (d+2)} \right. \hspace{-3mm} \nonumber \\
& & \left. -\frac{d+2}{64D} (2-D)^2 \left( (f^2+e^2-c^2) e^{-D} -(f^2+d^2-b^2) d^{-D} \right)^2 a^{D-\nu (d+2)} f^ {-\nu (d+2)} 
\right] \times \nonumber \\
& & \qquad \qquad \times \, \Theta(f<\max(a,d,e))
\eea
or when $a$ is contracted first:
\bea
C_2 &=&\left[ \frac{2-D}{8D}\left( b^2+e^2-c^2-d^2\right) 
\left( (a^2+b^2-c^2) b^{-D} -(a^2+d^2-e^2) d^{-D} \right) 
a^{-\nu (d+2)} f^ {-\nu (d+2)} \right. \hspace{-3mm} \nonumber \\
& & \left. -\frac{d+2}{64D} (2-D)^2 \left( (a^2+b^2-c^2) b^{-D} -(a^2+d^2-e^2) d^{-D} \right)^2 f^{D-\nu (d+2)} a^{-\nu (d+2)} 
\right]  \times \nonumber \\
& & \qquad \qquad \times \, \Theta(a<\max(b,d,f)) \\
\tilde C_2 &=&\left[ \frac{2-D}{8D}\left( b^2+e^2-c^2-d^2\right) 
\left( (a^2+e^2-d^2) e^{-D} -(a^2+c^2-b^2) c^{-D} \right) 
a^{-\nu (d+2)} f^ {-\nu (d+2)} \right. \hspace{-3mm} \nonumber \\
& & \left. -\frac{d+2}{64D} (2-D)^2 \left( (a^2+e^2-d^2) e^{-D} -(a^2+c^2-b^2) c^{-D} \right)^2 f^{D-\nu (d+2)} a^{-\nu (d+2)} 
\right]  \times \nonumber \\
& & \qquad \qquad \times\,\Theta(a<\max(c,e,f))
\eea
Two equivalent versions are given as one e.g.\ may put the counterterm
for $f\to 0$ on either endpoint of the distance $f$.

The last class of counterterms, equation \eq{e: CT D},
appears when either
$(c,d) \to 0$ or $(b,e) \to 0$. 
For the first contraction, the two equivalent versions are:
\bea
D_1&=&-\frac1{2D} a^{D-\nu d} \left(c^{2\nu}+d^{2\nu} \right)^{-d/2}
\Theta(d<a) \Theta(c<a) \nn \\
\tilde D_1&=&-\frac1{2D} f^{D-\nu d} \left(c^{2\nu}+d^{2\nu} \right)^{-d/2}
\Theta(d<f) \Theta(c<f) 
\eea
For the second contraction they are:
\bea
D_2&=& -\frac1{2D} a^{D-\nu d} \left(e^{2\nu}+b^{2\nu} \right)^{-d/2}
\Theta(b<a) \Theta(e<a)  \nn \\
\tilde D_2&=& -\frac1{2D} f^{D-\nu d} \left(e^{2\nu}+b^{2\nu} \right)^{-d/2}
\Theta(b<f) \Theta(e<f)
\eea
We now define the integrand which has to be taken in \eq{e:F3 star}
\be
F(a,b,c,d,e,f):=A-B-\half\left( C_1+\tilde C_1 +C_2+\tilde C_2+ D_1
+\tilde D_1 +D_2 +\tilde D_2\right)
\ee
We used the symmetric version of the counterterms, but could
have also taken $\lambda C_1+(1-\lambda) \tilde C_1$ instead of
$\half(C_1+\tilde C_1)$.  We will use this freedom later in order
to simplify the calculations. 
As usual this expression has to be integrated
over all distances restricted to
be smaller than $L$. Applying $L\partial/\partial L$ and mapping
onto $c=L=1$ results in 
\be \label{e:Jeps}
J(L)=L\frac{\partial}{\partial L} I(L)=
\int_{c=1; \,a,b,d,e,f} F(a,b,c,d,e,f) \max(a,b,c,d,e,f)^{-2\E}
\ee
The integral $J(L)$ is convergent as can be seen from a somehow
tedious generalized Taylor expansion for the domains of possible
divergences ($a\to0$; $f\to 0$; $(c,d) \to 0$, or as $c$ is fixed for
all the other distances to $\infty$; $(b,e) \to 0$).
So the term $\max(a,b,c,d,e,f)^{-2\E}$ can be dropped and the limit
$\E \to 0$, $d\to d_c$ can be performed.

We furthermore have checked that the subtracted terms $C_1$, 
$\tilde C_1$, $C_2$, $\tilde C_2$, 
$D_1$ and $D_2$ are up to subdominant contributions 
in $\E$ equivalent to the terms given in equation
(\ref{e:F3 star}), i.e. that the changes in the domain of integration
do not matter (the so-called ``two-loop miracle'').
This seems to be familiar by now, but nevertheless
has to be checked.

\subsection{Analytic continuation of the measure}
Now $F(a,b,c,d,e,f)$ will be integrated numerically.
The main problem is again, that the measure already used in 
section \ref{Coupling constant renormalization, second graph}
 and defined in \eq{all43} has to be improved 
by partial integration. As in 
subsection~\ref{s:Improvement of the measure} we have
to calculate the partial derivative applied to $A$, $B$, $C_1$, 
$\tilde C_1$, $C_2$, $\tilde C_2$, 
$D_1$ and $D_2$, i.e.\ $\mbox{\bf R}$, given by
equation \eq{mass4pnt1}  applied to
these terms. Hereby $\mbox{\bf R}$ will also act on the 
$\Theta$-functions yielding terms proportional to the $\delta$-distribution.
We will first discuss some ingenious methods to analytically
calculate these ``non-diagonal'' terms, which work well for
$D_1$, $D_2$ and $C_1$, but fail for $C_2$ or equivalently 
$\tilde C_2$. 
We then discuss another method to analytically continue the integral,
which cannot be used for the numerics neither.

The problem is finally solved by brute force:
As in section~\ref{Coupling constant renormalization, second graph}
we change the prescription for the sectors 
so that the corresponding $\Theta$-functions commute with
$\mbox{\bf R}$. The corrections are calculated numerically. 

To analyze the problem, we first write down $\mbox{\bf R}$ applied
to each term. We study especially the contributions proportional to 
the $\delta$-distributions, which cannot be calculated numerically.
In order to simplify these formulas,
we use the identity $\nu d_c=2D$. 
\bea
\mbox{\bf R} A &=& -\frac{1}{2D} \left[(2-D)a^2f^{-D} +2 a^{2\nu} 
+\frac{2-D}2 \left(e^2+b^2-c^2-d^2\right) \left(d^{-D}-e^{-D}\right)\right] \times
\nonumber\\
& & \qquad \qquad\times\left[ a^{2 \nu} f^{2 \nu} -\frac{1}{4} \left( c^{2\nu}-e^{2 \nu} -b^{2 \nu}
	+d^{2 \nu} \right)^2 \right]^{-d_c/2-1} \nonumber \\
& & +\frac{D+2}{2D}
\bigg[ f^{2 \nu}a^2 +a^{2 \nu}f^2 
+\half(e^{2}+b^{2}-c^2-d^2) \left(c^{2\nu}-e^{2 \nu} -b^{2 \nu}
	+d^{2 \nu}\right)
\bigg] \times\nonumber \\
& & \qquad \qquad \times
\left[a^{2\nu} f^{-D} -\frac1{2}\left( c^{2\nu}-e^{2 \nu} -b^{2 \nu}
	+d^{2 \nu} \right) \left(d^{-D}-e^{-D}\right)
\right] \times \nonumber \\
& & \qquad \qquad \times
\left[ a^{2 \nu} f^{2 \nu} -\frac{1}{4} \left( c^{2\nu}-e^{2 \nu} -b^{2 \nu}
	+d^{2 \nu} \right)^2 \right]^{-d_c/2-2} \\
\mbox{\bf R} B &=& \half a^{-2D} f^{-D-2} +a^{-D} f^{-2D-2} \\
\mbox{\bf R} C_1 &=& \left[-\frac{(2-D)(2+D)}{8D}\left( b^2+e^2-c^2-d^2\right) 
\left( (f^2+b^2-d^2) b^{-D} -(f^2+c^2-e^2) c^{-D} \right)  \times \nonumber \right. \\
& & \qquad \qquad \times\, a^{-D-2} f^{-D-4} \nonumber \\
& & \left. \ +\frac{(2-D)(2+D)^2}{32 D} \left( (f^2+b^2-d^2) b^{-D} -(f^2+c^2-e^2) c^{-D} \right)^2
a^{-2} f^{-D-4} \right] \times \nonumber \\
& & \qquad \qquad \times \, \Theta(f<\max(a,b,c)) \nonumber \\
& & -\left[ \frac{2-D}{8D}\left( b^2+e^2-c^2-d^2\right) 
\left( (f^2+b^2-d^2) b^{-D} -(f^2+c^2-e^2) c^{-D} \right) 
a^{-D-2} f^ {-D-2} \right. \hspace{-1cm} \nonumber \\
& & \quad \ \left. -\frac{(2-D)(2+D)}{32D} \left( (f^2+b^2-d^2) b^{-D} -(f^2+c^2-e^2) c^{-D} \right)^2 a^{-2} f^ {-D-2} 
\right] \times \nonumber \\
& & \qquad \qquad \times \, \frac1f \left[ 
\delta(f-a) \Theta(f>b)\Theta(f>c) + \right .\nonumber \\
& & \left. \hspace{3cm} \delta(f-b) \Theta(f>a)\Theta(f>c)+
\delta(f-c) \Theta(f>a)\Theta(f>b) \right]  \label{e:R C1} \\
\mbox{\bf R} \tilde C_1 &=& \left[ -\frac{2-D}{8D} \left( b^2+e^2-c^2-d^2\right)
 \right. a^{-D-2} f^{-D-2}\times \nonumber \\
& &  \qquad \qquad \times \left( D(f^2+e^2-c^2)e^{-D-2} -D(f^2+d^2-b^2) d^{-D-2} 
	-4e^{-D}+4d^{-D}
\right)  \nonumber \\ 
& & \ -\frac{(2-D)(2+D)}{8D}\left( b^2+e^2-c^2-d^2\right) 
\left( (f^2+e^2-c^2) e^{-D} -(f^2+d^2-b^2) d^{-D} \right)  \times \nonumber \\
& & \qquad \qquad \times\, a^{-D-2} f^{-D-4} \nonumber \\
& & \  +\frac{(2-D)(2+D)}{16D} 
	\left( (f^2+e^2-c^2) e^{-D} -(f^2+d^2-b^2) d^{-D} \right)
 a^{-2} f^{-D-2}  \times\nonumber \\
& & \qquad \qquad \times
\left(D (f^2+e^2-c^2) e^{-D-2} -D(f^2+d^2-b^2) d^{-D-2} -4e^{-D}+4d^{-D}
\right) \nonumber \\
& & \left. \ +\frac{(2-D)(2+D)^2}{32 D} \left( (f^2+e^2-c^2) e^{-D} -(f^2+d^2-b^2) d^{-D} \right)^2
a^{-2} f^{-D-4} \right] \times \nonumber \\
& & \qquad \qquad \times \, \Theta(f<\max(a,d,e)) \nonumber \\
& & -\left[ \frac{2-D}{8D}\left( b^2+e^2-c^2-d^2\right) 
\left( (f^2+e^2-c^2) e^{-D} -(f^2+d^2-b^2) d^{-D} \right) 
a^{-D-2} f^ {-D-2} \right. \hspace{-1cm} \nonumber \\
& & \quad \ \left. -\frac{(2-D)(2+D)}{32D}\left( (f^2+e^2-c^2) e^{-D} -(f^2+d^2-b^2) d^{-D} \right)^2 a^{-2} f^{-D-2} 
\right] \times \nonumber \\
& &\qquad \qquad \times \frac1f\delta(a-f)\Theta(f>d)\Theta(f>e)  \\
\mbox{\bf R} C_2 &=& \left[ \frac{2-D}{8} \left( b^2+e^2-c^2-d^2\right)
 \right. (a^2+d^2-e^2)d^{-D-2} a^{-D-2} f^{-D-2} \nonumber \\ 
& & \ -\frac{(2-D)(2+D)}{8D}\left( b^2+e^2-c^2-d^2\right) 
\left( (a^2+b^2-c^2)b^{-D} -(a^2+d^2-e^2) d^{-D}  \right)  \times \nonumber \\
& & \qquad \qquad \times \, a^{-D-2} f^{-D-4} \nonumber \\
& & \  -\frac{(2-D)(2+D)}{16} \left( (a^2+b^2-c^2)b^{-D} -(a^2+d^2-e^2) d^{-D}  \right)
 a^{-D-2} f^{-2}  \times\nonumber \\
& & \qquad \qquad \times
(a^2+d^2-e^2) d^{-D-2} \nonumber \\
& & \left. \ +\frac{(2-D)(2+D)}{16D} \left( (a^2+b^2-c^2)b^{-D} -(a^2+d^2-e^2) d^{-D}  \right)^2
a^{-D-2} f^{-4} \right] \times \nonumber \\
& & \qquad \qquad \times \Theta(a<\max(b,d,f)) \nonumber \\
& & + \left[ \frac{2-D}{8D}\left( b^2+e^2-c^2-d^2\right) 
\left( (a^2+b^2-c^2) b^{-D} -(a^2+d^2-e^2) d^{-D} \right) 
a^{-D-2} f^ {-D-2} \right. \nonumber \\
& & \quad \ \left. -\frac{(2-D)(2+D)}{32D} \left( (a^2+b^2-c^2) b^{-D} -(a^2+d^2-e^2) d^{-D} \right)^2 f^{-2} a^{-D-2} 
\right]  \times \nonumber \\
& & \qquad \qquad \times\,\left[ \frac1d \delta(a-d) \Theta(a>b) 
	\Theta(a>f) + \frac1f\delta(a-f) \Theta(a>b) \Theta(a>d) \right] \\
\mbox{\bf R} \tilde C_2 &=& \left[ -\frac{2-D}{8} \left( b^2+e^2-c^2-d^2\right)
 \right. (a^2+e^2-d^2)e^{-D-2} a^{-D-2} f^{-D-2} \nonumber \\ 
& & \ -\frac{(2-D)(2+D)}{8D}\left( b^2+e^2-c^2-d^2\right) 
\left( (a^2+e^2-d^2)e^{-D} -(a^2+c^2-b^2) c^{-D}  \right)  \times \nonumber \\
& & \qquad \qquad \times \, a^{-D-2} f^{-D-4} \nonumber \\
& & \  +\frac{(2-D)(2+D)}{16} \left( (a^2+e^2-d^2)e^{-D} -(a^2+c^2-b^2) c^{-D}  \right)
 a^{-D-2} f^{-2}  \times\nonumber \\
& & \qquad \qquad \times
(a^2+e^2-d^2) e^{-D-2} \nonumber \\
& & \left. \ +\frac{(2-D)(2+D)}{16D} \left( (a^2+e^2-d^2)e^{-D} -(a^2+c^2-b^2) c^{-D}  \right)^2
a^{-D-2} f^{-4} \right] \times \nonumber \\
& & \qquad \qquad \times \Theta(a<\max(c,e,f)) \nonumber \\
& &+\left[ \frac{2-D}{8D}\left( b^2+e^2-c^2-d^2\right) 
 \left( (a^2+e^2-d^2)e^{-D} -(a^2+c^2-b^2) c^{-D}  \right) 
a^{-D-2} f^ {-D-2} \right. \hspace{-1cm} \nonumber \\
& & \quad \ \left. -\frac{(2-D)(2+D)}{32D} \left( (a^2+e^2-d^2)e^{-D} -(a^2+c^2-b^2) c^{-D}  \right)^2 a^{-D-2} f^{-2} 
\right] \times \nonumber \\
& & \qquad \qquad \times\,\left[ \frac1e \delta(a-e) \Theta(a>c) 
	\Theta(a>f) + \frac1f\delta(a-f) \Theta(a>c) \Theta(a>e) \right] 
	\hspace{-1cm} \\
\mbox{\bf R} D_1 &=& a^{-D} (c^{2\nu}+d^{2\nu})^{-d_c/2-1} d^{-D} 
\Theta (d<a) \Theta(c<a) \nonumber\\
& &  +\frac{1}{2D} a^{-D} (c^{2\nu}+d^{2\nu})^{-d_c/2} \frac{1}{d} 
\delta (d-a) \Theta(c<a) \label{e:wave3} \\
\mbox{\bf R} \tilde D_1 &=& \left( 
\half f^{-D-2}(c^{2\nu}+d^{2\nu})^{-d_c/2-1}
+f^{-D} (c^{2\nu}+d^{2\nu})^{-d_c/2-1} d^{-D} \right)
\Theta (c<f) \Theta(d<f) \nonumber\\
& &  -\frac{1}{2D} f^{-D} (c^{2\nu}+d^{2\nu})^{-d_c/2} \frac{1}{f} 
\delta (c-f) \Theta(d<f) \label{e:wave3'} \\
\mbox{\bf R} D_2 &=& a^{-D} (e^{2\nu}+b^{2\nu})^{-d_c/2-1} e^{-D} 
\Theta (b<a) \Theta(e<a) \nonumber\\
& &  +\frac{1}{2D} a^{-D} (e^{2\nu}+b^{2\nu})^{-d_c/2} \frac{1}{e} 
\delta (e-a) \Theta(b<a) \label{e:wave4} \nn \\
\mbox{\bf R} \tilde D_2 &=& 
\left( \half f^{-D-2} (e^{2\nu}+b^{2\nu})^{-d_c/2}
	+f^{-D} (e^{2\nu}+b^{2\nu})^{-d_c/2-1} e^{-D} \right) 
\Theta (b<f) \Theta(e<f) \nonumber\\
& &  -\frac{1}{2D} f^{-D} (e^{2\nu}+b^{2\nu})^{-d_c/2} \frac{1}{f} 
\delta (e-a) \Theta(b<a) \label{e:wave4'}
\eea
First we show that in this parametrization the terms proportional to the $\delta$-distribution
in (\ref{e:wave3}) and (\ref{e:wave4}) can be calculated analytically.
We demonstrate that for (\ref{e:wave3}).

We use the standard measure of equation \eq{all43}. 
The vectors over which the integration will be performed are $a$ and 
$d$.
The integration over $d_3$ has been improved by partial integration.
Let us apply that to the last term in (\ref{e:wave3}), i.e. to 
\be
\frac{1}{2D} a^{-D} (c^{2\nu}+d^{2\nu})^{-d_c/2} \frac{1}{d} 
\delta (d-a) \Theta(c<a) \ .
\ee
The integration over $a$ is standard and can easily be performed, resulting in:
\be
\frac{1}{2D}  (c^{2\nu}+d^{2\nu})^{-d_c/2} \frac{1}{d^2} \Theta(c<d)
\ee
The former integral over $\vec d=(x,y,z,0,\ldots)$ was
\be
\frac{S_{D-2}}{S_D} \int_{-\infty}^{\infty}\rmd x \,  \int_{-\infty}^{\infty}\rmd y \,
 \int_{0}^{\infty}\rmd z \, z^{D-3}\, F(a,b,c,d,e,f)   
\ee
and is changed to
\be 
\frac{1}{2-D} \frac{S_{D-2}}{S_D} \int_{-\infty}^{\infty}\rmd x \,  \int_{-\infty}^{\infty}\rmd y \,
 \int_{0}^{\infty} \rmd z \, z^{D-1} \,\mbox{\bf R} F(a,b,c,d,e,f) \ .  
\ee
Reversing the derivation of the measure we recognize the invariant 
measure in $D+2$ dimensions:
\be 
	\frac1{2-D} \frac{S_{D-2} S_{D+2}}{S_D^2} \int_{d \in \Rscript^{D+2}} =
	-\frac1{D} \int_{d \in \Rscript^{D+2}}
\ee
This can be summarized in the following formula, valid as long 
as the integral over $\vec d$ factorizes from the integral over
$a$, $b$ and $c$:
\be
\int_{d \in \Rscript^{D}} F(a,b,c,d,e,f) = -\frac1{D} \int_{d \in \Rscript^{D+2}}
\mbox{\bf R} F(a,b,c,d,e,f)
\ee
The final result is given by the integral
\be
-\frac{1}{2D^2} \int_1^{\infty} \frac{\rmd d}{d} d^{D} \left(1+d^{2\nu}\right)^{-d_c/2}
=-\frac{1}{4D^2(2-D)} \mbox{B} \left(\frac{D}{2-D},\frac D{2-D} \right) \,
\ee
where $\mbox{B}$ is the standard Beta-function.
It is worth to mention that this is equivalent to 
\be
\frac{1} D \, \half \, \Bigg<\GH \Bigg| \GO \Bigg>_{\E} \Bigg<\GM \Bigg| \GB \Bigg>_{\E}
\ee
The factor $1/D$ is due to the measure, the factor $1/2$ due to the fact that
the {\bf R}-operation is acting on half of the bounded distances only.
The concerned reader will be able to reconstruct the factor $1/D$ from
the identity
\be
\int_d \frac{d_3^2}{d^2} F(d)=\frac 1 D \int_d F(d) \ .
\ee
The analogous term in (\ref{e:wave4}) gives the same result. It nevertheless
is somehow harder to calculate. As a first step a mapping has to be performed,
which exchanges $b$ with $c$ and 
$d$ with $e$. This mapping is possible, although the measure 
is not invariant, as this mapping does not affect the height of the 
tetrahedron.

Equivalently it is possible to calculate the diagonal term of $\mbox{\bf R}C_1$,
equation(\ref{e:R C1}). We find:
\be
\frac1D  \, \Bigg< \GH \Bigg| \GO \Bigg>_\E \Bigg< \GZ \Bigg| \GO \Bigg>_\E
\ee
This time no factor $1/2$ appears, since {\bf R} was acting on all the
distances.

The non-diagonal part of $C_2$ or $\tilde C_2$  {\em cannot}
be calculated by these methods. (The concerned reader is encouraged
to try this himself.) Nor is it possible to arrange the counterterms 
in such a way that the non-diagonal terms cancel. (Warning: The measure
is not invariant by the conformal mapping, as the partial integration
by $f_3$ has been performed.)

Regarding the problems with the {\bf R}-operation one may ask, 
whether it is not better to use an alternative prescription for
the analytic continuation. One might think of replacing
the crucial integral
\be
\int \rmd f_3  f_3^{D-3} F(a,b,c,d,e,f)
\ee
by
\be 
\int \rmd f_3  f_3^{D-3} F(a,b,c,d,e,f) -f_3^{D-3} F(a,b,c,d_p,e_p,f_p)
\ee
where $d_p$, $e_p$ and $f_p$ are the projections of $d$, $e$ and $f$ onto
the plane spanned by $f_1$ and $f_2$ in the standard parametrization 
of the measure in  equation \eq{all43}. 
Here the problem arizes that even if the length of
$\vec f=(f_1,f_2,f_3)$ is large, the length of 
$\vec f_p=(f_1,f_2,0)$ may tend to 0. This is a new divergence, which is 
 integrable,
but not handable by the standard measure in the numerical integration
procedure. Thus this method cannot be used.

\subsection{Numerical integration}
We finally solve the problem by modifying as in 
section~\ref{Coupling constant renormalization, second graph}
the $\Theta$-functions and by calculating the corrections later. 
We integrate  $\mbox{\bf R} \tilde F(a,b,c,d,e,f)$, 
where 
\be
\tilde F=A-B-\tilde C_1 -\tilde C_2 -D_1 -D_2
\ee 
but with modified prescriptions for the sectors: $f<d$ for $\tilde C_1$, $a<c$ for $\tilde C_2$, $c<a$ for 
$D_1$ and $b<a$ for $D_2$.
We therefore have to integrate numerically:
\begin{figure}[tb] 
\centerline{
\epsfxsize=12cm \parbox{12cm}{\epsfbox{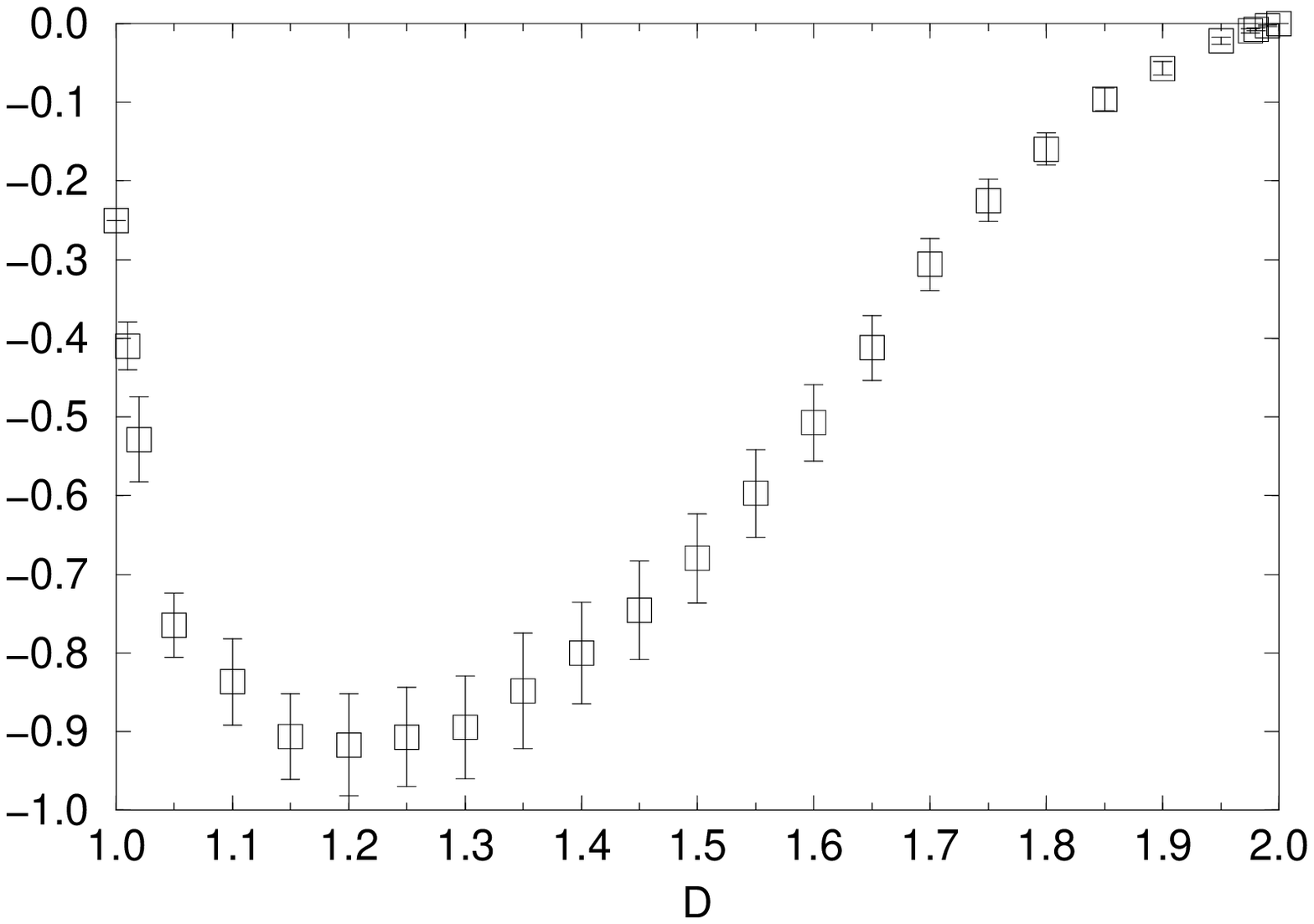}}
}
\vspace{2mm}
{\tabcolsep2.5mm
\begin{tabular}[t]{|l|c|} \hline
$D$ & 
\\ \hline \hline
1.00 &	$-0.25$ \\ \hline
1.01 &	$-0.315  \pm 0.028-0.072$ \\ \hline
1.02 &	$-0.446  \pm 0.053 -0.070$\\ \hline
1.05 &	$-0.727 \pm 0.041 -0.036$\\ \hline
1.10 &	$-0.831 \pm 0.055 -0.006$ \\ \hline
1.15 &	$-0.905  \pm 0.055$ \\ \hline
1.20 &	$-0.917  \pm 0.065$ \\ \hline
1.25 &	$-0.907  \pm 0.063$ \\ \hline
\end{tabular}\hfill
\begin{tabular}[t]{|l|c|} \hline
$D$ & \\ \hline \hline
1.30 &	$-0.895  \pm0.066$ \\ \hline
1.35 &	$-0.848  \pm0.074$ \\ \hline
1.40 &	$-0.800  \pm0.065$ \\ \hline
1.45 &	$-0.746  \pm0.063$ \\ \hline
1.50 & 	$-0.680  \pm0.057$ \\ \hline
1.55 &	$-0.597  \pm0.056$ \\ \hline
1.60 &	$-0.507  \pm0.049$ \\ \hline
1.65 &	$-0.412  \pm0.041$ \\ \hline
\end{tabular}\hfill
\begin{tabular}[t]{|l|c|} \hline
$D$ & \\ \hline \hline
1.70 &	$-0.306 \pm 0.034$ \\ \hline
1.75 &	$-0.225 \pm 0.027$ \\ \hline
1.80 &	$-0.159 \pm 0.021$ \\ \hline
1.85 &	$-0.096 \pm 0.015$ \\ \hline
1.90 &	$-0.056 \pm 0.009$ \\ \hline
1.95 &	$-0.022 \pm 0.005$ \\ \hline
1.975 &	$-0.0091 \pm 0.0024$ \\ \hline
2.00	 &	0\\ \hline
\end{tabular}}
\caption{Numerical results for (\protect{\ref{e:wfcor1}}). 
The first error is the statistical error, the second the correction 
for the systematic error
as discussed in the text. The latter is only given, when it is not
negligible.}
\label{f:F1 brut}
\end{figure}\bea
& & \hspace{-0.5cm} \mbox{\bf R} \tilde F(a,b,c,d,e,f)=
\nonumber \\
& & -\frac{1}{2D} \left[(2-D)a^2f^{-D} +2 a^{2\nu} 
+\frac{2-D}2 \left(e^2+b^2-c^2-d^2\right) \left(d^{-D}-e^{-D}\right)\right] \times
\nonumber\\
& & \qquad \qquad\times\left[ a^{2 \nu} f^{2 \nu} -\frac{1}{4} \left( c^{2\nu}-e^{2 \nu} -b^{2 \nu}
	+d^{2 \nu} \right)^2 \right]^{-d_c/2-1} \nonumber \\
& & +\frac{D+2}{2D}
\bigg[ f^{2 \nu}a^2 +a^{2 \nu}f^2 
+\half(e^{2}+b^{2}-c^2-d^2) \left(c^{2\nu}-e^{2 \nu} -b^{2 \nu}
	+d^{2 \nu}\right)
\bigg] \times\nonumber \\
& & \qquad \qquad \times
\left[a^{2\nu} f^{-D} -\frac1{2}\left( c^{2\nu}-e^{2 \nu} -b^{2 \nu}
	+d^{2 \nu} \right) \left(d^{-D}-e^{-D}\right)
\right] \times \nonumber \\
& & \qquad \qquad \times
\left[ a^{2 \nu} f^{2 \nu} -\frac{1}{4} \left( c^{2\nu}-e^{2 \nu} -b^{2 \nu}
	+d^{2 \nu} \right)^2 \right]^{-d_c/2-2} \nonumber \\
& & -\half a^{-2D} f^{-D-2} -a^{-D} f^{-2D-2} \nonumber \\
& & +\left[ \frac{2-D}{8D} \left( b^2+e^2-c^2-d^2\right)
 \right. a^{-D-2} f^{-D-2}\times \nonumber \\
& &  \qquad \qquad \times \left( D(f^2+e^2-c^2)e^{-D-2} -D(f^2+d^2-b^2) d^{-D-2} 
	-4e^{-D}+4d^{-D}
\right)  \nonumber \\ 
& & \quad\ +\frac{(2-D)(2+D)}{8D}\left( b^2+e^2-c^2-d^2\right) 
\left( (f^2+e^2-c^2) e^{-D} -(f^2+d^2-b^2) d^{-D} \right)  \times \nonumber \\
& & \qquad \qquad \times\, a^{-D-2} f^{-D-4} \nonumber \\
& & \quad\  -\frac{(2-D)(2+D)}{16D} 
	\left( (f^2+e^2-c^2) e^{-D} -(f^2+d^2-b^2) d^{-D} \right)
 a^{-2} f^{-D-2}  \times\nonumber \\
& & \qquad \qquad \times
\left(D (f^2+e^2-c^2) e^{-D-2} -D(f^2+d^2-b^2) d^{-D-2} -4e^{-D}+4d^{-D}
\right) \nonumber \\
& & \left. \quad\ -\frac{(2-D)(2+D)^2}{32 D} \left( (f^2+e^2-c^2) e^{-D} -(f^2+d^2-b^2) d^{-D} \right)^2
a^{-2} f^{-D-4} \right] \times \nonumber \\
& & \qquad \qquad \times \, \Theta(f<d) \nonumber \\
& & +\left[\frac{2-D}{8} \left( b^2+e^2-c^2-d^2\right)
 \right. (a^2+e^2-d^2)e^{-D-2} a^{-D-2} f^{-D-2} \nonumber \\ 
& & \quad\ +\frac{(2-D)(2+D)}{8D}\left( b^2+e^2-c^2-d^2\right) 
\left( (a^2+e^2-d^2)e^{-D} -(a^2+c^2-b^2) c^{-D}  \right)  \times \nonumber \\
& & \qquad \qquad \times \, a^{-D-2} f^{-D-4} \nonumber \\
& & \quad\  -\frac{(2-D)(2+D)}{16} \left( (a^2+e^2-d^2)e^{-D} -(a^2+c^2-b^2) c^{-D}  \right)
 a^{-D-2} f^{-2}  \times\nonumber \\
& & \qquad \qquad \times
(a^2+e^2-d^2) e^{-D-2} \nonumber \\
& & \left. \quad\  -\frac{(2-D)(2+D)}{16D} \left( (a^2+e^2-d^2)e^{-D} -(a^2+c^2-b^2) c^{-D}  \right)^2
a^{-D-2} f^{-4} \right] \times \nonumber \\
& & \qquad \qquad \times \Theta(a<c) \nonumber \\
& & -a^{-D} (c^{2\nu}+d^{2\nu})^{-d_c/2-1} d^{-D} 
	\Theta(c<a) \nonumber\\
& & -a^{-D} (e^{2\nu}+b^{2\nu})^{-d_c/2-1} e^{-D} 
\Theta (b<a) 
\label{e:wave5}
\eea
The variable transformations are the same as discussed in 
section \ref{all42} and \ref{s:decomp}. 
The numerical calculations are very difficult. 
Due to the complexity of the integrand, the numerical errors
induced by the limited precision of the workstation 
became important too and limited the reduction of 
the statistical error.
The total CPU-time was about 1000 hours on a workstation.
 We obtained the numerical results 
summarized in figure \ref{f:F1 brut}.

The systematic error was corrected 
using table \ref{f:cor}. The error-bars represent the
statistical error of the AMC-integration, which could not
be reduced due to the lack of performance of the work-station,
both in speed and in precision. 
The numerical results are in agreement with 
the analytically calculated value $ -\frac14$ for $D \to 1$, discussed later. 

\subsection{The correction for the first marginal counterterm}
\begin{figure}[htb] 
\centerline{
\epsfxsize=12cm \parbox{12cm}{\epsfbox{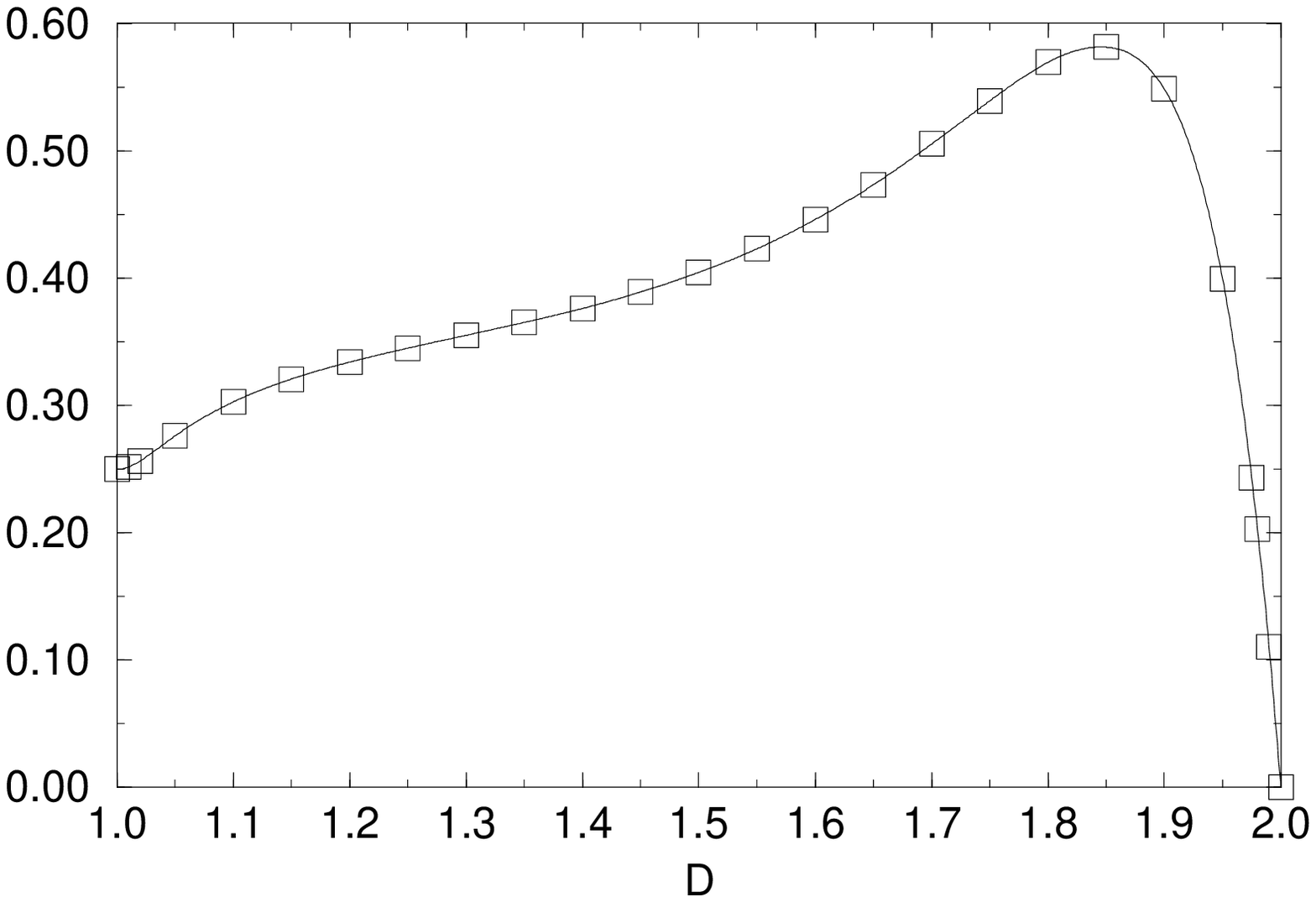}}
}
\vspace{2mm}
\centerline{\tabcolsep3.0mm
\begin{tabular}[t]{|l|c|} \hline
$D$ & \\ \hline \hline
1.00 &	$1/4$ \\ \hline
1.01 &	$0.252$\\ \hline
1.02 &	$0.256$ \\ \hline
1.05 &	$0.276$ \\ \hline
1.10 &	$0.303$ \\ \hline
\end{tabular}\hfill\begin{tabular}[t]{|l|c|} \hline
$D$ & \\ \hline \hline
1.15 &	$0.321$ \\ \hline
1.20 &	$0.334$ \\ \hline
1.25 &	$0.345$ \\ \hline
1.30 &	$0.355$ \\ \hline
1.35 &	$0.365$ \\ \hline
\end{tabular}\hfill\begin{tabular}[t]{|l|c|} \hline
$D$ & \\ \hline \hline
1.40 &	$0.376$ \\ \hline
1.45 &	$0.389$ \\ \hline
1.50 & 	$0.405$ \\ \hline
1.55 &	$0.423$ \\ \hline
1.60 &	$0.446$ \\ \hline
\end{tabular}\hfill\begin{tabular}[t]{|l|c|} \hline
$D$ & \\ \hline \hline
1.65 &	$0.473$ \\ \hline
1.70 &	$0.505$ \\ \hline
1.75 &	$0.539$ \\ \hline
1.80 &	$0.570$ \\ \hline
1.85 &	$0.581$ \\ \hline
\end{tabular}\hfill\begin{tabular}[t]{|l|c|} \hline
$D$ & \\ \hline \hline
1.90 &	$0.548$ \\ \hline
1.95 &	$0.399$ \\ \hline
1.975 &	$0.244$ \\ \hline
1.99 &	$0.110$ \\ \hline
2.00	 &	0\\ \hline
\end{tabular}}
\caption{Numerical results for (\protect{\ref{e:wfcor1}}). 
The relative statistical error is $10^{-3}$.}
\label{f:wfcor1}
\end{figure}We arranged the counterterms $\tilde C_1$ and $\tilde C_2$ so that they both
have the same correction. 
This correction is up to terms of ${\cal O} (\E)$ (for the 
notation cf.\ figure \ref{f:abcforient}):
\begin{figure}[htb] 
\centerline{
    \epsfxsize=3.5cm \epsfbox{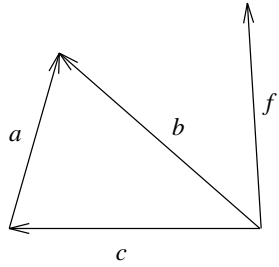}
}
\caption{The orientation of the vectors $\vec a$, $\vec b$, $\vec c$ and $\vec f$. }
\label{f:abcforient}
\end{figure}\bea
& & \hspace{-0.5cm}\int_{c=1;\ a,b,d,e,f} \left[ \frac{2-D}{2D} \, \vec a \vec f \, \left( \vec b \vec f b^{-D} -\vec c \vec f c^{-D} \right)
a^{ -\nu (d+2) } f^{-\nu (d+2)}\right. \nonumber \\ 
& & \hspace{1.3cm} \left .-\frac{d+2}{16 D} (2-D)^2 \left(\vec b \vec f b^{-D} -\vec c \vec f c^{-D} \right)^2 
a^{D-\nu (d+2)} f^{-\nu (d+2)} \right]\times \nonumber \\
& & \hspace{3cm} \times\left[ \Theta(f<c) - \Theta(f<\max(a,b,c)) \right] \nn \\
&=&  \int_{a=1; \, b,c}
	\left [ \frac{2-D}{2D^2} \left( \vec a \vec b b^{-D} -\vec a \vec c c^{-D} \right)
		a^{-\nu(d+2)}
	-\frac{(2-D)^2(2+d)}{16 D^2} \left( \vec b b^{-D} - \vec c c^{-D} \right)^2 
	a^{D-\nu (d+2)}
\right] \times
\weiter \hspace{3cm}\times  \frac1\E\left[c^\E- \max(a,b,c)^\E\right]
\label{64646} 
\eea
Expanding $\frac1\E \left[c^\E- \max(a,b,c)^\E\right] =
\ln(c) -\ln(\max(a,b,c))+{\cal O}(\E)$
and using the symmetry of \eq{64646} yields:
\bea
& &  \int_{a=1; \, b,c}
	\left [ \frac{2-D}{4D^2} \left( \vec a \vec b b^{-D} -\vec a \vec c c^{-D} \right)
		a^{-D-2}
	-\frac{(2-D)(2+D)}{16 D^2} \left( \vec b b^{-D} - \vec c c^{-D} \right)^2 
	a^{-2}
\right] \times
\weiter \hspace{3cm}\times  \left[\ln(c)+\ln(b)-2 \ln(\max(a,b,c))\right] +{\cal O}(\E) \nn \\
&=&  \int_{a=1; \, b,c}
	\left [ \frac{2-D}{8D^2} \left( (a^2+b^2-c^2) b^{-D} +(a^2+c^2-b^2) c^{-D} \right)
		a^{-D-2}\right.
\weiter \hspace{1.5cm} \left.	-\frac{(2-D)(2+D)}{16 D^2} \left( b^{2-2D} +c^{2-2D} +(a^2-b^2-c^2)b^{-D}c^{-D}\right) \right] a^{-2}\times
\weiter \hspace{3cm}\times  \left[\ln(c)+\ln(b)-2 \ln(\max(a,b,c))\right] +{\cal O}(\E)
\label{e:wfcor1}
\eea
This integral is calculated numerically using the measure given 
by \eq{e:m1} to \eq{e:m2} in 
section~\ref{The correction for the unusual marginal counterterm}.
We find the results in figure \ref{f:wfcor1}.

For $D\to 1$ an analytical calculation gives:
\be
\half \int_0^1 \rmd b \, \ln(b) -\frac 34 \int_0^1 \rmd b \, \ln (b) = \frac 14
\ee
This is in agreement with the numerical data in figure \ref{f:wfcor1}.

\subsection{The correction for the second marginal counterterm}
\begin{figure}[htb] 
\centerline{
\epsfxsize=12cm \parbox{12cm}{\epsfbox{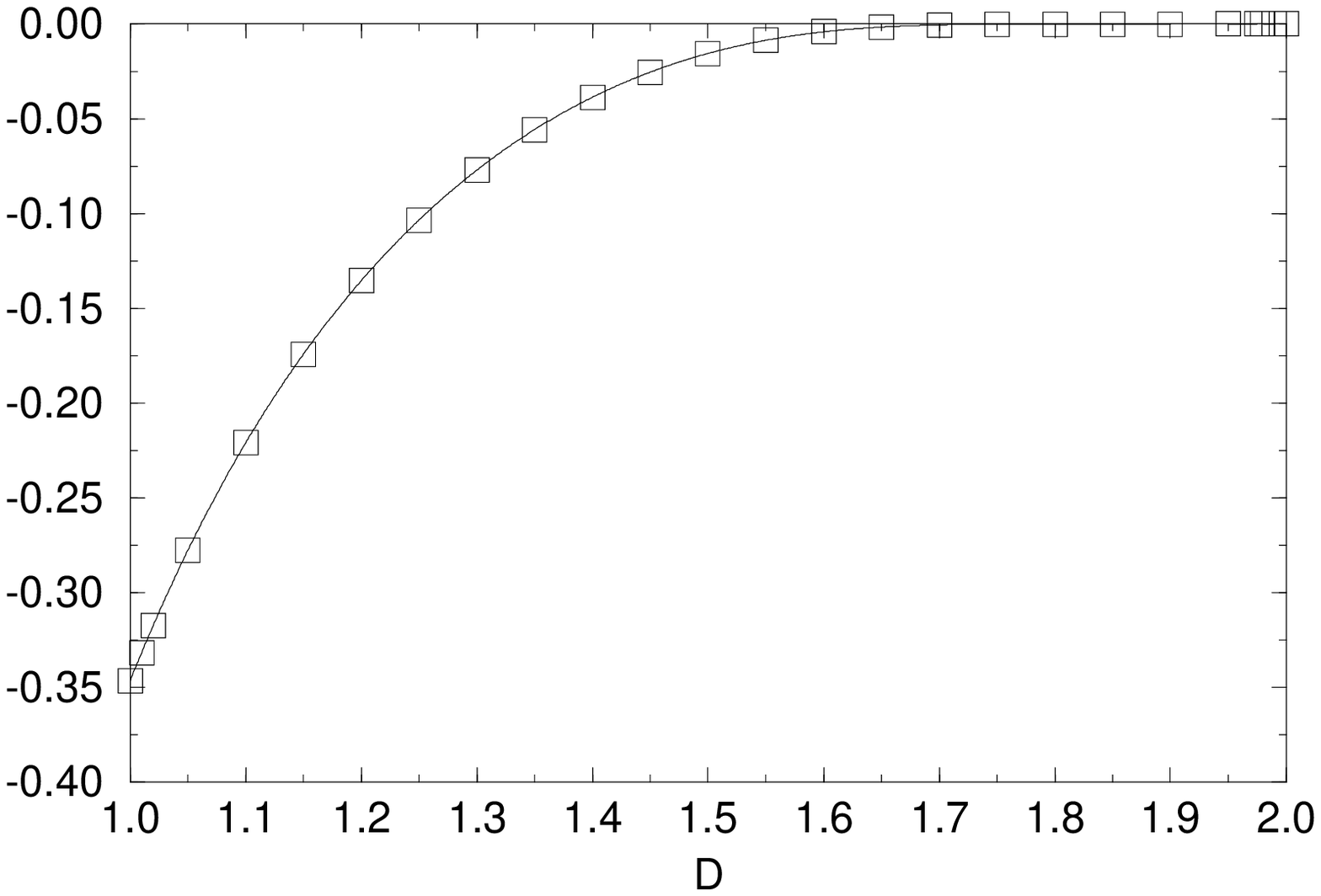}}
}
\vspace{2mm}
{\tabcolsep3.0mm
\begin{tabular}[t]{|l|c|} \hline
$D$ & \\ \hline \hline
1.00 &	$-3.4657\times 10^{-1}  $ \\ \hline
1.01 &	$-3.3167\times 10^{-1}$ \\ \hline
1.02 &	$-3.1736\times 10^{-1}  $ \\ \hline
1.05 &	$-2.7766\times 10^{-1}  $ \\ \hline
1.10 &	$-2.2098\times 10^{-1} $ \\ \hline
1.15 &	$-1.7416\times 10^{-1}$ \\ \hline
1.20 &	$-1.3541\times 10^{-1}  $ \\ \hline
1.25 &	$-1.0338\times 10^{-1} $ \\ \hline
\end{tabular}\hfill\begin{tabular}[t]{|l|c|} \hline
$D$ & \\ \hline \hline
1.30 &	$-7.7076\times 10^{-2} $ \\ \hline
1.35 &	$-5.5699\times 10^{-2}  $ \\ \hline
1.40 &	$-3.8640\times 10^{-2} $ \\ \hline
1.45 &	$-2.5397\times 10^{-2}  $ \\ \hline
1.50 &	$-1.5529\times 10^{-2}  $ \\ \hline
1.55 &	$-8.6017\times 10^{-3}  $ \\ \hline
1.60 &	$-4.1478\times 10^{-3}  $ \\ \hline
1.65 &	$-1.6350\times 10^{-3}  $ \\ \hline
\end{tabular}\hfill\begin{tabular}[t]{|l|c|} \hline
$D$ & \\ \hline \hline
1.70 &	$-4.7420\times 10^{-4}  $ \\ \hline
1.75 &	$-8.3621\times 10^{-5}  $ \\ \hline
1.80 &	$-6.1097\times 10^{-6}  $ \\ \hline
1.85 &	$-7.5149\times 10^{-8}  $ \\ \hline
1.90 &	$-1.0256\times 10^{-11} $ \\ \hline 
1.95 &	$-1.7552\times 10^{-23} $ \\ \hline 
1.975 &	$-2.8185\times 10^{-47}$ \\ \hline  
2.00 &	$0 $ \\ \hline 
\end{tabular}}
\caption{Numerical results for (\protect{\ref{e:wfcor2}}). 
The relative statistical error is $10^{-5}$.}
\label{f:wfcor2}
\end{figure}
Up to order $\E$ for $D_1$ and $D_2$ each, the following correction has to be added:
\bea & & \hspace{-1cm}\frac1{2D} \int_{c=1;\ a,b,d,e,f} a^{D-\nu d} \left(c^{2 \nu} + d^{2 \nu}\right)^{-d/2} \Theta(c<a)
	(\Theta(d<a)-1) \nn \\ 
&=& - \frac{1}{2D} \int_{c=1; \, d>c}  \left(c^{2 \nu} + d^{2 \nu}\right)^{-d/2} \frac1\E \left(d^\E-c^\E\right)
\nn \\
&=& \frac1{2D} \frac{1}{(2-D)^2} \int_0^1
\frac{\rmd x}x x^{D/(2-D)} (1+x)^{-2D/(2-D)} \ln(x) + {\cal O}(\E)
\label{e:wfcor2}
\eea
For $D\to 1$ an analytical calculation gives:
\be
\frac 12 \int_0^1 \rmd x \, \frac{\ln (x)}{(1+x)^2} =
 -\half \ln 2 =-0.3465735903
\ee
The numerical data are given in figure \ref{f:wfcor2}.
Numerical and analytical data fit nicely together.

\subsection{Analytical calculation for $D \to 1$}
\begin{figure}[t]
$$
\vbox{\epsfbox{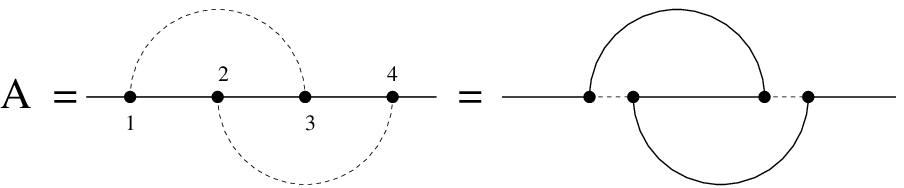}}
$$
$$
\vbox{\epsfbox{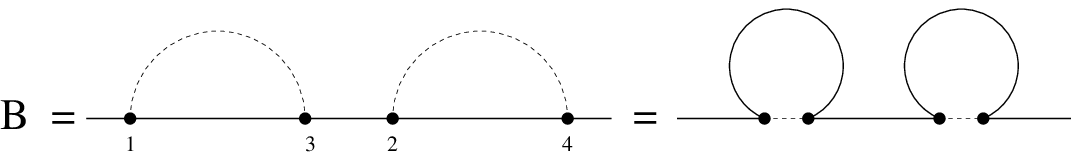}}
$$
$$
\vbox{\epsfbox{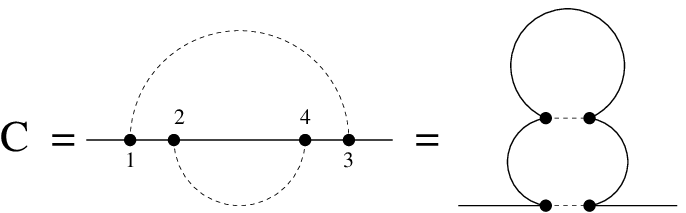}}
$$
\caption{The 3 topological inequivalent classes A, B and C and the 
corresponding Feynman-diagrams of scalar $\varphi^4$ theory.}
\label{f:wavetop}
\end{figure}
We calculate now $J(L=1)$, equation \eq{e:Jeps}, for D=1 and $\E=0$.
We give the integrals and the numerical results for
the different regions. The combinatorical factor 4 for 
every pair of integrals compensates against the factor $1/4$ from the measure.
The symmetrized version of the counterterms has to be used.
Otherwise the 4 integrals in one equivalence class may not 
coincide. So we have to integrate
$A-B-\half( C_1+\tilde C_1 +C_2+\tilde C_2 + D_1+\tilde D_1+ D_2 +\tilde D_2)$.
\bea
\newlength{\temp} \settowidth{\temp}{\epsfbox{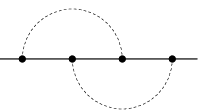}}
\parbox{\temp}{\epsfbox{./eps/waveline4points1p.eps}}&=&\int_1^\infty \rmd a
\int_1^\infty \rmd f \left[-\half (fa^2+af^2-2af) (af-1)^{-3}
+\half (a^{-2}f^{-1} +a^{-1}f^{-2}) \right. \nonumber \\
& & \hspace{2.75cm} \left.+\frac14 a^{-2}f^{-1} \Theta(a-f)
+\frac14 a^{-1}(a+f-2)^{-2} \Theta(a-f+1) \right.\nonumber \\
& & \hspace{2.75cm} \left.+\frac14 a^{-1}f^{-2} \Theta(f-a)
+\frac14 f^{-1}(a+f-2)^{-2} \Theta(f-a+1)\right] \nonumber \\
&=& -\half \ln(2) +\frac34 
\\
\settowidth{\temp}{\epsfbox{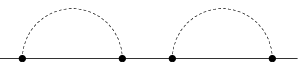}}
\parbox{\temp}{\epsfbox{./eps/waveline4points2p.eps}}&=&\int_0^\infty \rmd a
\int_0^\infty \rmd f \, 0 \nonumber \\
&=& 0 \\
\settowidth{\temp}{\epsfbox{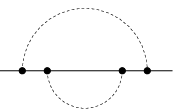}}
\parbox{\temp}{\epsfbox{./eps/waveline4points3p.eps}}
&=&\int_1^\infty \rmd a
\int_0^{a-1} \rmd f \left[	-\half a f^{-2} (a-f)^{-2}+\half \left(a^{-2}f^{-1}+f^{-2}a^{-1}\right)\right. \nonumber \\
& & \hspace{3.0cm} \left.+\half a^{-2}f^{-1}
	+\frac14 \left(a^{-1}(a-f)^{-2}+a^{-1}(a+f)^{-2} \right) \right. \nonumber \\
& & \hspace{3.0cm} \left.+\frac14 f^{-1}(a-f)^{-2} \Theta(f-1) \Theta(2f+1-a) \right] \nonumber \\
&=& -\half \ln(2) -\half 
\eea
Together this gives:
\be
\settowidth{\temp}{\epsfbox{./eps/waveline4points1p.eps}}
\parbox{\temp}{\epsfbox{./eps/waveline4points1p.eps}}
+
\settowidth{\temp}{\epsfbox{./eps/waveline4points2p.eps}}
\parbox{\temp}{\epsfbox{./eps/waveline4points2p.eps}}
+
\settowidth{\temp}{\epsfbox{./eps/waveline4points3p.eps}}
\parbox{\temp}{\epsfbox{./eps/waveline4points3p.eps}}
=-\ln(2)+\frac14 = -0.4431\ldots
\ee

\subsection{The complete diagram}
\begin{figure}[h] 
\centerline{
\epsfxsize=12cm \parbox{12cm}{\epsfbox{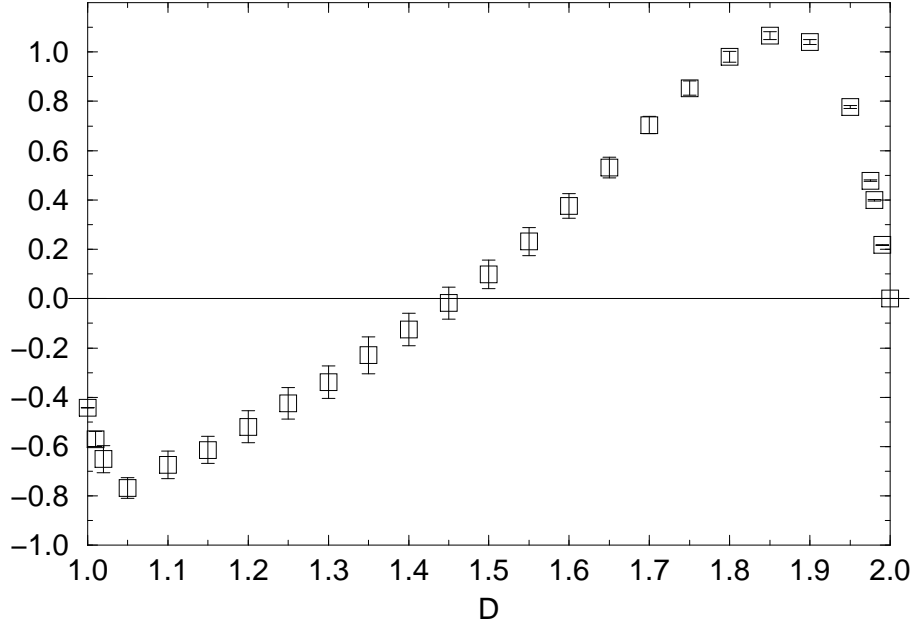}}
}
\caption{Numerical results for $J(1)$, equation (\ref{e:Jeps}). 
Only the statistical error is given.}
\label{f:F2 complet}
\end{figure}
In figure \ref{f:F2 complet} we  show the complete
results for $J(1)\ts_{\E=0}$, equation (\ref{e:Jeps}). We can verify that
the limit $D\to 1$ is correctly reproduced within the 
error-bars.

\section{Second Contribution for the Renormalization of the Wave Function}
\begin{figure}[htb] 
\centerline{
    \epsfxsize=3.5cm \epsfbox{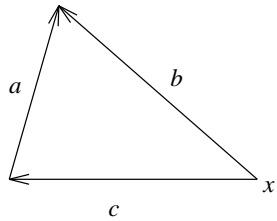}
}
\caption{The orientation of the vectors $\vec a$, $\vec b$ and $\vec c$ and
the position of the point $x$. }
\label{f:abcorient}
\end{figure}The second contribution to the renormalization of the 
wave-function is, see \eq{e:F2}:
\bea
\hspace{-1cm}{\cal F}_2&=&  -\half \Bigg< \GH \Bigg| \GO \Bigg>_{\E} \times 
\weiter \times \left(  \Bigg< \GZ \Bigg| \GO \Bigg>_L +
	\nu(d+2)  \Bigg< \GH \Bigg| \GO \Bigg>_L
	-\Bigg< \GH \Bigg| \GO \Bigg>_L \Bigg< \GW \Bigg| \GO \Bigg>_L \right) \nn
\eea
In order to determine ${\cal F}_2$, we calculate:
\be \label{e:s0}
 \Bigg< \GZ \Bigg| \GO \Bigg>_L +\nu (d+2)
  \Bigg< \GH \Bigg| \GO \Bigg>_L - \Bigg< \GH \Bigg| \GO \Bigg>_L
 \Bigg< \GW \Bigg| \GO \Bigg>_L
\ee
We recall that 
\bea \label{e:s1}
\hspace{-1cm}\Bigg< \GZ \Bigg| \GO \Bigg>_L &=&\frac{\nu (d+2)}{2D} \frac{2-D}{2} 
\int_{a,b,c<L}  \left( \vec b b^{-D} - \vec c c^{-D} \right)^2
a^{D-\nu (d+2)} 
\weiter -\frac{2-D}D \int_{a,b,c<L} \left(\vec a \vec b b^{-D} -\vec a \vec c c^{-D} \right) a^{-\nu (d+2)} \ , \\
\Bigg< \GH \Bigg| \GO \Bigg>_L \Bigg< \GW \Bigg| \GO \Bigg>_L
 &=& -\frac{2-D}{D} \int_{a,c<L} a^{-\nu(d+2)} \left(a^2 -D 
	\left(\vec a \vec c \right)^2 c^{-2} \right) c^{-D} \label{e:s3}
\eea
and 
\be \label{e:s2}
\nu(d+2) \Bigg< \GH \Bigg| \GO \Bigg>_L=-\frac{\nu(d+2)}{2D}\int_{a<L} a^{D-\nu d}
\ee
First of all one easily convinces oneself that the integral equals 0 for 
$D=1$. 
In the following we give two independent derivations, 
which were both integrated numerically and which were found to 
coincide. This gives an additional check that the global prefactor,
which cannot be checked analytically, is correct. 

We start with the first derivation:
In order to subtract (\ref{e:s2}) from the first term in (\ref{e:s1}) we use the fact that
\be
\frac{2-D}2 \int_{a=\mbox{\scriptsize fixed};\, b,c} \left( \vec b b^{-D} - \vec c c^{-D} \right)^2 = a^{2-D} \ .
\ee
Therefore the first contribution is:
\be
\frac{\nu (d+2)}{2D} \frac{2-D}{2} \left[ \int_{a,b,c<L} -\int_{a<L}\right] 
	\left( \vec b b^{-D} - \vec c c^{-D} \right)^2 a^{D-\nu (d+2)}
\ee
Applying $\frac1\E L \frac \partial {\partial L}$ and mapping onto $a=1$ yields:
\be
\frac{\nu (d+2)}{2D} \frac{2-D}{2} \int_{a=1;\, b,c}
\left( \vec b b^{-D} - \vec c c^{-D} \right)^2 a^{D-\nu (d+2)} 
\frac 1 \E \left( \max(a,b,c)^{-\E} -a^{-\E} \right)
\ee
In the limit $\E \to 0 $ this expression simplifies to: 
\be	\label{e:sw1}
	- \frac{(2-D)(2+D)}{4D} \int_{a=1; \, b,c} 
	\left( \vec b b^{-D} - \vec c c^{-D} \right)^2 a^{-D-2}
	 \left( \ln(\max(a,b,c))-\ln(a) \right) +{\cal O}(\E)
\ee
The second contribution is:
\bea
& & \hspace{-0.9cm}-\frac{2-D}D \int_{a,b,c<L}
	\left(\vec a \vec b b^{-D} -\vec a \vec c c^{-D} \right) a^{-\nu (d+2)}
+ \frac{2-D}{D} \int_{a,c<L} 
  \left(a^2 -D 
	\left(\vec a \vec c \right)^2 c^{-2} \right) c^{-D}a^{-\nu(d+2)} \nn \\
&=& -\frac{2-D}D \int_{a,b,c<L} \left[
	\left(\vec a \vec b b^{-D} -\vec a \vec c c^{-D} \right)  
-  \left(a^2 -D 
	\left(\vec a \vec c \right)^2 c^{-2} \right) c^{-D} \right] a^{-\nu(d+2)}
\weiter +\frac{2-D}D \left[ \int_{a,c<L} -\int_{a,b,c<L} \right]\left(a^2 -D 
	\left(\vec a \vec c \right)^2 c^{-2} \right) c^{-D}  a^{-\nu(d+2)}
\label{e:smile}
\eea
The last term becomes after application of $\frac 1 \E L \frac \partial {\partial L}$
and mapping onto $a=1$:
\be \label{e:star}
\frac{2-D}D   \int_{a=1; \, b,c}
\left( a^2 -D \left(\vec a \vec c\right)^2 c^{-2} 
\right) c^{-D} a^{-\nu (d+2)}
\frac 1 \E \left( \max(a,c)^{-\E} -\max(a,b,c)^{-\E} 
\right)
\ee
In the limit $\E \to 0$ it reduces to
\be \label{e:star'}
\frac{2-D}D   \int_{a=1; \, b,c}
\left( a^2 -D \left(\vec a \vec c\right)^2 c^{-2} 
\right) c^{-D} a^{-D-2}
\Big( \ln(\max(a,b,c)) -\ln(\max(a,c)) \Big) +{\cal O}(\E)
\ee
The other term in (\ref{e:smile}) is:
\be
-\frac{2-D}D \int_{a,b,c<L} \left[
	\left(\vec a \vec b b^{-D} -\vec a \vec c c^{-D} \right)  
-  \left(a^2 -D 
	\left(\vec a \vec c \right)^2 c^{-2} \right) c^{-D} \right] a^{-\nu(d+2)}
\label{e:smile2}
\ee
By power-counting, a pole in $\frac 1 \E$ can appear for $a \to 0$. In 
this limit however the terms in the square-brackets cancel and  make
the integral finite.
Let us apply $L \frac{\partial }{\partial L} $ and map onto  $a=1$:
\be
-\frac{2-D}D \int_{a=1;\, b,c} \left[
	\left(\vec a \vec b b^{-D} -\vec a \vec c c^{-D} \right)  
-  \left(a^2 -D 
	\left(\vec a \vec c \right)^2 c^{-2} \right) c^{-D} \right] a^{-D-2}
	a^{\E} \max(a,b,c)^{-\E}
\ee
This integral is a function of  $\E$, which has the form:
\be
c_1 \E +{\cal O}(\E^2)
\ee
Especially it vanishes for $\E=0$. The integral thus does not change
if we subtract its  value for $\E=0$. Equation (\ref{e:smile2}) then 
becomes -- remember that a  factor  $\frac 1 \E$ has to be reintroduced:
\bea
& & \hspace{-1cm}-\frac{2-D}D \int_{a=1;\, b,c} \left[
	\left(\vec a \vec b b^{-D} -\vec a \vec c c^{-D} \right)  
-  \left(a^2 -D 
	\left(\vec a \vec c \right)^2 c^{-2} \right) c^{-D} \right] a^{-D-2}\times
\weiter \qquad
	\times \frac1\E \left( a^{\E} \max(a,b,c)^{-\E} -1 \right)
\eea
In the limit $\E \to 0$ this simplifies  to:
\bea
& & \hspace{-1cm} \frac{2-D}D \int_{a=1;\, b,c} \left[
	\left(\vec a \vec b b^{-D} -\vec a \vec c c^{-D} \right)  
-  \left(a^2 -D 
	\left(\vec a \vec c \right)^2 c^{-2} \right) c^{-D} \right] a^{-D-2}
	\times
\weiter \qquad \times \left( \ln(\max(a,b,c))- \ln(a) \right) +{\cal O}(\E)
\label{e:smile3}
\eea
The final result is the sum of (\ref{e:sw1}), (\ref{e:star'}) and (\ref{e:smile3}):
\bea
\lefteqn{\hspace{-.7cm}\Bigg< \GZ \Bigg| \GO \Bigg>_L +\nu (d+2)
  \Bigg< \GH \Bigg| \GO \Bigg>_L - \Bigg< \GH \Bigg| \GO \Bigg>_L
 \Bigg< \GW \Bigg| \GO \Bigg>_L }\nn \\
 =	&-& \frac{(2-D)(2+D)}{4D} \int_{a=1; \, b,c} 
	\left( b^{2-2D} +  c^{2-2D}+(a^2-b^2-c^2) b^{-D}c^{-D} \right) a^{-D-2}\times 
\weiter \hspace{5cm} \times
	 \Big( \ln(\max(a,b,c))-\ln(a) \Big) \nn \\
&+& \frac{2-D}{2D}  \int_{a=1; \, b,c} 
\weiter \hspace{1.0cm} \bigg[
\,\left( a^2 -D/4 \left(a^2+c^2-b^2\right)^2 c^{-2} 
\right) c^{-D} \Big( \ln(a) -\ln(\max(a,c)) \Big) 
\weiter 
\hspace{1.0cm} + \!\left( a^2 -D/4 \left(a^2+b^2-c^2\right)^2 b^{-2} 
\right) b^{-D}
\Big( \ln(a)  -\ln(\max(a,b))\Big) \nn \\
 & & \hspace{1.0cm}+\!\left((a^2+b^2-c^2) b^{-D} +(a^2+c^2-b^2) c^{-D} \right)
\Big( \ln(\max(a,b,c))- \ln(a) \Big)  
\bigg]\times
\weiter \hspace{3.5cm} \times a^{-D-2}
  +{\cal O}(\E) \label{e:sw-res}
\eea

\begin{figure}[tb] 
\centerline{
\epsfxsize=12cm \parbox{12cm}{\epsfbox{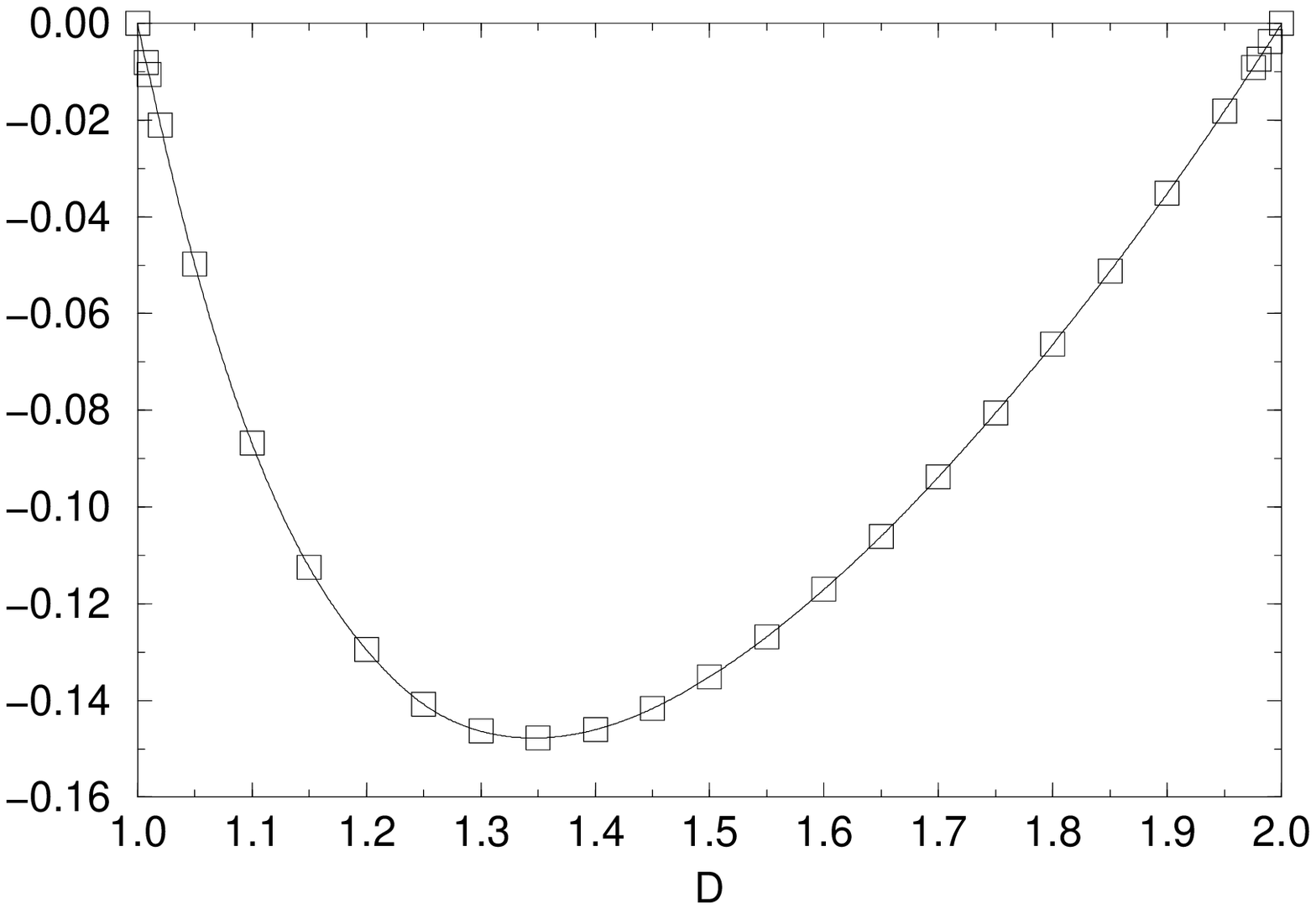}}
}
\vspace{2mm}
{\tabcolsep5.0mm
\begin{tabular}[t]{|l|c|} \hline
$D$ & 
\\ \hline \hline
1.0	& $		0	$ \\ \hline	
1.007& $ 		-8.15 \times 10^{-3}  $ \\ \hline
1.01 & $  	-1.06 \times 10^{-2}  $ \\ \hline 1.02 & $  	-2.10 \times 10^{-2}  $ \\ \hline 1.05 & $  	-4.98 \times 10^{-2}  $ \\ \hline 1.10 & $  	-8.68 \times 10^{-2}  $ \\ \hline 1.15 & $  	-1.12 \times 10^{-1}  $ \\ \hline 1.20 & $  	-1.30 \times 10^{-1}  $ \\ \hline 1.25 & $  	-1.41 \times 10^{-1}  $ \\ \hline \end{tabular}\hfill\begin{tabular}[t]{|l|c|} \hline
$D$ & 
\\ \hline \hline
1.30 & $  	-1.46 \times 10^{-1}  $ \\ \hline 1.35 & $  	-1.48 \times 10^{-1}  $ \\ \hline 1.40 & $  	-1.46 \times 10^{-1}  $ \\ \hline 1.45 & $  	-1.42 \times 10^{-1}  $ \\ \hline 1.50 & $  	-1.35 \times 10^{-1}  $ \\ \hline 1.55 & $  	-1.27 \times 10^{-1}  $ \\ \hline 1.60 & $  	-1.17 \times 10^{-1}  $ \\ \hline 1.65 & $  	-1.06 \times 10^{-1}  $ \\ \hline 1.70 & $  	-9.38 \times 10^{-2}  $ \\ \hline \end{tabular}\hfill\begin{tabular}[t]{|l|c|} \hline
$D$ & 
\\ \hline \hline
1.75 & $  	-8.05 \times 10^{-2}  $ \\ \hline 1.80 & $  	-6.63 \times 10^{-2}  $ \\ \hline 1.85 & $  	-5.12 \times 10^{-2}  $ \\ \hline 1.90 & $  	-3.52 \times 10^{-2}  $ \\ \hline 1.95 & $  	-1.81 \times 10^{-2}  $ \\ \hline 1.975& $	-9.19 \times 10^{-3 } $ \\ \hline 1.98 & $  	-7.37 \times 10^{-3}  $ \\ \hline 1.99 & $  	-3.71 \times 10^{-3}  $ \\ \hline 2.00 & $ 0 $ \\ \hline
\end{tabular}}
\caption{Numerical results for (\protect{\ref{e:sw-res}}). 
The relative statistical error is $10^{-3}$.}
\label{f:F2}
\end{figure}Another way to treat the second term in
(\ref{e:s1}) is to 
perform a partial integration:
\bea
&& \hspace{-1cm} \frac1D \int_{a<L} \int_x \,\Theta(L-b) \Theta(L-c)\, \vec a
\vec\nabla_{\!\!x} \left(b^{2-D} -c^{2-D} \right) a^{-\nu (d+2)}
\weiter
=-\frac1D \int_{a<L} \int_x \vec a \vec\nabla_{\!\!x} \left[\Theta(L-b) \Theta(L-c) \right]
\left(b^{2-D} -c^{2-D} \right) a^{-\nu (d+2)}
\weiter= -\frac 1 D \int_{a<L} \int_{x} \vec a \left( \delta(L-b) \vec b b^{-2}
\Theta(L-c) 
+ \Theta(L-b)
\delta(L-c) \vec c c^{-2} \right) \times 
\weiter \hspace{3cm} \times \left(b^{2-D} -c^{2-D} \right) a^{-\nu (d+2)}
\weiter = -\frac 1 D \int_{a,c<b=L} \vec a \vec b b^{-2} 
\left(b^{2-D} -c^{2-D} \right) a^{-\nu (d+2)}
\weiter \quad  -\frac 1 D \int_{a,b<c=L} \vec a \vec c c^{-2} 
\left(b^{2-D} -c^{2-D} \right) a^{-\nu (d+2)}
\weiter = -\frac 2 D \int_{a,b<c=L} \vec a \vec c c^{-2} 
\left(b^{2-D} -c^{2-D} \right) a^{-\nu (d+2)} \label{e:blablu}
\eea
The pole in (\ref{e:s3}) has to be subtracted. It 
can be written as 
\be 
 -\frac{2-D}{D} \int_{a<L; c=L}
\left(\vec a \vec c \right)^2 a^{-\nu (d+2)} c^{-2-D} 
\ee  
and is split into two parts
\be
-\frac {2-D} D \left[ \int_{a,b<c=L} +\int_{a<c=L<b} \right]
\left(\vec a \vec c\right)^2 a^{-\nu (d+2)} c^{-2-D} 
\ee
which are mapped onto the same sector as in (\ref{e:blablu}):
\be
	-\frac{2-D} D\int_{a,b<c=L} \left(\vec a \vec c \right)^2 a^{-\nu (d+2)} c^{-2-D}  
+ \left(\vec a \vec b \right)^2 a^{-\nu (d+2)} b^{-2-D} (b/c)^{-\E}
\Theta(a<b) \label{e:blablo}
\ee
The complete expression, i.e.\  $(\ref{e:blablu})-(\ref{e:blablo})$ is:
\be
-\frac 1 D \int_{a,b<c=L} \Bigg\{ 2 \vec a \vec c c^{-2} 
\left(b^{2-D} -c^{2-D} \right) -(2-D) \left[\left(\vec a \vec c \right)^2
c^{-2-D} +\left(\vec a \vec b \right)^2 b^{-2-D} (b/c)^{-\E}\Theta(a<b) \right]
\Bigg\} a^{-\nu(d+2)}
\ee
One checks that the integral is locally convergent  and that the
limit $\E \to 0$ can be taken. Of course this derivation is not
systematic but the final result is easier to integrate numerically
and will therefore be used in the following. It is:
\bea
&&\hspace{-1.2cm}
-\frac 1 D \int_{a,b<c=L} \Bigg\{ 2 \vec a \vec c c^{-2} 
\left(b^{2-D} -c^{2-D} \right) 
\weiter \hspace{1.1cm}-(2-D) \left[\left(\vec a \vec c \right)^2
c^{-2-D} +\left(\vec a \vec b \right)^2 b^{-2-D} \Theta(a<b) \right]
\Bigg\} a^{-D-2} \nn \\
&=& \frac 1 D \int_{a,b<c=L} \Bigg\{ (a^2+c^2-b^2) c^{-2} 
\left(b^{2-D} -c^{2-D} \right) +\frac{(2-D)}4\times
\weiter \hspace{1.1cm} \times\left[\left( a^2+c^2-b^2 \right)^2
c^{-2-D} +\left(a^2+b^2-c^2\right)^2 b^{-2-D} \Theta(a<b) \right]
\Bigg\} a^{-D-2}\quad
\eea
The numerical integration is performed using the measure
given in equations \eq{e:m1} to \eq{e:m2} in 
section~\ref{The correction for the unusual marginal counterterm}.
We find the results in figure \ref{f:F2}. We verify that for $D\to1$ and
for $D\to 2$ the analytically predicted value $0$ is correctly 
reproduced.

\section{Contribution to the Wavefunction Renormalization from
the 1-loop Coupling Constant Renormalization}
\begin{figure}[tb] 
\centerline{
\epsfxsize=12cm \parbox{12cm}{\epsfbox{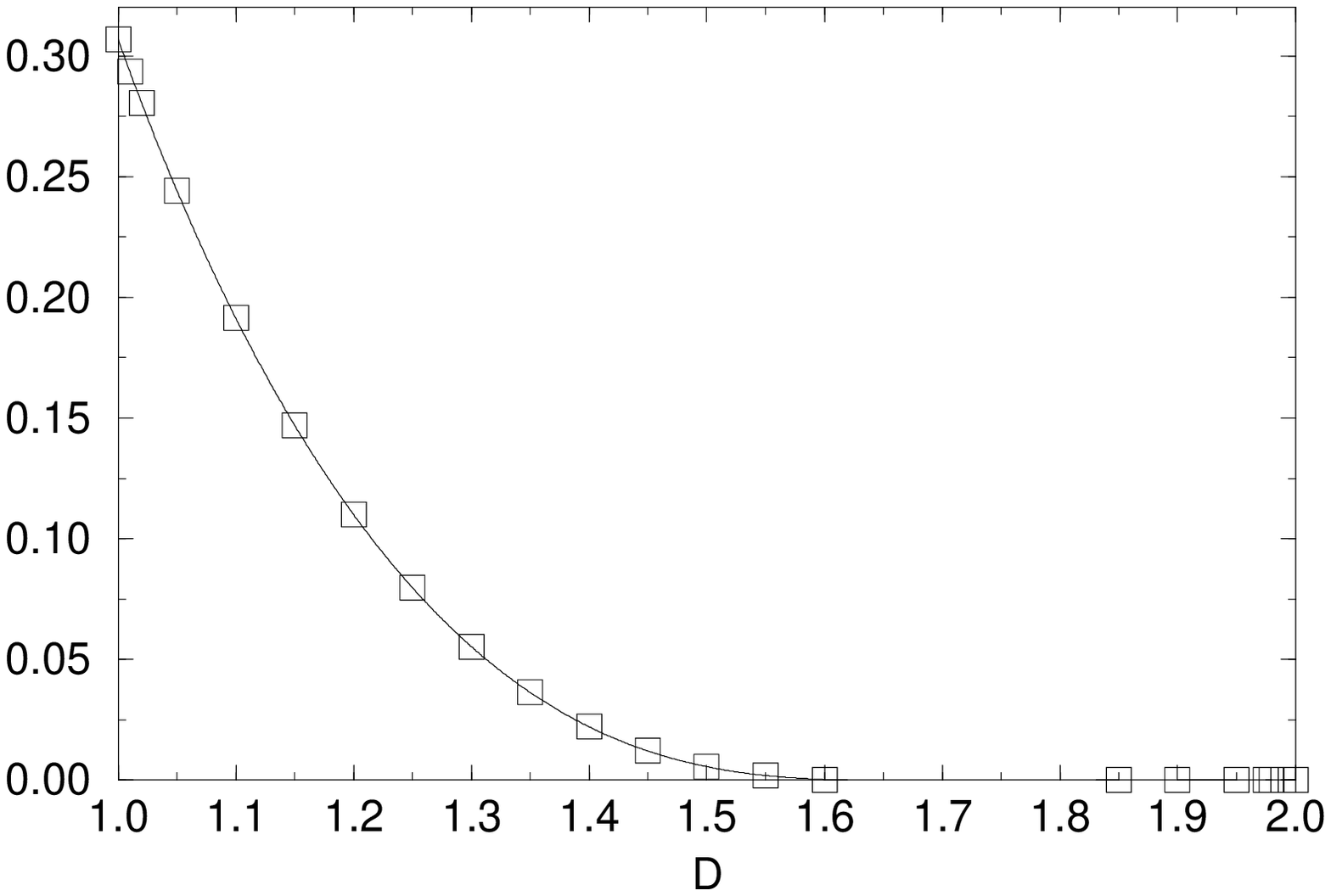}}
}
$$ \Bigg< \GM \ \Bigg| \ \GB \Bigg>_{\E^0}$$
{\tabcolsep1.0mm
\begin{tabular}[t]{|l|c|} \hline
$D$ & \\ \hline \hline
1.00 & $	3.0685 \times 10^{-1} $ \\ \hline 1.01 & $	2.9345 \times 10^{-1} $  \\ \hline 1.02 & $	2.8050 \times 10^{-1} $  \\ \hline 1.05 & $	2.4419 \times 10^{-1} $  \\ \hline 1.10 & $	1.9136 \times 10^{-1} $  \\ \hline 1.15 & $	1.4701 \times 10^{-1} $  \\ \hline \end{tabular}\hfill\begin{tabular}[t]{|l|l|} \hline
$D$ & \\ \hline \hline
1.20 & $	1.1009 \times 10^{-1} $  \\ \hline 1.25 & $	7.9770 \times 10^{-2} $  \\ \hline 1.30 & $	5.5395 \times 10^{-2} $  \\ \hline 1.35 & $	3.6373 \times 10^{-2} $  \\ \hline 1.40 & $	2.2143 \times 10^{-2} $  \\ \hline 1.45 & $	1.2116 \times 10^{-2} $  \\ \hline \end{tabular}\hfill\begin{tabular}[t]{|l|r|} \hline
$D$ & \\ \hline \hline
1.50 & $	5.6359 \times 10^{-3} $  \\ \hline 1.55 & $	1.9633 \times 10^{-3} $  \\ \hline 1.60 & $	2.8999 \times 10^{-4} $  \\ \hline 1.65 & $	-1.8953 \times 10^{-4} $  \\ \hline 1.70 & $	-1.5855 \times 10^{-4} $  \\ \hline 1.75 & $	-4.9540 \times 10^{-5} $  \\ \hline \end{tabular}\hfill\begin{tabular}[t]{|l|c|} \hline
$D$ & \\ \hline \hline
1.80 & $	-5.5019 \times 10^{-6} $  \\ \hline 1.85 & $	-9.5846 \times 10^{-8} $  \\ \hline 1.90 & $	-1.7933 \times 10^{-11} $  \\ \hline 1.95 & $	-4.2091 \times 10^{-23} $  \\ \hline 1.975& $	-8.0880 \times 10^{-47} $  \\ \hline 2.00 & 0 \\ \hline
\end{tabular}}
\caption{Numerical results for equation (\protect{\ref{e:int21fin}}). 
The relative statistical error is $10^{-5}$. }
\label{f:int21fin}
\end{figure}
We will calculate the last diagram, equation \eq{e:F3}, 
\be
{\cal F}_3=\half \bigg< \GH \bigg| \GO \bigg>_{\E^{-1}} 
\bigg< \GM \bigg| \GB \bigg>_{\E^{0}}. 
\ee
We expand up to first order in $\E$ the following term:
\begin{equation}
\Bigg< \GM \Bigg| \GB \Bigg>_L =\int_{s<L} \int_{t<L} \left(s^{2\nu}+t^{2\nu}\right)^{-d/2}  
\end{equation}
We apply $L {\partial \over \partial L}\ts_{L=1}$ and map onto $s=L=1$: 
\begin{eqnarray}
 & &\hspace{-1cm} \int_0^{\infty} \frac{dt}t t^D \left(1 + t^{2 \nu} \right)^{-d/2} \max(1,t)^{-\E}\nonumber 
\\
& & = \int_0^{\infty} \frac{dt}t t^D \left(1 + t^{2 \nu} \right)^{-d/2} 
+ \int_1^{\infty} \frac{dt}t t^D \left(1 + t^{2 \nu} \right)^{-d/2} (t^{-\E}-1) \nonumber \\
& & =\frac1{2-D} \frac{\Gamma\left(\frac{D}{2-D} \right)
\Gamma\left(\frac{D-\E}{2-D} \right) }{\Gamma\left(\frac{2D-\E}{2-D} \right)}
+ \frac\E{(2-D)^2} \int_0^1 \frac{dt}t t^{D/(2-D)} (1+t)^{-2D/(2-D)} \ln(t) 
\end{eqnarray}
This expression can still be expanded as:
\bea
&&	\frac1{2-D} \frac{\Gamma\left(\frac{D}{2-D} \right)^2
 }{\Gamma\left(\frac{2D}{2-D} \right)}
\left[1+  \E\left( \Psi\left({2 D\over 2-D}\right)
-\Psi\left({D \over 2-D}\right) \right)\right] \nn \\
&& \qquad \qquad + \frac\E{(2-D)^2} \int_0^1 \frac{dt}t t^{D/(2-D)} (1+t)^{-2D/(2-D)} \ln(t) \quad
\eea
The final result is 
\bea
\bigg< \GM \bigg| \GB \bigg>_{\E^{0}}&=&\frac1{2-D} \frac{\Gamma\left(\frac{D}{2-D} \right)^2}{\Gamma\left(\frac{2D}{2-D} \right)}
\left( \Psi\left({2 D\over 2-D}\right)
-\Psi\left({D \over 2-D}\right) \right) \nn \\
&& \qquad + \frac1{(2-D)^2} \int_0^1 \frac{dt}t t^{D/(2-D)} (1+t)^{-2D/(2-D)} \ln(t) \quad
\label{e:int21fin}
\eea
The numerical results are given in figure~\ref{f:int21fin}.
For $D=1$, equation \eq{e:int21fin} is analytically calculated to be
\be
1-\ln(2)=0.30685281944\ldots
\ee

\section{Extrapolations and Calculation of Critical Exponents}
\newcommand{\rvec}{{\vec r}}
\newcommand{\nuflory}{\nu_{\mbox{\scriptsize Flory}}}
\newcommand{\nuvar}{\nu_{\mbox{\scriptsize var}}}
\newcommand{\Dd}[2]{\parbox{0mm}{\raisebox{-3.02cm}[0mm][0mm]{\tiny\hspace{4.34cm}$0$}}\parbox{0mm}{\raisebox{2.6cm}[0mm][0mm]{\scriptsize\hspace{0.725cm}$(D,d)=(#1,#2)$}}}

\subsection{The renormalization-group functions}
\label{s:The renormalization group functions}
As we already discussed in section \ref{s:Renormalization at 1-loop order},
the renormalization group $\beta$-function and the anomalous scaling
dimension $\nu$ of $\vec r$ are in the MS-scheme that we use
obtained from the  variation of the 
coupling constant 
and the field with respect to the renormalization scale $\mu$,
keeping the bare couplings fixed. Written in terms of
$Z$ and $Z_b$ they are:
\begin{eqnarray}
\label{e:beta'}
\beta(b) &=&
\mu \frac{\partial}{\partial \mu }\lts_{b_0} b =
\frac{-\E b} {1+ b\frac{\partial}{\partial b } \ln Z_b +
\frac{d}2 b \frac{\partial}{\partial b} \ln Z} 
\\
\label{e:nu'}
\nu (b) &=&
\frac{2-D}{2}-\half \mu \frac{\partial}{\partial \mu }\lts_{b_0} \ln Z =
\frac{2-D}2 -\half \beta(b) \frac{\partial}{\partial b} \ln Z 
\end{eqnarray}
We recall the form of $Z$ and $Z_b$ from equations \eq{e:ZZb2},
\eq{e:f1c1tilde} and \eq{e:fc1fc1tilde}:
\begin{eqnarray}
Z&=&1+{e_1\over\varepsilon}b+\left({\tilde f_1\over\varepsilon}+
{\tilde f_2(\E)\over\varepsilon^2}\right) b^2+{\cal O}(b^3)
\nonumber\\
Z_b&=&1+{a_1\over\varepsilon}b+\left({\tilde c_1\over\varepsilon}+
{\tilde c_2(\E) \over\varepsilon^2}\right) b^2+{\cal O}(b^3) 
\end{eqnarray}
Using equations \eq{e:c1}, \eq{e:a1}, \eq{e:Z2lead} and 
\eq{e:Zb2lead} we obtain
\be
\beta(b)= -\E \,b + \left(a_1 +1 -\frac{\E}{2D} \right) b^2
+2\frac{2 D\tilde f_1-D\tilde c_1-\tilde f_1\E+2\tilde c_1}{2-D}\,b^3 + {\cal O}(b^4) 
\ee
and
\be \label{e:nu(b)}
\nu(b)=\frac{2-D}2 + \frac{2-D}{4D}\,b + \tilde f_1 \,b^2 + {\cal O}(b^3)
\ee
For $\E>0$, the $\beta$-function has a non-trivial IR-attractive fixed point
for $\beta(b^*)=0$. Up to second order in $\E$, $b^*$ is:
\be \label{e:bstar}
b^*(\E)= \frac{1}{1+a_1}\E+ \frac{(2-D)(1+a_1)-4D(2-D) \tilde c_1-8D^2\tilde f_1}{2D(2-D)(1+a_1)^3}
\E^2 +{\cal O} (\E^3)
\ee
Plugging in \eq{e:bstar} in \eq{e:nu(b)} yields:
\be
\nu^*=
 \frac{2-D}2 + \frac{2-D}{4D(1+a_1)}\E+ \frac{8D^2a_1 \tilde f_1-4D(2-D)
\tilde c_1 +(2-D)(1+a_1)}{8D^2(1+a_1)^3} \E^2 +{\cal O}(\E^3)
\label{e:nustar}
\ee
$\tilde c_1$ and $\tilde f_1$ were both calculated numerically as a function
of $D$ in the interval $1\le D\le 2$.

\begin{figure}[t] \label{f:nu12}
\centerline{
\epsfxsize=13cm \parbox{13cm}{\epsfbox{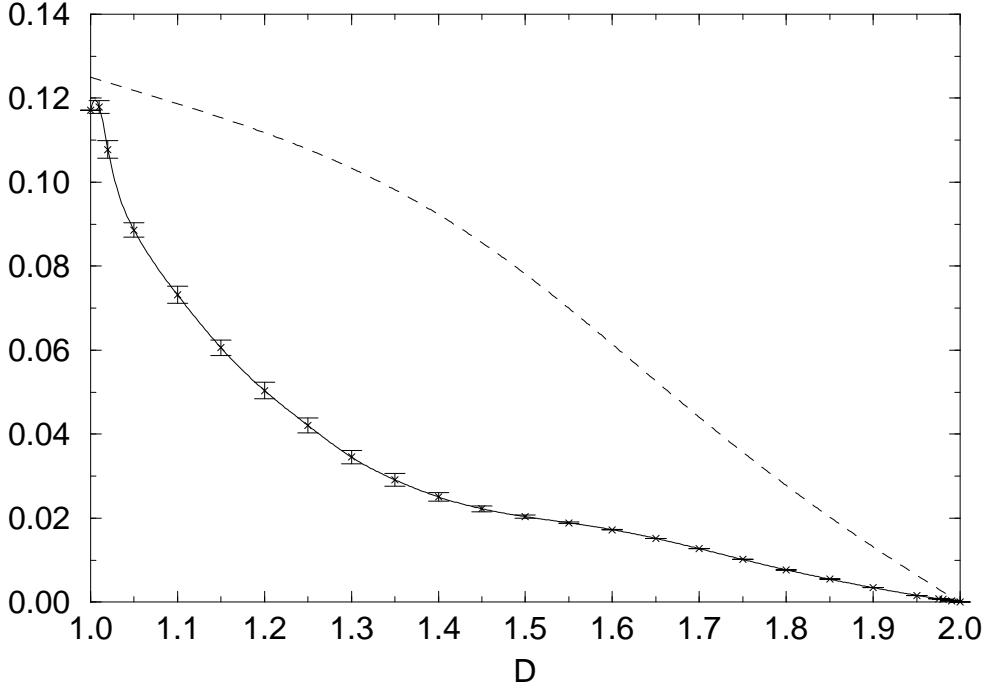}}
}
\caption{The functions $\nu_1(D)$ (dashed line) and $\nu_2(D)$. The latter
is given with the corresponding error-bars of the statistical error. }
\end{figure}\subsection{The exponent $\nu $}
\begin{figure}[tb] 
\centerline{
\Dd{1}{3}%
\epsfxsize=8cm \parbox{8cm}{\epsfbox{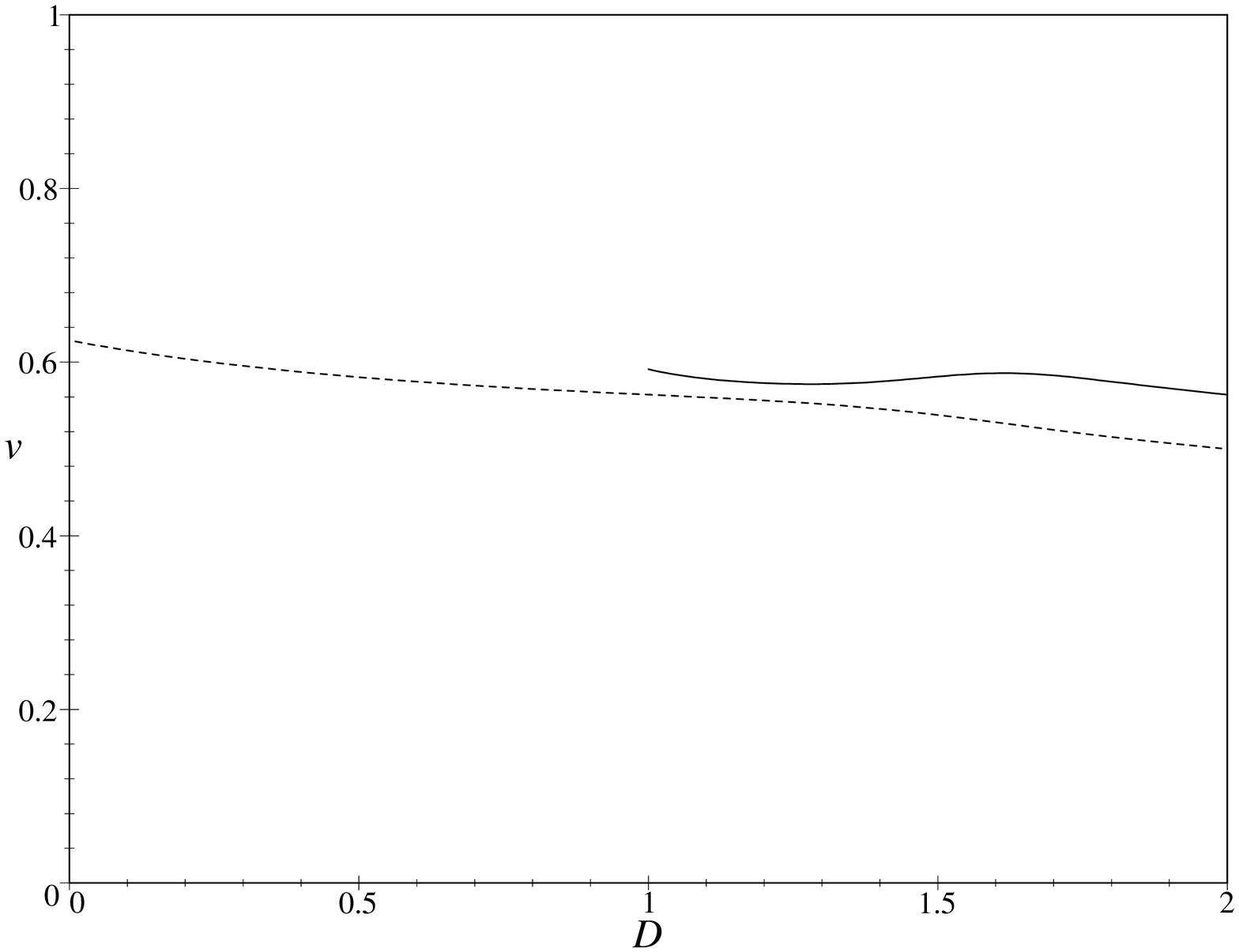}}
\Dd{2}{3}%
\epsfxsize=8cm \parbox{8cm}{\epsfbox{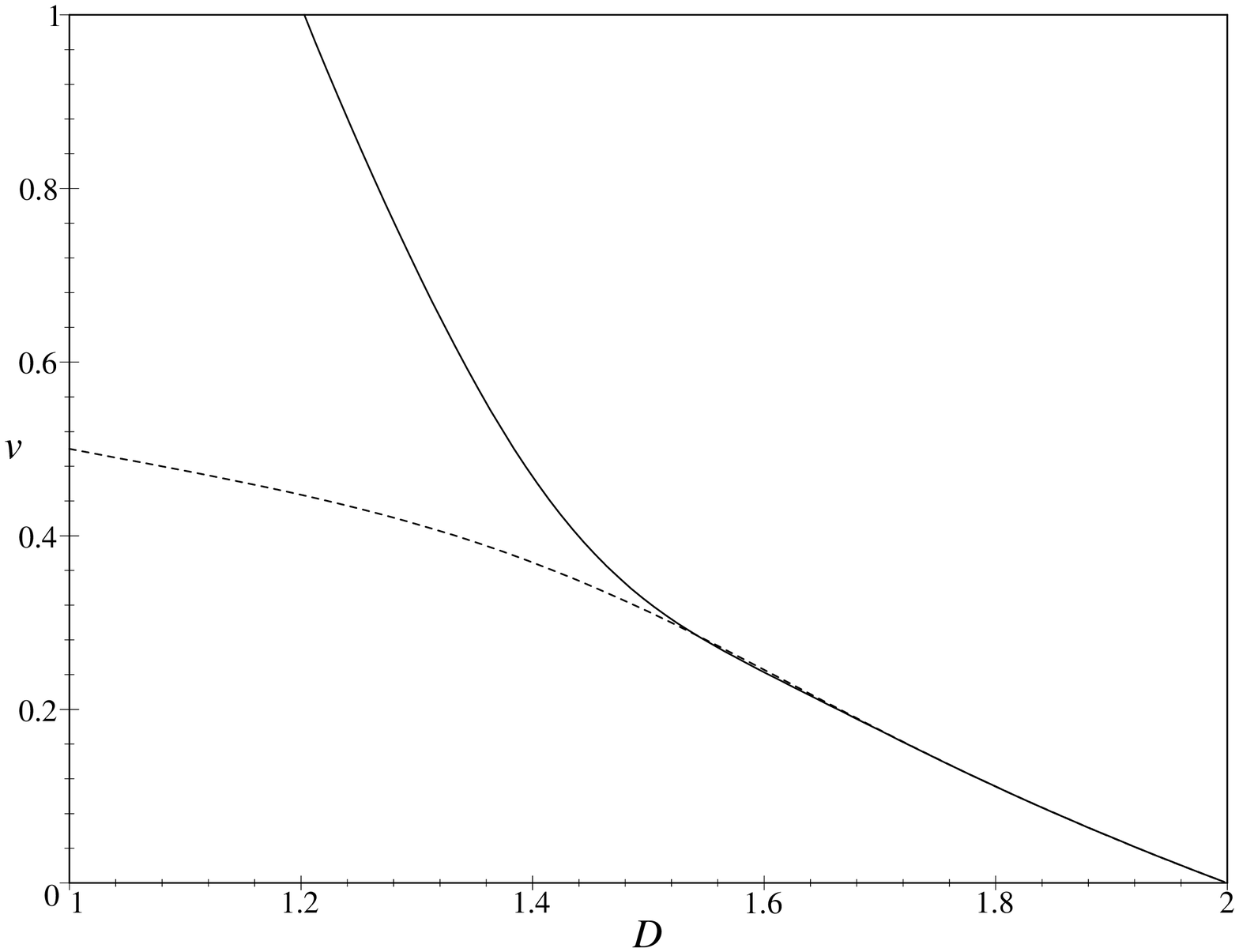}}
}
\caption{Extrapolation of $\nu$ in $D$ and $\E$ to $(D,d)=$ (1,3) and (2,3).
The dashed line is the 1-loop result,
the full line the 2-loop result. The exponent $\nu$ is plotted as function of the expansion-point
$D_0$. }
\label{fig: nu-D-eps}
\end{figure} 

\begin{figure}[htb] 
\centerline{
\Dd{1}{3}%
\epsfxsize=8cm \parbox{8cm}{\epsfbox{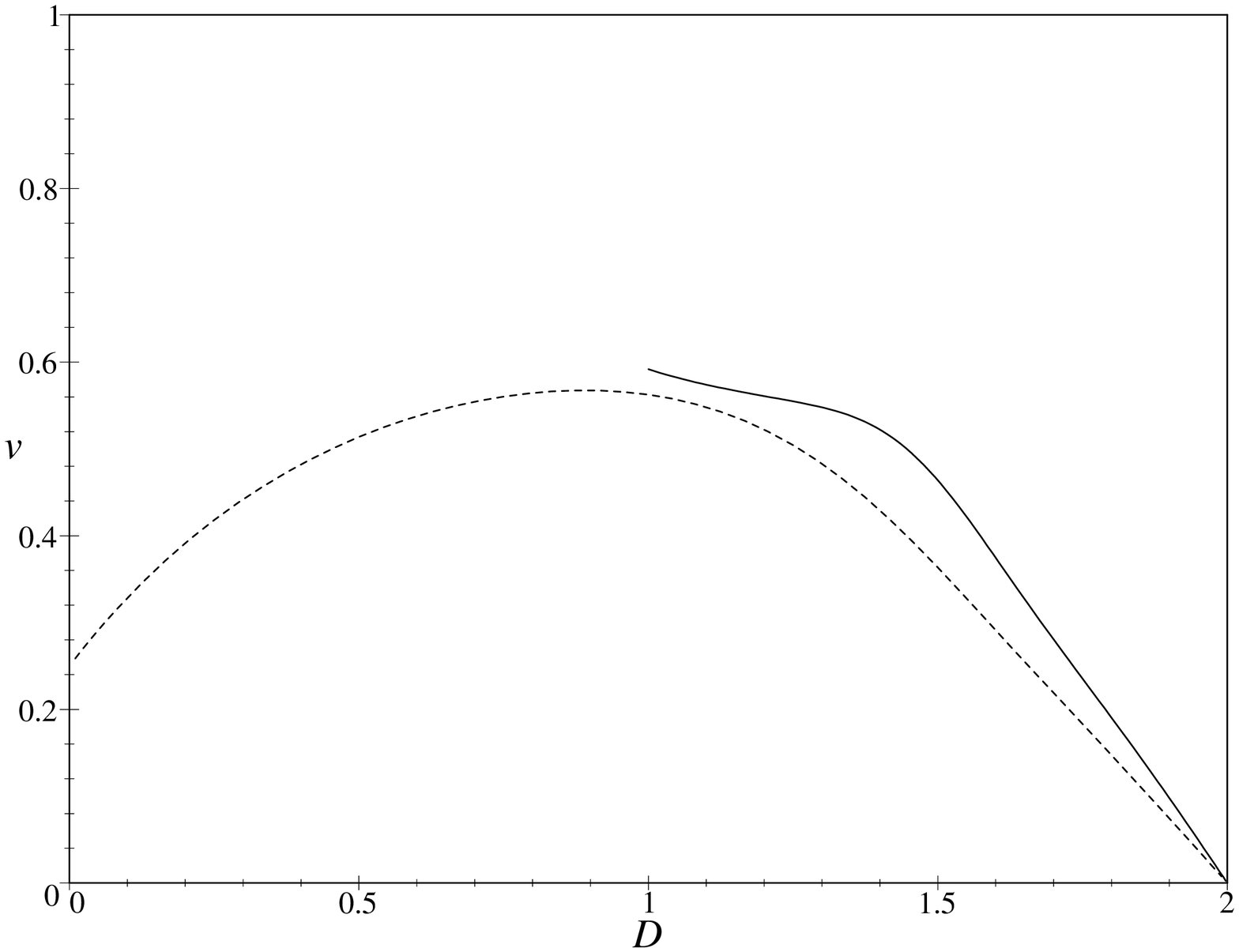}}
\Dd{2}{3}%
\epsfxsize=8cm \parbox{8cm}{\epsfbox{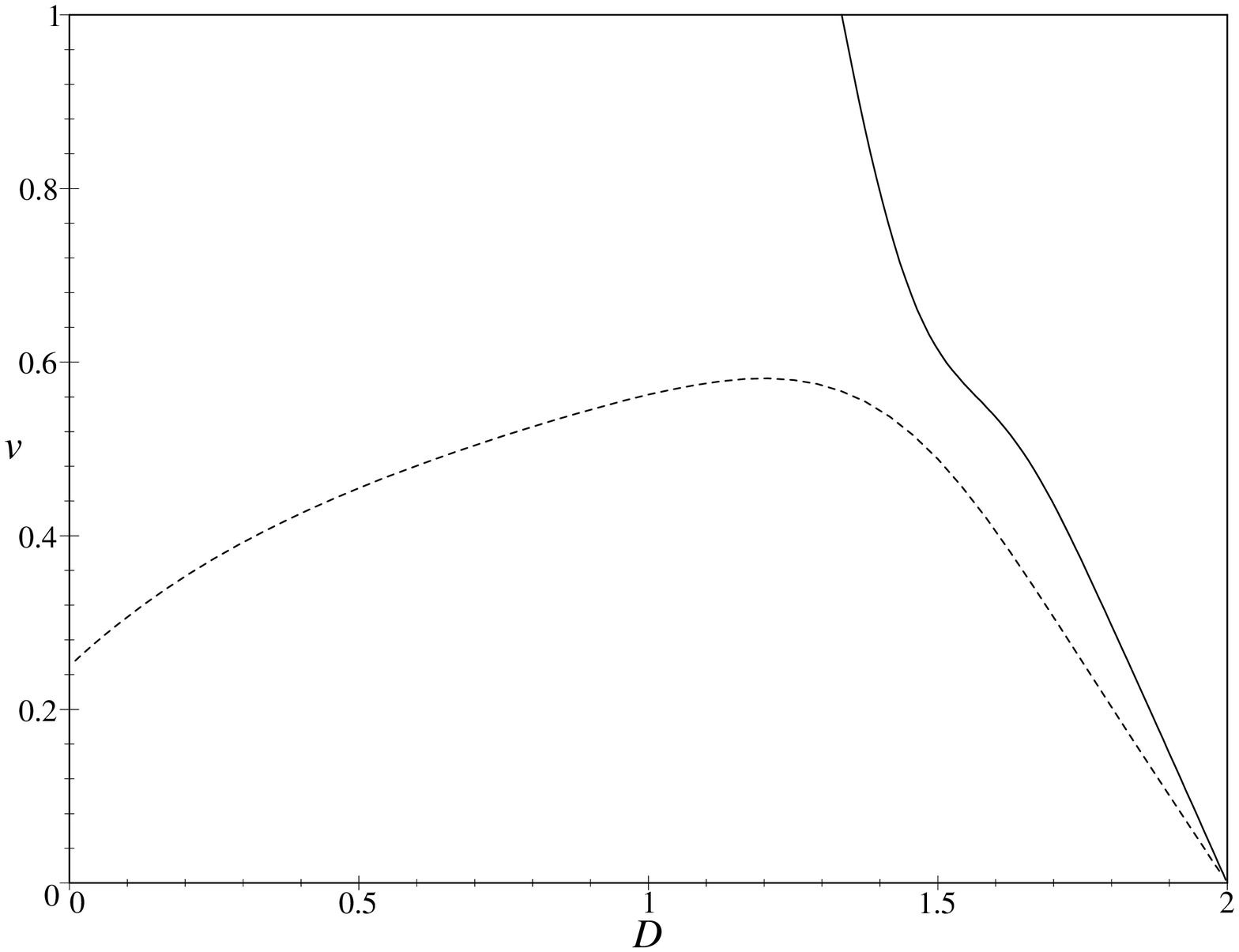}}
}
\caption{Extrapolation in $D$ and $d$ to $(D,d)=$ (1,3) and (2,3).}
\label{f: nu-D-d}
\end{figure}
\begin{figure}[htb] 
\centerline{
\Dd{1}{1}%
\epsfxsize=8cm \parbox{8cm}{\epsfbox{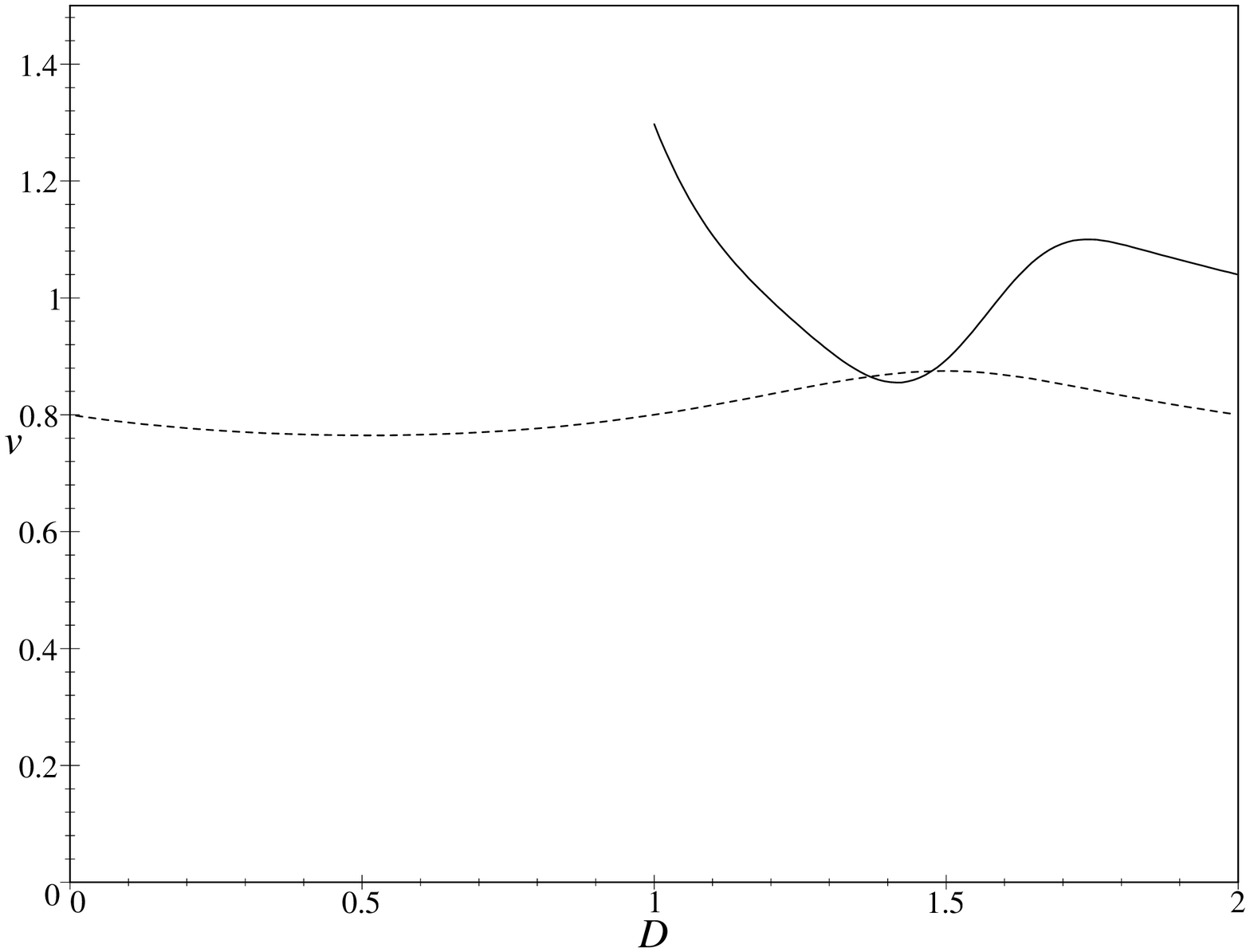}}
\Dd{1}{2}%
\epsfxsize=8cm \parbox{8cm}{\epsfbox{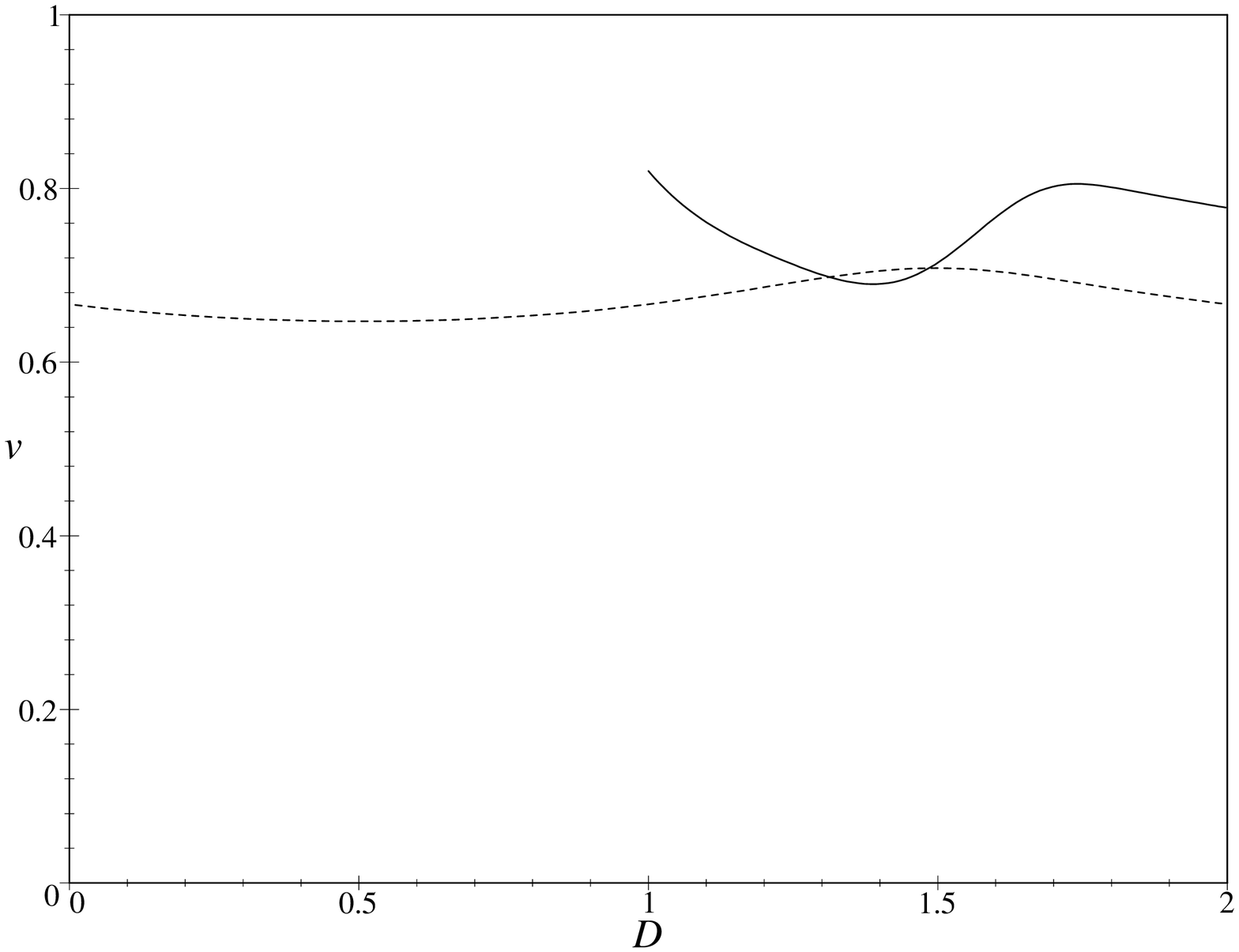}}
}
\centerline{
\Dd{1}{3}%
\epsfxsize=8cm \parbox{8cm}{\epsfbox{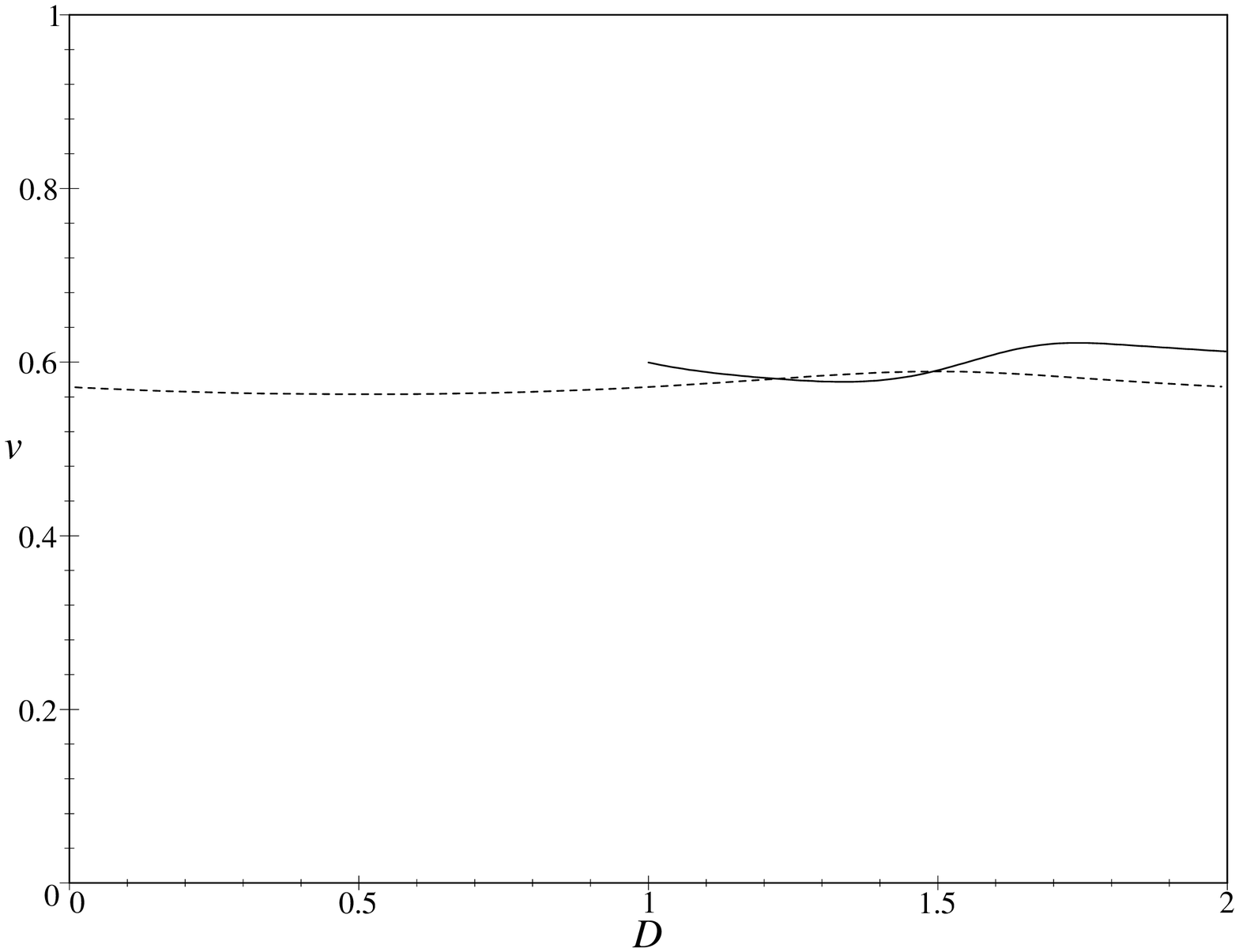}}
}
\caption{Extrapolation for polymers
in $D$ and $D_c(d)$ to $(D,d)=$ (1,1), (1,2), (1,3)}
\label{f: nu1-D-Dc(d)}
\end{figure} 
\begin{figure}[htb] 
\centerline{
\Dd{2}{2}%
\epsfxsize=8cm \parbox{8cm}{\epsfbox{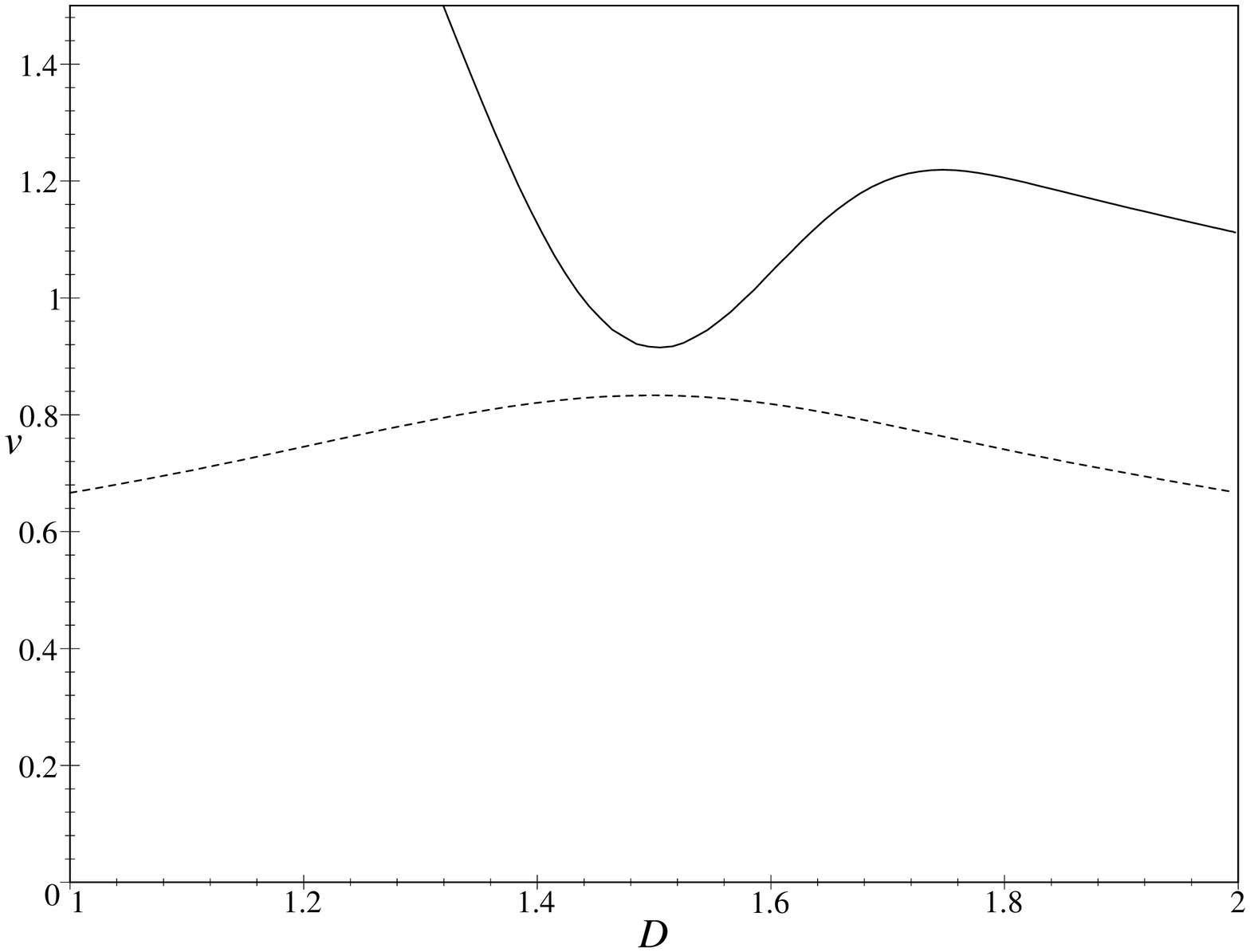}}
\Dd{2}{3}%
\epsfxsize=8cm \parbox{8cm}{\epsfbox{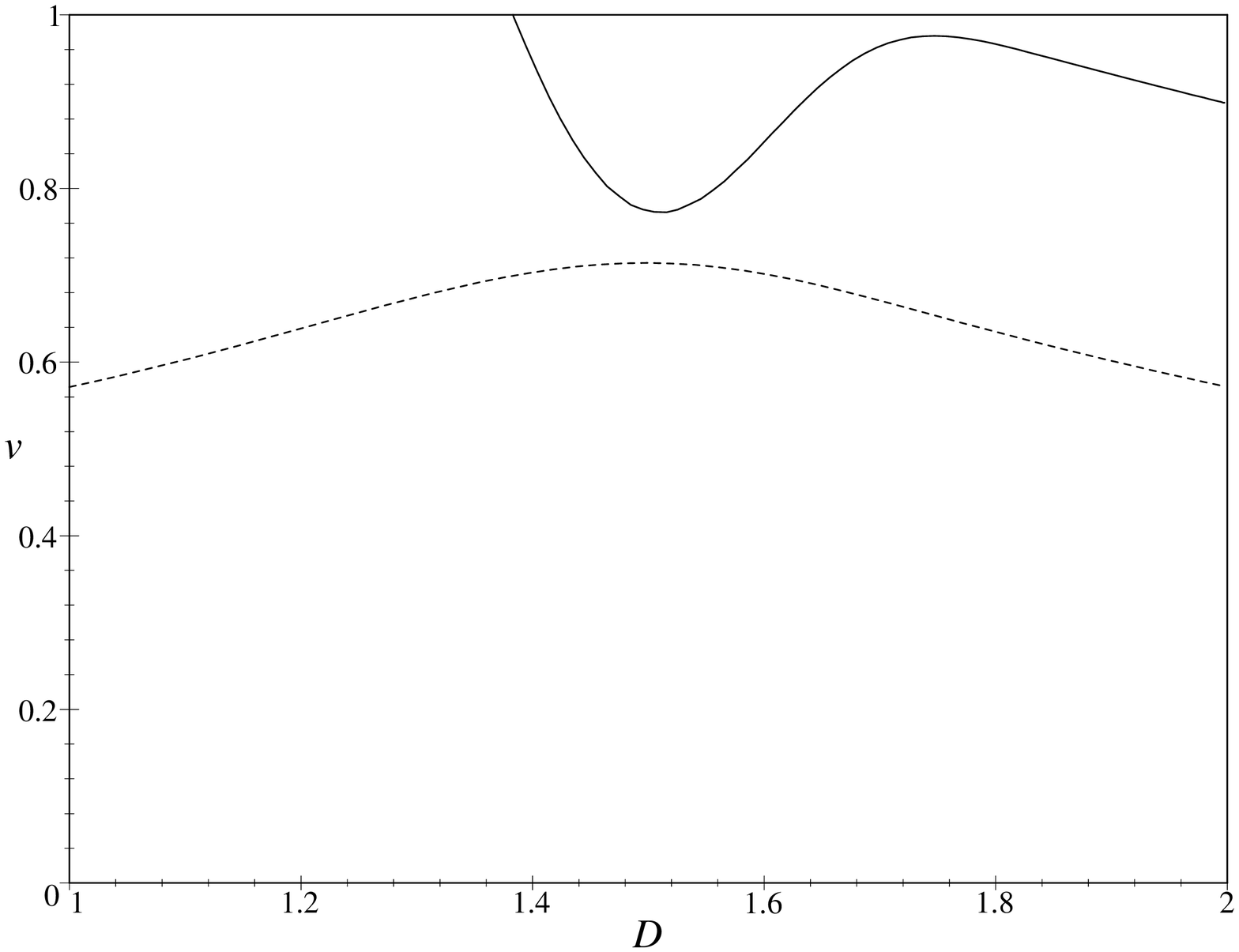}}
}
\centerline{
\Dd{2}{4}%
\epsfxsize=8cm \parbox{8cm}{\epsfbox{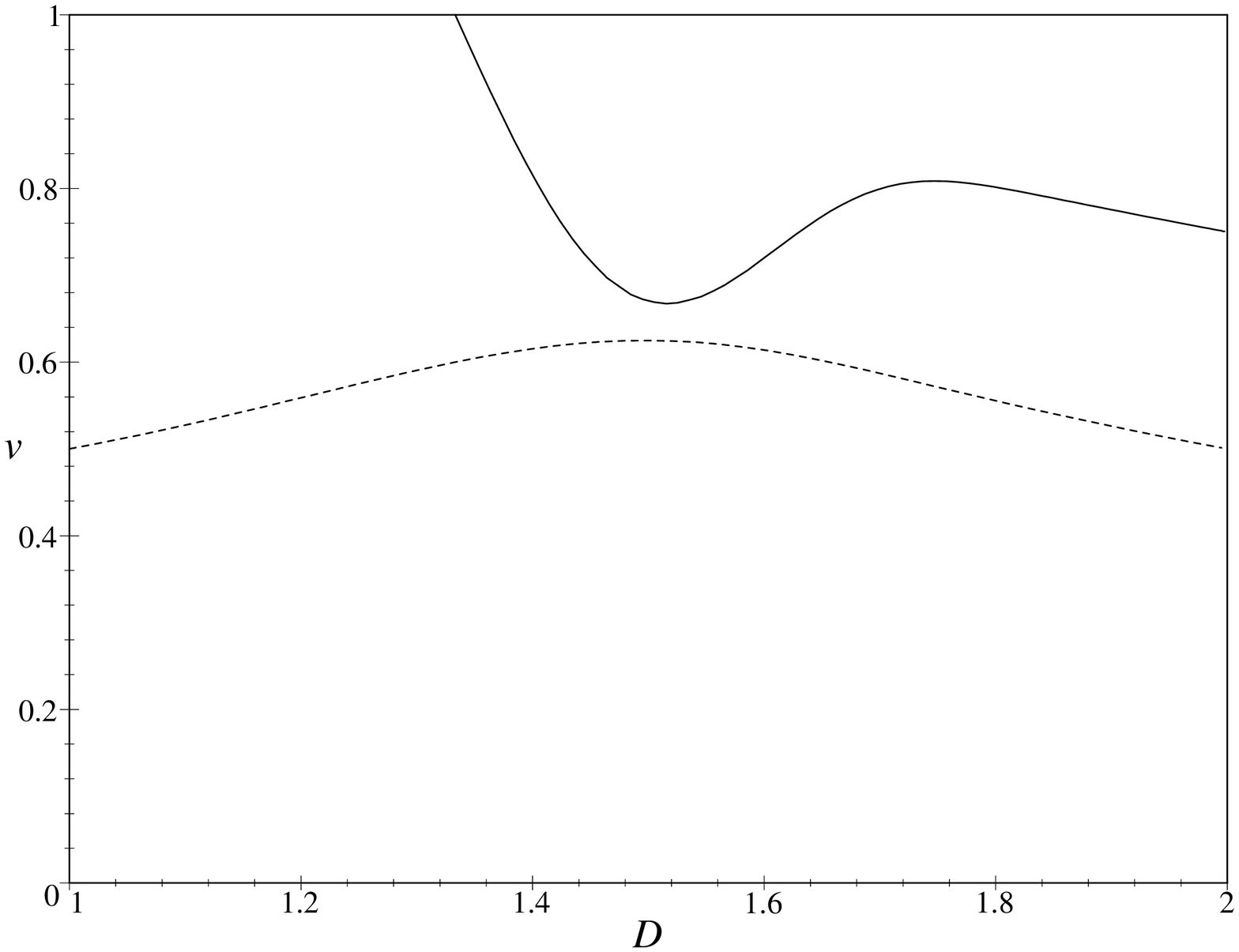}}
\Dd{2}{8}%
\epsfxsize=8cm \parbox{8cm}{\epsfbox{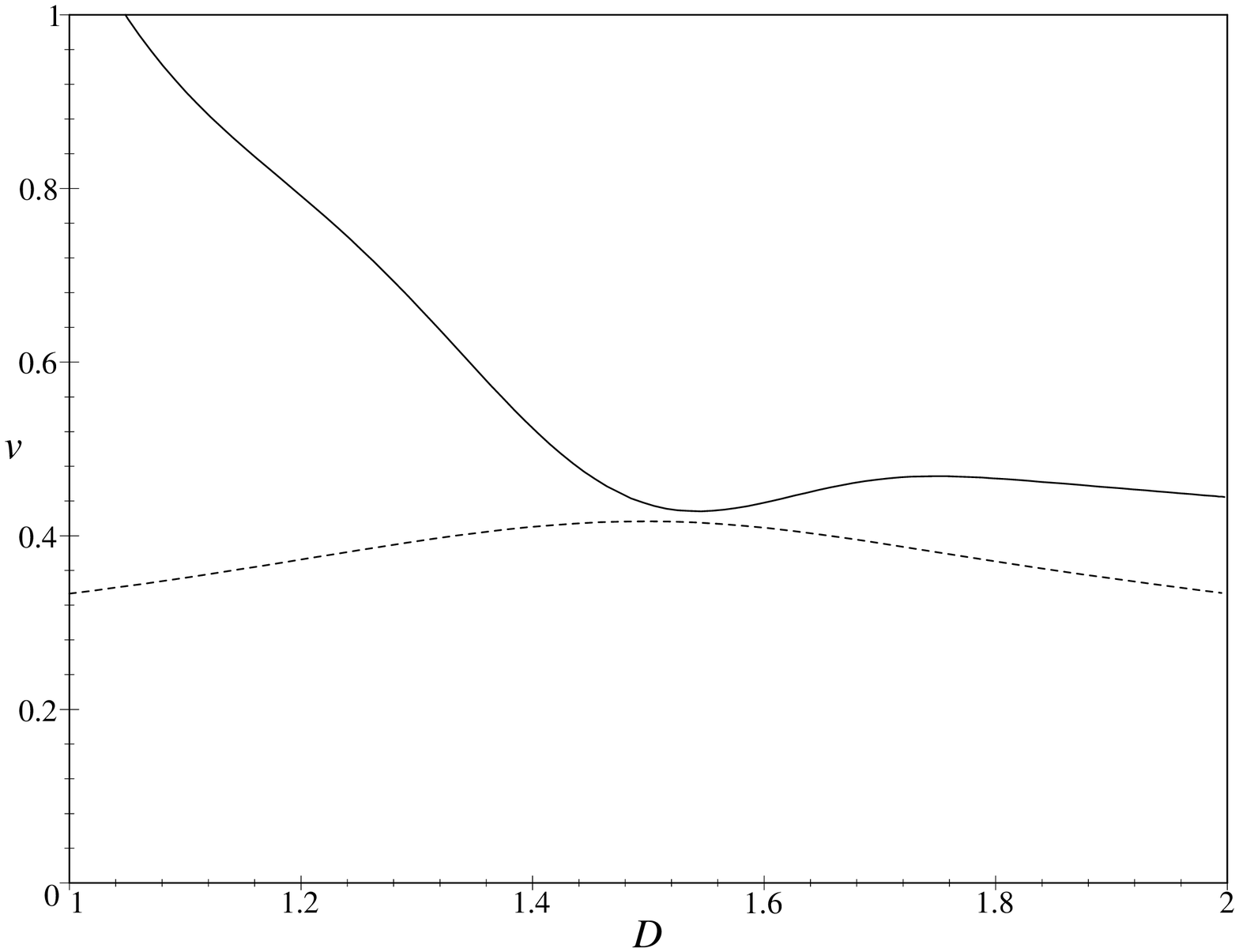}}
}
\centerline{
\Dd{2}{15}%
\epsfxsize=8cm \parbox{8cm}{\epsfbox{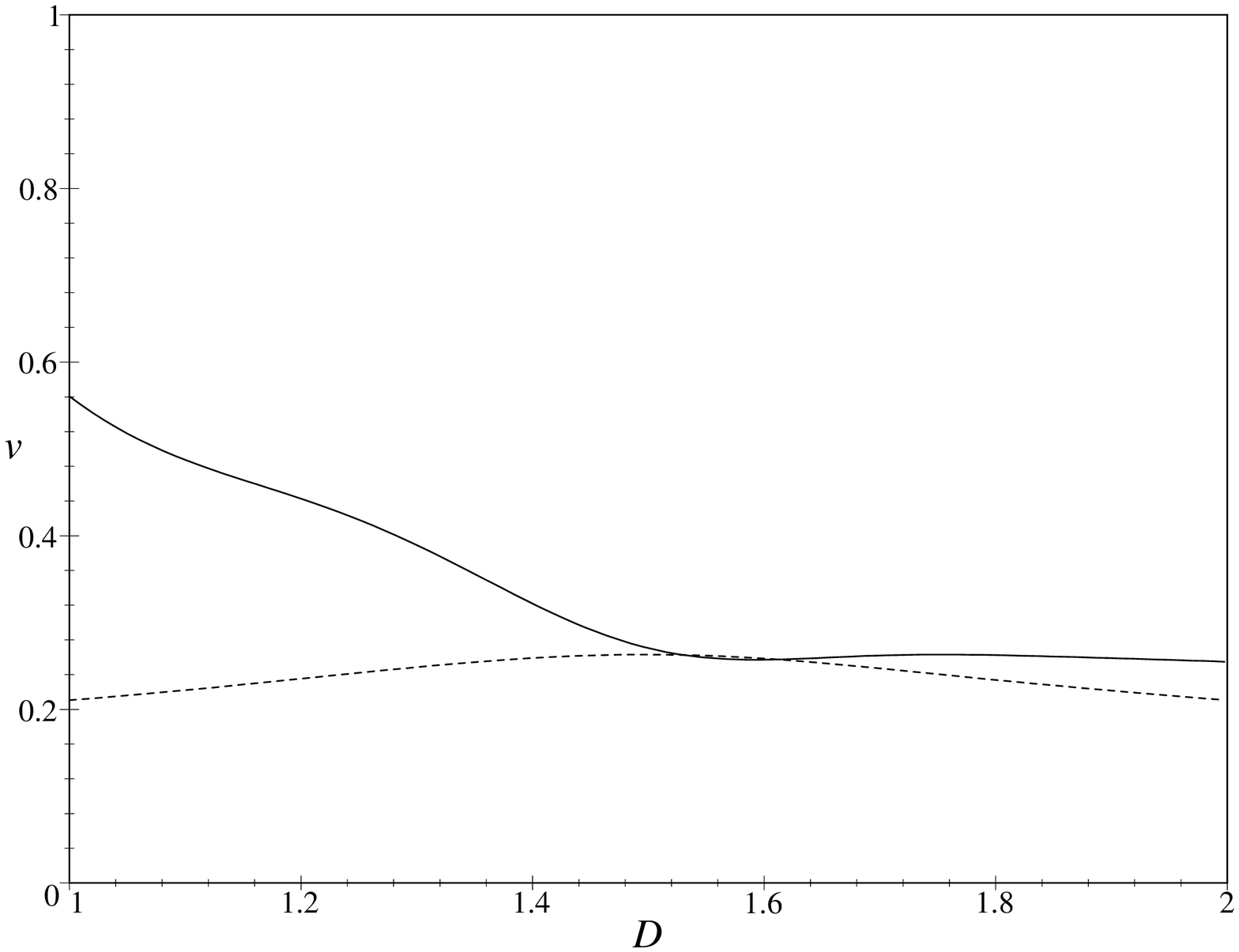}}
\Dd{2}{20}%
\epsfxsize=8cm \parbox{8cm}{\epsfbox{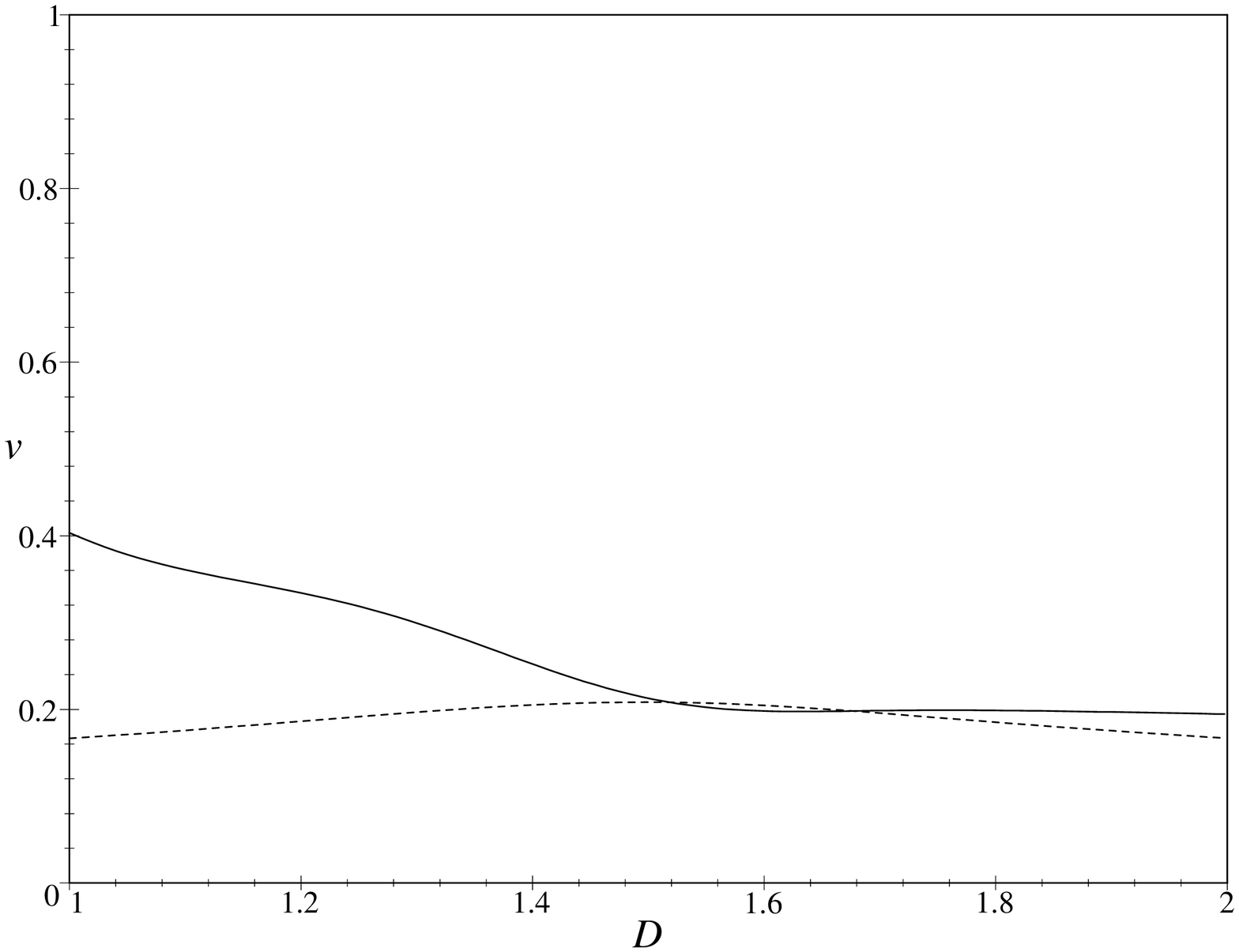}}
}
\caption{Extrapolation for membranes
in $D$ and $D_c(d)$ to $(D,d)=$ (2,2), (2,3), (2,4),
(2,8), (2,15) and (2,20)}
\label{f: nu-D-Dc(d)}
\end{figure} 
\begin{figure}[htb] 
\centerline{
\Dd{1}{1}%
\epsfxsize=8.0cm \parbox{8.0cm}{\epsfbox{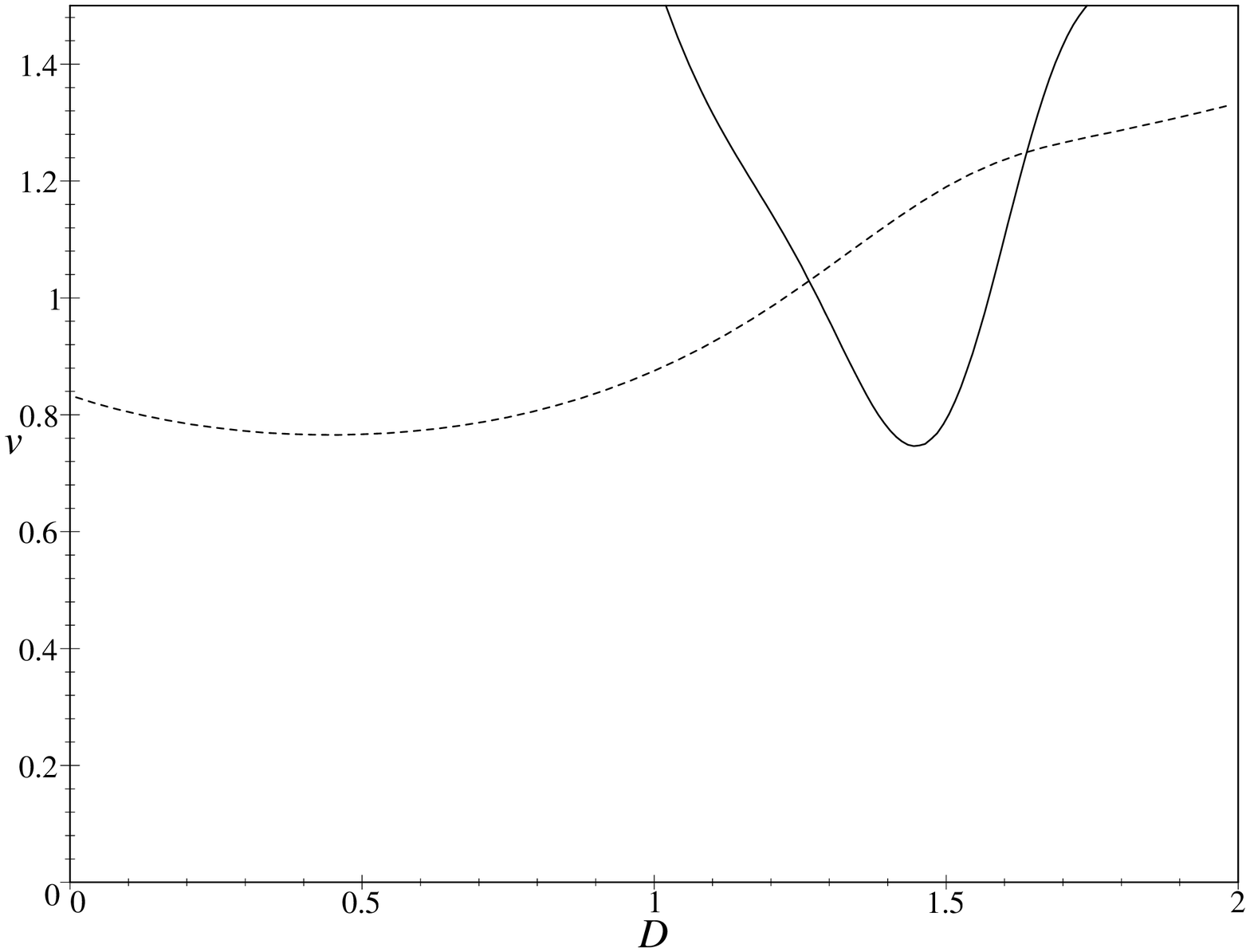}}
\hspace{0.0cm}
\Dd{1}{2}%
\epsfxsize=8.0cm \parbox{8.0cm}{\epsfbox{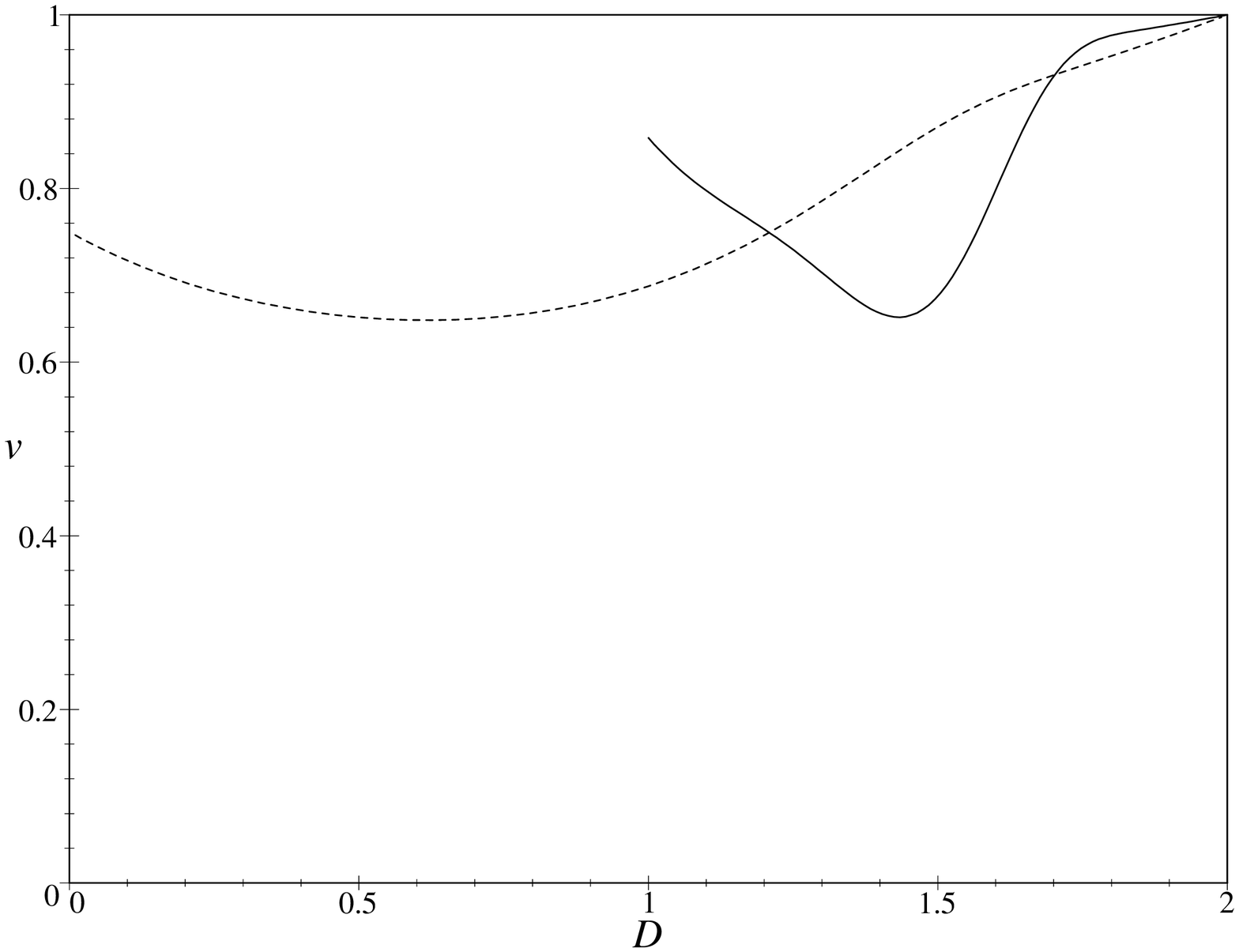}}
}
\centerline{
\Dd{1}{3}%
\epsfxsize=8cm \parbox{8.0cm}{\epsfbox{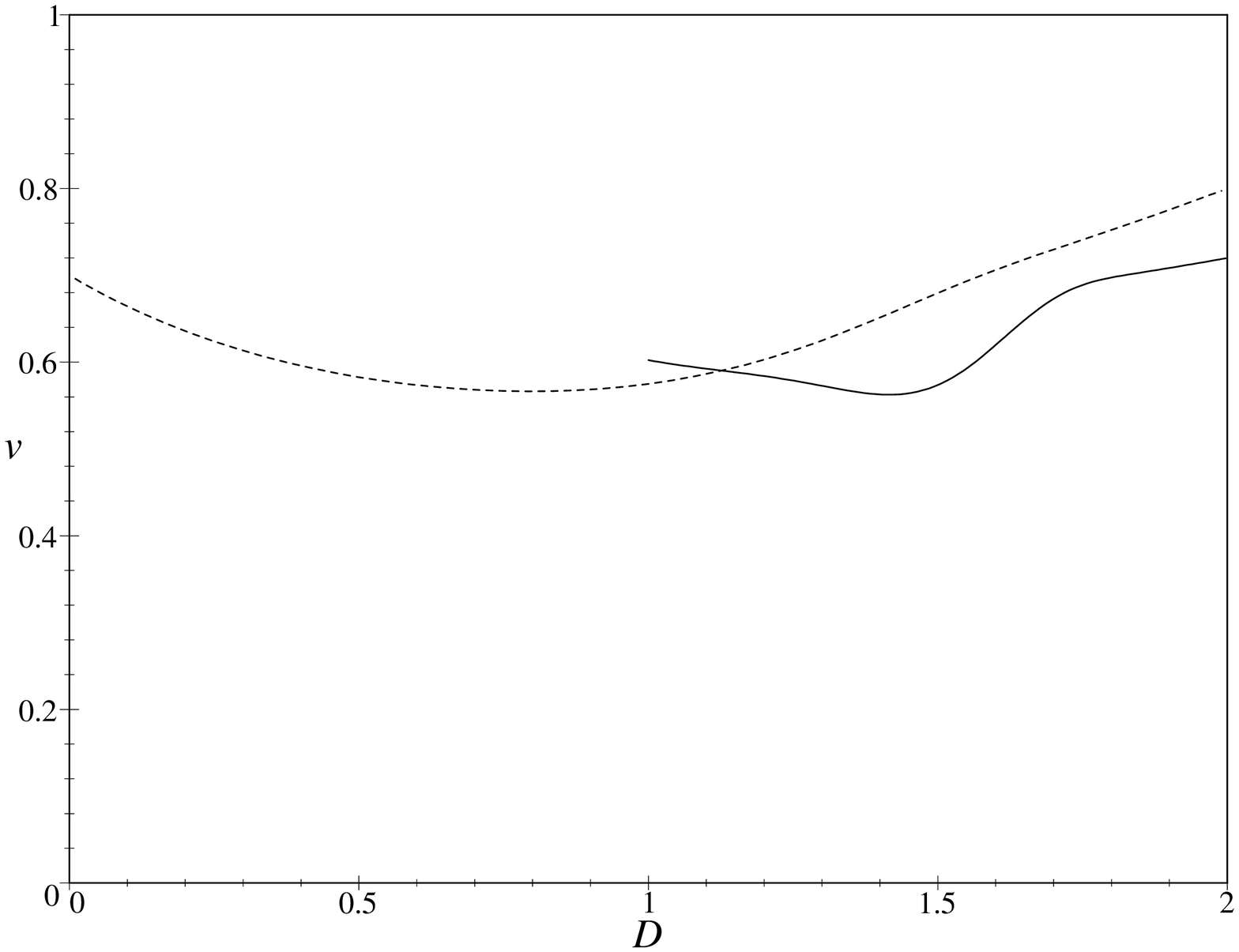}}
\hspace{0.0cm}
\Dd{2}{2}%
\epsfxsize=8cm \parbox{8cm}{\epsfbox{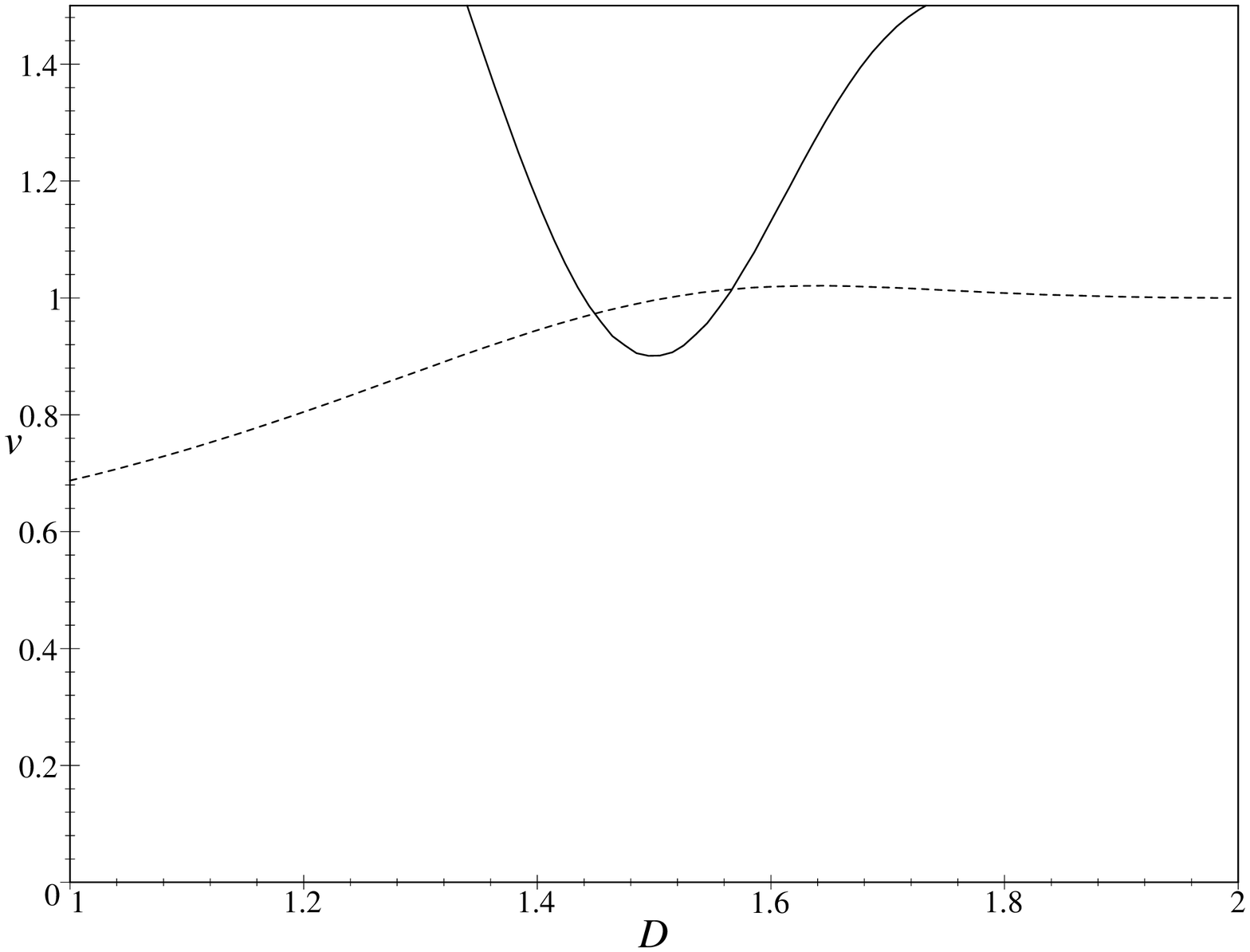}}
}
\centerline{
\Dd{2}{3}%
\epsfxsize=8cm \parbox{8cm}{\epsfbox{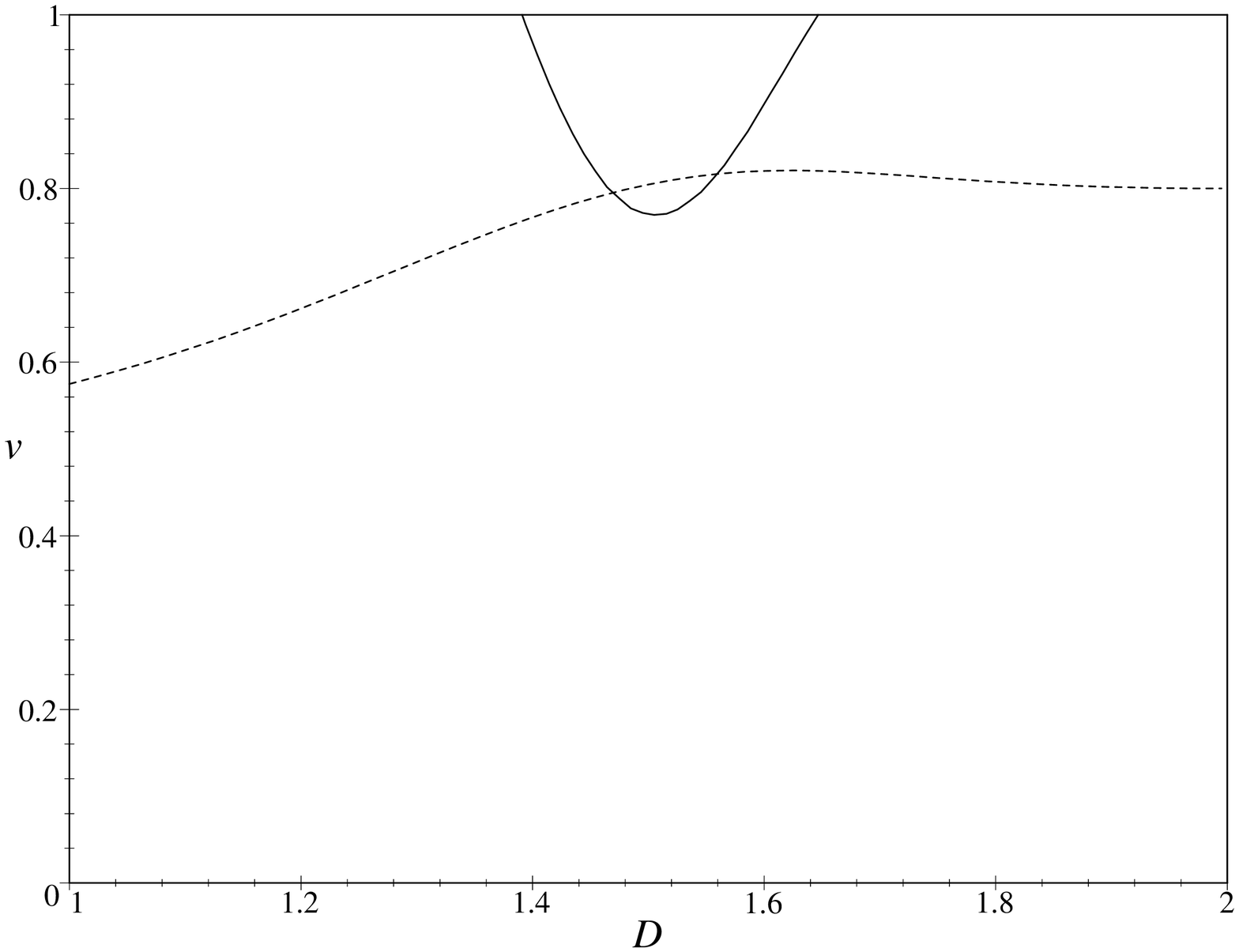}}
\hspace{0cm}
\Dd{2}{20}%
\epsfxsize=8cm \parbox{8cm}{\epsfbox{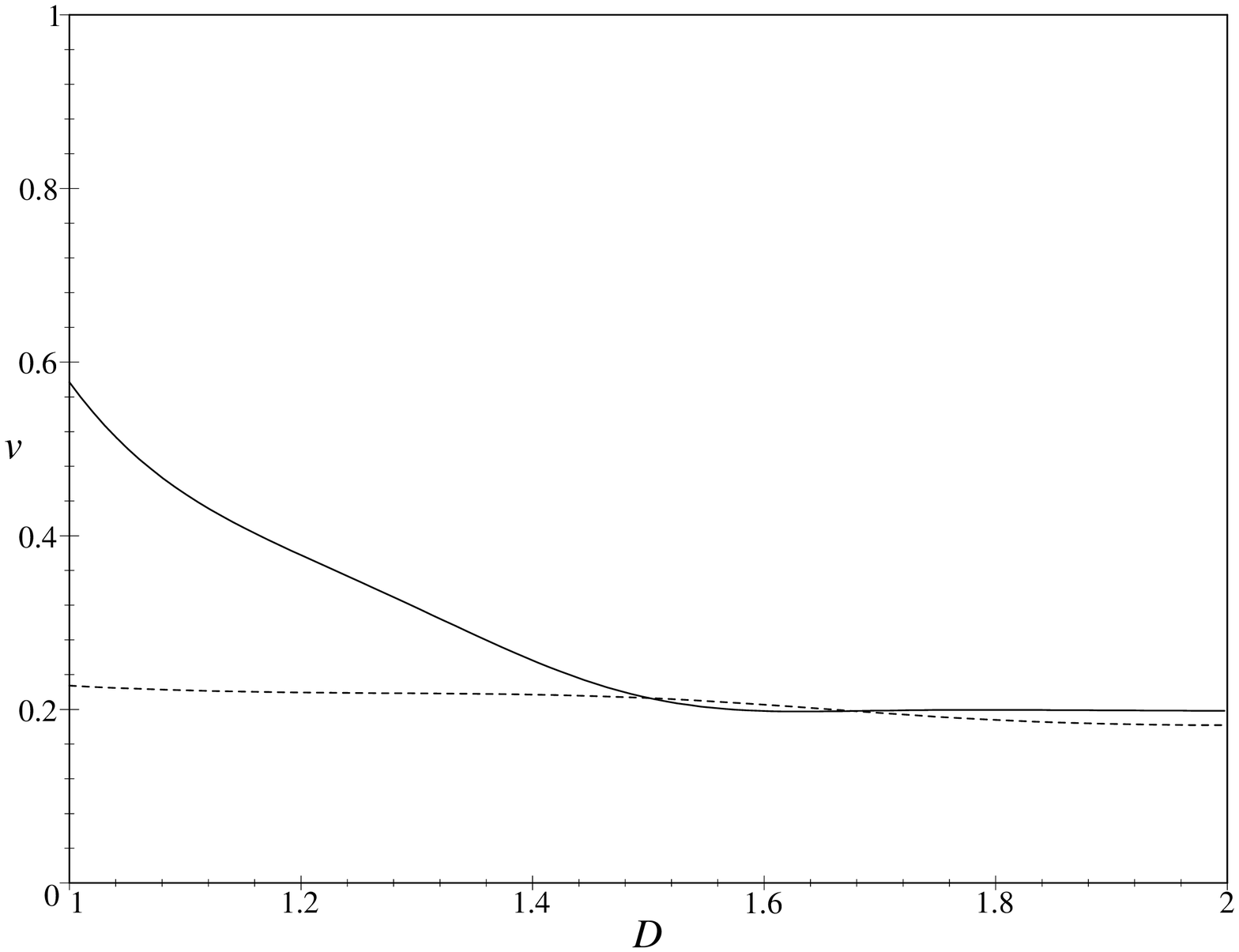}}
}
\caption{Extrapolation in $D$ and $1/(d+2)$ to $(D,d)=$ (1,1), (1,2), (1,3),
(2,2), (2,3) and (2,20)}
\label{f: nu-D-1/(d+2)}
\end{figure}
\begin{figure}[htb] 
\centerline{
\Dd{1}{1}%
\epsfxsize=8cm \parbox{8cm}{\epsfbox{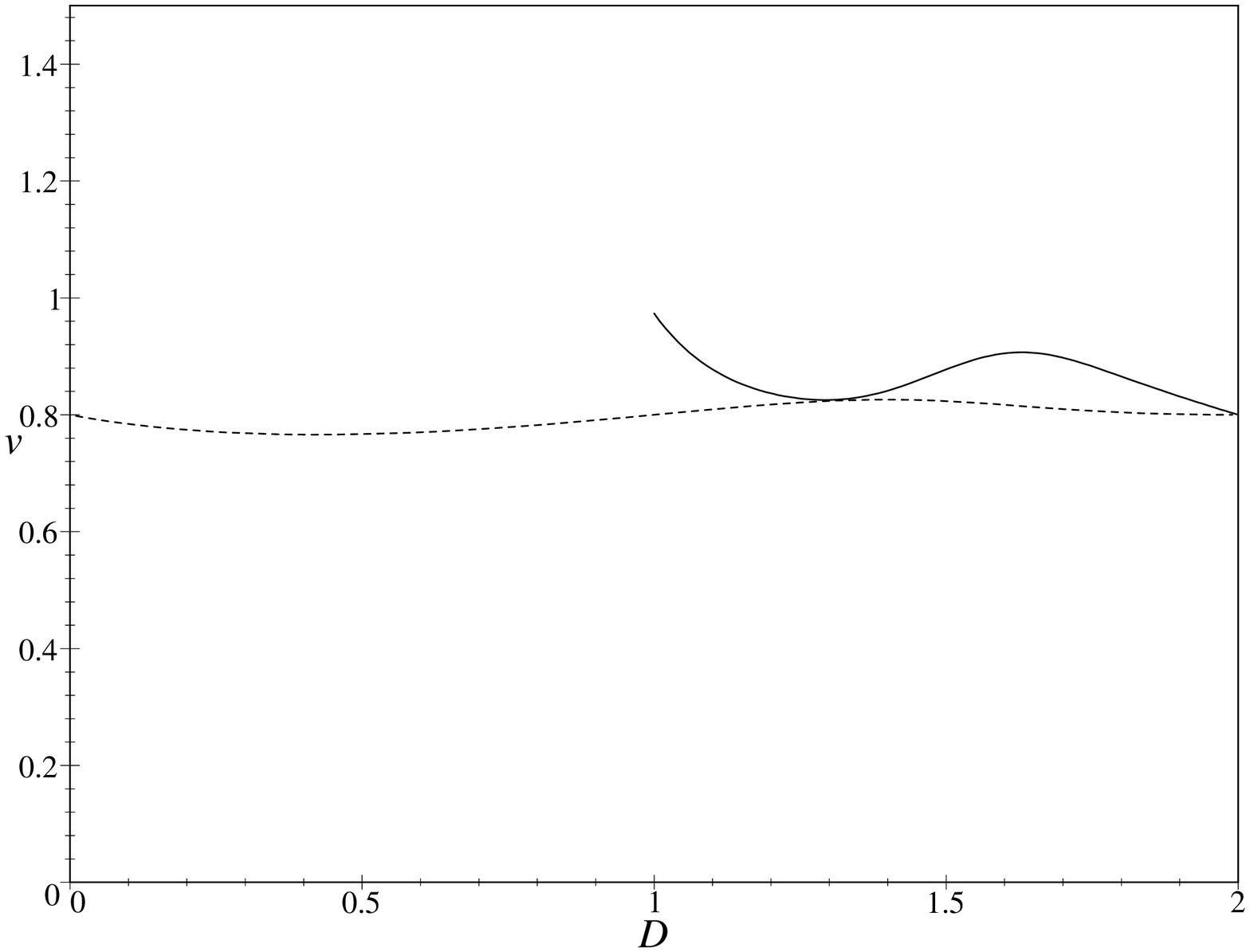}}
\Dd{1}{2}%
\epsfxsize=8cm \parbox{8cm}{\epsfbox{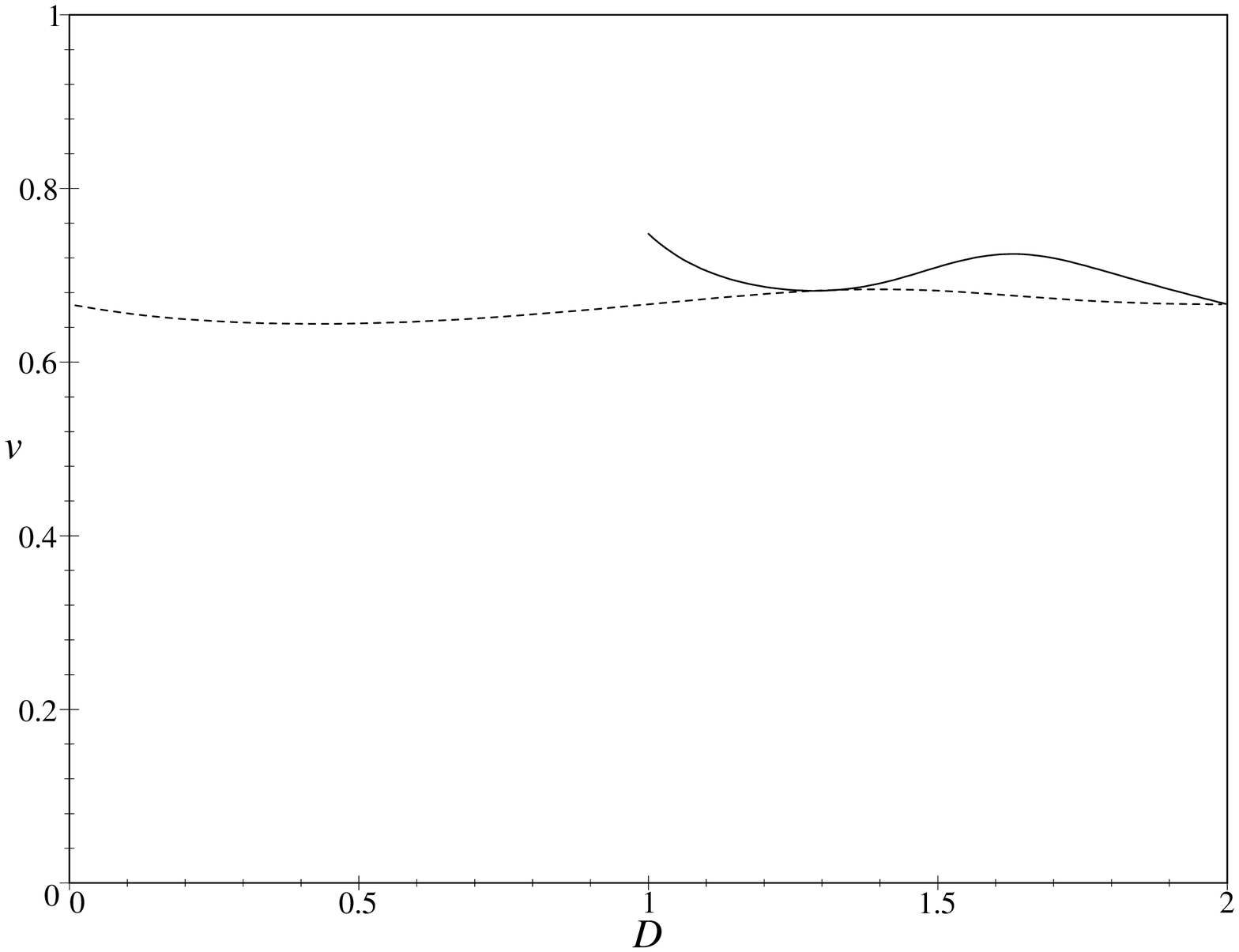}}
}
\centerline{
\Dd{1}{3}%
\epsfxsize=8cm \parbox{8cm}{\epsfbox{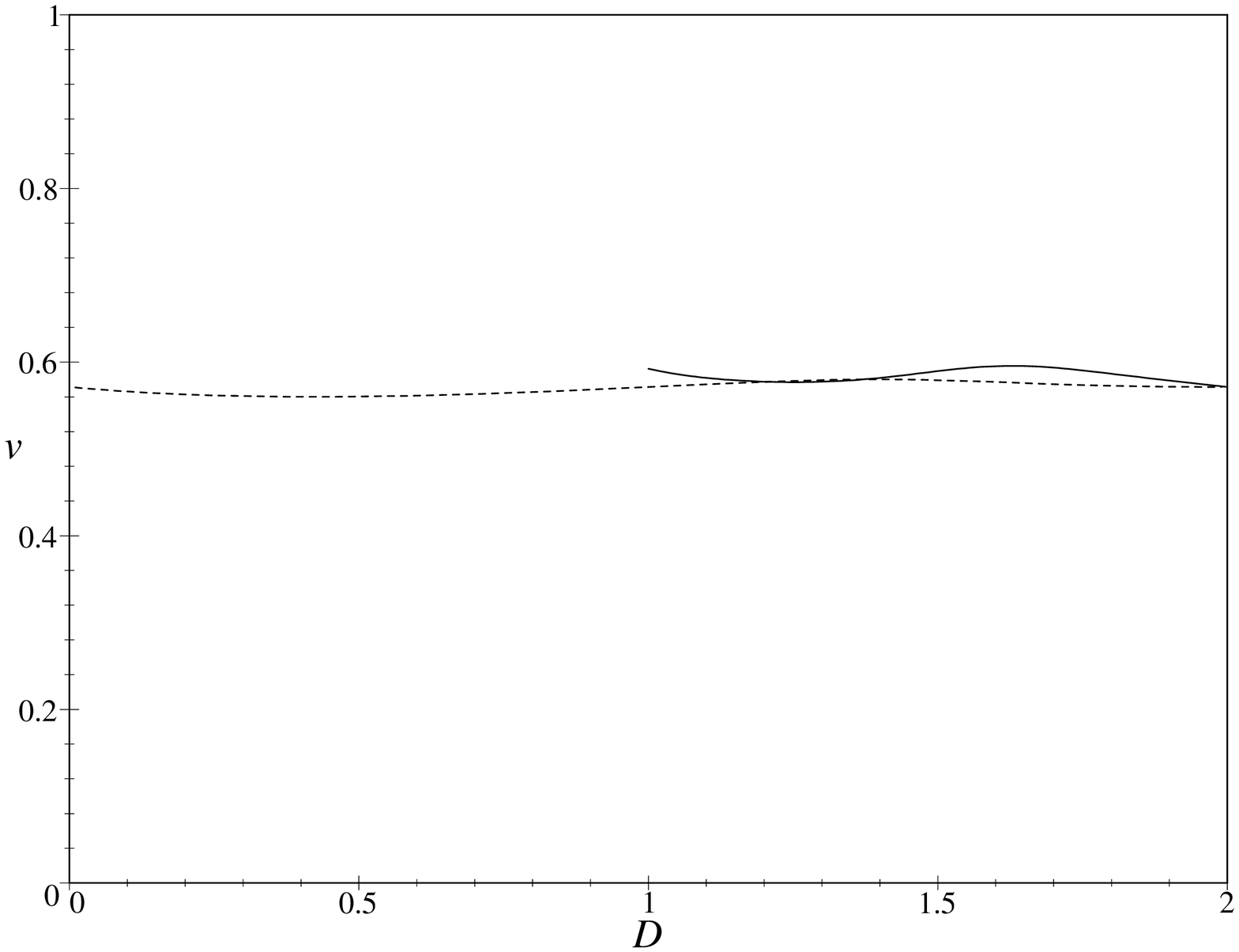}}
 \parbox{8cm}{}
}
\caption{Extrapolation for polymers in $D_c(d)$ and $\E$ to $(D,d)=$ (1,1), (1,2) and (1,3)}
\label{nu1-Dc(d)-eps}
\end{figure} 
\begin{figure}[htb] 
\centerline{
\Dd{2}{2}%
\epsfxsize=8cm \parbox{8cm}{\epsfbox{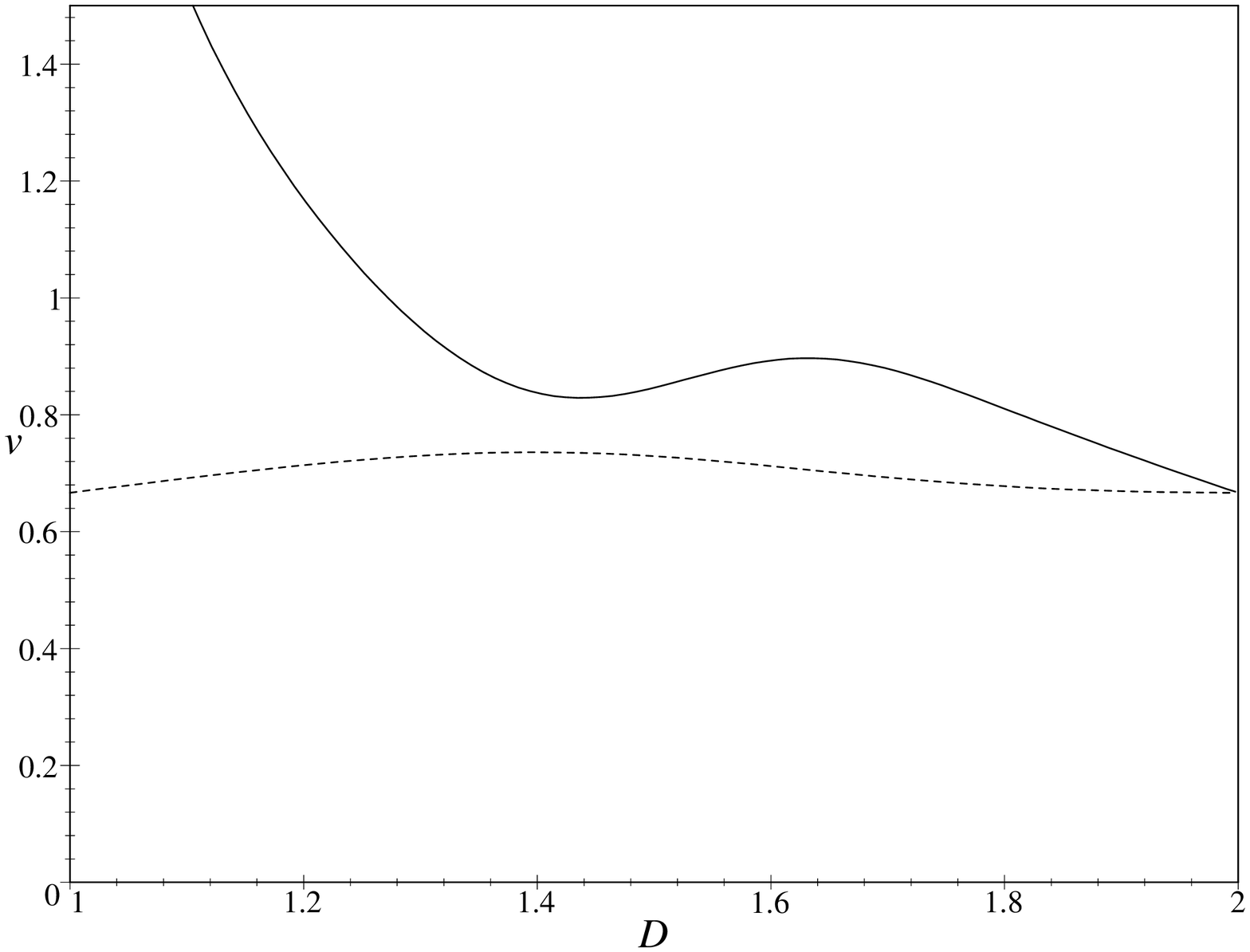}}
\Dd{2}{3}%
\epsfxsize=8cm \parbox{8cm}{\epsfbox{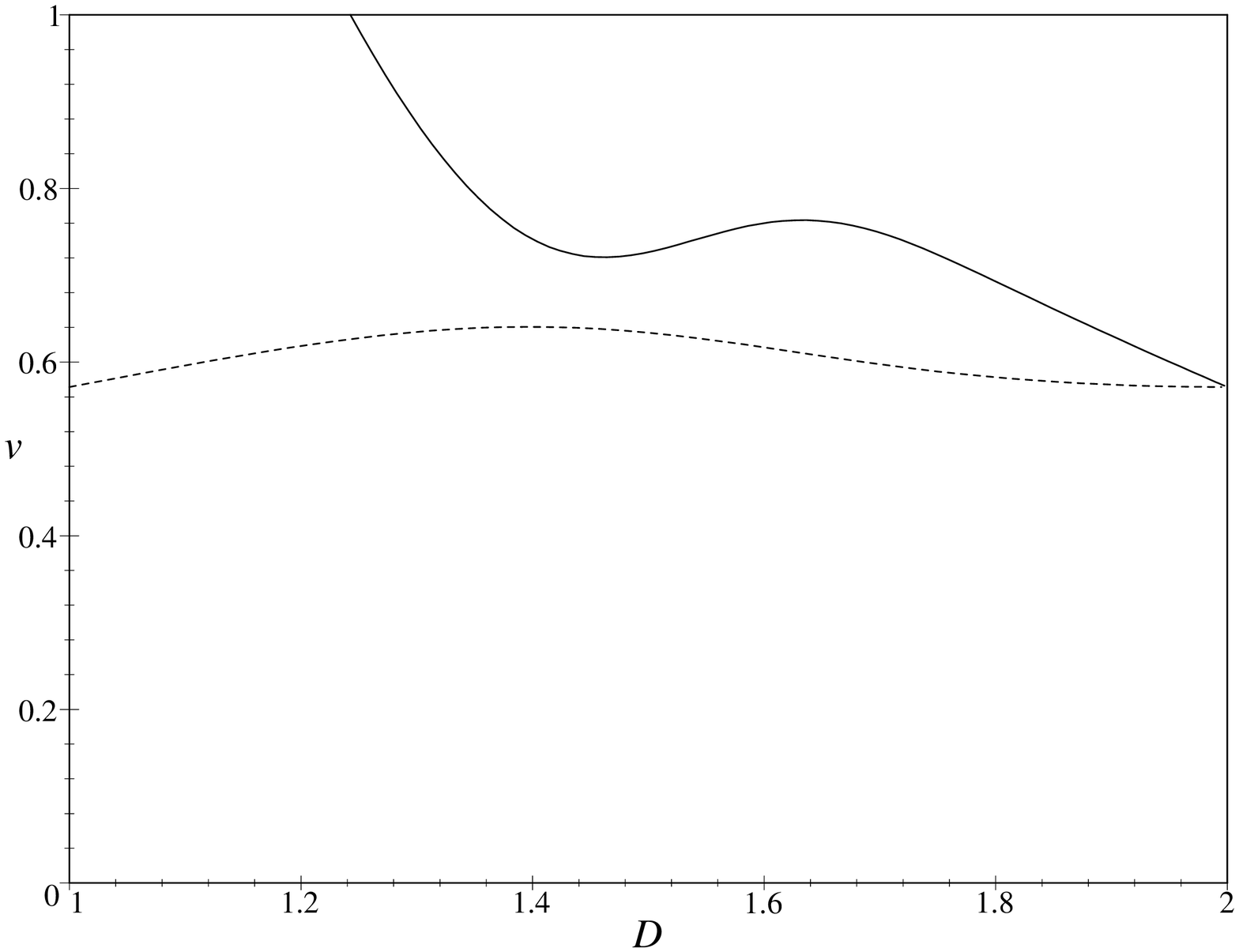}}
}
\centerline{
\Dd{2}{4}%
\epsfxsize=8cm \parbox{8cm}{\epsfbox{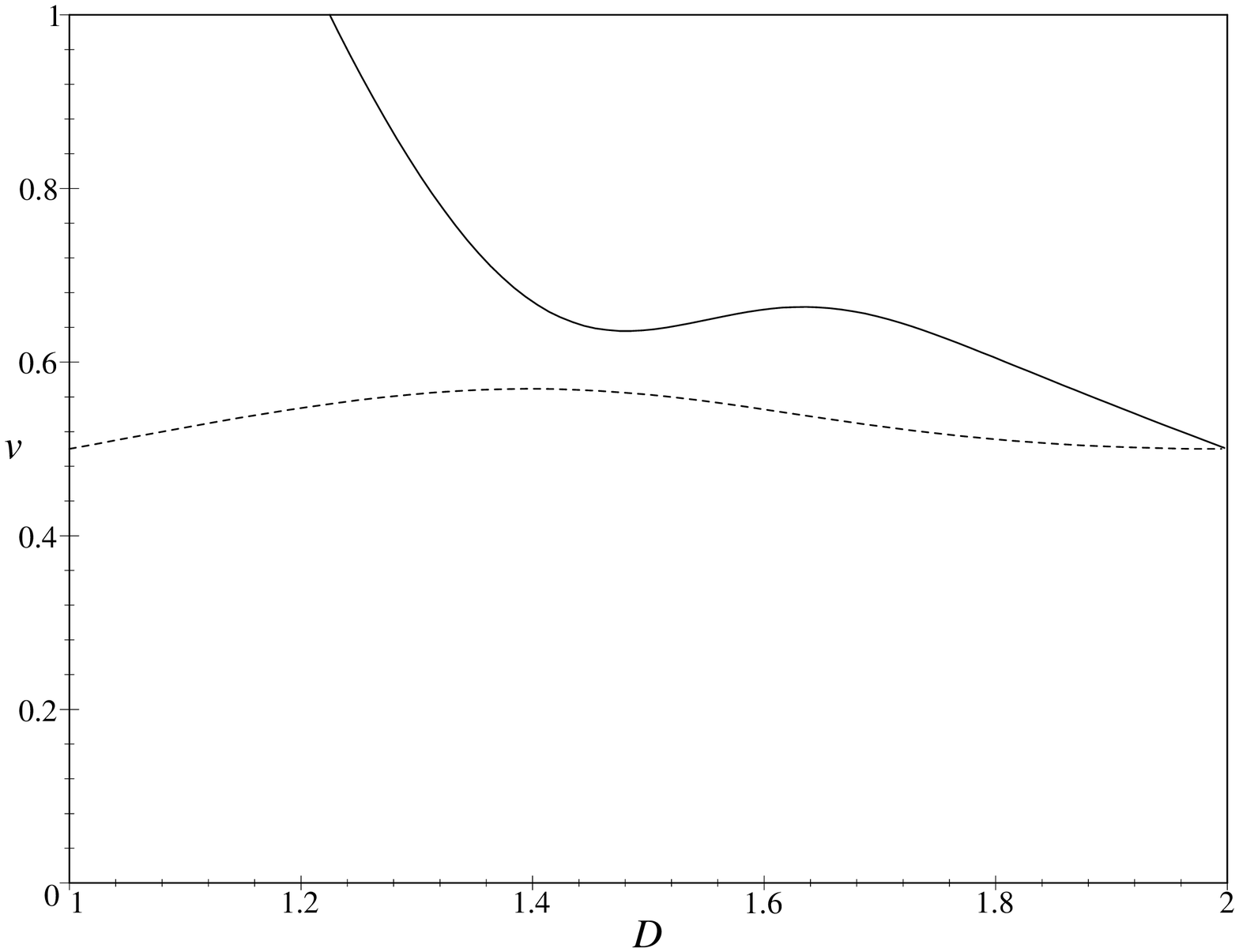}}
\Dd{2}{8}%
\epsfxsize=8cm \parbox{8cm}{\epsfbox{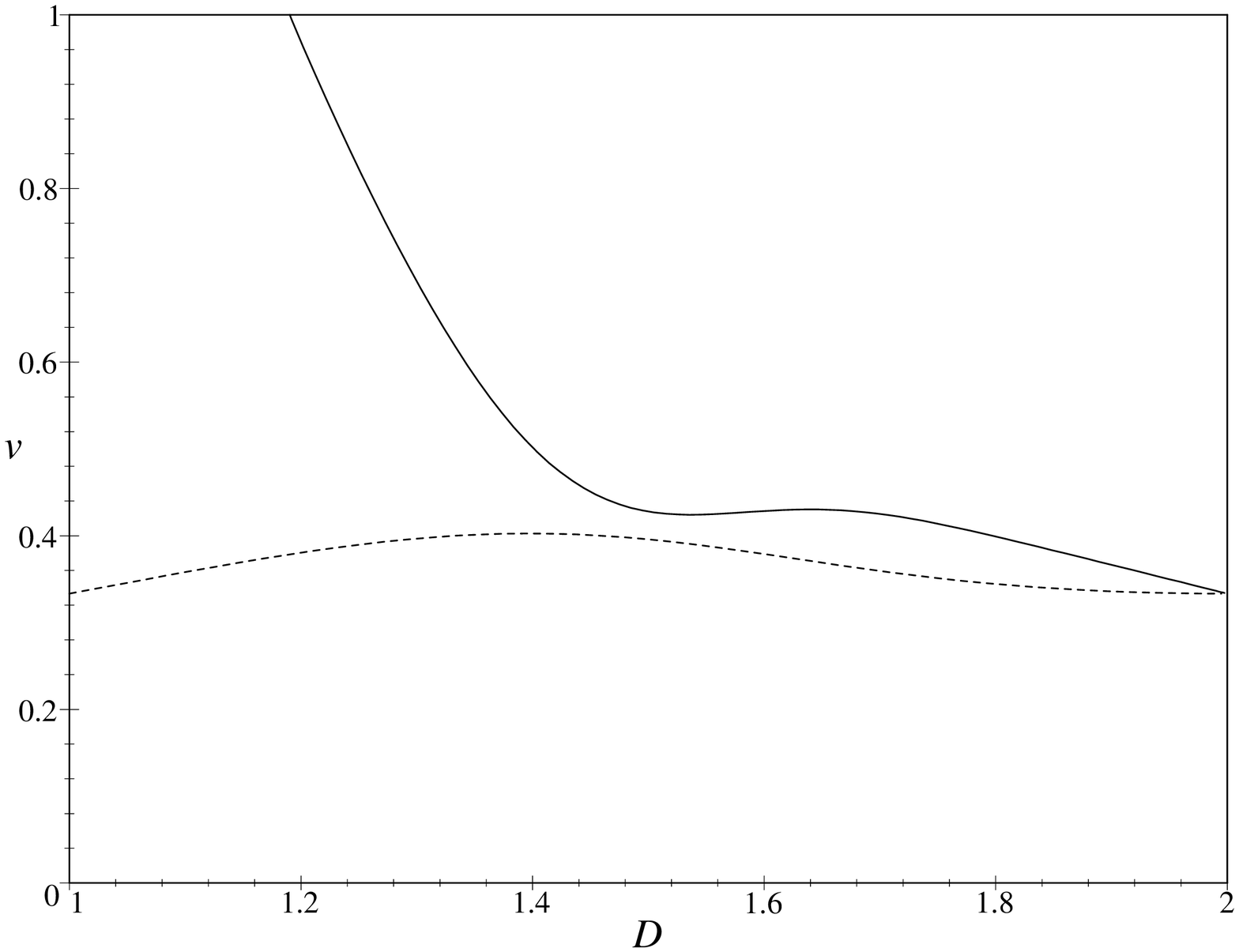}}
}
\centerline{
\Dd{2}{15}%
\epsfxsize=8cm \parbox{8cm}{\epsfbox{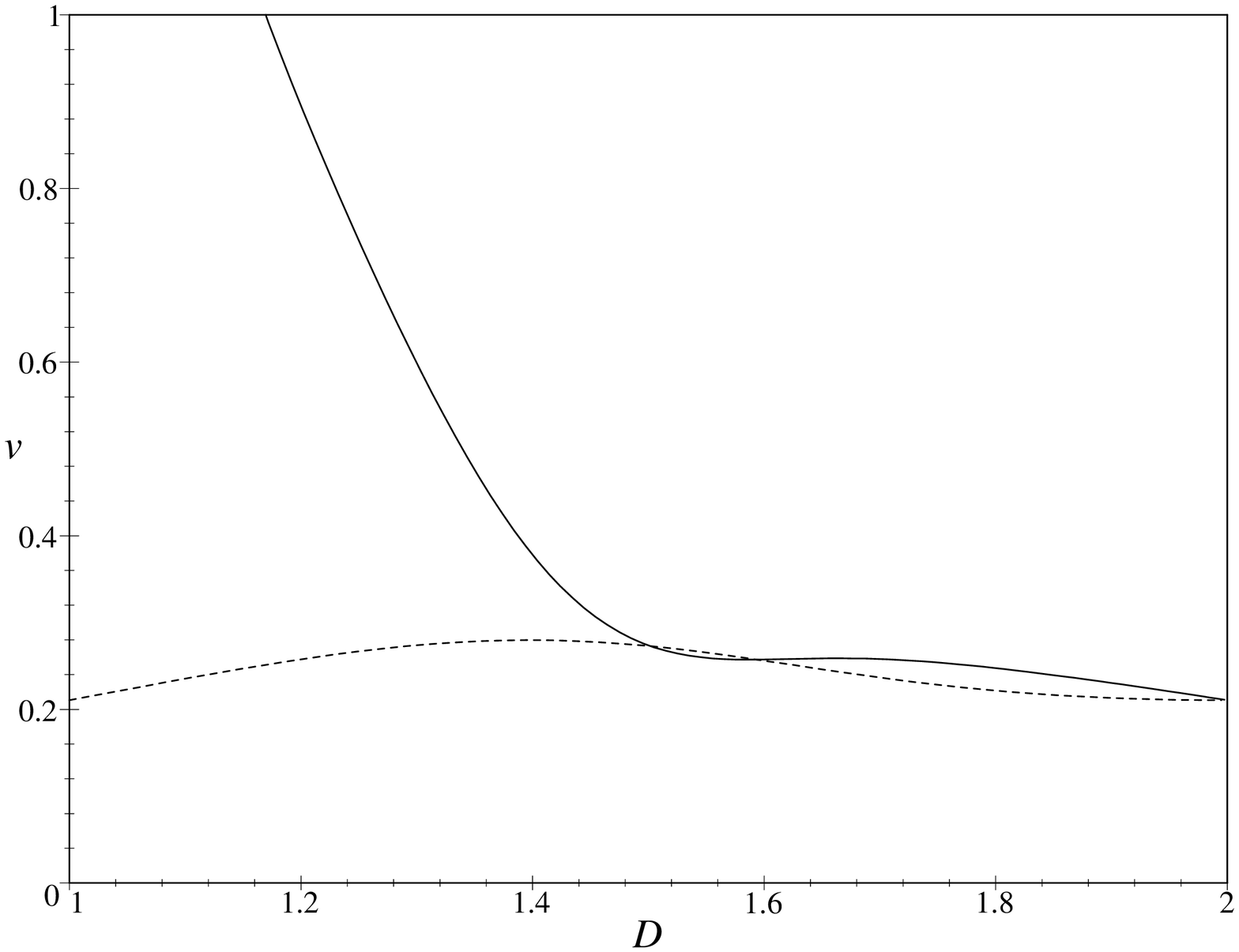}}
\Dd{2}{20}%
\epsfxsize=8cm \parbox{8cm}{\epsfbox{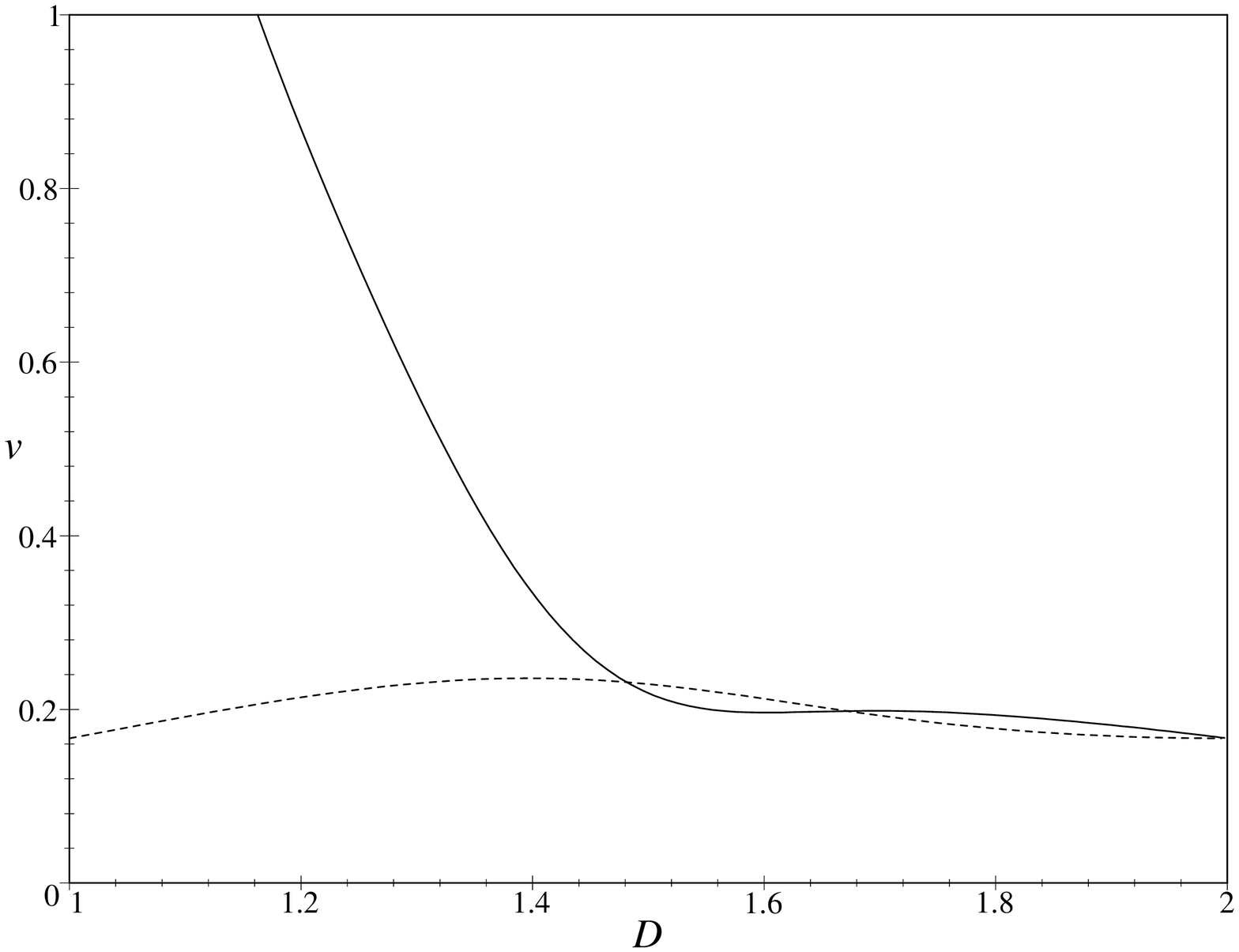}}
}
\caption{Extrapolation in $D_c(d)$ and $\E$ to $(D,d)=$ (2,2), (2,3), (2,4),
(2,8), (2,15) and (2,20).}
\label{nu-Dc(d)-eps}
\end{figure} 
We can rewrite \eq{e:nustar} as an $\E$-expansion for the critical exponent
$\nu$ in terms of $\E$ and $D$.
\bea \label{e:nuuu}
\nu^*=\nu(D,d) &=& \nu_0(D) + \nu_1(D) \E (D,d) + \nu_2(D) \E^2(D,d) + \ldots \\
\E(D,d) &=& 2D - \frac{2-D}2 d 
\eea
The result for the coefficients is given in figure \ref{f:nu12}. 
In order to proceed, we shall use a polynomial interpolation for 
$\tilde f_1(D)$ and for $\tilde c_1(D)/a_1(D)$. 
(A polynomial interpolation for $\tilde c_1(D)$ is bad as this
term vanishes exponentially with $1/(2-D)$.)

We now have to use these 2-loop results to calculate the critical
exponent $\nu$ for self-avoiding two-dimensional membranes $(D=2)$
as a function of the dimension $d$ of space and to compare the results
with the previous 1-loop results for $\nu$.

As already stressed in previous works, the $\E$-expansion
given by \eq{e:nuuu} cannot be used directly for membranes, as it can be done
for polymers.
There one fixes $D=1$ and uses sophisticated resummation
methods to evaluate $\nu$ for $d=3$ ($\E=1/2$) and even for $d=2$ ($\E=1)$.
Indeed, directly setting $D=2$ in the $\E$-expansion for $\nu$ gives a trivial,
but absurd, result, since all the terms $\nu_n(2)$ of the $\E$-expansion
vanish!
Moreover when $D=2$, $\E=4$ irrespective of the value of $d$.
This simply means that the point $D=2$, $\E=0$ (which corresponds to
$d_c=\infty$) is a singular point and that it is not possible to
perform a direct $\E$-expansion around it.
Instead, one may perform a similar expansion, starting from another point
$(D_0, d_0=\frac{4D_0}{2-D_0})$ on the critical curve ($\E=0$).
This approach has already been used in \cite{Hwa} to extrapolate the
1-loop results.
It introduces two different kinds of arbitrariness in the extrapolations.
Firstly one is a priori free to chose any starting point on the
critical curve $\E=0$ (or at least any point in some subset of this curve).
Secondly there is an arbitrariness in the ``path of extrapolation" 
which goes from the critical curve to the physical point $(D=2, d)$.
Of course if one knows an exact resummation procedure of the full series
in $\E$, one should obtain the same result for any starting point and for any 
path of extrapolation.
However, if one has a finite number of terms of the $\E$-expansion, even for an
adequate resummation method, the final result will depend on this
arbitrariness.

In \cite{Hwa} a minimal sensitivity method was introduced by Hwa to analyze the
1-loop results.
He chose a  given extrapolation method and selected the starting point
$(D_0,d_0)$
on the critical curve which gives an estimate for $\nu$, $\nu(D_0)$
which is the less sensitive to the choice of the starting point.
This method works well for polymers and gives interesting results for
membranes.
However, when one goes to 2-loops, it delivers largely varying
or even diverging estimates for $\nu(D_0)$ and in addition
his choice for the extrapolation path is somewhat arbitrary.

We shall use a generalization of the methods introduced in \cite{Hwa}.
Note that the expansion (\ref{e:nuuu}) 
is exact in $D$ and of order 2 in $\E$, thus it can be expanded up to 
order 2 both in $D-D_0$ and $\E$.
Now we can change our extrapolation path through any invertible transformation 
$\{ x,y \} = \{x(D,\E), y(D,\E) \}$.
One can express $D$ and $\E$ as function
of $x$ and $y$ and re-expand $\nu$ up to order 2 in $x$ and $y$
around the point $(x_0,y_0)$ on the critical curve.
\bea
\nu(D,\E)&&=\tilde\nu(x,y)= 
\tilde\nu_{0,0}(x_0,y_0)+\Delta x\tilde\nu_{1,0}(x_0,y_0)
+\Delta y\tilde\nu_{0,1}(x_0,y_0)
\nonumber\\
&&\hskip-2.em+(\Delta x)^2\tilde\nu_{2,0}(x_0,y_0)
+2\Delta x\Delta y\tilde\nu_{1,1}(x_0,y_0)
+(\Delta y)^2\tilde\nu_{0,2}(x_0,y_0)
+\cdots
\nonumber\\
&&
\hskip 5em \Delta x=x-x_0\ ,\ \ \ \Delta y=y-y_0
\label{e:nuxy}
\eea
The goal is to find an optimal choice of variables $\{x,y\}$.
Our guidelines for such a choice are the following:
({\it i}\/) the estimate for $\nu$ should depend ``the least" on the choice of 
the expansion point on the critical curve,
({\it ii}\/) it should reproduce well the known result for polymers
$(D=1)$,
({\it iii}\/) for membranes $(D=2)$ the $d\to \infty$ limit should not be singular
and we should get results close to those obtained by a Gaussian variational
approximation. This last point is not arbitrary and will be justified below.
It turns out to be quite stringent.

Finally, we must choose some resummation procedure to extrapolate $\nu$ from
the know\-ledge of the series (\ref{e:nuxy}) up to order 2.
Since we have only a few terms and since we do not have insight in the
large order behavior of these
series or in the analytical structure of the resummed series, we cannot use
sophisticated resummation methods
(for instance those based on Borel transforms).
Therefore we shall always use the truncated series at order 2 and boldly sum
its terms.

Let us now discuss possible extrapolation variables.
The simplest choice is to take $D$ and $\E$.
This works well for polymers $(D=1)$, since both at 1- and 2-loop order we 
get results which are quite stable with respect to $D_0$.
However this gives very poor results for membranes $(D=2)$, since both
at 1- and 2-loop order the results depend very much on $D_0$.
This could be expected, since in this case $\E=4$ independent of $d$.
See figure
\ref{fig: nu-D-eps}. 
In the case of the membrane, no prediction is possible. 

Another possibility is to expand in $D$ and $d$.
This expansion gives reasonable results for polymers, but poor results for
membranes.
This can be seen by looking on the results of figure~\ref{f: nu-D-d} and
is not surprising, since if we apply this extrapolation method to the 
Gaussian variational estimate $\nuvar=2D/d$ it  also gives poor results,
although the result $\nuvar$ is expected to be close to the exact $\nu$ at
large $d$ (as argued in subsection~\ref{Variational method and perturbation expansion}).

A more interesting choice is to use $D$ and $D_c(d)= 2d/(4+d)$ or equivalently
$D$  and $1/(d+4)$.
This expansion has the advantage to represent the critical curve $D_0$,
$d_c(D_0)$ as a straight line.
For polymers in 3 dimensions this method delivers a remarkable broad plateau,
i.e. the extrapolated value of $\nu$ is relatively independent of
the expansion-point $D_0$.
For polymers in 2 and 1 dimensions we still have a plateau when $D_0\to 2$,
which delivers 2-loop extrapolations close to the exact results ($\nu=0.75$ and
$\nu=1$ respectively)
(see figure~\ref{f: nu1-D-Dc(d)}).
For membranes, this method also delivers  interesting results (see figure~\ref{f: nu-D-Dc(d)}).
For large $d$, we find a stable plateau at 2-loops when $D_0\to 2$, which
gives for $\nu$ a result very close to the variational estimate
$\nuvar=4/d$, while 
at smaller $D_0$, the plateau stops and the 2-loop estimate for $\nu$
increases sharply as $D_0\to 1$.
For small $d$, the $D_0\to 2$ plateau becomes an oscillatory region,
still followed by a sharp increase for $D_0\to 1$.
In this case, to calculate $\nu$, we use the minimum and the maximum 
of $\nu(D_0)$.
Whereas the first is expected to be an underestimation, the second is expected 
to be an overestimation.
We also give their mean-value, cf.\ figures \ref{tab: nu} 
and \ref{f: nu-D-Dc(d)}.
One might think of developing in $D$ and $1/(d+c)$, with $c\ne 4$.
$c=0$ is suggested by the variational ansatz.
In fact we prefer to take $c=2$, which is suggested by the prediction of
the Flory argument (see subsection\ref{s:Flory}).
We obtain similar estimates as above for $d$ large, but larger variations
for smaller $d$.
For this case we evaluated $\nu$ by the request that the second
order corrections should vanish.
We find the results in figure \ref{tab: nu} and \ref{f: nu-D-1/(d+2)}.
The predictions for $\nu$ are good in the known cases and reasonable for
membranes in 3 dimensions. 

Another promising method is the expansion in $\E$ and $D_c(d)$.
This expansion is also regular for $D\to 2$ and is perhaps more in the spirit
of an $\E$-expansion.
Let us discuss the features of this expansion in more detail.
See figure \ref{nu-Dc(d)-eps}.  
For polymers in 3 dimensions we find the flattest plateau of all extrapolation
methods.
For membranes in 3 dimensions the prediction at 1-loop order (dashed line) 
is essentially independent of the expansion point.
For membranes in large dimension, at 2-loop order the 
estimate starts from the 1-loop result at $D=2$, grows until it
reaches a plateau, where $\nu\approx\nu_{var}$ and then grows rapidly again.
For smaller $d$, there is still a plateau and in order to extract $\nu$ from
figure~\ref{nu-Dc(d)-eps}, one uses the  maximum and the minimum of the plateau.
Their mean is an estimate for $\nu$, their difference an estimate of the error
in {\em this} expansion scheme.

The results for $\nu$ from the various extrapolations are summarized in
figure~\ref{tab: nu}.
\begin{figure}[tb]%
\centerline{ \renewcommand{\arraystretch}{1.2}
\begin{tabular}[t]{|l|c|c|c|c|c|c|c|c|} \hline
$(x,y)\ ;\ (D,d)$ & $(1,1)$ & $(1,2)$ & $(1,3)$ & $(2,2)$ & $(2,3)$ & (2,4) & (2,8) & $(2,20)$
\\ \hline \hline
${\mbox{exact}}$ & 1.0 & 0.75 & $0.586(4)$ & 1  & --- & --- & --- & ---
\\ \hline
${\mbox{Flory}}$ & 1.0 & 3/4 & 3/5 & 1 & 4/5 & 2/3 & 2/5 & 2/11 
\\ \hline
${\mbox{variational}}$ & 2 & 1 & 2/3 & 2 & 4/3 & 1 & 1/2 & 1/5 
\\ \hline
${D,\,(d+2)^{-1}}$ & 1.0 & 0.75 & 0.59 & 0.99 &0.81 & 0.68 & 0.43 & 0.198
\\ \hline 
${D,\,D_c(d)}\ {\mbox{min}}$ & 0.85 & 0.69 & 0.57 & 0.91 & 0.77 & 0.67 & 0.43 & 0.198
\\ \hline 
${D,\,D_c(d)}\ {\mbox{max}}$ & 1.10 & 0.81 & 0.62 & 1.22 & 0.98 & 0.81 & 0.47 & 0.199
\\ \hline 
${D,\,D_c(d)}\ {\mbox{mean}}$ & 0.98 & 0.75 & 0.60 & 1.08 & 0.88 & 0.78 & 0.45 & 0.198
\\ \hline
${D_c(d),\,\E}\ {\mbox{min}}$ & 0.83 & 0.68 & 0.58 & 0.83 & 0.72 & 0.64 & 0.42 & 0.196
\\ \hline
${D_c(d),\,\E}\ {\mbox{max}}$ & 0.91 & 0.73 & 0.60 & 0.90 & 0.76 & 0.66 & 0.42 & 0.198
\\ \hline  
${D_c(d),\,\E}\ {\mbox{mean}}$ & 0.87 & 0.71 & 0.59 & 0.87 & 0.74 & 0.65 & 0.43 & 0.197
\\ \hline 
\end{tabular}  \renewcommand{\arraystretch}{1.0} }
\caption{Results of the numerical extrapolations for $\nu$. 
If not stated otherwise  the error is $\pm 1$ in the last digit.
$ (x,y)$ indicates the expansion-parameters $x$ and $y$ as discussed 
in the text. ${\mbox{min}}$,  ${\mbox{max}}$ 
and  ${\mbox{mean}}$ are the minimum and the maximum of the
plateau and their mean-value respectively.}
\label{tab: nu}
\end{figure}%

\clearpage

\subsection{Variational method and perturbation expansion}
\label{Variational method and perturbation expansion}

In the various extrapolation schemes that we used, we have seen that the
plateau structure for $D_0\to 2$ becomes clearer when the space dimension $d$
is large and that the corresponding estimates for $\nu$ are close to the value
$\nu_{\mbox{\scriptsize var}}=4/d$ obtained from a variational ansatz.
In order to understand this phenomenon and to have a better understanding of
the plateau structure of the extrapolations, we shall discuss the status of the
variational method.

Using a Gaussian variational ansatz \cite{Guitter variational},
the exponent $\nu$ for the crumpled phase of a self-avoiding $D$-dimensional
tethered membrane is found to be
\be \label{nu var}
	\nu_{\mbox{\scriptsize var}} = \frac{2D}{d} \ .
\ee
It was noticed in \cite{r:DDG3} that in the case of membranes with long range
interactions the variational estimate for $\nu$  is exact and can
be reproduced easily by simply  assuming that there is no coupling constant
renormalization, i.e.\ that one can take $Z_b=1$ in the renormalized 
Hamiltonian (\ref{e:Ham1}).
This last assumption can be proven for membranes with long range forces
\cite{r:DDG3,r:DDG4}.
Let us rewrite the full dimension $\nu(b)$ in a way which makes this
point clear.
Starting from the definitions of the RG-functions $\beta(b)$ \eq{e:beta'} 
and $\nu(b)$
\eq{e:nu'}, we reexpress $\frac{\partial}{\partial b} \ln Z$ in terms of
$\beta(b)$ and $\frac{\partial}{\partial b} \ln Z_b $, 
\be
\label{e:dlZdlZb}
\frac{\partial}{\partial b} \ln Z = -\frac2{db} \left(
\frac{\E b}{\beta(b)} + 1 + b \frac{\partial}{\partial b} \ln Z_b 
 \right)
\ee
Inserting this into (\ref{e:nu'}) gives
\be
\label{e:nudZb}
\nu(b) = \frac{2D}{d} +\frac{\beta(b)}{db} +\frac{ \beta(b)} d \frac{\partial}{\partial b} \ln Z_b
\ee
At the IR fixed point $b=b^*$, the second term of the r.h.s. of \eq{e:nudZb}
vanishes. The last one does not vanish in general, but vanishes if $Z_b=1$.
In this case, we get the variational result (\ref{nu var}).

Using this observation,
one can understand why the 1-loop and 2-loop extrapolations for $\nu$ 
coincide with $\nu_{\mbox{\scriptsize var}}$ for large $d$.
Taking $d_0\to\infty$ on the critical line amounts to take $D_0\to 2$.
The limit $d\to\infty$ corresponds thus to the limit $D\to 2$ for the
counterterms. 
The 1-loop wave-function counterterm is given by the residue
\be
\label{e:1lWFDto2}
\bigg< \GH \bigg| \GO \bigg>_\E \approx 
1\ \ \mbox{as}\ \ D\to 2
\ee
while the 1-loop coupling constant counterterm is given by the residue
\be
\label{e:1lCCDto2}
\bigg< \GM \bigg| \GB \bigg>_\E \approx 2^{-2D/(2-D)}
\ \ \mbox{as}\ \ D\to 2
\ee
which is exponentially smaller than (\ref{e:1lWFDto2}) when $D\to 2$.
A similar exponential factor appears for the 2-loop coupling constant
counterterm compared to the 2-loop wave-function counterterm
(this can be checked from the analytical expressions and the numerical
results).

When looking at the general structure of the divergent diagrams at $N$-loop
order, we can argue that this phenomenon will persist, but we have no 
rigorous proof.
However, if this exponential bound $\ln Z_b\ll \ln Z$ as $D\to 2$ is correct,
this means that $\nu-\nu_{\mbox{\scriptsize var}}\sim\exp(-\mbox{cst.}/d)$
when $d\to\infty$.

\begin{figure}[htb] 
\centerline{
\Dd{2}{3}%
\parbox{0mm}{\raisebox{0.0cm}[0mm][0mm]{\scriptsize\hspace{4.25cm}$1$}}%
\parbox{0mm}{\raisebox{0.3cm}[0mm][0mm]{\scriptsize\hspace{4.9cm}$2$}}%
\parbox{0mm}{\raisebox{0.5cm}[0mm][0mm]{\scriptsize\hspace{5.6cm}$3$}}%
\parbox{0mm}{\raisebox{1.0cm}[0mm][0mm]{\scriptsize\hspace{6cm}$4$}}%
\epsfxsize=8cm \parbox{8cm}{\epsfbox{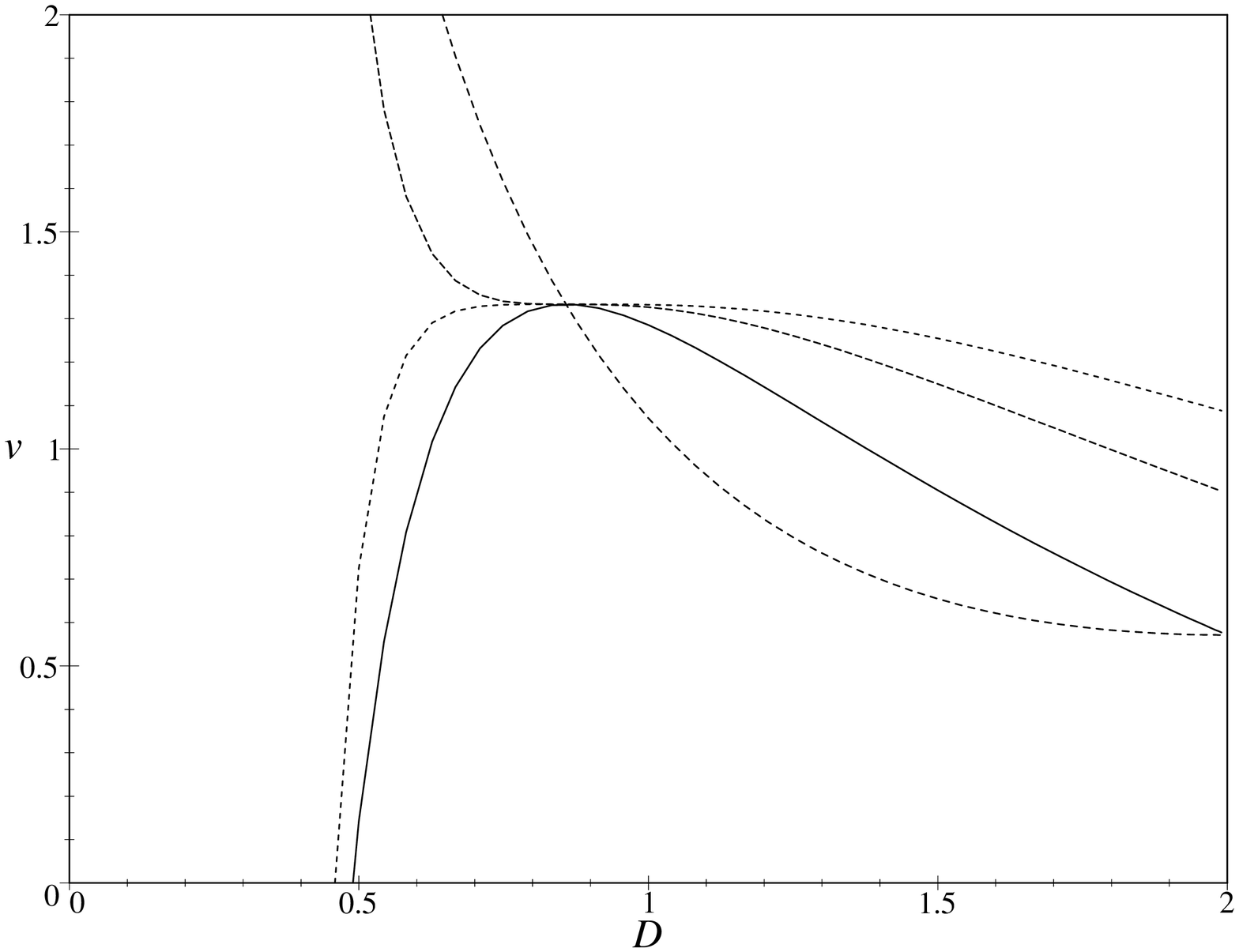}}
\Dd{2}{3}%
\epsfxsize=8cm \parbox{8cm}{\epsfbox{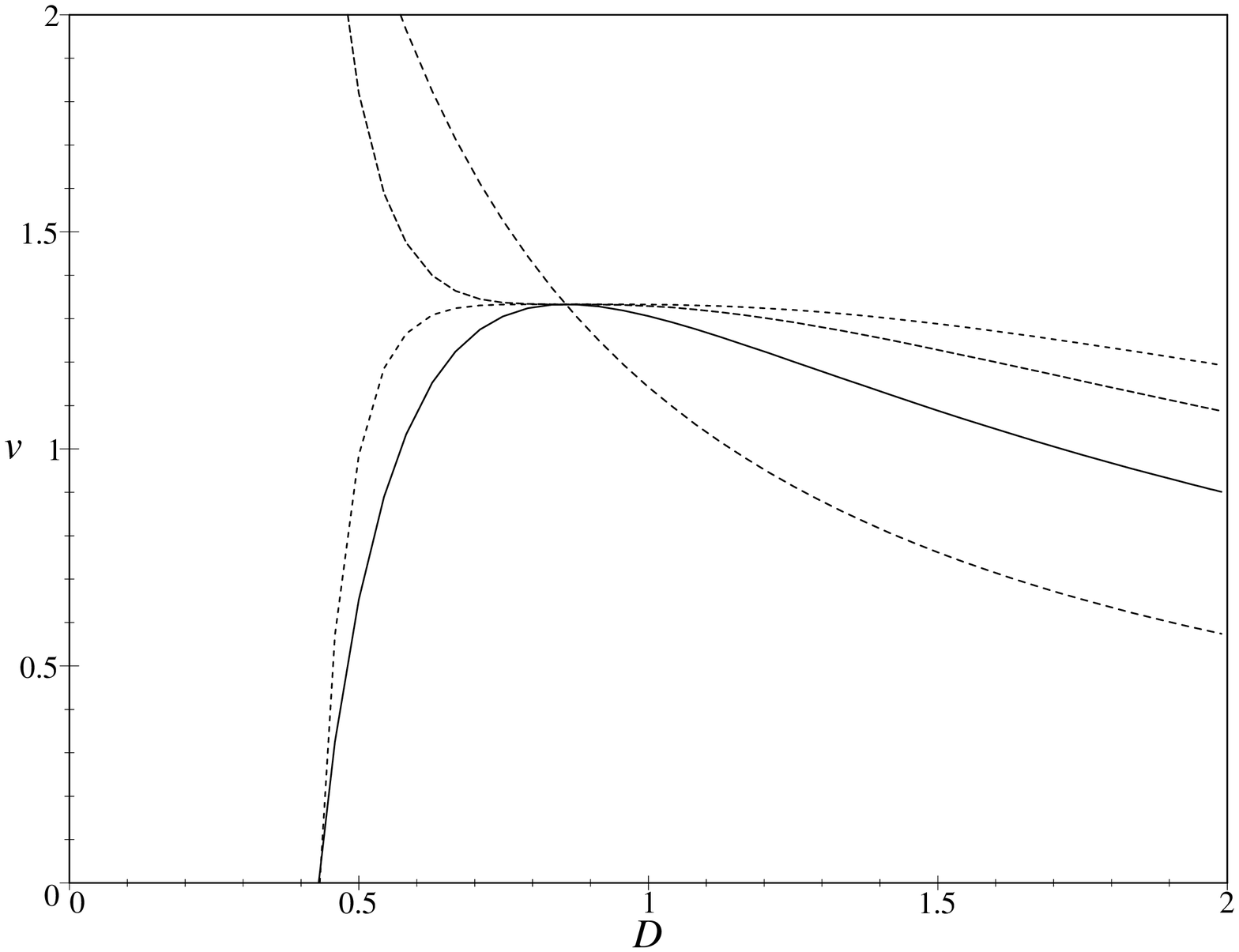}}
}
\centerline{
\Dd{2}{10}%
\epsfxsize=8cm \parbox{8cm}{\epsfbox{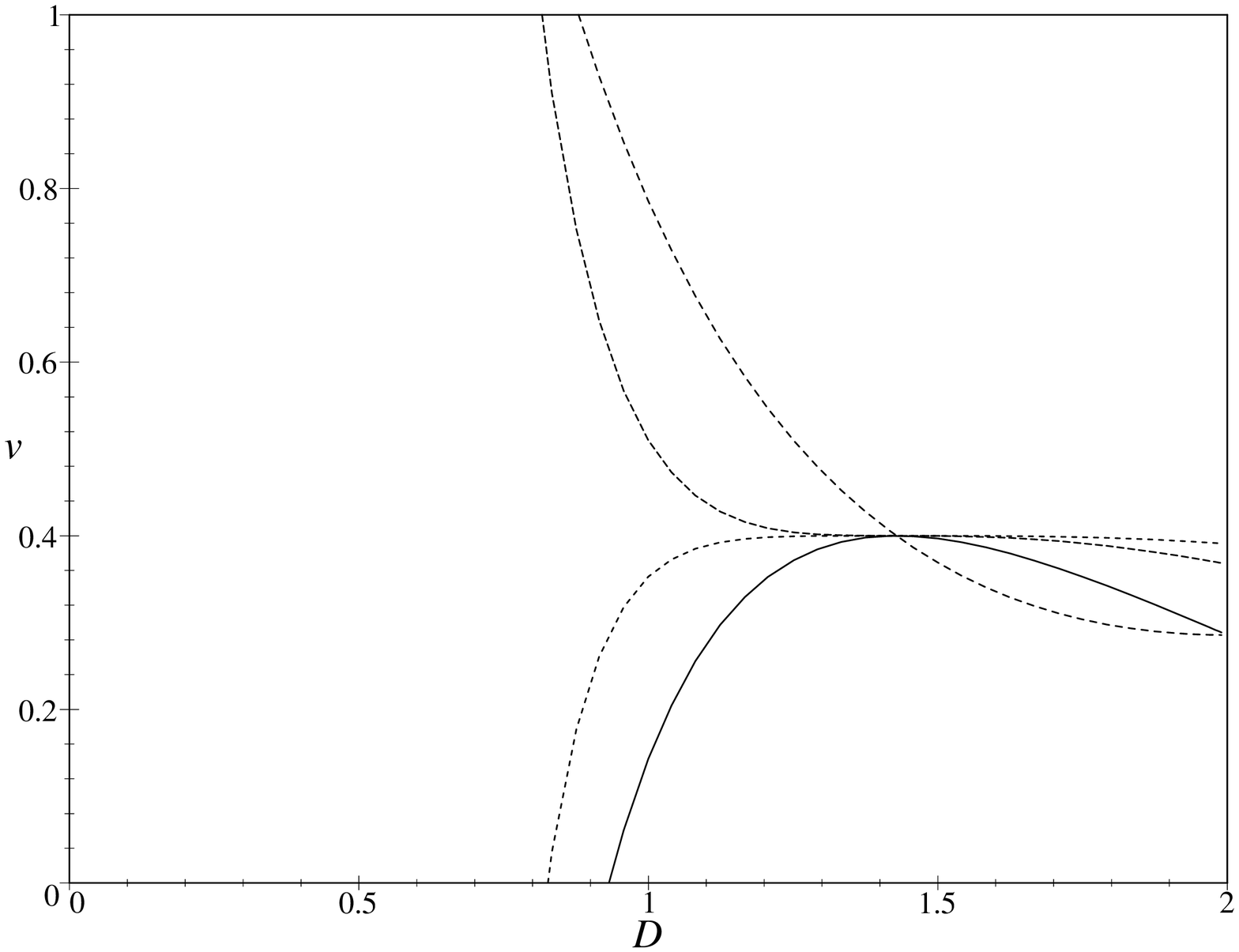}}
\Dd{2}{10}%
\epsfxsize=8cm \parbox{8cm}{\epsfbox{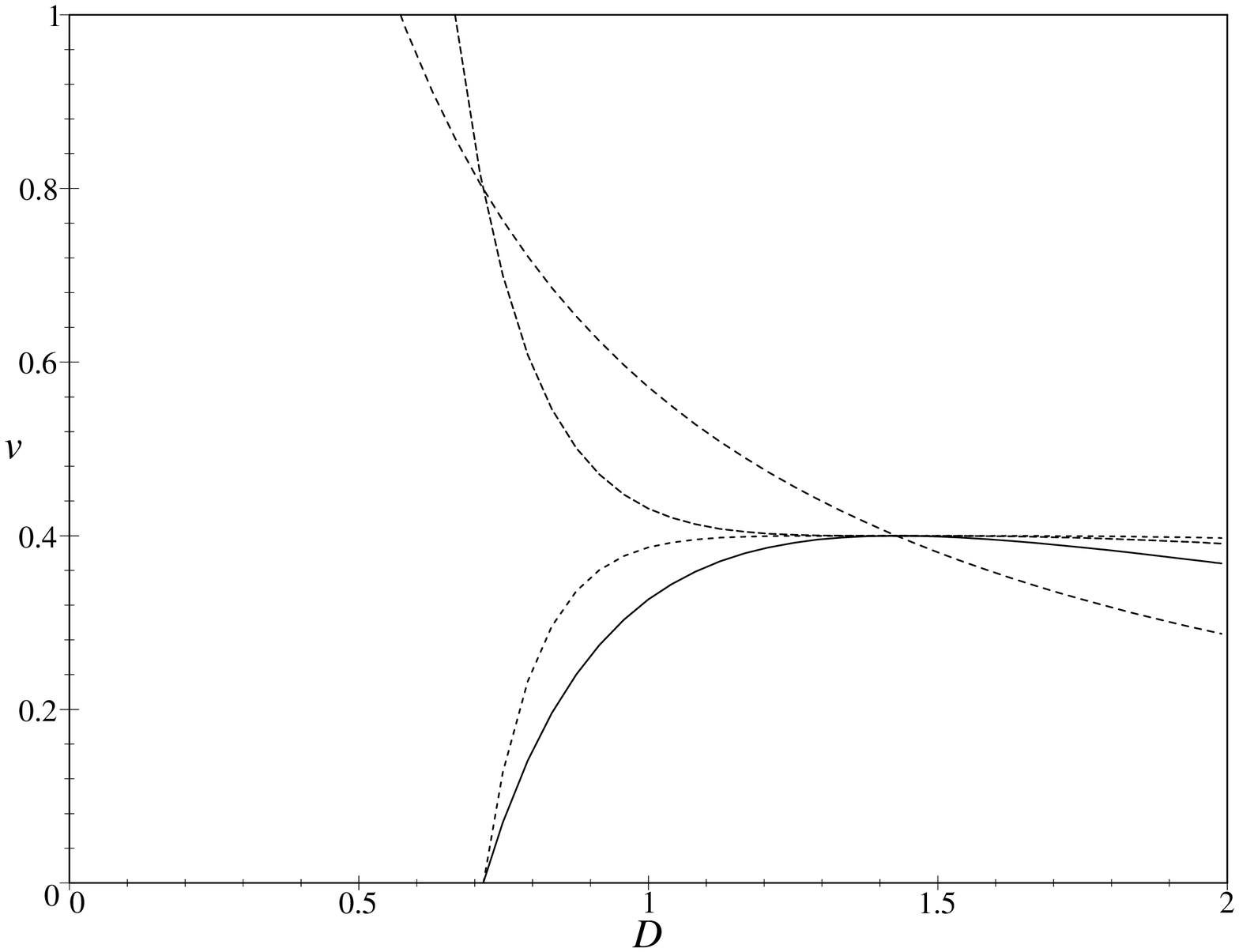}}
}
\centerline{
\Dd{2}{20}%
\epsfxsize=8cm \parbox{8cm}{\epsfbox{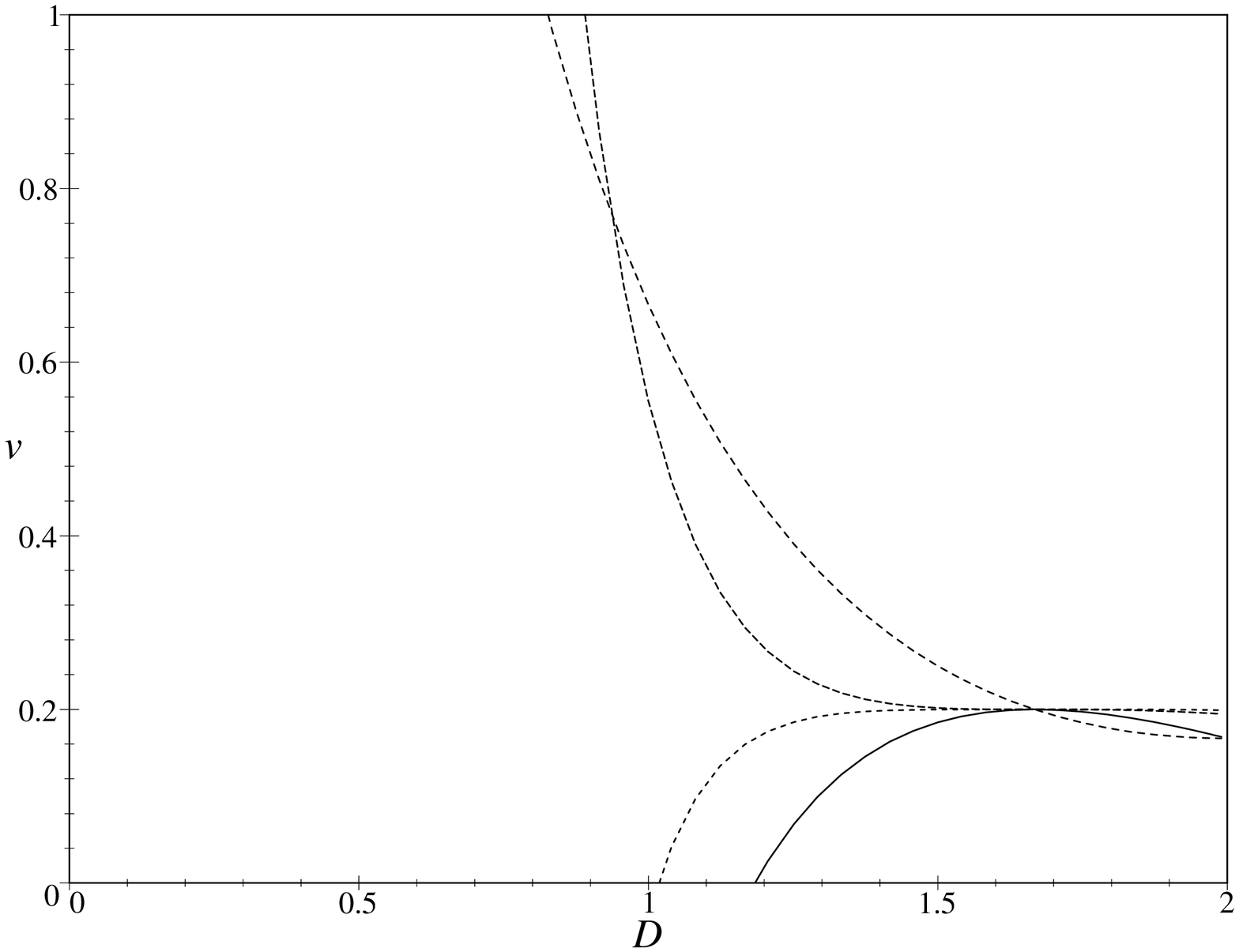}}
\Dd{2}{20}%
\epsfxsize=8cm \parbox{8cm}{\epsfbox{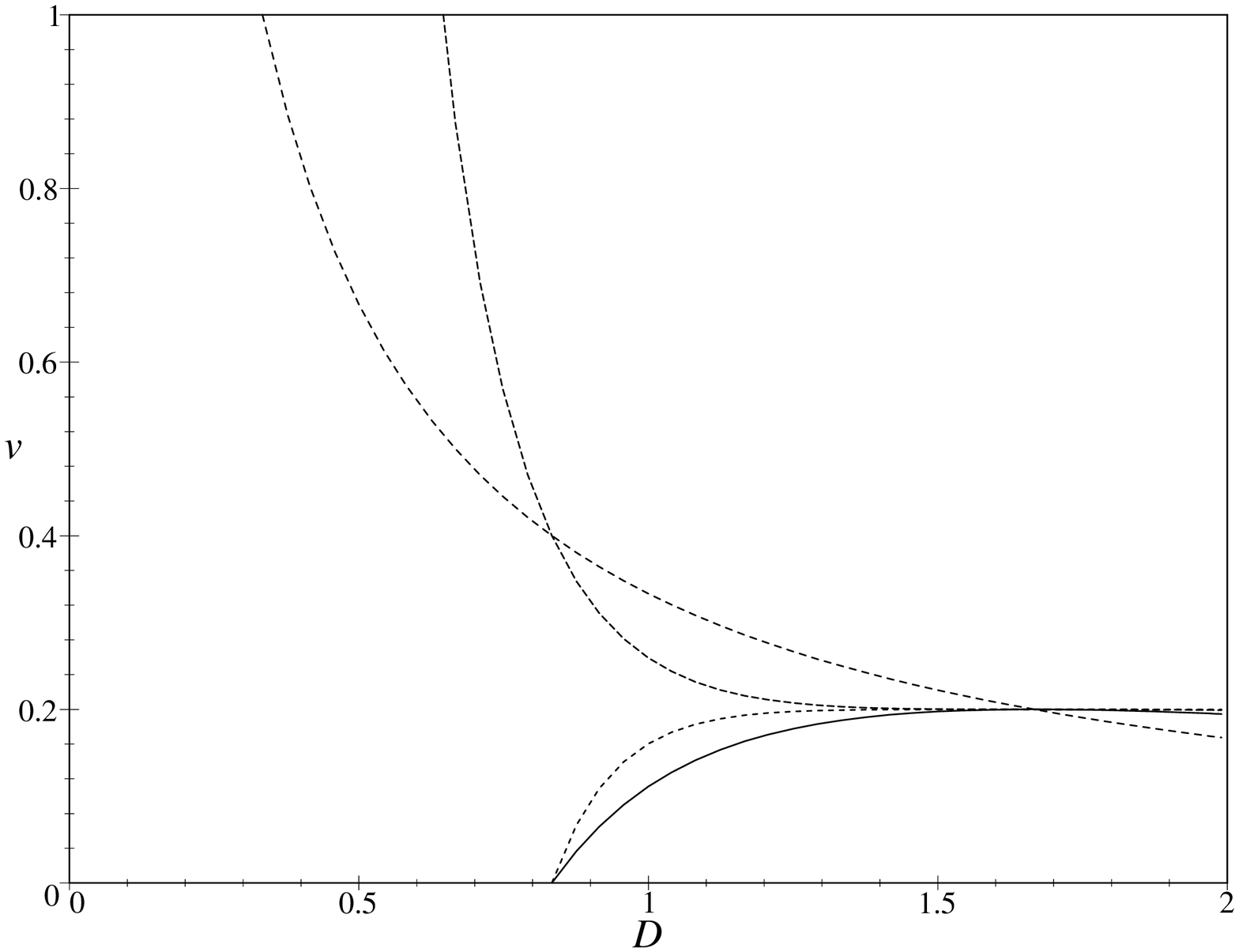}}
}
\caption{Extrapolation of $\nuvar=2D/d$ for membranes in
$\{D_c(d),\E$\} (left),
and $\{D_c(D), D\}$ (right), to $(D,d)=$ (2,3), (2,10) and (2,20).
On the first figure the labels $1$ -- $4$  indicate the order of the
$\E$-expansion i.e.\ ${\cal O}(\E^1)$ -- ${\cal O}(\E^4$). On all figures, 
the full line is the result at order  $\E^2$, the other lines are 
the results at order $\E$, $\E^3$ and $\E^4$.}
\label{f:nuvarextrap}
\end{figure}

Assuming the variational estimate (\ref{nu var})
to be a good approximation for large $d$,
it is interesting to test the various extrapolation methods that we have
used to get the 2-loop results.
The principle is  to start from the exact formula for
$\nu_{\mbox{\scriptsize  var}}$, to write an $\E$-like-expansion
around the critical curve $\E=0$, to truncate it at a fixed order and to
resum the result as done previously.
The results of such a resummation, using the extrapolation in $\{x,y\}$,
are presented on figure~\ref{f:nuvarextrap}.
Let us note the following points:
\begin{enumerate}
\item
 The extrapolations at 1- and 2-loop orders of $\nu_{\mbox{\scriptsize  var}}$
are indeed very similar to the extrapolations of $\nu$ at large $d$.
In particular one recovers the same plateau structure and it should be noted
that already the 2-loop optimal estimate for $\nu_{\mbox{\scriptsize  var}}$,
obtained by the minimal sensitivity method, gives the exact
$\nu_{\mbox{\scriptsize  var}}$!
This is a strong point for this method, when applied for smaller $d$.

\item
 One should note that the $\E$-expansion for
$\nu_{\mbox{\scriptsize  var}}$ is convergent, but that is has a finite
radius of convergence.
As a consequence, one can show that the $N$-loop extrapolations for
$\nu_{\mbox{\scriptsize  var}}$ converge towards the exact result
when $N\to\infty$ only in a finite range of starting points $D_0$ on the
critical curve.
This range is explicitly
\be
\label{e:RangeD}
{2d\over 8+d}<D_0<2
\ee
\item
The large variation of the 2-loop estimates for $\nu$ as $D_0$ becomes small
reflects the fact that we are outside of the range of convergence of the 
$\E$-expansion.
This is clear when one compares the 2-loop estimate with the 3- and
4-loop estimates on figure~\ref{f:nuvarextrap}.
We expect that this is still true for smaller $d$ and that the optimal
values for $D_0$ are still in the domain of confidence of the $\E$-expansion.
\end{enumerate}

\medskip
{\tabcolsep1.9mm
\begin{figure}[tb]%
\centerline{ \renewcommand{\arraystretch}{1.20}
\begin{tabular}[t]{|l|c|c|c|c|c|c|c|c|} \hline
$(x,y)\ ;\ (D,d)$ & $(1,1)$ & $(1,2)$ & $(1,3)$ & $(2,2)$ & $(2,3)$ & $(2,4)$ & $(2,8)$ & $(2,20)$
\\ \hline \hline 
${\mbox{exact}}$ & 1 & 3/4 & 0.586(4) & 1  & --- & --- & --- & ---
\\ \hline
${\mbox{Flory}}$ & 1 & 3/4 & 3/5 & 1 & 4/5 & 2/3 & 2/5 & 2/11 \\ \hline
${\mbox{variational}}$ & 2 & 1 & 2/3 & 2 & 4/3 & 1 & 1/2 & 1/5 
\\ \hline
${D,\,D_c(d)}\ {\mbox{left crossing}}$ & 1.09 & 0.69 & 0.58 & 0.97 &
0.797 & 0.80 & 0.45 & 0.20
\\ \hline
${D,\,D_c(d)}\ {\mbox{right crossing}}$ & 2.08 & 0.76 & 0.60 & 1.05 & 0.82 & 1.00 & 0.50 & 0.20
\\ \hline
${D,\,D_c(d)}\  {\mbox{mean crossing}}$ & 1.59 & 0.73 & 0.59 & 1.01 & 0.80 & 0.90 & 0.48 & 0.20
\\ \hline
\end{tabular}  \renewcommand{\arraystretch}{1.0} }
\caption{Results for $\nu$ using the numerical extrapolations for $\nu d$.
If not stated otherwise  the error is $\pm 1$ in the last digit.}
\label{nudtab}
\end{figure}}%

\begin{figure}[htb]
\centerline{
\Dd{1}{2}%
\epsfxsize=8cm \parbox{8cm}{\epsfbox{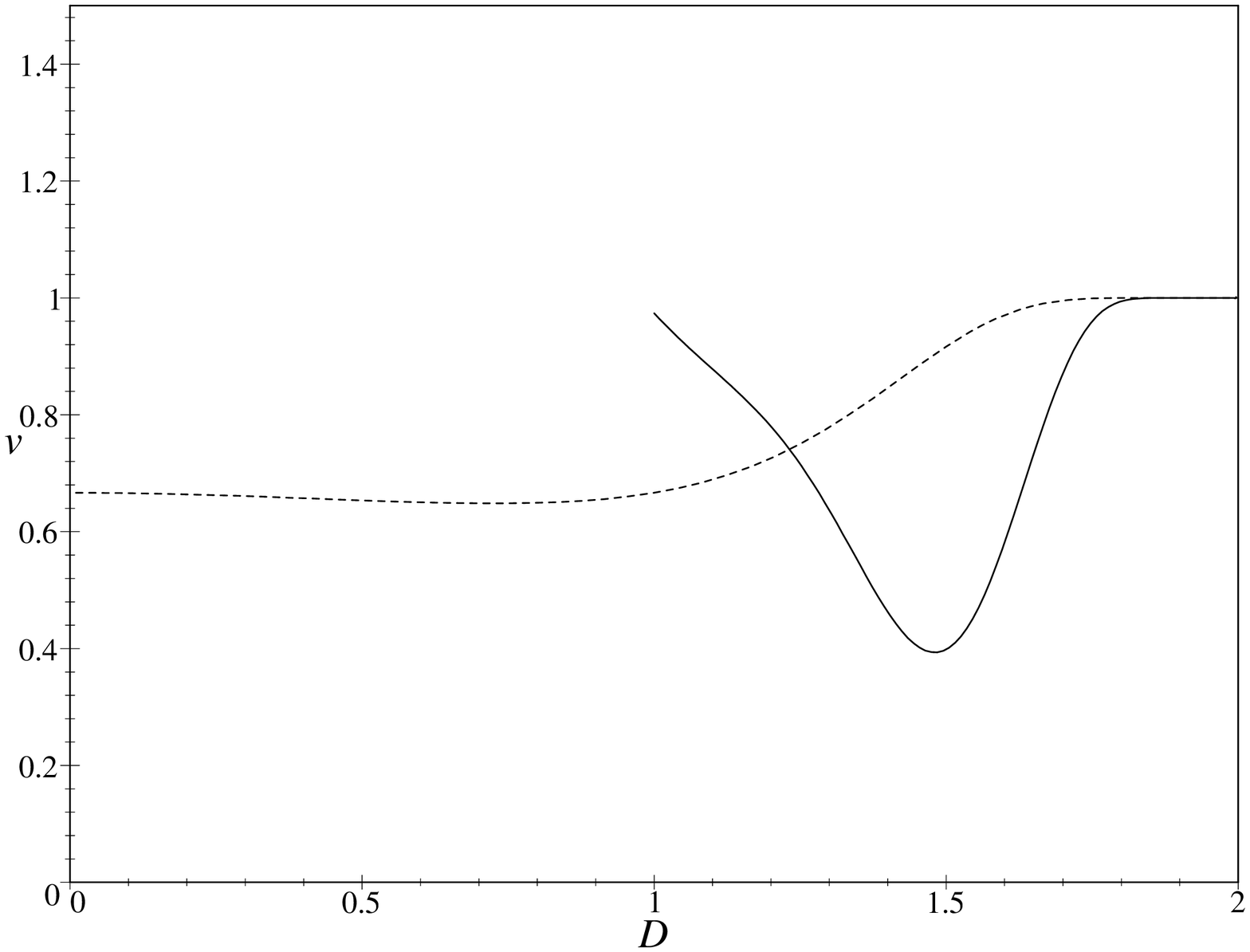}}
\Dd{1}{3}%
\epsfxsize=8cm \parbox{8cm}{\epsfbox{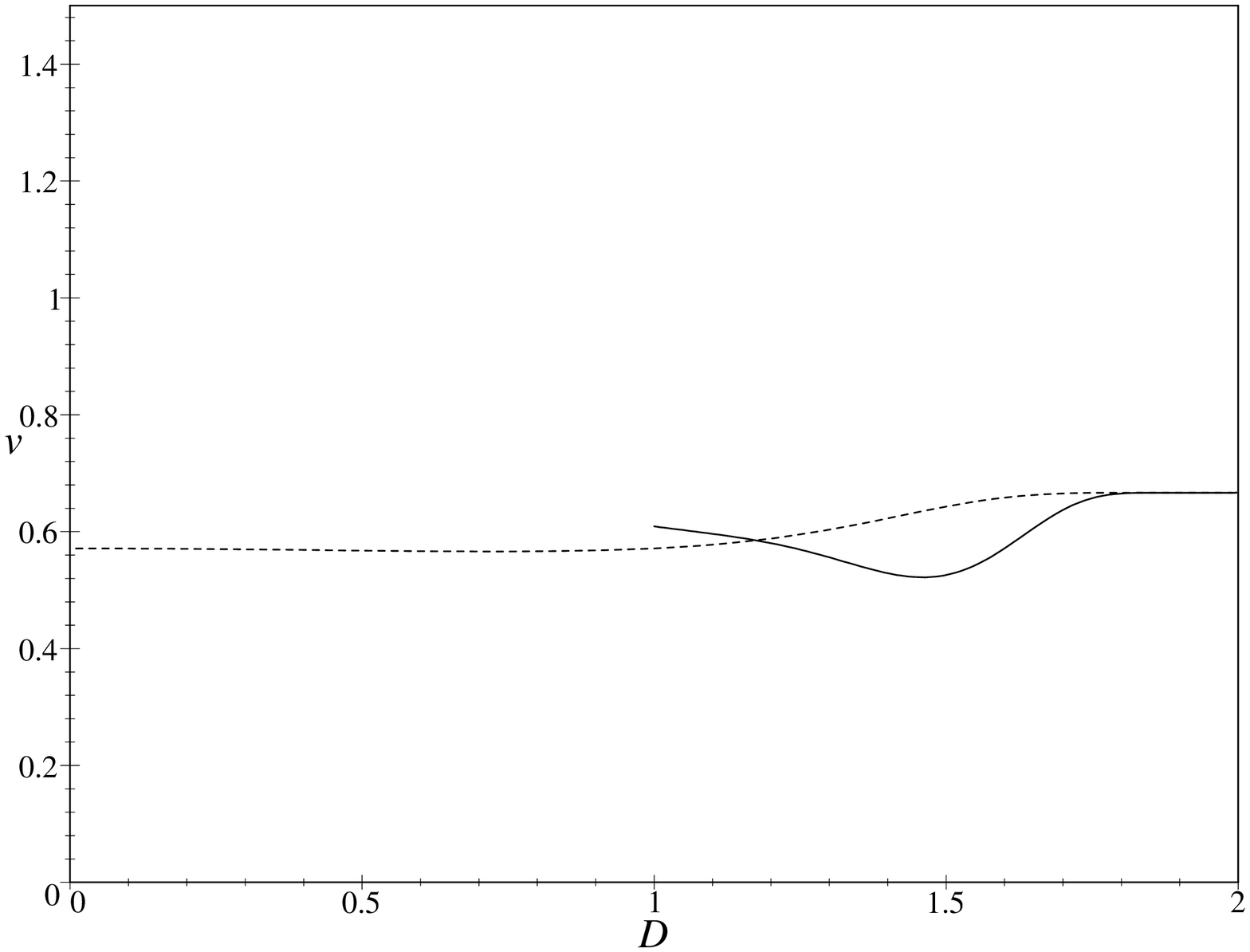}}
}
\centerline{
\Dd{2}{2}%
\epsfxsize=8cm \parbox{8cm}{\epsfbox{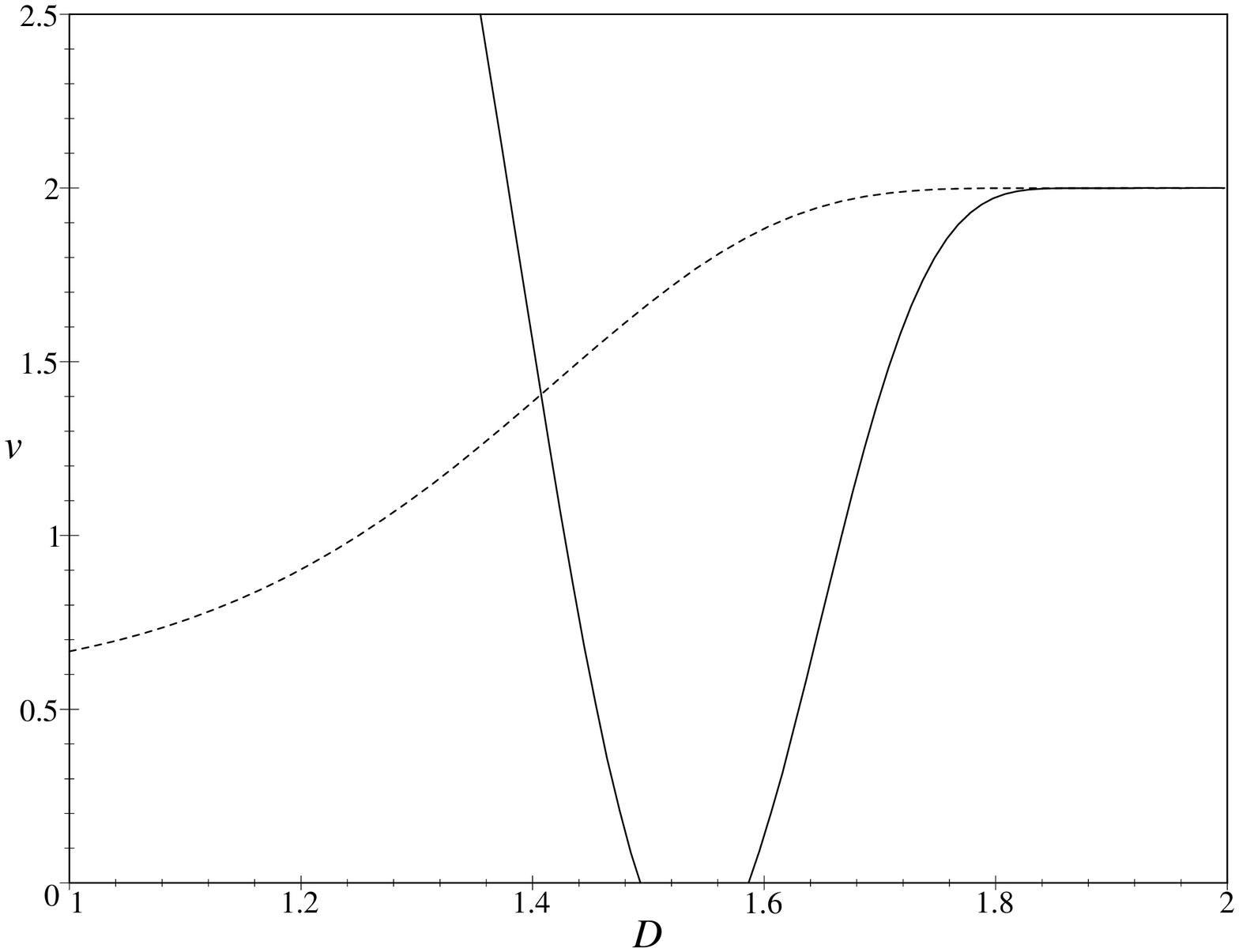}}
\Dd{2}{3}%
\epsfxsize=8cm \parbox{8cm}{\epsfbox{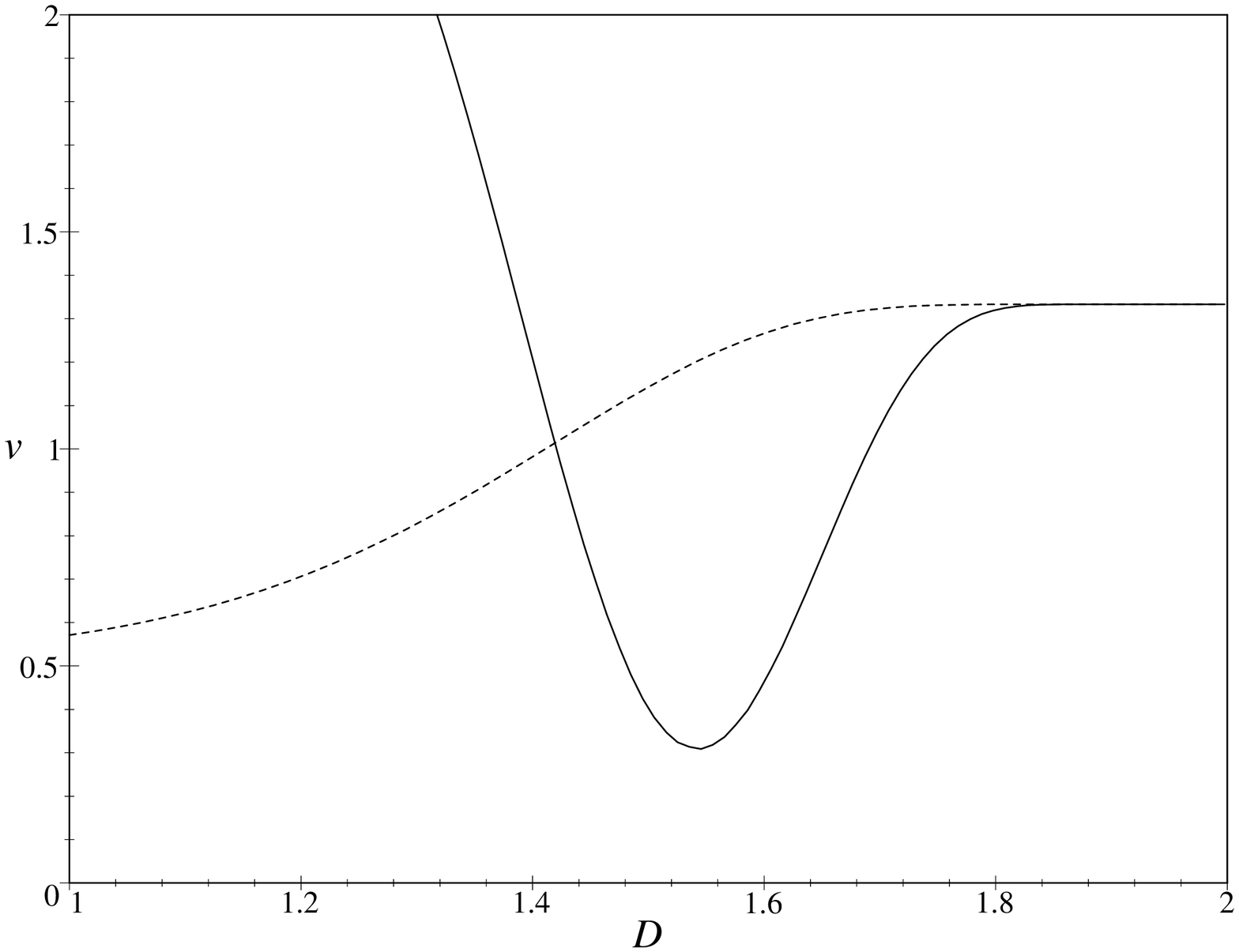}}
}
\centerline{
\Dd{2}{6}%
\epsfxsize=8cm \parbox{8cm}{\epsfbox{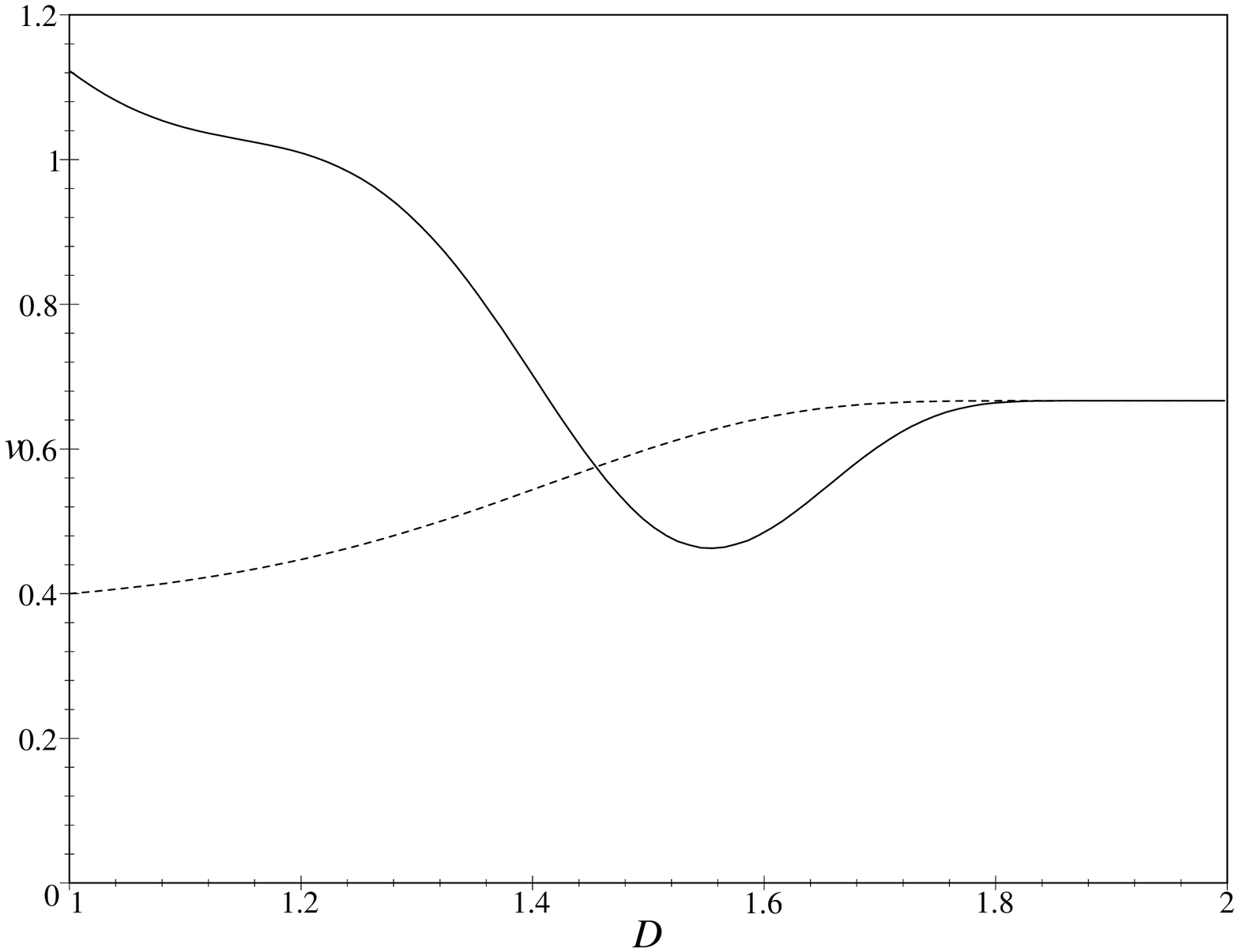}}
\Dd{2}{20}%
\epsfxsize=8cm \parbox{8cm}{\epsfbox{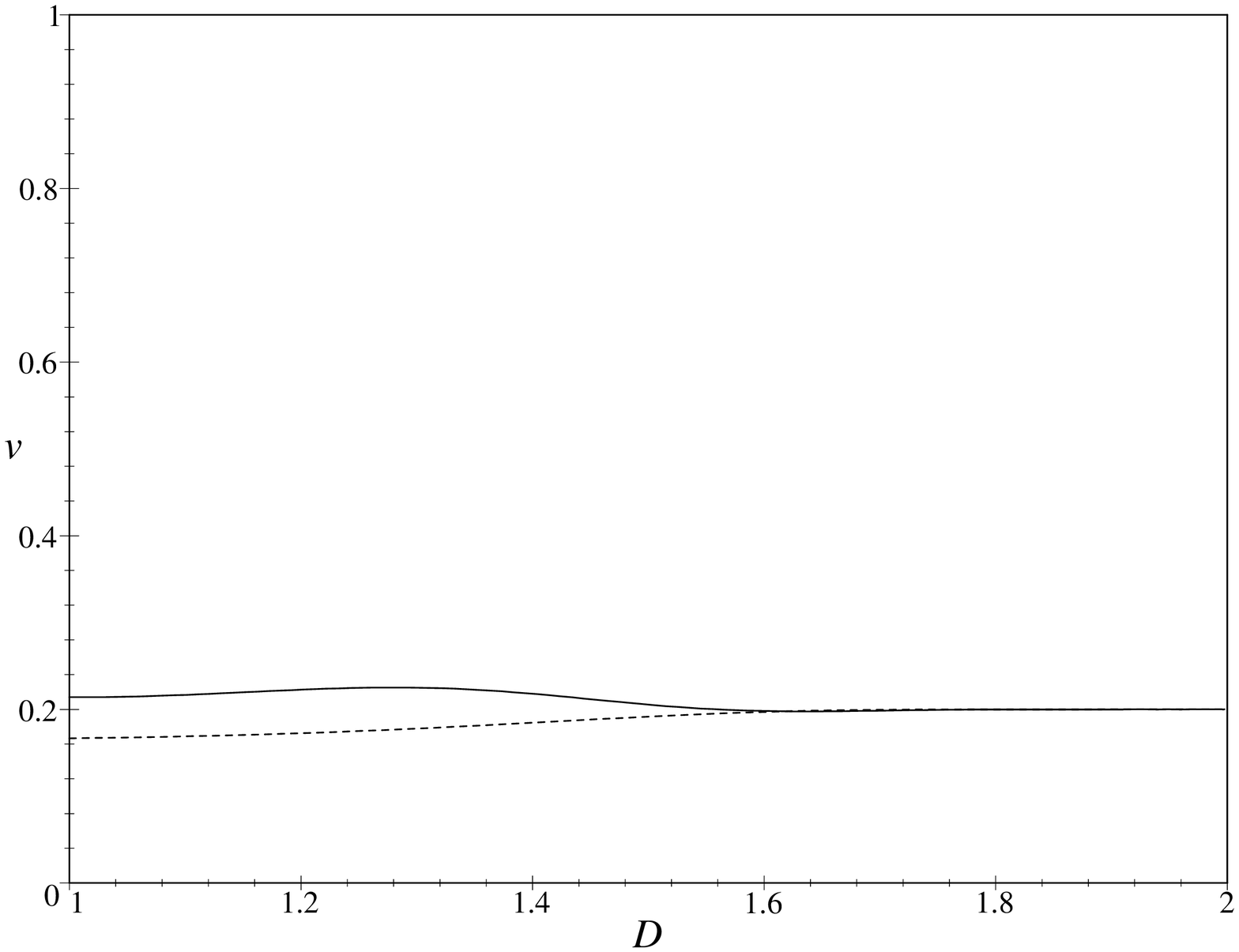}}
}
\caption{Extrapolation of $\nu d$ in $D$ and $D_c(d)$ to $(D,d)=$ (1,2), (1,3),
(2,2),
(2,3), (2,6), (2,20).}
\label{nudfig}
\end{figure}%

We now present a new $\E$-expansion, well suited to large $d$, which
is suggested by the expression \eq{e:nudZb} for $\nu(b)$.
Indeed, when evaluating the exponent $\nu^*=\nu(b^*)$ by \eq{e:nudZb}, we
see that  what
we really expand in $\E$ is not $\nu$ but $\nu d$:
\bea
\nu^* d &&
= 2D + \left( \beta(b) \frac{\partial}{\partial b} \ln Z_b(b)\right)\lts_{b=b^*}
\nonumber\\
&&= 2D+{\cal O}(\E)
\eea
Thus we may perform the $\E$-expansion for $\nu d$ rather than for $\nu$.
This new expansion has the advantage that it starts at order $\E^0$
from the result predicted by the variational ansatz.
The same extrapolation methods used for $\nu$ can be used for $\nu d$.

The results are given in figures~\ref{nudtab} and \ref{nudfig}.
We find that for polymers $(D=1)$, when using the $(D, D_c(d))$ variables, the
minimal sensitivity method gives poor results, but that the criterion to
take as optimal $D_0$ the point where the second order correction vanishes 
gives good results. 
We use the same criterion for membranes ($D=2$).

\subsection{Expansion around Flory's estimation}
\label{s:Flory}

In the last section we have seen that the $\E$-expansion for $\nu d$
is in fact an $\E$-expansion around the variational ansatz,
$\nu d = 2D +{\cal O} (\E)$.
Another stimulating idea is to perform a similar expansion around the prediction
made by Flory's argument.
It is well known that the Flory result for polymers $\nuflory=3/(2+d)$
is simply obtained by assuming that the elastic term and the contact interaction
term in the Edwards Hamiltonian scale in the same way with the internal size
of the polymer. 
The same scaling assumption for membranes leads to the prediction
\be
\label{e:nuFlory}
\nuflory\ =\ {2+D\over 2+d}
\ee
It is possible to perform an $\E$-expansion around $\nuflory$ by
simply expanding $\nu (d+2)$.
Indeed,  we can set
\be
Z_{\mbox{\scriptsize Flory}}\ =\ \sqrt{Z\,Z_b}\quad,\quad Z_\E\ =\ \sqrt{Z/Z_b}
\label{e:ZFlory}
\ee
and eliminate $Z_{\mbox{\scriptsize Flory}}$ from the system of equations
(\ref{e:beta'}) and (\ref{e:nu'}).
We obtain
\be
\nu^* (d+2) = D+2 + \left( \beta(b) \frac{\partial}{\partial b}
\ln\!\left( \frac{Z_b}{Z} \right) \right)\lts_{b=b^*}
\label{e: nu(d+2)=}
\ee
This makes clear that if the wave-function and coupling-constant
renormalizations are the same (more precisely if $Z_b/Z$ stays finite at the
IR fixed point $b^*$) the Flory result becomes exact.
Moreover, the $\E$-expansion of $\nu(d+2)$ is clearly an $\E$-expansion 
around $\nuflory$.

This expansion seems to be the most satisfying numerically.
In particular, the method of minimal sensitivity and that of minimizing
the second order term give generally close results.
We therefore give in the following plots for all interesting combinations of
variables and dimensions $(D,d)$.
Let us stress that good expansion-parameters for
$\nu$ are not necessarily good for $\nu (d+2)$ and vice versa.
E.g.\ the expansion in $D$ and $d$ is bad for $\nu$ but works quite well 
(although not optimal) for $\nu (d+2)$.
We study the expansion in $(D,d)$, $(d,\E)$, $(D,D_c(d))$ and
$(D_c(d), \E)$.
The results of the extrapolations are collected in figure \ref{tab: nu(d+2)}.
{\tabcolsep1.9mm
\begin{figure}[tb]%
\centerline{ \renewcommand{\arraystretch}{1.2}
\begin{tabular}[t]{|l|c|c|c|c|c|c|c|c|} \hline
$(x,y)\ ;\ (D,d)$ & $(1,1)$ & $(1,2)$ & $(1,3)$ & $(2,2)$ & $(2,3)$ & $(2,4)$ & $(2,8)$ & $(2,20)$
\\ \hline \hline
${\mbox{  exact}}$ & 1 & 3/4 & $0.586(4)$ & 1  & ---  & --- & --- & ---
\\ \hline
${\mbox{  Flory}}$ & 1 & 3/4 & 3/5 & 1 & 4/5 & 2/3 & 2/5 & 2/11 
\\ \hline
${\mbox{  variational}}$ & 2 & 1 & 2/3 & 2 & 4/3 & 1 & 1/2 & 1/5
\\ \hline
${D,\,d}\  {\mbox{  min}}$ &  0.93 & 0.71 & 0.58 & 1.02 & 0.83 & 0.70 & 0.43 &  0.20
\\ \hline 
${D,\,d}\  {\mbox{  max}}$ & 1.09 & 0.82 & 0.65 & 1.20 & 0.95 & 0.79 & 0.46 &  0.19
\\ \hline 
${D,\,d}\  {\mbox{   mean}}$ & 1.01 & 0.76 & 0.62 & 1.11 & 0.89 & 0.75 & 0.45 & 0.20
\\ \hline 
${D,\,d}\  {\mbox{  left crossing}}$ & 0.95 & 0.72 & 0.59 & 1.12 & 0.90 & 0.75 &  0.44 & 0.20 
\\ \hline 
${D,\,d}\  {\mbox{  right  crossing}}$ & 1.00 & 0.75 & 0.60 & 1.16 & 0.93 & 0.78 & 0.45 & 0.20
\\ \hline 
${D,\,d}\  {\mbox{   mean  crossing}}$ & 0.97 & 0.73 & 0.59 & 1.14 & 0.91 & 0.76 & 0.44 & 0.20
\\ \hline  
${d,\,\E}\  {\mbox{  min}}$ & 0.98 & 0.74 & 0.59 & 0.98 & 0.80 & 0.68 & 0.42 & 0.19 
\\ \hline 
${d,\,\E}\  {\mbox{  max}}$ & --- & --- & --- & 1.10 & 0.88 & 0.73 & 0.44 & 0.20
\\ \hline 
${d,\,\E}\  {\mbox{   mean}}$ & --- & --- & --- & 1.04 & 0.88 & 0.73 & 0.44 & 0.20
\\ \hline 
${d,\,\E}\  {\mbox{  left  crossing}}$ & 1.00 & 0.74 & 0.59 & 0.99 & 0.81 & 0.73 & 0.44 & 0.20 
\\ \hline 
${d,\,\E}\  {\mbox{  right  crossing}}$ & --- & --- & --- & 1.03 & 0.83 & 0.68 &  0.42 & 0.20
\\ \hline 
${d,\,\E}\  {\mbox{   mean  crossing}}$ & --- & --- & --- & 1.01 & 0.88 & 0.69 & 0.42 & 0.20 
\\ \hline 
${D_c(d),\,\E}\  {\mbox{  min}}$ & 0.75 & 0.56 & 0.57 & 0.83 & 0.74 & 0.64 & 0.43 & 0.20 
\\ \hline 
${D_c(d),\,\E}\  {\mbox{  max}}$ & 1.24 & 0.86 & 0.64 & 1.40 & 1.08 & 0.86 & 0.48 & 0.20 
\\ \hline 
${D_c(d),\,\E}\  {\mbox{   mean}}$ & 0.99 & 0.76 & 0.60 & 1.12 & 0.91 & 0.75 & 0.45 & 0.20 
\\ \hline 
${D_c(d),\,\E}\  {\mbox{  left  crossing}}$ & 0.84 & 0.69 & 0.58 & 0.86 & 0.74 & 0.64 & --- & 0.21  
\\ \hline 
${D_c(d),\,\E}\  {\mbox{  right  crossing}}$ & 0.91 & 0.72 & 0.60 & 0.63 & 0.75 & 0.64 & --- & 0.20
\\ \hline 
${D_c(d),\,\E}\  {\mbox{   mean  crossing}}$ & 0.88 & 0.70 & 0.59 & 0.87 & 0.74 & 0.64 & --- & 0.20 
\\ \hline 
${D,\,D_c(d)}\  {\mbox{  min}}$ & 0.35 & 0.55 & 0.55 & --- & 0.49 & 0.52 & 0.41 &  0.20
\\ \hline 
${D,\,D_c(d)}\  {\mbox{  max}}$ & 1.39 & 0.90 & 0.65 & --- & 1.18 & 0.85 & 0.50 & 0.20
\\ \hline 
${D,\,D_c(d)}\  {\mbox{   mean}}$ & 0.87 & 0.72 & 0.60 & --- & 0.84 & 0.68 &  0.45 & 0.20
\\ \hline 
${D,\,D_c(d)}\  {\mbox{  left  crossing}}$ & 0.95 & 0.72 & 0.58 & 1.12 & 0.90 & 0.74 & 0.45 & 0.20 
\\ \hline 
${D,\,D_c(d)}\  {\mbox{  right  crossing}}$ & 1.27 & 0.84 & 0.63 & 1.38 & 1.06 & 0.85 & 0.47 & 0.20
\\ \hline 
${D,\,D_c(d)}\  {\mbox{   mean  crossing}}$ & 1.06 & 0.78 & 0.61 & 1.25 & 0.98 & 0.80 & 0.46 & 0.20 
\\ \hline 
\end{tabular}  \renewcommand{\arraystretch}{1.0} }
\caption{Results for $\nu$ from the numerical extrapolations for $\nu(d+2)$ 
at second order.
``min", ``max" and "mean" denote respectively the estimate at the minimum of
the plateau, at the maximum of the plateau and their mean value.
``left crossing", ``right crossing" and ``mean crossing" denote respectively the estimate at the leftmost point where the second order correction vanishes, at 
the rightmost point and their mean value.
If not stated otherwise the error is $\pm 1$ in the last digit.}
\label{tab: nu(d+2)}
\end{figure}
}

\begin{figure}[htb] 
\centerline{
\Dd{1}{2}%
\epsfxsize=8.0cm \parbox{8.0cm}{\epsfbox{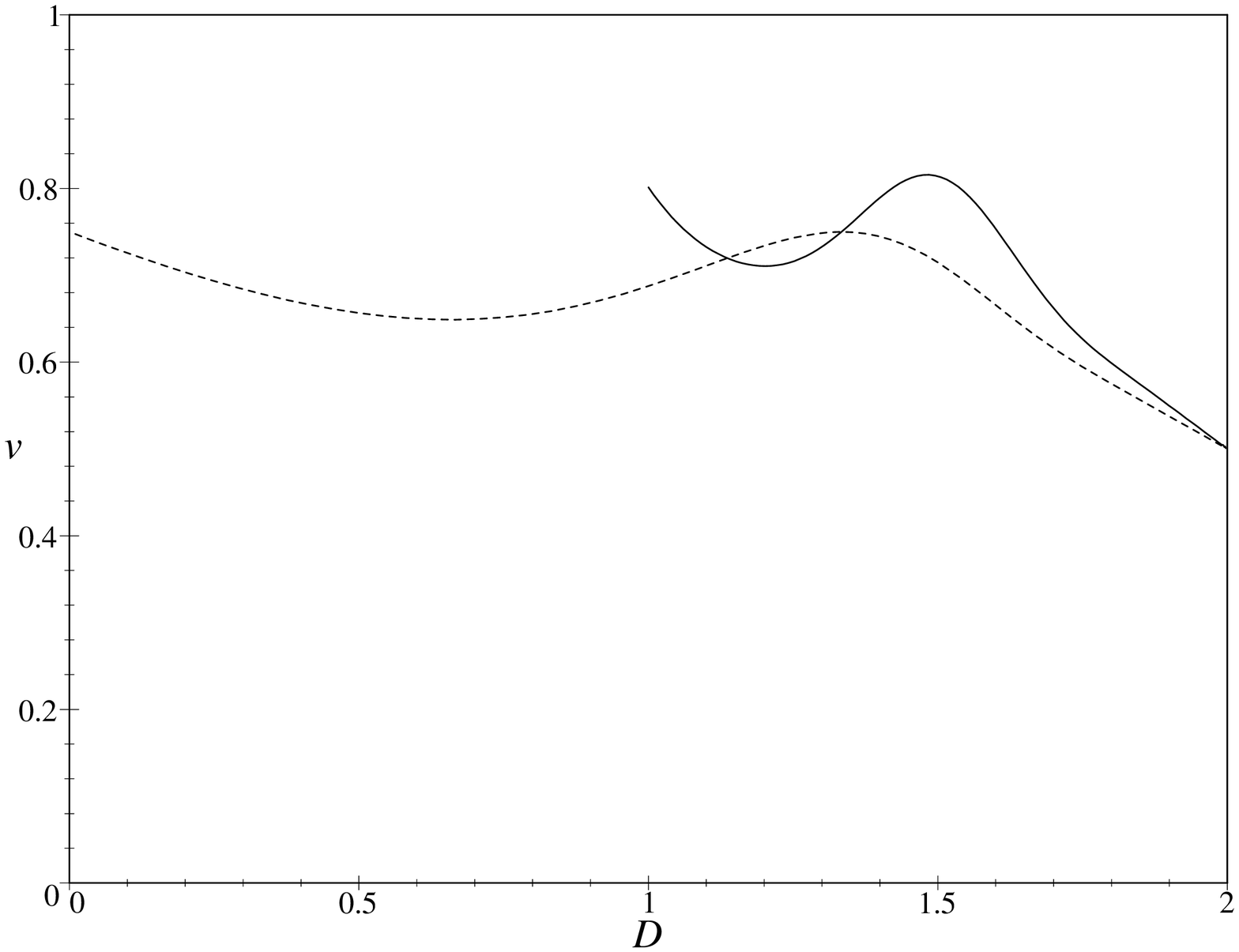}}
\hspace{0.0cm}
\Dd{1}{3}%
\epsfxsize=8.0cm \parbox{8.0cm}{\epsfbox{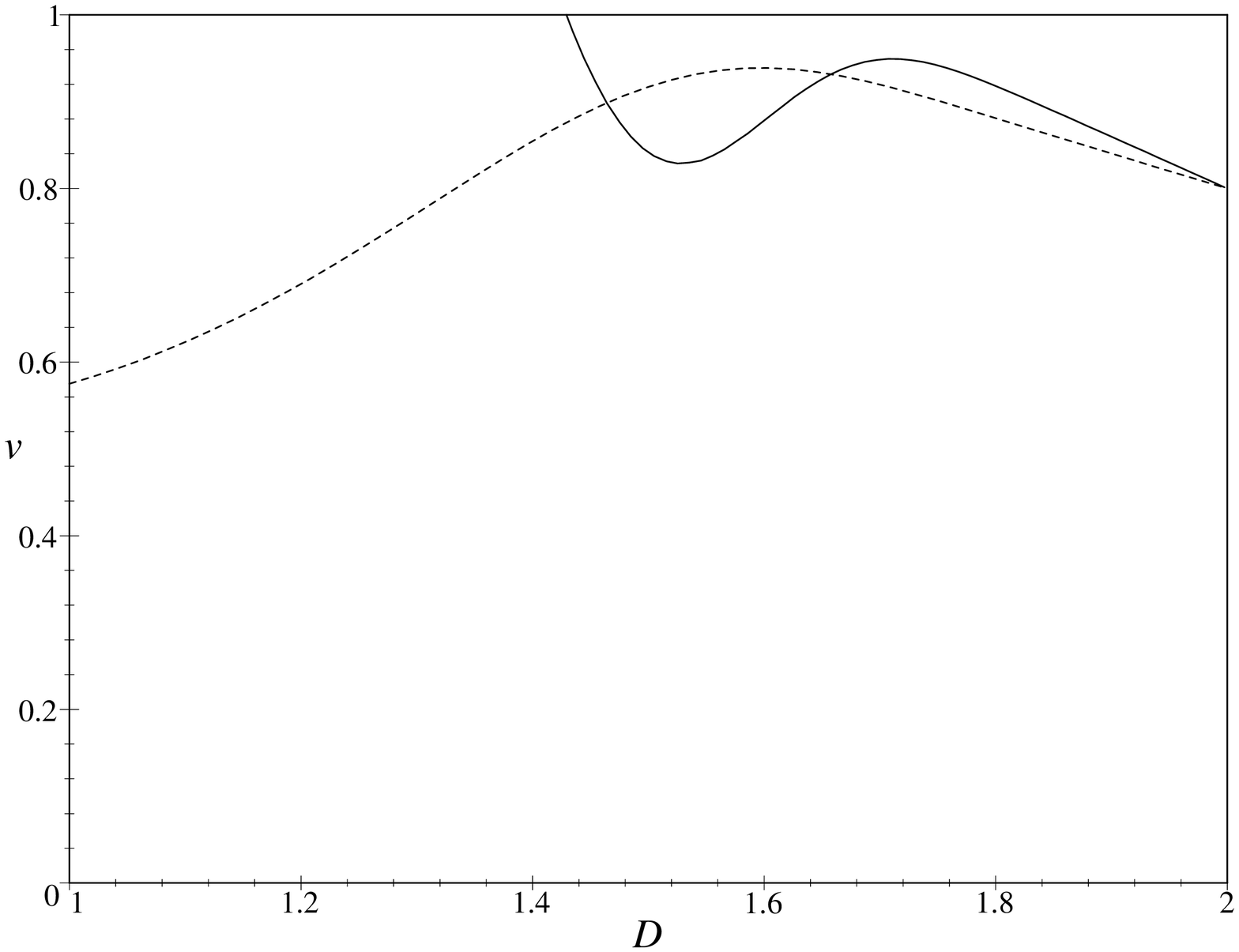}}
}
\centerline{
\Dd{2}{2}%
\epsfxsize=8.0cm \parbox{8.0cm}{\epsfbox{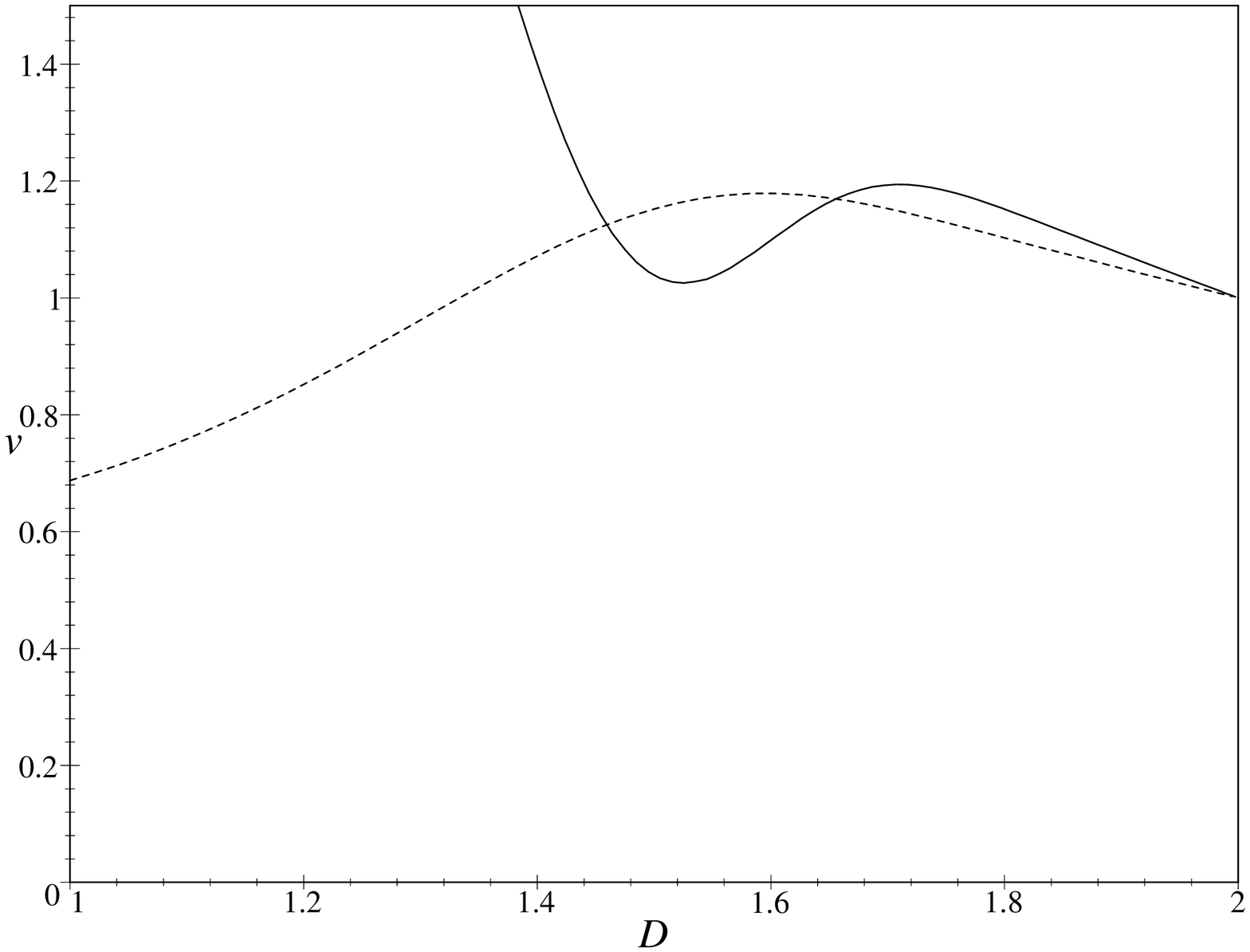}}
\hspace{0.0cm}
\Dd{2}{3}%
\epsfxsize=8.0cm \parbox{8.0cm}{\epsfbox{./eps/nudp2_D_d_2_3.eps}}
}
\centerline{
\Dd{2}{6}%
\epsfxsize=8.0cm \parbox{8.0cm}{\epsfbox{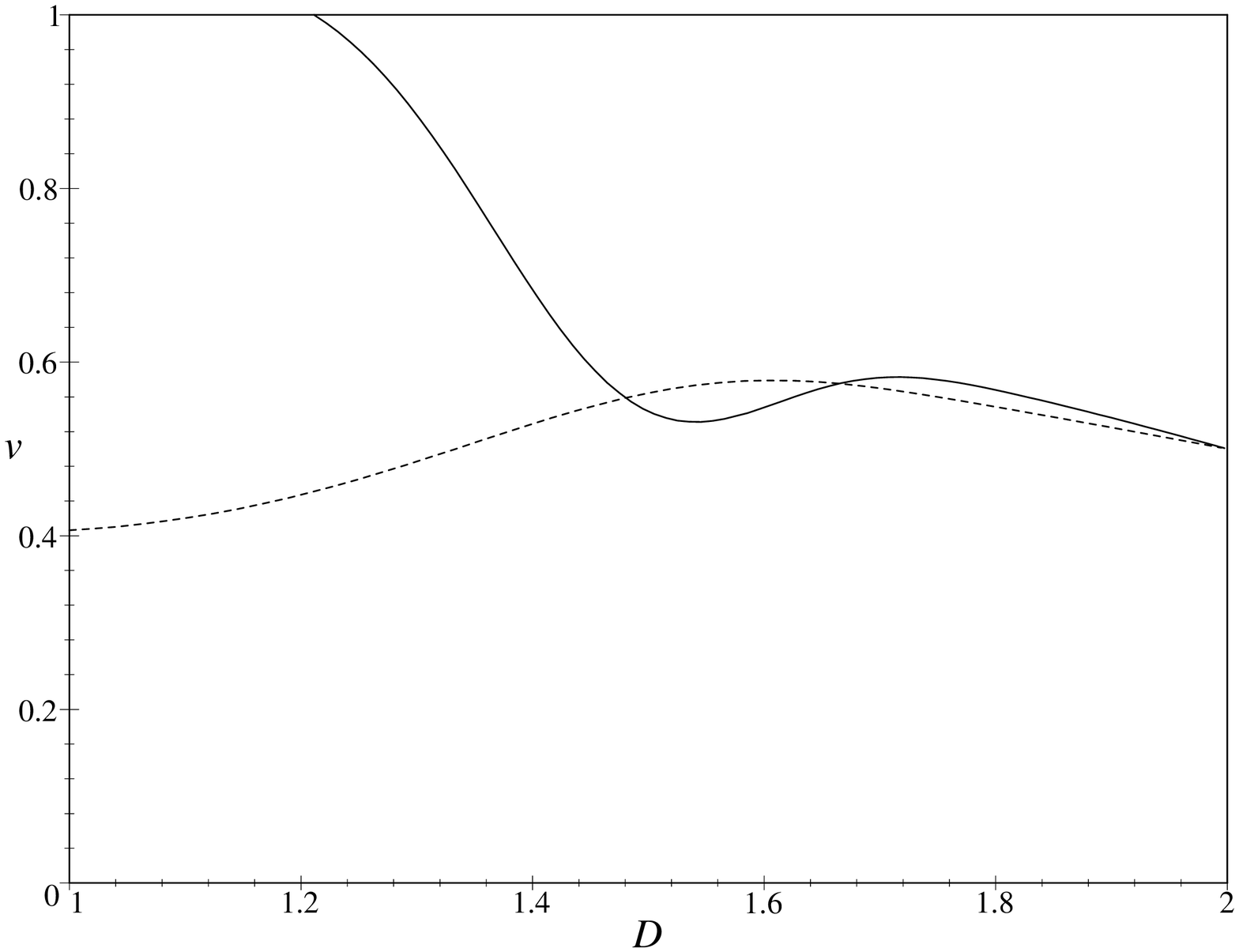}}
\hspace{0.0cm}
\Dd{2}{20}%
\epsfxsize=8.0cm \parbox{8.0cm}{\epsfbox{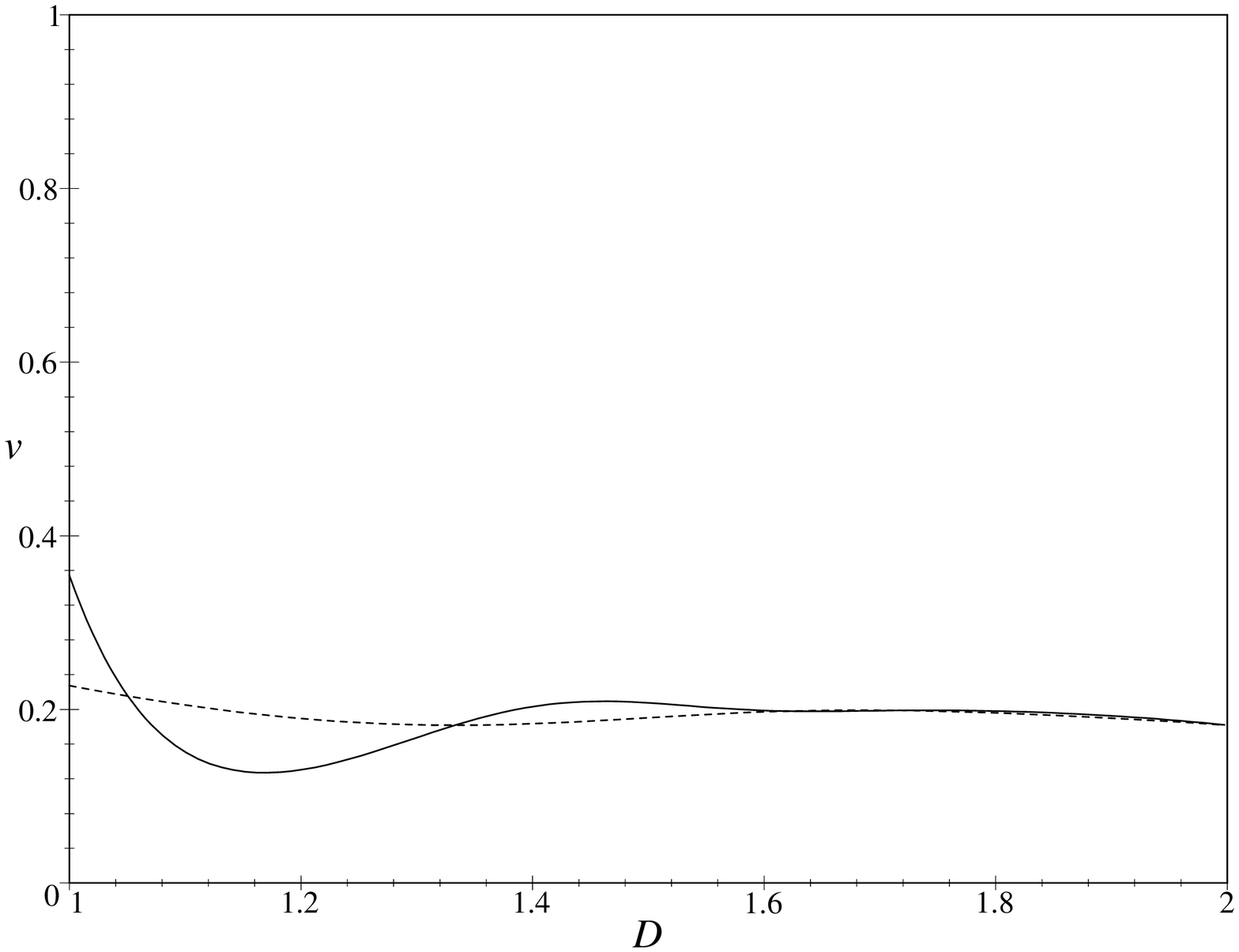}}
}
\caption{Extrapolation of $\nu (d+2)$ in $D$ and $d$ to $(D,d)=$ (1,2), (1,3), (2,2),
(2,3), (2,6), (2,20).}
\label{nu(d+2)-D-d}
\end{figure}

\begin{figure}[htb] 
\centerline{
\Dd{1}{2}%
\epsfxsize=8.0cm \parbox{8.0cm}{\epsfbox{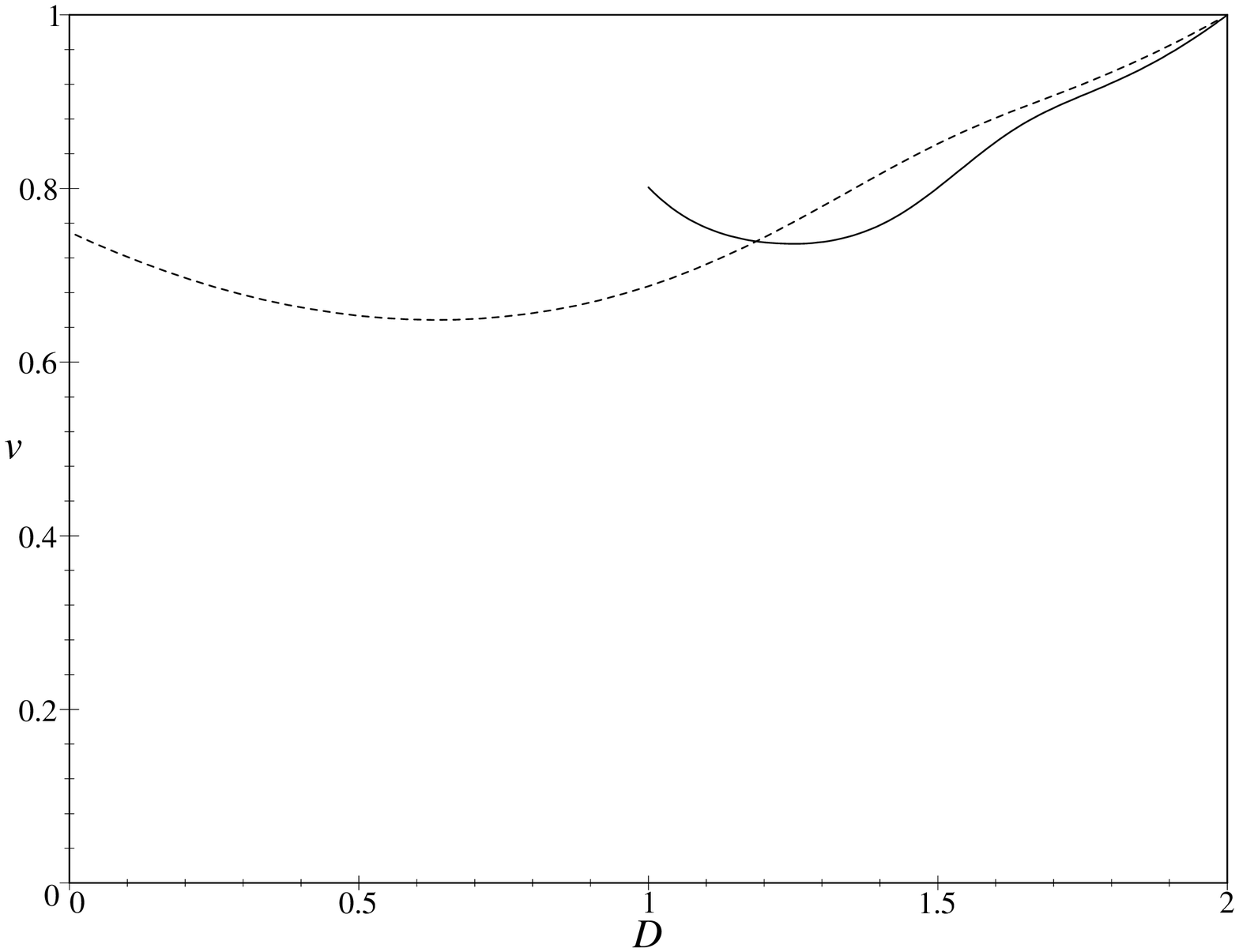}}
\hspace{0.0cm}
\Dd{1}{3}%
\epsfxsize=8.0cm \parbox{8.0cm}{\epsfbox{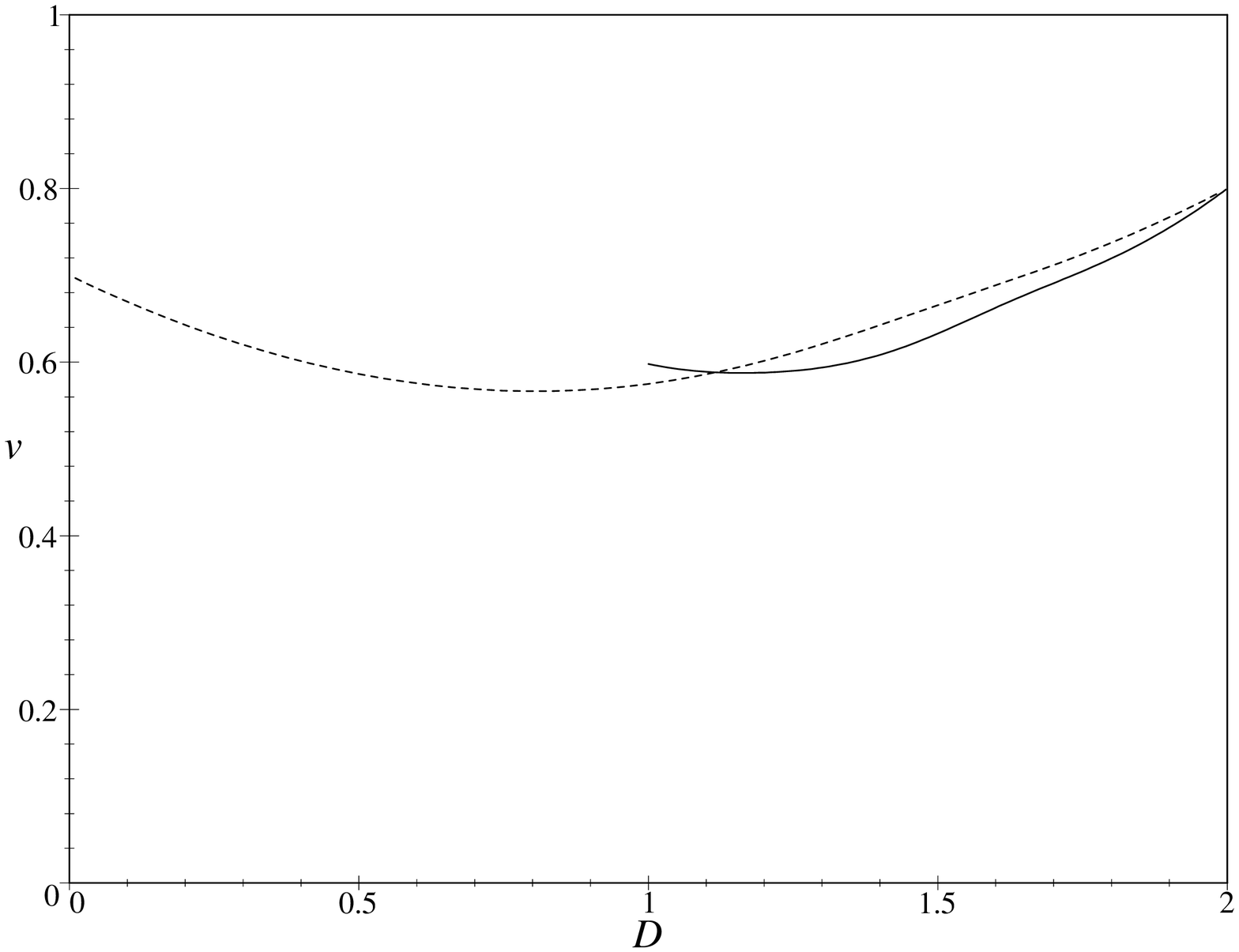}}
}
\centerline{
\Dd{2}{2}%
\epsfxsize=8.0cm \parbox{8.0cm}{\epsfbox{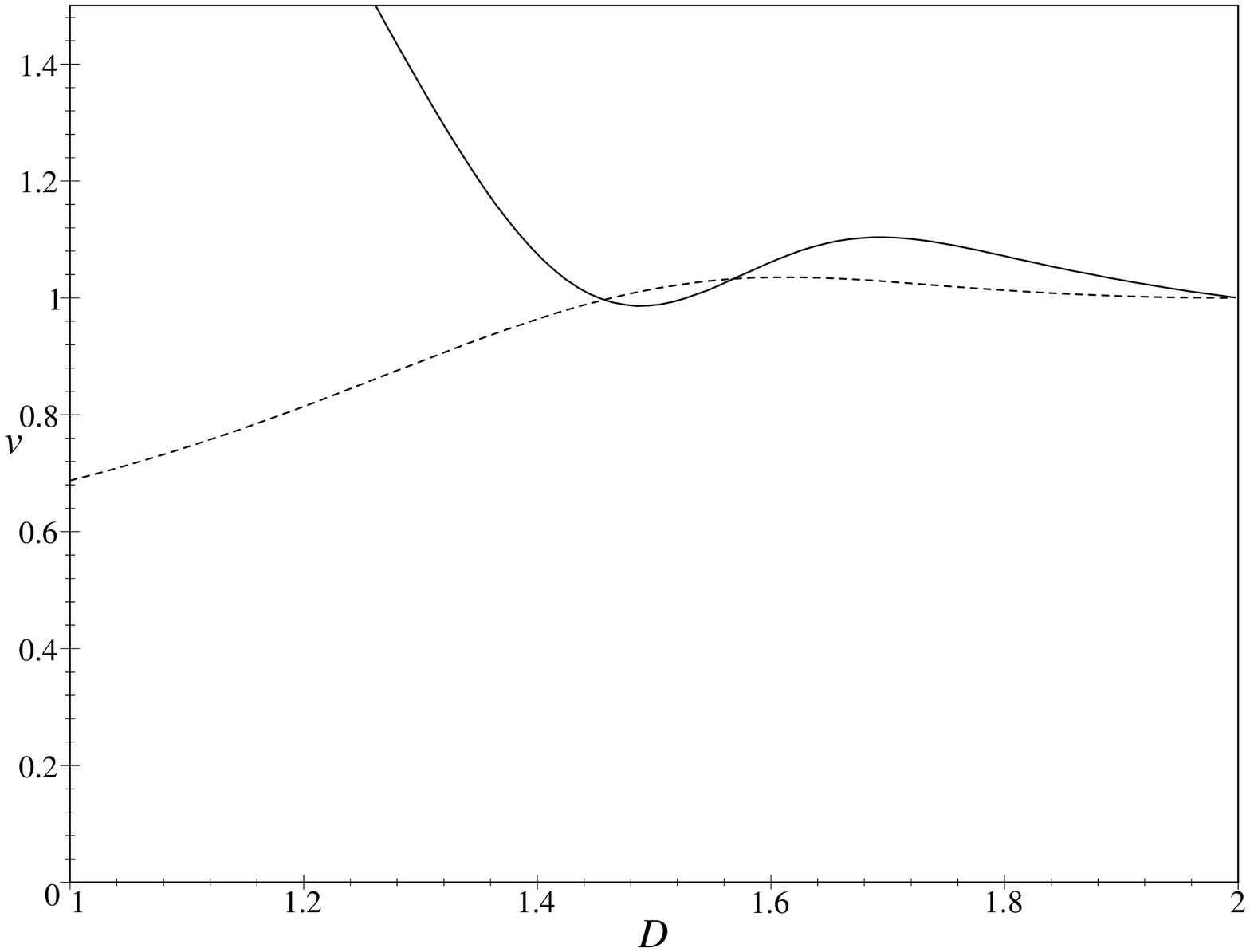}}
\hspace{0.0cm}
\Dd{2}{3}%
\epsfxsize=8.0cm \parbox{8.0cm}{\epsfbox{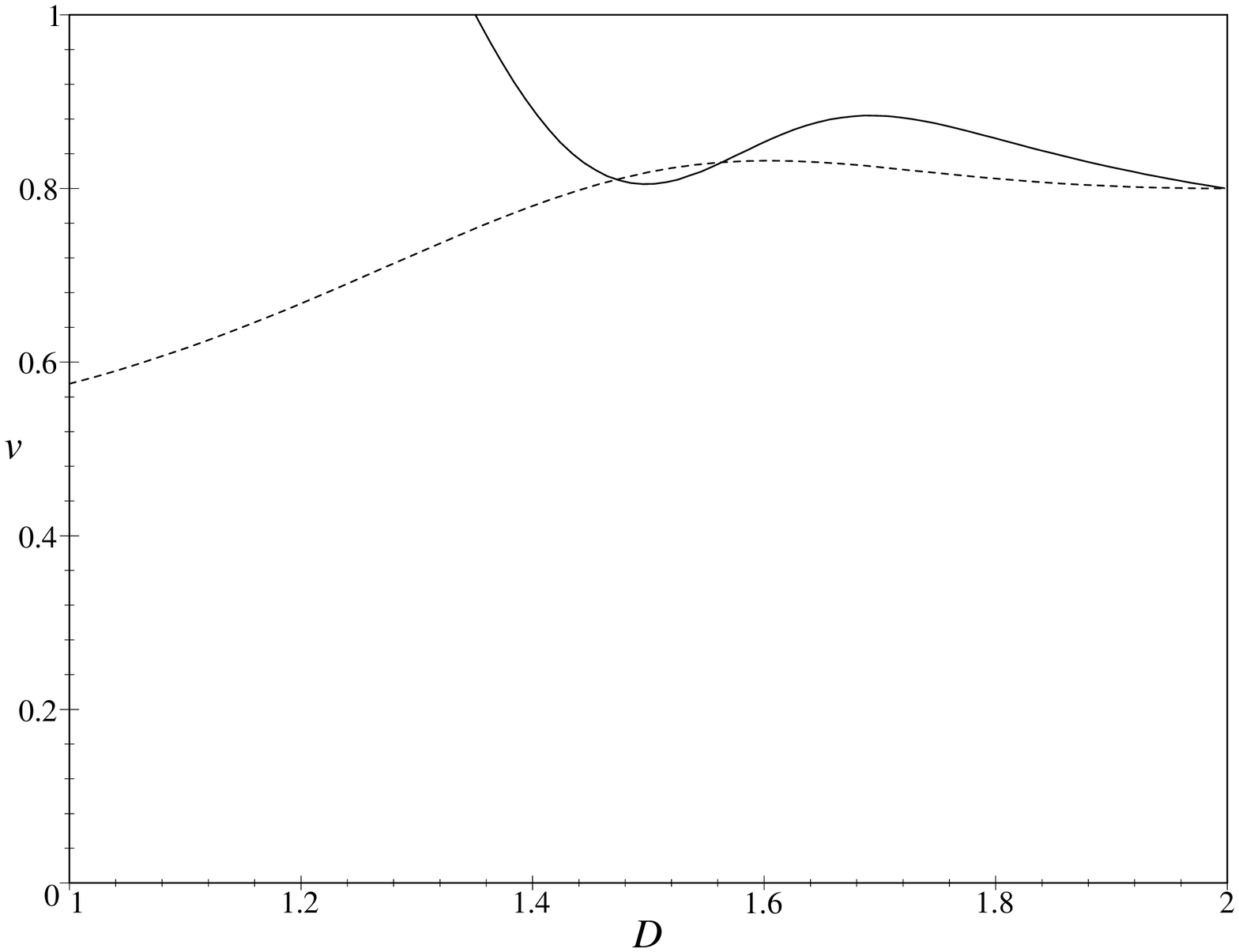}}
}
\centerline{
\Dd{2}{6}%
\epsfxsize=8.0cm \parbox{8.0cm}{\epsfbox{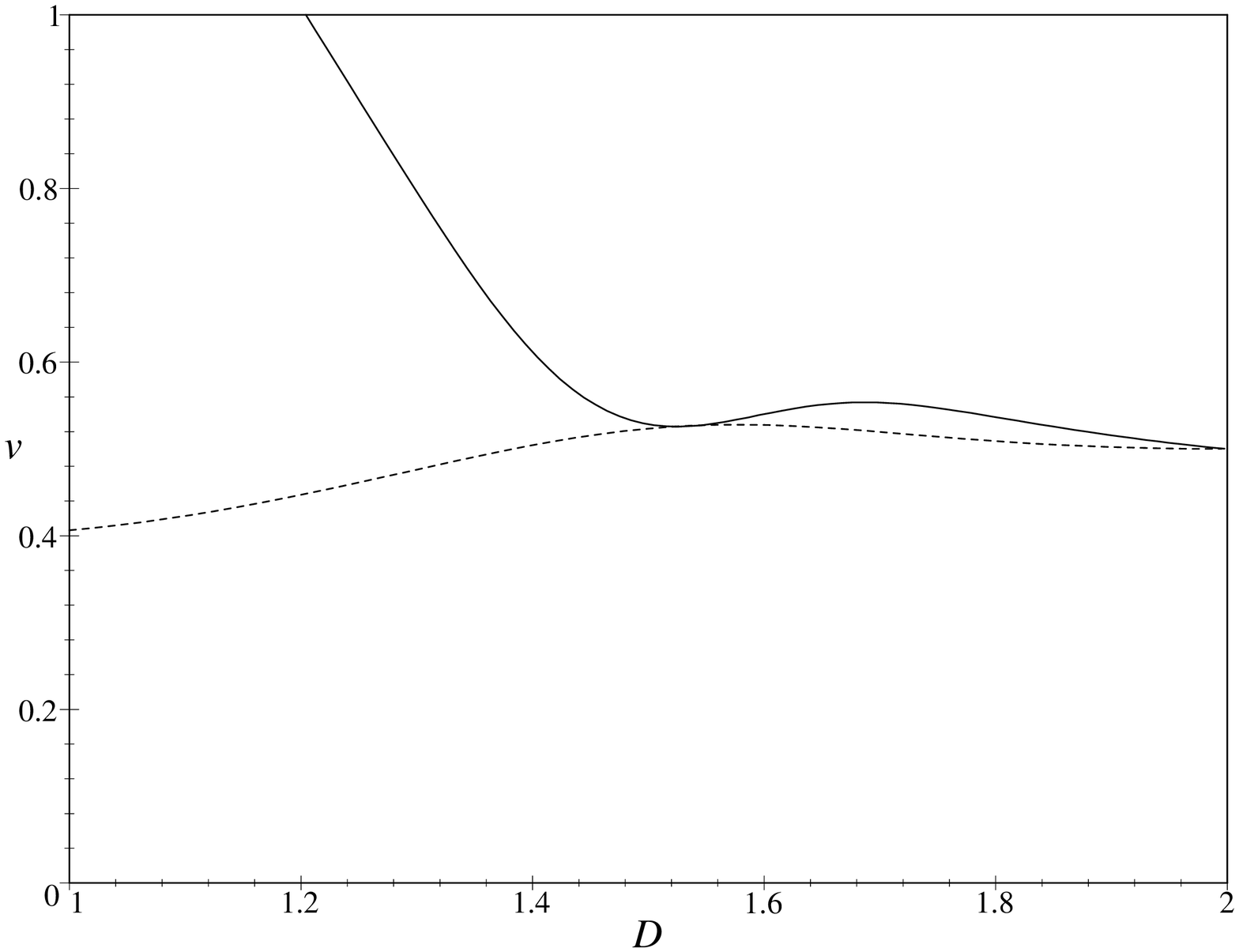}}
\hspace{0.0cm}
\Dd{2}{20}%
\epsfxsize=8.0cm \parbox{8.0cm}{\epsfbox{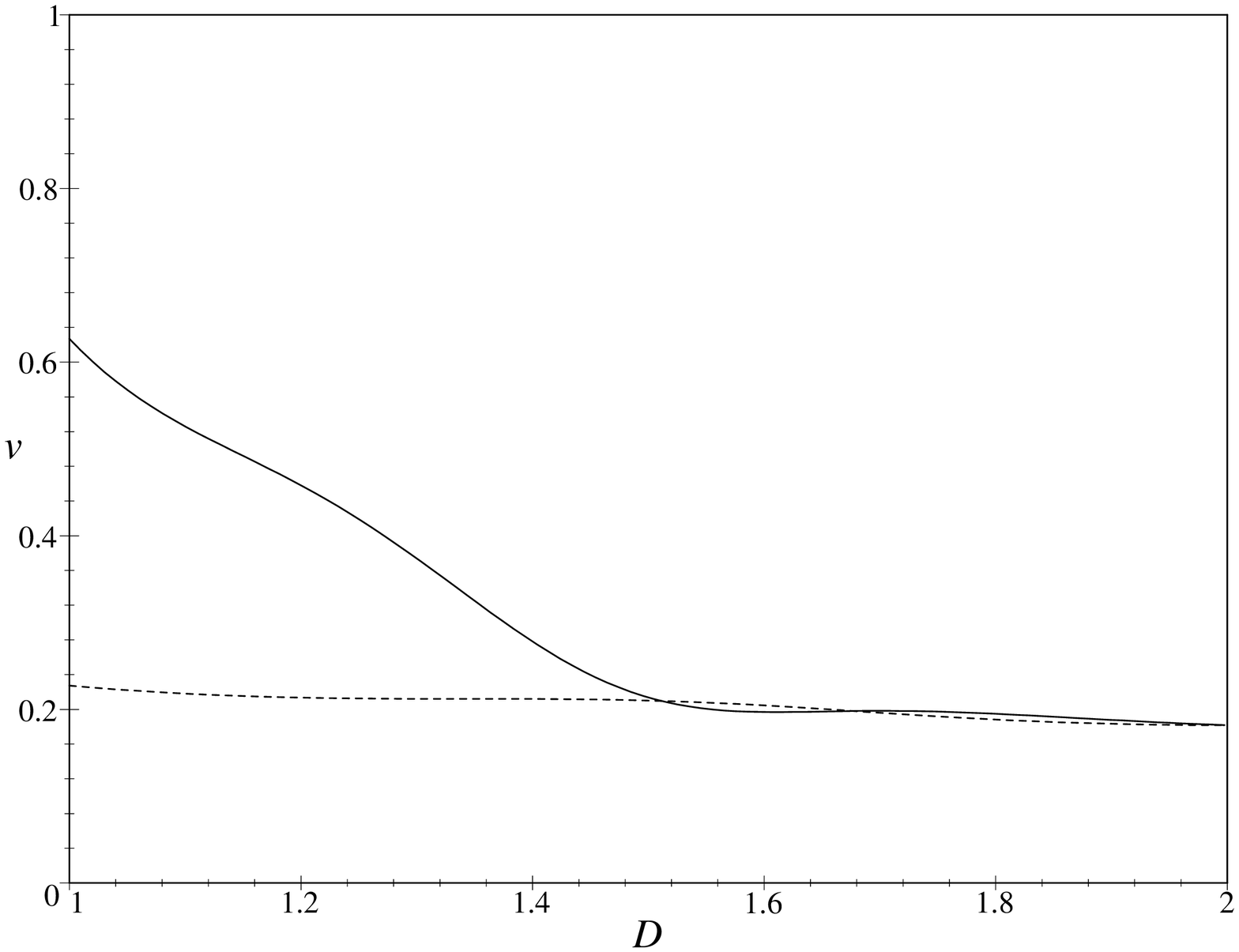}}
}
\caption{Extrapolation of $\nu(d+2)$ in $d$ and $\E$ to  $(D,d)=$ (1,2), (1,3), (2,2),
(2,3), (2,6), (2,20).}
\label{f:nudp2dE}
\end{figure}

\begin{figure}[htb] 
\centerline{
\Dd{1}{2}%
\epsfxsize=8.0cm \parbox{8.0cm}{\epsfbox{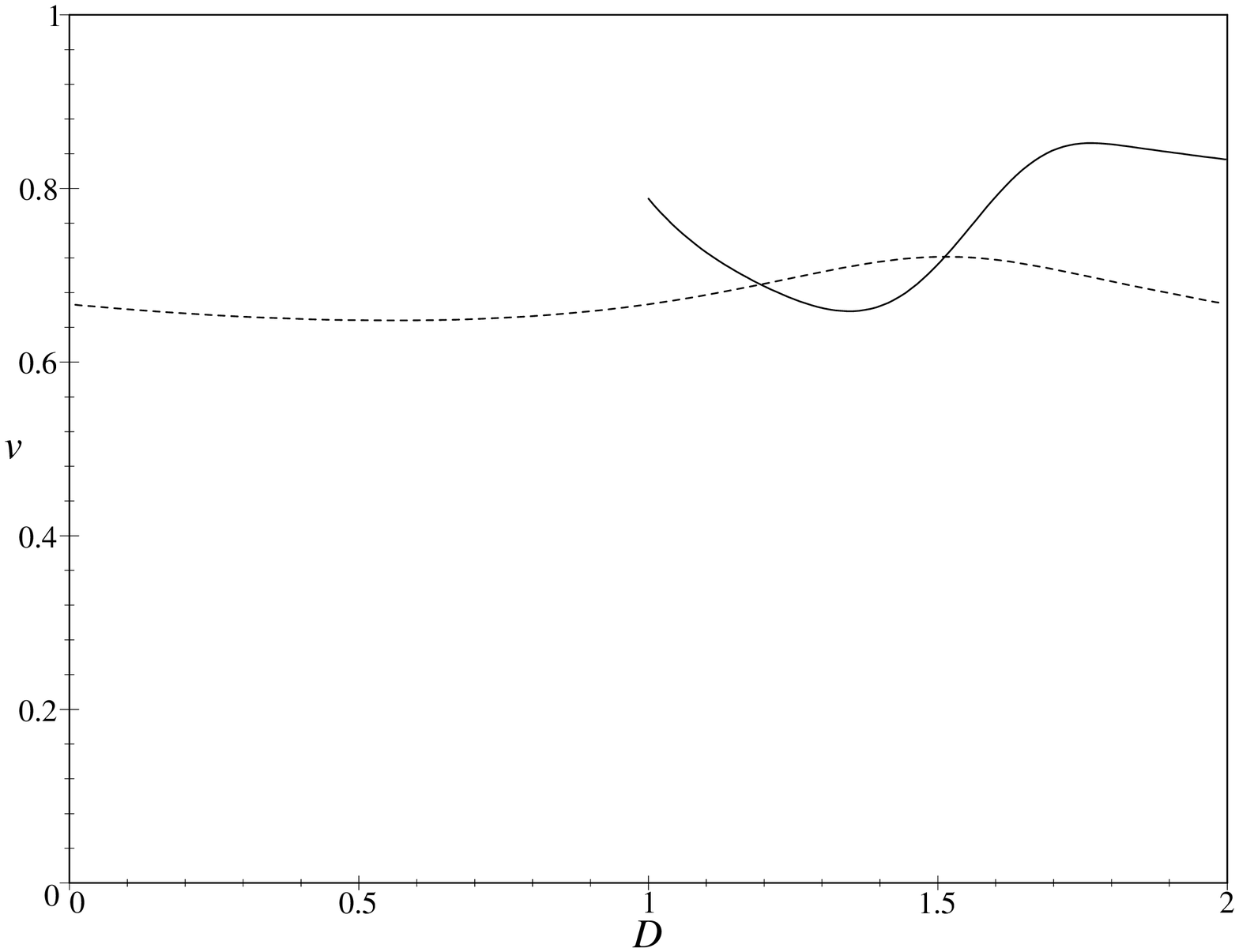}}
\hspace{0.0cm}
\Dd{1}{3}%
\epsfxsize=8.0cm \parbox{8.0cm}{\epsfbox{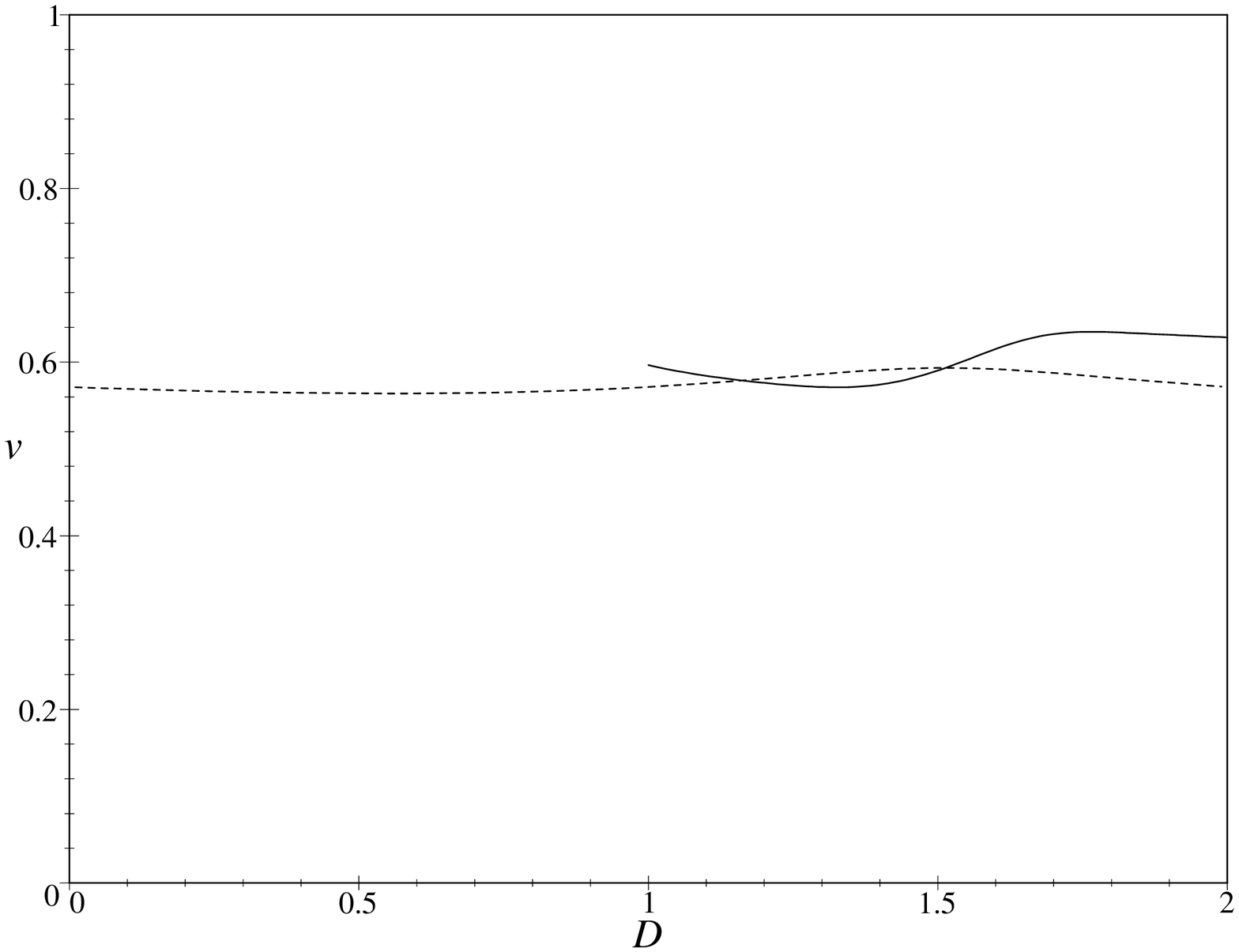}}
}
\centerline{
\Dd{2}{2}%
\epsfxsize=8.0cm \parbox{8.0cm}{\epsfbox{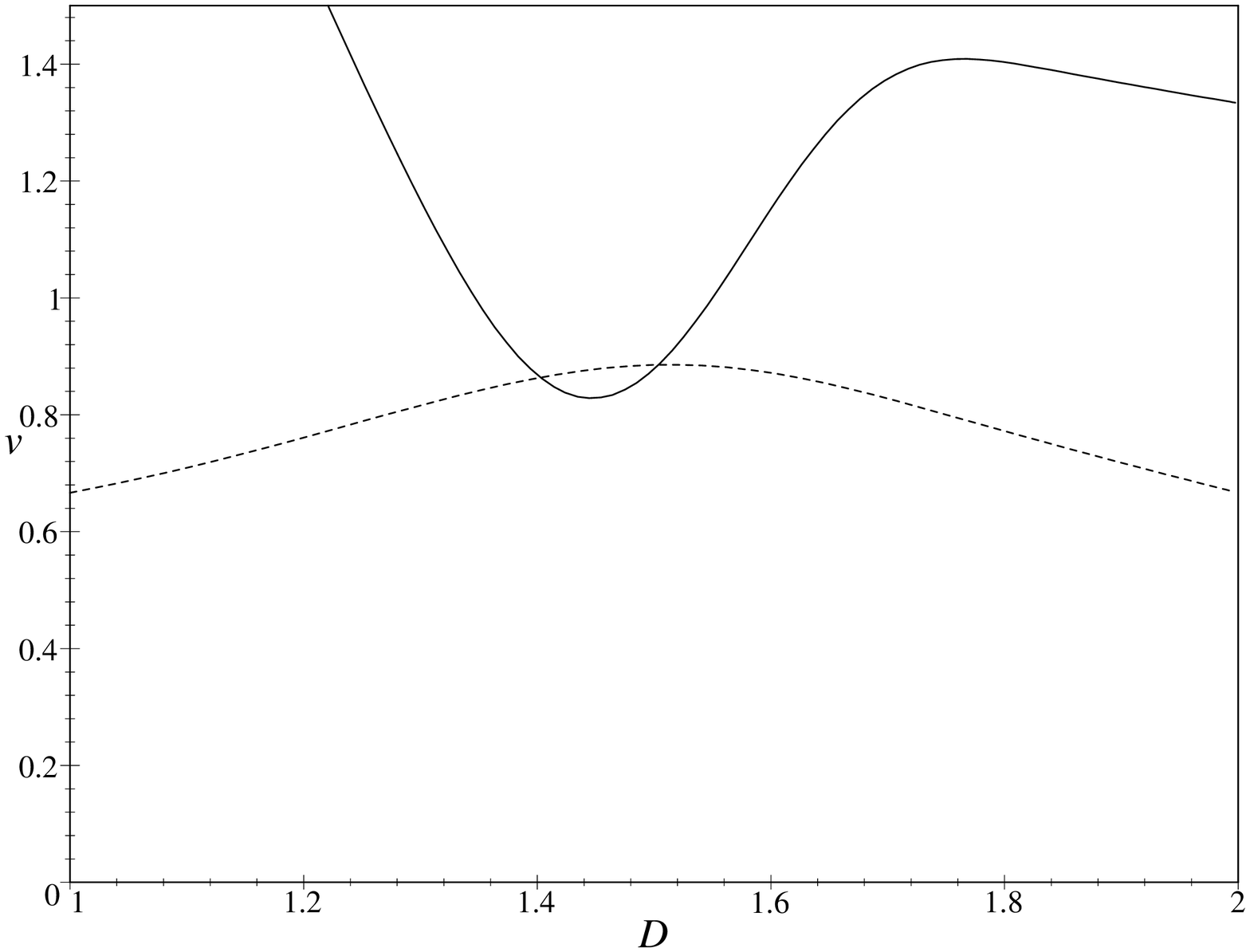}}
\hspace{0.0cm}
\Dd{2}{3}%
\epsfxsize=8.0cm \parbox{8.0cm}{\epsfbox{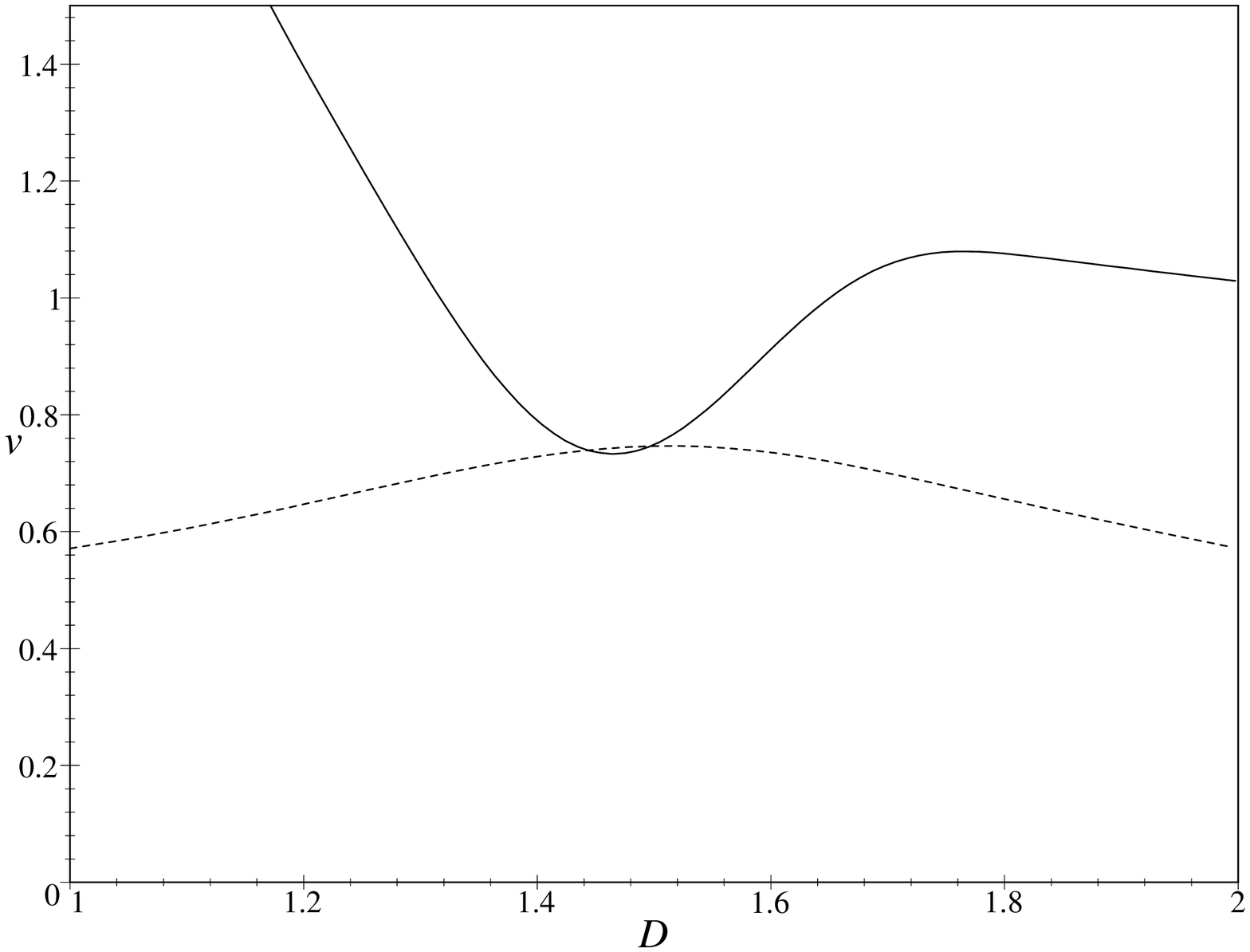}}
}
\centerline{
\Dd{2}{6}%
\epsfxsize=8.0cm \parbox{8.0cm}{\epsfbox{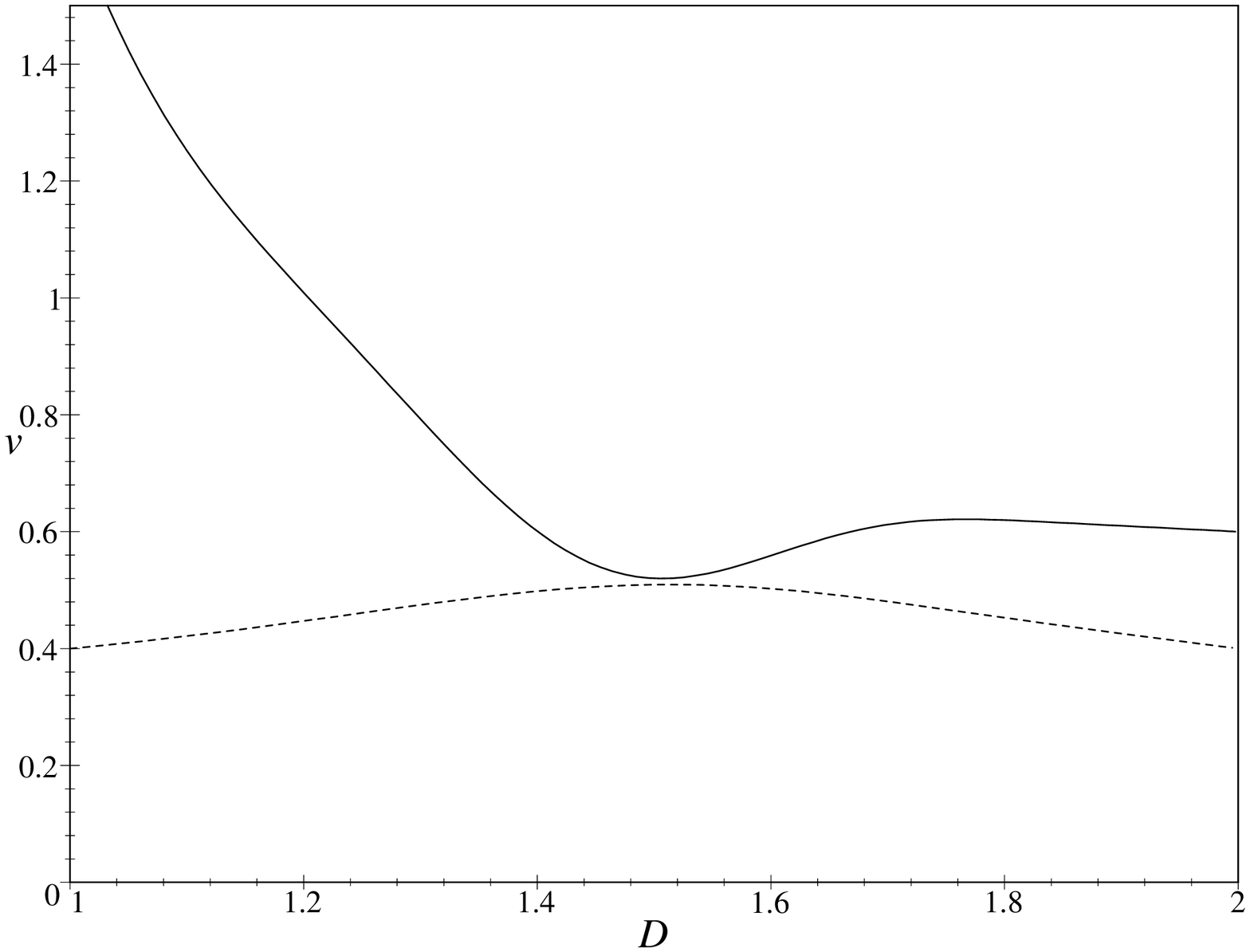}}
\hspace{0.0cm}
\Dd{2}{20}%
\epsfxsize=8.0cm \parbox{8.0cm}{\epsfbox{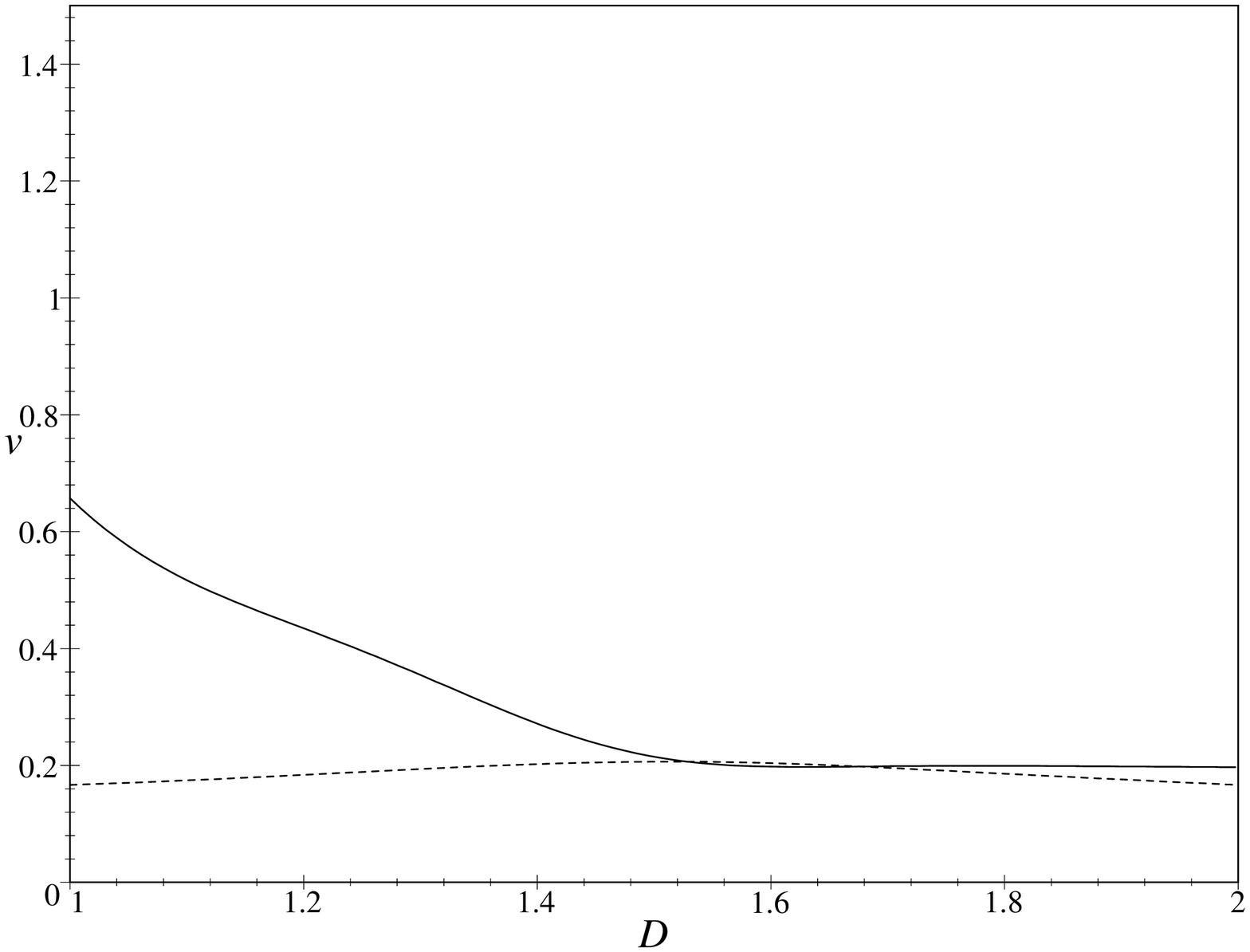}}
}
\caption{Extrapolation of $\nu(d+2)$ in $D_c(d)$ and $\E$ to $(D,d)=$ (1,2), (1,3), (2,2),
(2,3), (2,6), (2,20).}
\label{f:nudp2DcE}
\end{figure}

\begin{figure}[htb] 
\centerline{
\Dd{1}{2}%
\epsfxsize=8.0cm \parbox{8.0cm}{\epsfbox{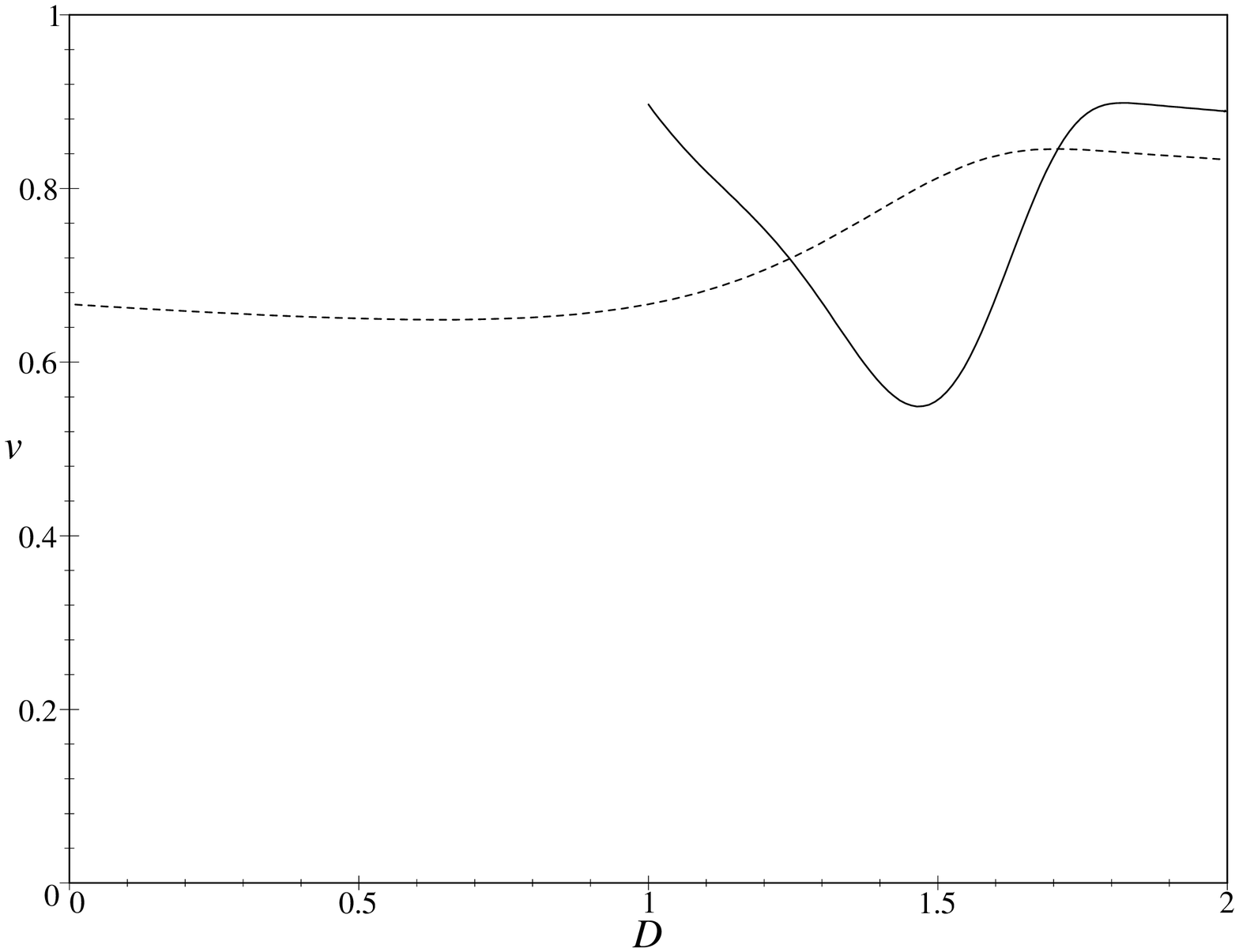}}
\hspace{0.0cm}
\Dd{1}{3}%
\epsfxsize=8.0cm \parbox{8.0cm}{\epsfbox{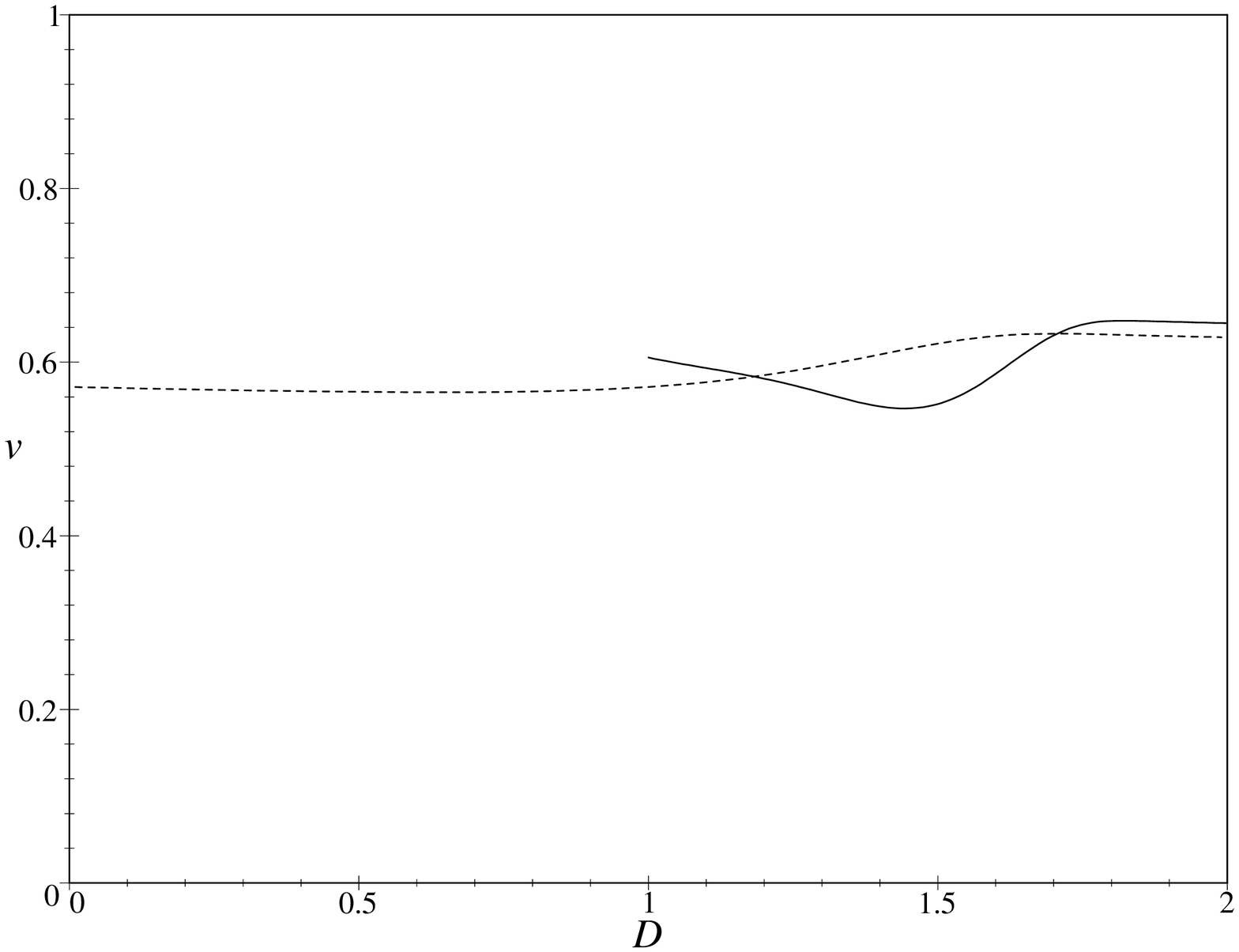}}
}
\centerline{
\Dd{2}{2}%
\epsfxsize=8.0cm \parbox{8.0cm}{\epsfbox{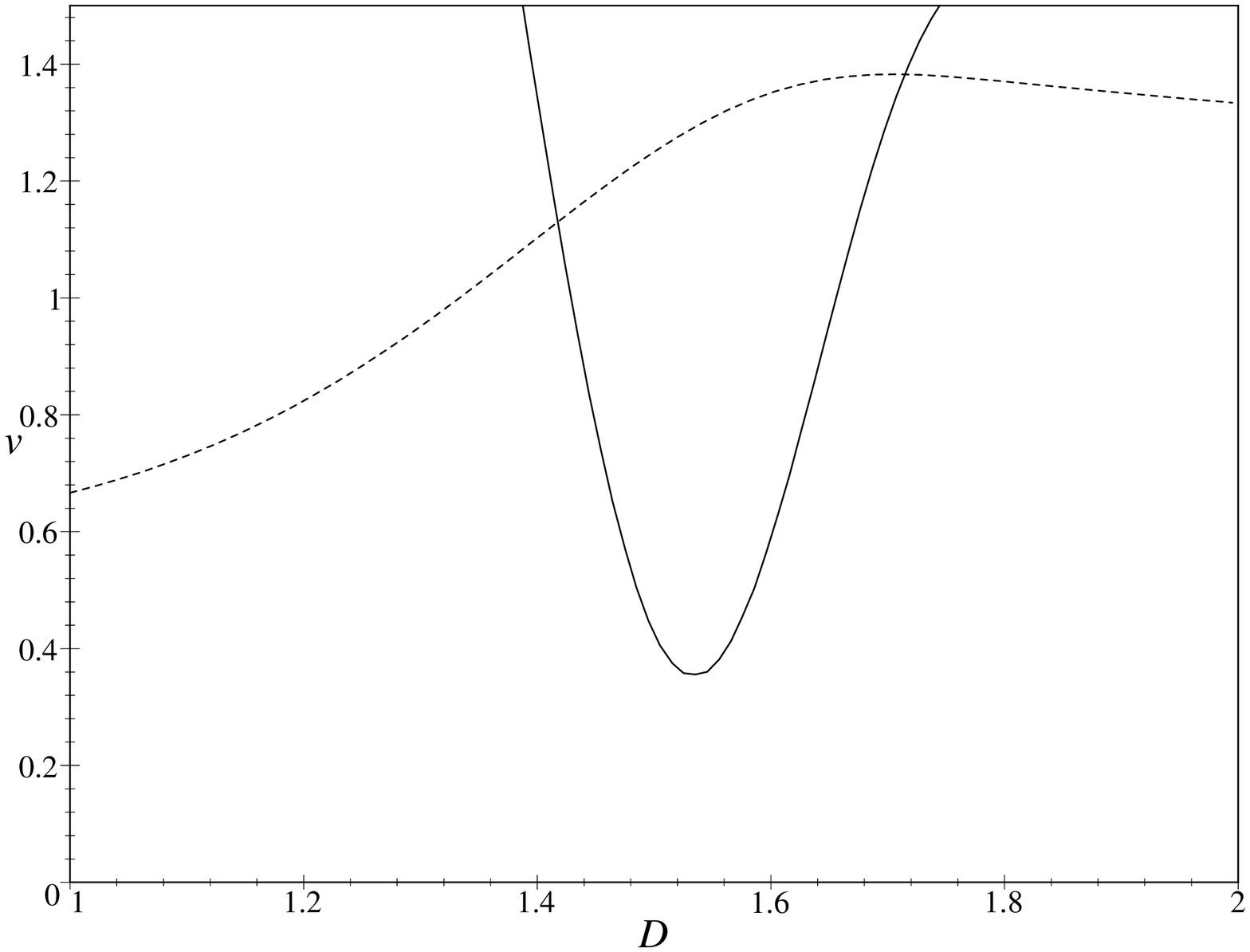}}
\hspace{0.0cm}
\Dd{2}{3}%
\epsfxsize=8.0cm \parbox{8.0cm}{\epsfbox{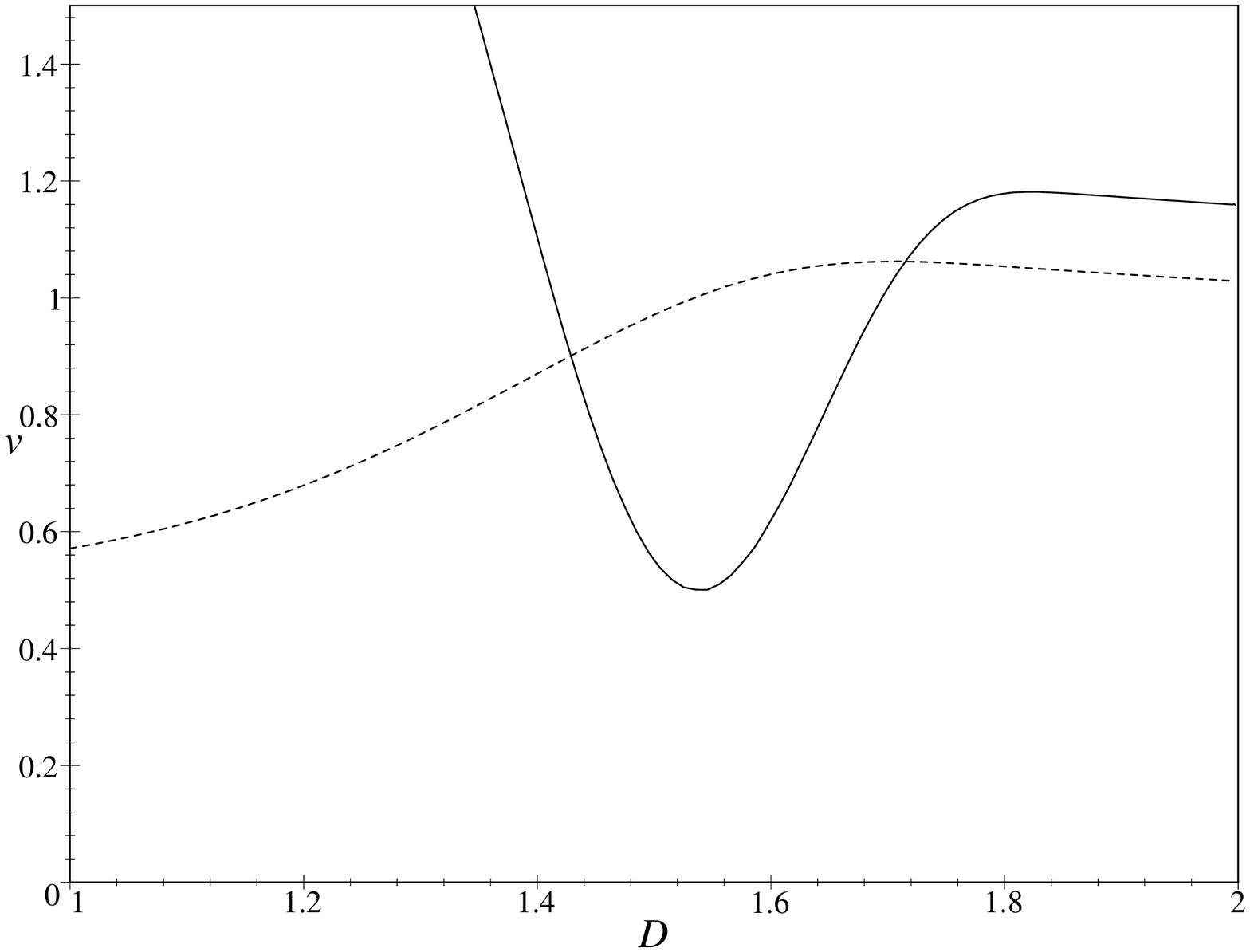}}
}
\centerline{
\Dd{2}{6}%
\epsfxsize=8.0cm \parbox{8.0cm}{\epsfbox{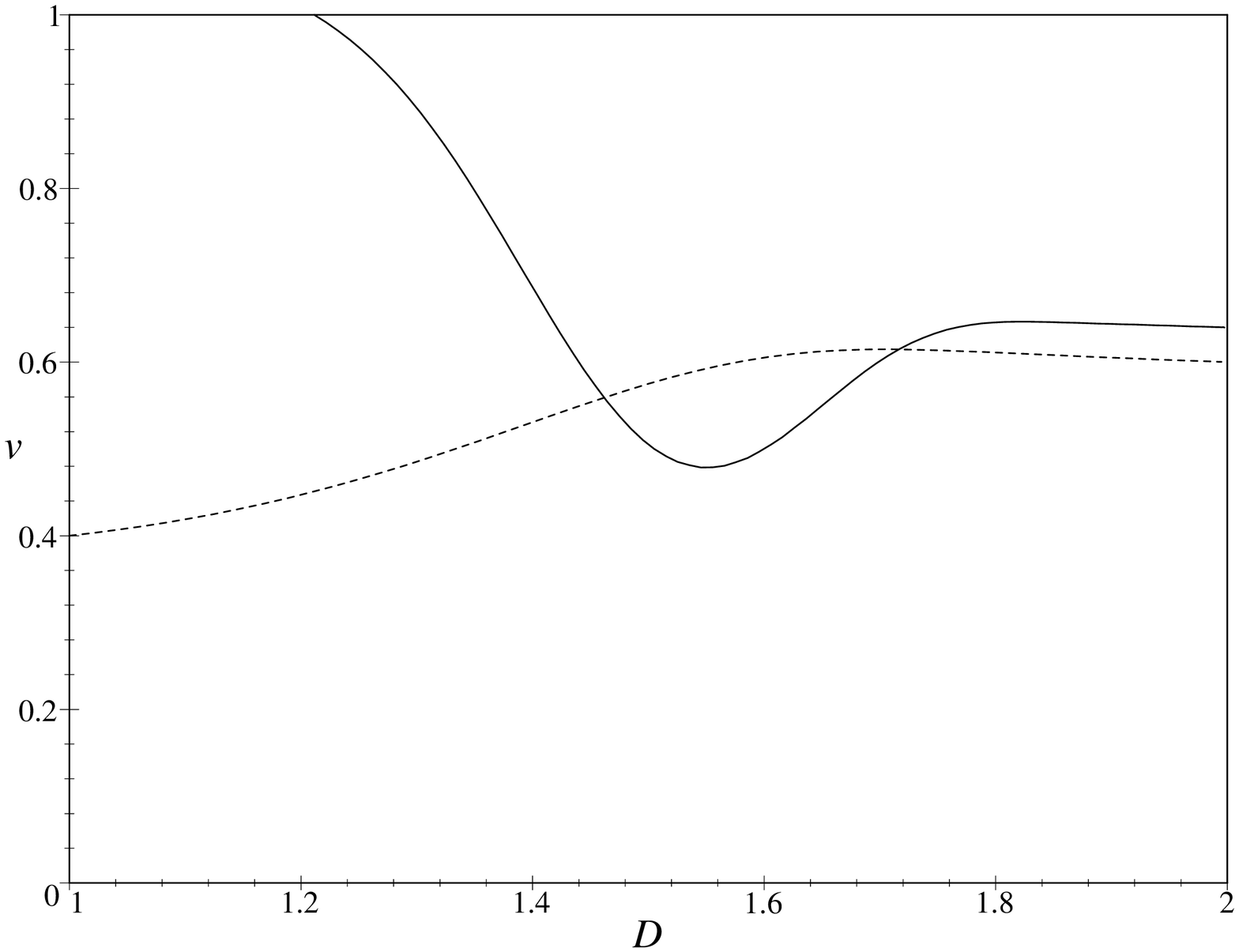}}
\hspace{0.0cm}
\Dd{2}{20}%
\epsfxsize=8.0cm \parbox{8.0cm}{\epsfbox{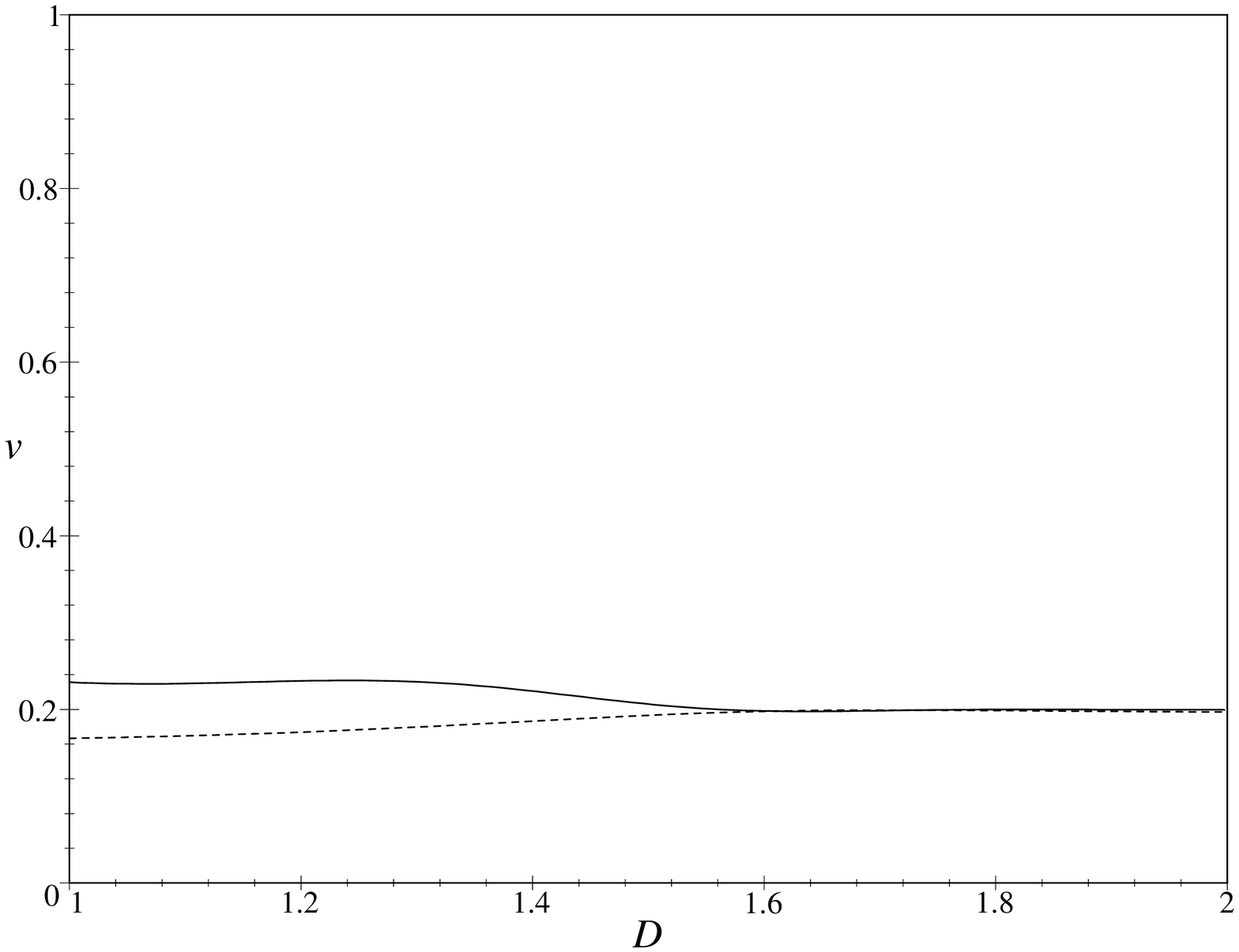}}
}
\caption{Extrapolation of $\nu(d+2)$ in $D$ and $D_c(d)$ to $(D,d)=$ (1,2), (1,3), (2,2),
(2,3), (2,6), (2,20).}
\label{f:nudp2DDc}
\end{figure} 

\clearpage

\subsection{Summary of the 2-loop extrapolations for $\nu$.}

\begin{figure}[h] 
\centerline{
\epsfxsize=13cm \parbox{13cm}
{\epsfbox{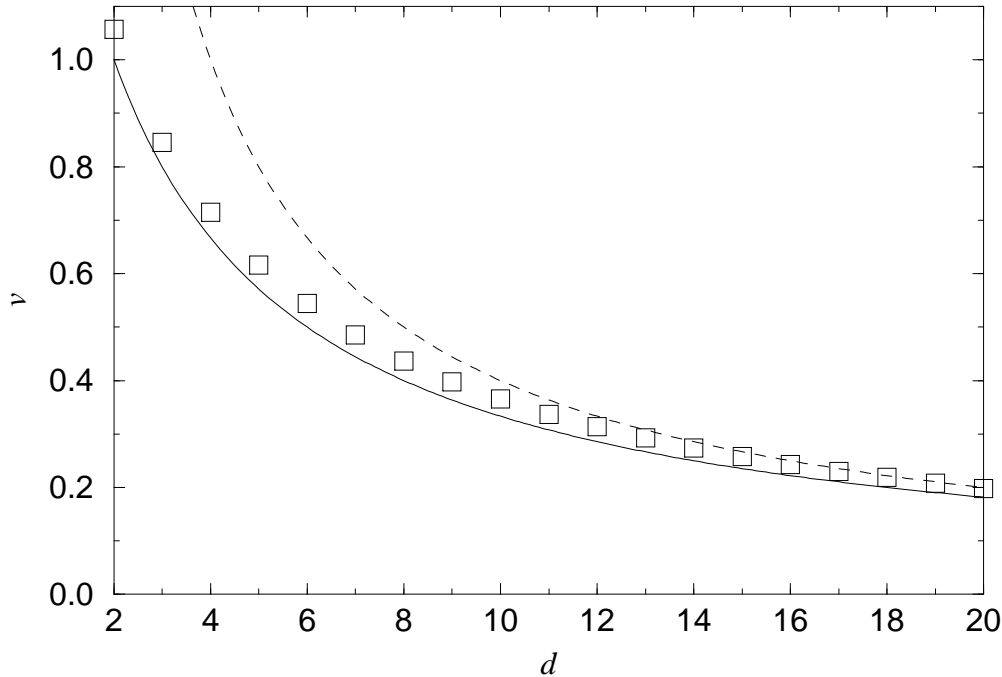}}}
\caption{Extrapolation of the 2-loop results in $d$ and $\E$ for membranes
$D=2$ in $d$ dimensions, using (\protect\ref{e: nu(d+2)=}) (squares). The solid line is the prediction made
by Flory's theory, the dashed line by the variational ansatz.}
\label{f: nu(d)}
\end{figure}

Let us summarize.
In figure \ref{f: nu(d)} we represent the results of 
a 2-loop extrapolation for $\nu$ in the case of 
membranes ($D=2$) in $d$ dimensions ($2\le d\le 20)$.
We see that for $d\to\infty$ the prediction of the Gaussian variational
method becomes exact, as argued above.
For small $d$, the prediction made by Flory's argument is close to our
results.
This is a non-trivial statement, since the membrane case corresponds to
$\E=4$ and in comparison with polymers in $d=3$, where $\E=1/2$,  the 2-loop
corrections were expected to be large.
In fact we have found that the 2-loop corrections are small when one expands
around the critical curve $\E=0$ for an adequate range of $D\approx 1.5$
(depending slightly on $d$ and on the choice of variables) and a suitable
choice of extrapolation variables.
In this case the 2-loop corrections are even smaller than the 1-loop
corrections and allow for more reliable extrapolations to $\E=4$.

\subsection{Other critical exponents}
The 2-loop calculations presented in this paper allow in principle to
compute other scaling exponents for self-avoiding tethered membranes.
The first exponent is the so-called correction to scaling
exponent $\omega$ which governs the corrections to the large $L$ scaling
behaviour.
It is known that this exponent is given by the slope of the $\beta$-function
at the IR fixed point $b^*$
\be
\omega\ =\ \left.\frac{\partial}{\partial b} \beta(b)\right|_{b=b^*}
\label{e:omegac}
\ee
Its $\E$-expansion is given by
\be
\label{e:omeg2E}
\omega = \E + \frac{(2-D) \tilde c_1(D) + 2D \tilde f_1(D) }{(2-D) \left( 
\frac1{2-D} \frac{\Gamma^2\left(\frac D{2-D}\right)}
{\Gamma\left( \frac{2D}{2-D}\right)}\right)^2} \E^2 + {\cal O} (\E^3)
\ee
(For the definition of $\tilde c_1(D)$ and $\tilde f_1(D)$ cp.\ (\ref{c1_res})
and (\ref{f1_res})).

We have tried to use the extrapolation methods that we have developed
for $\nu$ to get 2-loop estimates for $\omega$.
Unfortunately, it turns out that the term of order $\E^2$ is always very
large compared to the term of order $\E$ and there is no
domain along the critical curve where a reliable estimate can be
extracted from the $\E$-expansion. 
Thus the situation for the exponent $\omega$ seems to be very different from 
that for $\nu$.

\medskip

Another scaling exponent which can be obtained from our calculations is
the so-called bulk contact exponent $\theta_2$.
For a general introduction to contact exponents for polymers and
membranes we refer to \cite{r:BDcontact}.
A detailed discussion of the calculation of contact exponents  within the
Edwards model for self-avoiding membranes will be given in \cite{r:DDG4}.
We simply recall the basic results here.
The contact exponent $\theta_2$ is related to the probability to find two
{\it fixed}
points $x_1$ and $x_2$ inside the membrane at a relative distance
$r=|\vec r|$ in external $d$-dimensional space:
\be
\label{e:ContactP}
P(r;x_1,x_2) = \left< \tilde \delta^d(\rvec-(\rvec(x_1)-\rvec(x_2))) \right>
\ee
For a large membrane, $P$ is expected to take the scaling form
\be
\label{e:ScalingP}
	P(r;x_1,x_2)=R_{12}^{-d} F(r/R_{12})
\ee
where $R_{12}$ is the mean distance between $x_1$ and $x_2$.
\be
\label{e:R12}
R_{12}^2=\half \left< (\rvec(x_1) -\rvec(x_2))^2 \right>
\ee
The contact exponent $\theta_2$ is given by the small $r$ behavior of the
scaling function $F$
\be
\label{e:defThe2}
F\left(\frac r {R_{12}}\right)\ \sim\ \left(\frac r {R_{12}}\right)^{\theta_2}\qquad\mbox{when}\qquad r\to 0
\ee
$\theta_2$ is simply related to the scaling dimension $\omega_{12}$
of the 2-membrane contact operator
\be
\label{e:delta12}
\delta_{12}(x_1,x_2)\ =\ \tilde \delta^d(\rvec_1(x_1)-\rvec_2(x_2))
\ee
in the model of two independent self-avoiding membranes.
This model is described by the Hamiltonian
\bea
\label{e:2MemHam}
{\cal H} &=& 
\left( \int_{x_1\in M_1} \half (\nabla \rvec_1(x_1))^2
+ \int_{x_2\in M_2} \half (\nabla \rvec_2(x_2))^2\right)
\nonumber\\
&& + b 
\left(
\int_{x_1\in M_1}\int_{y_1\in M_1}\tilde\delta^d(\rvec_1(x_1)-\rvec_1(y_1))
+
\int_{x_2\in M_2}\int_{y_2\in M_2}\tilde\delta^d(\rvec_2(x_2)-\rvec_2(y_2))
\right)
\nonumber\\
&&\,+\, 2\,t\int_{x_1\in M_1}\int_{y_2\in M_2}
\tilde\delta^d(\rvec_1(x_1)-\rvec_2(y_2))
\eea 
This model can be made UV-finite at $\E=0$ by the same renormalization factors
for $\rvec$ and $b$ as the 1-membrane model,
i.e.\ by replacing in \eq{e:2MemHam}
$\rvec\to Z(b)^{1/2}\rvec$ and $b\to b {Z_b(b)} Z(b)^{d/2}\mu^\E$,
but with an additional renormalization for the inter-membrane coupling
$t\to t Z_t(b,t) Z(b)^{d/2} \mu^\E$.
The new counter-term $Z_t$ can be calculated in terms of the same divergent
diagrams as those which contribute to $Z_b$, but with different numerical
factors.
In particular, one can show that when $t=b$, $Z_t(b,t=b)=Z_b(b)$, so that
the symmetric 2-membranes model reduces to the 1-membrane model.

As a consequence of this formalism, one can define a new RG function, $\beta_t$
\be
\label{e:Betat}
\beta_t(b,t)\ =\ 
\left.\mu{\partial\over\partial\mu}t\right|_{\mbox{\scriptsize bare $b,t$}\atop 
\mbox{\scriptsize fixed}}
\ee
calculate the RG-flow in the $(b,t)$ plane and check that
$(b,t)=(b^*,b^*)$ is the IR stable fixed point which governs the scaling
behavior of a large membrane.
It turns out that the $\theta_2$  contact exponent and the anomalous dimension
$\omega_{12}$ of the  contact operator $\delta_{12}$ are related to the
$t$-derivative of the $\beta_t$ function at the IR-fixed point by
\be
\omega_{12}\ =\ -\nu^*\theta_2\ =\ \left.
{\partial\over\partial t}\beta_t\right|_{b=t=b^*}
\label{e:omega12}
\ee
As a result, the $\E$-expansion for $\theta_2$ involves the same diagrams 
as $\omega$ and $\nu$, but in a different combination.
The result is somehow complicated, so let us simply write the counterterm
$Z_t$ and compare it with  $Z_b$ (already given in sections~3 and 4).
\bea
Z_b(b)\ &=&\ 1+{a_1\over\E}b+\left({a_1(a_1-1/2)\over\E^2}+{a_1/4D+{\cal C}_1+
{\cal C}_2+{\cal C}_3\over\E}\right) b^2+\cdots\nonumber\\
Z_t(b,t)\ & =&\ 1+{a_1\over \E}t+
\left({{a_1}^2\over\E}+{{\cal C}_1\over\E}\right)t^2
+\left(-{a_1/2\over\E^2}+{a_1/4D+{\cal C}_2+{\cal C}_3\over\E}\right)bt
+\cdots
\nonumber\\
&&
\label{e:ZbZt}
\eea

We have tried to use the extrapolation methods that we have developed
for $\nu$ to get 2-loop estimates for $\theta_2$.
As for $\omega$, it turns out that the term of order $\E^2$ is always very
large compared to the term of order $\E$ and that there is no
domain along the critical curve where a reliable estimate can be
extracted from the $\E$-expansion.

\subsection{Scaling for membranes at the $\theta$-point (tri-critical point)}
\begin{figure}[htb] 
\centerline{
\epsfxsize=8cm \parbox{8cm}{\epsfbox{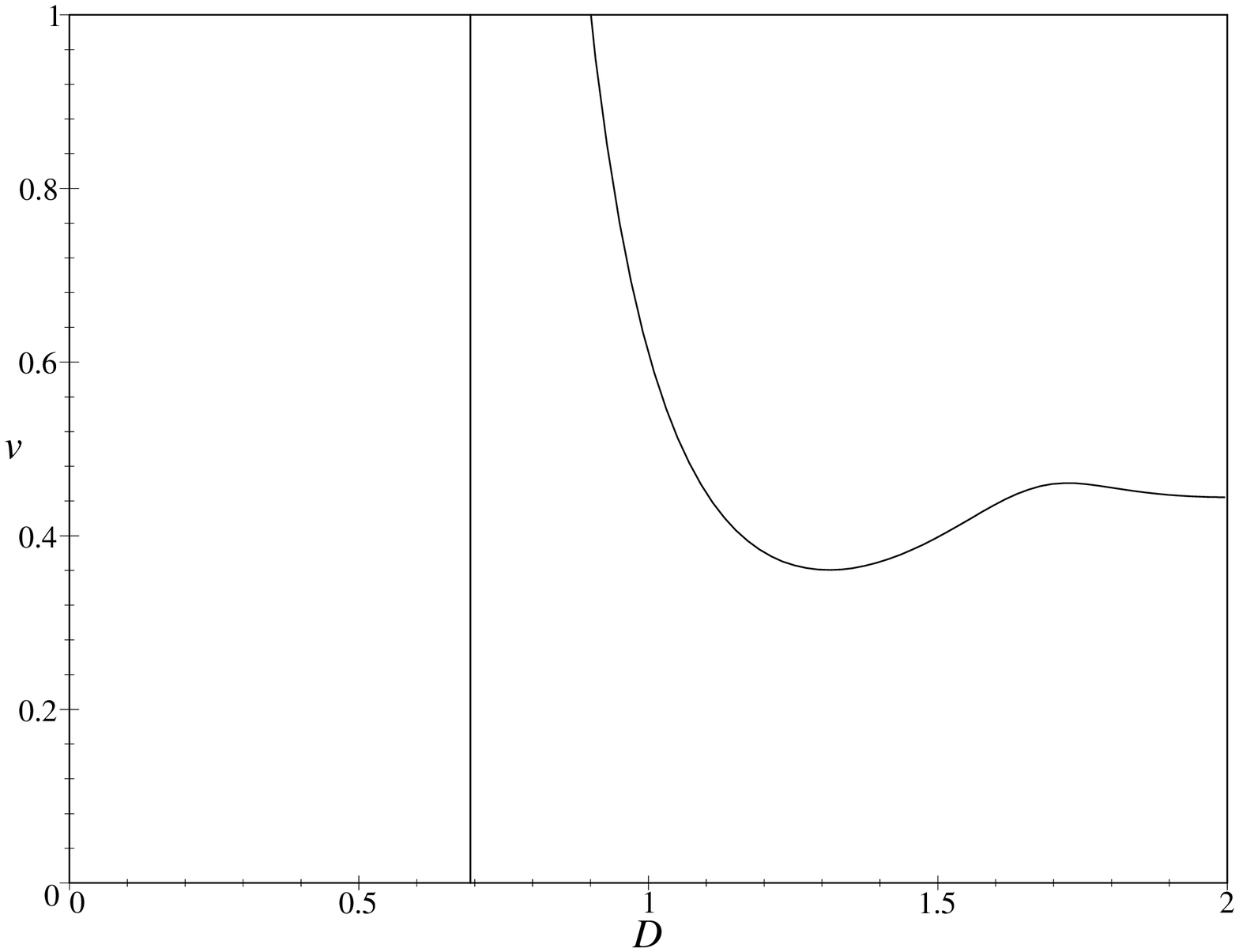}}
\epsfxsize=8cm \parbox{8cm}{\epsfbox{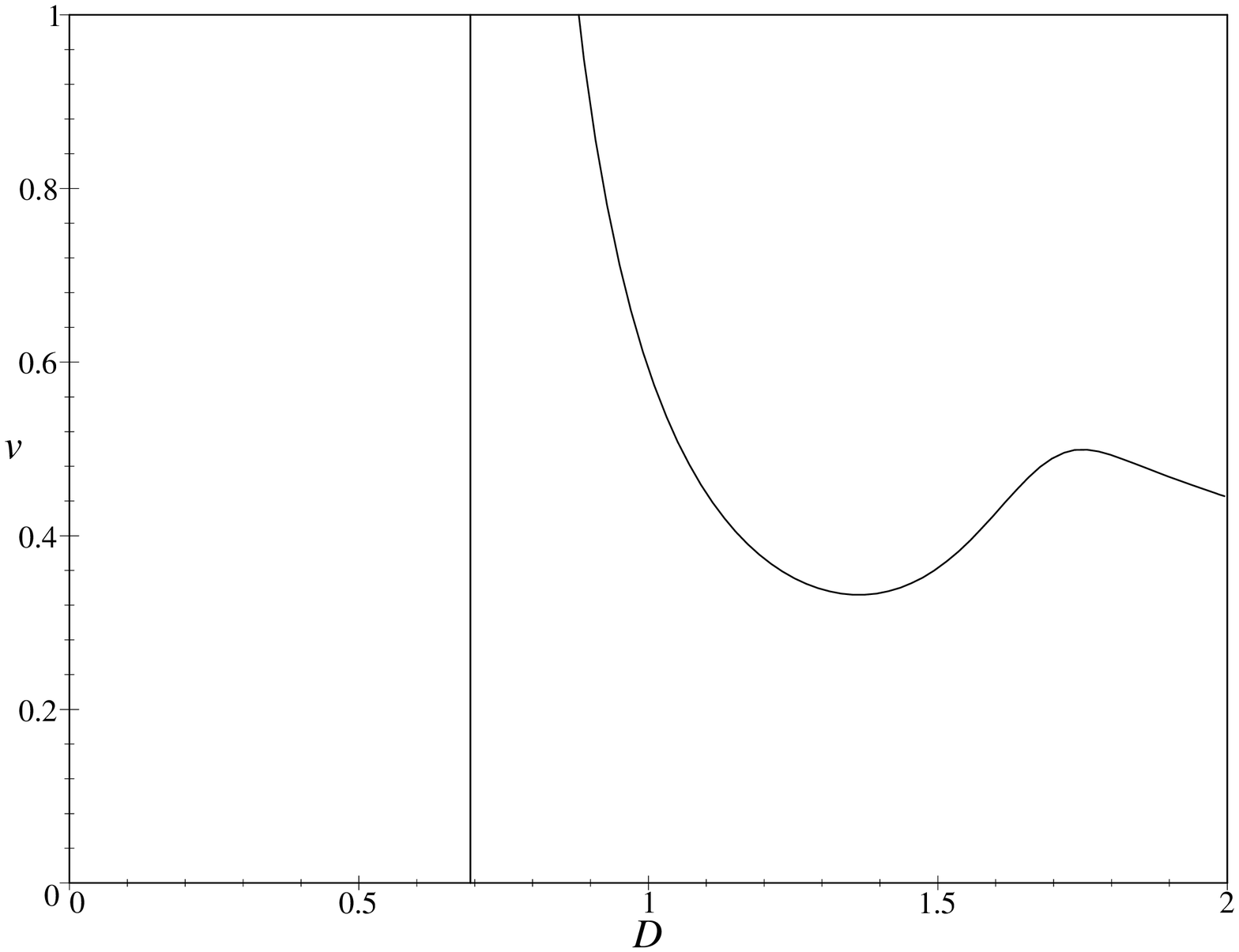}}
}
\caption{Extrapolation for $\nu $ at the tri-critical point in $(D_c(d),\E')$ 
and in $(D,D_c(d))$ to $D=2$ and 
$d=3$, 1-loop results.}
\label{fig: tricrit}
\end{figure}
In \cite{r:WieseDavid95} it was shown that the scaling behavior of
membranes at the tri-critical $\theta$-point is dominated by the
modified 2-point interaction ${\tilde\delta''}(\rvec(x)-\rvec(y))$, which is
repulsive at short distance but attractive at larger distance.
Membranes at the $\theta$-point are thus quite different from polymers, where
the 3-point repulsive interactions dominate.
As a consequence, the scaling exponent $\nu^\theta$ at the $\theta$-point has a
different $\E$-expansion, which can be computed analytically at first order
and is found to be \cite{r:DDG4,r:WieseDavid95}
\be 
\label{e:epsexptheta}
\nu^\theta(D, \E')= \frac{2-D}2 +\frac{2-D}{4D \left( 1+\half \frac{(10D-D^2-8)\Gamma^2\left( 
\frac{D}{2-D}\right)}{(2-D)^3 \Gamma\left( \frac{2D}{2-D} \right) }\right)} \E' + {\cal O}(\E'^2)
\ee
with 
\be
\label{e:eps3}
\E'=3D-2- d{(2-D)\over 2}
\ee
This expansion is again correct up to first order in $d$ or $\E'$ and exact in
$D$. 
We can use the extrapolation methods developed in this paper to evaluate the
exponent $\nu$ from the 1-loop results.
The extrapolation method proposed by Hwa \cite{Hwa} does not work in this case.
Two variants of our method deliver reasonable  results:
The expansion in $\E'$ and $D_c(d)$ and the expansion in $D$ and $D_c(d)$.
The 1-loop estimate for membranes in 3 dimensions is
(cf.\ figure \ref{fig: tricrit}):
\be
\label{e:nu3ext}
	\nu^\theta= 0.42 \pm 0.08
\ee 
Due to our experiences with the former extrapolations, we expect that the 1-loop
results will be an underestimation.
They may be compared to the Flory-result
\be
\label{e:nu3flo}
\nu_{\mbox{\scriptsize Flory}}^\theta
 = \frac{2+D}{4+d}  \ ,
\ee
which evaluates to 0.57 in the case of membranes in 3 dimensions.

\subsection{Comparison with numerical simulations and experiments}
The numerical study of two-dimensional self-avoiding tethered membranes 
imbedded into three dimensions 
was started in 86 by Kantor, Kardar and Nelson 
\cite{r:KantorKardarNelson1986a,r:KantorKardarNelson1986b}.
They found for the exponent $\nu$ a value close to the prediction made
by Flory's argument.
These simulations as most of the following
use a system composed of balls and springs (spring and bead model).
Self-avoidance is effective between the balls.
It was a surprise when it was found \cite{r:PlischkeBoal88,r:Abr&al89} 
that the simulations of larger membranes obtain a flat phase.
Abraham and Nelson\cite{r:AbrNel90} explained this result by suggesting that
an effective bending rigidity is induced by the geometric constraints
of the model (there is a maximal angle for which neighbouring 
faces can bend due to the finite size of the balls).
If this induced effective rigidity 
is larger than the critical rigidity, where
the crumpling transition (transition to a flat phase induced by the
bending rigidity {\em without} self-avoidance) occurs, the membrane
will always be in the flat phase.

After these studies, several attempts were made to reduce the effective
rigidity of the membrane in order to observe a crumpled phase,
by using smaller balls \cite{r:KantorKremer93}
or by bond-dilution \cite{r:GrestMurat1990,r:PlischkeFourcade91}.
None of these attempts was successful. 
Another possibility to reduce the rigidity
is to impose self-avoidance not between balls but between faces 
of the lattice.
Using this method Baumg\"artner et al.\ found a crumpled swollen phase.
Kroll and Gompper later declared that also in
this model the membrane is flat \cite{r:KrollGompper1993}.
Finally let us remark that a crumpled phase has been found by balancing
the induced rigidity by long-range attractive interactions
\cite{r:LiuPlischke92,r:GrestPetsche94}.
We still want to mention the simulations of self-avoiding tethered membranes
in a space of 4, 5, 6 and 8 dimensions.
Grest \cite{r:Grest91} found that in 4 dimensions, the 
membrane is flat, but is crumpled in dimensions higher than 4. 
Subsequent simulations by 
Barsky and Plischke \cite{r:BarskyPlischke94} confirm this conclusion.

To summarize: For 3 dimensions most numerical simulations 
indicate that the membrane is in the flat phase, but the 
results are still not fully conclusive.
The simulations which find a crumpled phase give values for $\nu$ which are
close to our analytical estimates.
The situation seems to be similar in 4 dimensions.

A few experimental studies of tethered membranes have also been performed,
but the situation is  not clear neither.
The best studied system is graphite oxide, i.e.\ a mono-layer of carbon-atoms.
The first experiments by Hwa et al.\ \cite{r:graphitefrac} found a 
phase with fractal dimension $d_f=2.4$.
This result is contested in \cite{r:graphiteflat} by direct electron-microscopy
methods, although the latter authors obtain within the error-bars the same data
from their diffraction experiments.
It is not clear if these membranes are sloppy enough or if their 
internal stiffness is sufficiently large to induce in itself the transition towards a
flat phase.
Another system studied is the spectrin network of red blood cells \cite%
{r:Membrane Skeletons}, but the
role of disorder (which may induce a wrinkling transition) seems here important. 
In summary, the experiments are not yet helpful to clarify the situation.

\subsection{A possible scenario for low dimensions}

The fact that most numerical simulations only find a flat phase (with $\nu=1$)
for self-avoiding tethered membranes in 3 dimensions, while our
analytical estimates leads to $\nu<1$, has still to be explained.
As mentioned above, a possible explanation is that the analytical calculations
apply to floppy membranes (with very small bending rigidity),
while ```real systems"
studied numerically or experimentally are rigid enough to cross the crumpling
transition barrier and to stay in the flat phase.

However, the fact that reducing the rigidity does not induce the crumpling
transition suggest another, more drastic possibility.
It is known that for phantom tethered membranes, the lower critical dimension
$d_l$, below which the crumpled phase does not exist, lies between 1 and 2
($1<d_l<2$).
It is possible that, when self-avoidance is taken into account, the
lower critical dimension $d_l$ for the crumpling transition moves
upwards to about 3. 
In this case self-avoiding tethered membranes are always flat, however small
the rigidity is.

Let us give heuristic arguments that this is indeed the case in 3
dimensions.
Let us start from phantom membranes (without self-avoidance).
The fractal exponent $\nu$ is then 0 in the crumpled phase, 1 in the flat phase,
and equal to
\be \label{e:nu-c}
	\nu_c=1-\frac 1 d + {\cal O} \left( \frac 1 {d^2} \right)
\ee 
at the crumpling transition \cite{DavidGuitter}.
This last estimate is the result of a large $d$ expansion.
Its applicability to low dimensions is thus not clear a priori, but numerical
simulations \cite{r:KaNe87} show that even in 3 dimensions this approximation
is reasonable ($\nu_c=2/3$).
Let us now ask whether self-avoidance is relevant at the crumpling transition.
By naive power-counting we find that this is the case if
\be
	{\cal D} = 2\cdot 2 -\nu_c \,d > 0
\ee
i.e.\ with \eq{e:nu-c} for 
\be
	d<5 \ .
\ee
For ${\cal D}>0$ we expect that the fractal exponent $\nu_c$ at the
crumpling transition is different with or without self-avoidance.
But let us assume that self-avoidance does not change drastically
the exponent $\nu_c$ at the crumpling transition,
i.e.\ that 
\be \label{nuc approx}
\nu_c \approx \nu_{c+\mbox{\scr SA}}
\ee
where the second exponent is the exponent $\nu$ at the 
crumpling transition in the presence of self-avoidance.
This is certainly true for ${\cal D}$ small. 
As we expect that the radius of gyration of a self-avoiding membrane with rigidity
is an increasing function of the bending rigidity modulus $\kappa$, until
we reach the crumpling transition, it is simple to deduce the inequality 
between the fractal exponent of self-avoiding membranes in the crumpled phase
$\nu_{\mbox{\scr SA}}$ and at the crumpling transition $\nu_{c+\mbox{\scr SA}}$
\be
\nu_{c+\mbox{\scr SA}} \le \nu_{\mbox{\scr SA}} \ .
\ee
Assuming that \eq{nuc approx} holds, using the estimate \eq{e:nu-c} for
$\nu_{c+{\mbox{\scr SA}}}$ and our two-loop estimates for $\nu_{\mbox{\scr SA}}$, we find that this
basic inequality is violated if 
\be \label{3.8}
d< d_l\approx 3.8
\ee
Our interpretation is that $d_l$ is nothing but the lower critical dimension
of the crumpling transition with self-avoidance, and that for $d<d_l$ the
membrane is always in the flat, rigid phase.

We want to emphasize that this line of arguments is still somehow speculative.
It is however tempting to compare our crude estimate of $d_l\approx 3.8$
with the numerical simulations in 3, 4 and 5 dimensions which 
yield $4\le d_l \le 5$.
To confirm or disapprove this scenario, one must take into account the 
effect of the bending rigidity
in our renormalization group calculations.

\section{Conclusions}
In this article we presented  the first renormalization
group calculation at 2-loop order
for self-avoiding flexible tethered membranes. These second
order corrections were found to be surprisingly small if one 
uses an adequate extrapolation scheme.  

We were able to clarify the status of the Gaussian variational
method and to show that it becomes exact for $d\to \infty$.
For low dimensions the 2-loop results are in good agreement
with the prediction made by Flory's approximation, although
{\em systematically} slightly larger. 

In order to improve these results, one should understand  if the
plateau phenomenon observed at 2-loop order persists to higher orders, and
one should
control the general large order behavior of perturbation theory for this
model.
Another important issue is whether the IR-fixed point studied here is stable
towards perturbation by bending rigidity.
Indeed, for small enough $d$ this
might destabilize the crumpled phase and explain why numerical simulations
in $d=3$ normally see a flat phase.
We gave additional arguments which corroborate this scenario.

\subsection*{Acknowledgments}
We thank E. Guitter, G.S. Grest, T.Hwa and J. Zinn-Justin for useful discussions.

\begin{appendix}
\addtocontents{toc}{\protect\contentsline{section}
{\protect\numberline{ }\protect\Large \protect\bf Appendix}{ }}
\section{Normalizations}
\label{s:Normalizations}

We use peculiar normalizations in order to simplify the calculations.
First of all, we normalize the integration measure of the 
internal space as ($S_D $ is the volume of the 
$ D $-dimensional unit-sphere)
\begin{equation} \int^{ }_ x= {1 \over S_D} \int^{ }_{ } \mbox{d}^Dx\ ,\ \ \ \ S_D = 2 {\pi^{
D/2} \over \Gamma( D/2)} \label{s:Normalizations 2} 
\end{equation}
This provides 
\begin{equation}
	\int_x |x|^{\E-D} \Theta(|x|-L) =\frac1{\E} L^\E
\end{equation}
The $\delta$-distribution is normalized according to
\begin{equation} \tilde \delta ^ d(r(x)-r(y)) = (4\pi)^{ d/2}\delta ^ d(r(x)-r(y)) = \int^{ }_ p \mbox{e}^{ip(r(x)-r(y))} \label{24} \end{equation}
with
\begin{equation} \int_p= \pi^{ -d/2} \int^{ }_{ } \mbox{d}^dp \label{25} \end{equation}
to have
\begin{equation} \int^{ }_ p {\mbox{e}}^{-p^2a} = a^{-d/2} \ . \label{26} \end{equation}
Using for the free Hamiltonian
\be
	{\cal H}_0 = \frac{1}{2-D}\int_x \half \left(\nabla r(x)\right)^2
\ee
delivers
\be
\left< \half (r_i(x)-r_j(y))^2 \right>_0 = \delta_{ij} |x-y|^{2-D} \ .
\ee

\section{Example of the MOPE}
\label{s:Example of the MOPE}
\begin{figure}[htb]
\centerline{
	\epsfxsize=8.0cm \parbox{8.0cm}{\epsfbox{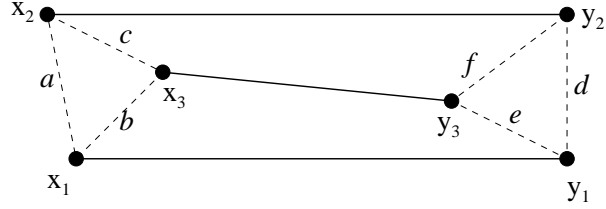}}
}
\caption{The distances and the points in (\protect{\ref{e:MOPE1}})}
\label{f:MOPE}
\end{figure}
We give as an explicit example of the MOPE the derivation of
\eq{e:ccr1:1}.
\bea \label{e:MOPE1}
\lefteqn{\hspace{-1cm} _{x_1}\!\!\GB_{y_1} \ 
 _{x_2}\!\!\GB_{y_2} \ 
 _{x_3}\!\!\GB_{y_3} \ } \nn \\
&=& \int_k\int_p\int_q 
	:\rme^{ikr(x_1)}:\, :\rme^{ipr(x_2)}:\, :\rme^{iqr(x_3)}:\ 
	:\rme^{-ikr(y_1)}:\, :\rme^{-ipr(y_2)}:\, :\rme^{-iqr(y_3)}:\,
\eea
These exponentials shall be contracted like
\be
\GB \GB \GB \ \longrightarrow	\ \GJ
\ee
We therefore use the OPE for the points $x_1$, $x_2$ and $x_3$,
supposed the differences between these points become small:
\be \label{e:MOPE2}
	:\rme^{ikr(x_1)}:\, :\rme^{ipr(x_2)}:\, :\rme^{iqr(x_3)}:
	\ =\ :\rme^{ikr(x_1)+ipr(x_2)+iqr(x_3)}:
	\rme^{kp\,a^{2\nu}+kq \,b^{2\nu} + pq\, c^{2\nu} }
\ee
The new variables for the distances between the points are given
in figure \ref{f:MOPE}.
An analogous expansion is valid for  $y_1$, $y_2$ and $y_3$. 
In order to retain only the most important contribution,
we expand
\be
	:\rme^{ikr(x_1)+ipr(x_2)+iqr(x_3)}: \ =\ 
:\rme^{i(k+p+q)r\left((x_1+x_2+x_3)/3\right)}
\left( 1+ {\cal O}(\nabla r) \right):
\ee
and neglect the contributions of order ${\cal O}(\nabla r)$ because
they are proportional to  irrelevant operators.
After a shift in the integration-variable $q$,
$$
 q \longrightarrow q-k-p \ ,
$$
equation \eq{e:MOPE1} becomes:
\bea
&&\hspace{-1cm} \int_q :\rme^{iq r\left((x_1+x_2+x_3)/3\right) }:\,
: \rme^{-iq r\left( (y_1+y_2+y_3)/3 \right)}: \times
\weiter
\times \int_k\int_p
\rme^{kp (a^{2\nu}+d^{2\nu}) +k(q-k-p) (b^{2\nu}+e^{2\nu}) + 
p(q-k-p) (c^{2\nu}+f^{2\nu}) }
\eea
The integral over $q$ yields the $\delta$-distribution plus higher 
derivatives of this distribution. The latter are irrelevant operators
and can be neglected. As they come from the expansion 
of the last exponential factor in $q$, we only have to retain 
the last factor, evaluated at $q=0$. This gives:
\bea
&&\hspace{-1cm}	\int_k\int_p
\rme^{kp (a^{2\nu}+d^{2\nu}) -k(k+p) (b^{2\nu}+e^{2\nu})  
-p(k+p) (c^{2\nu}+f^{2\nu}) } \nn \\ 
&& = \left[ (b^{2\nu}+e^{2\nu})(c^{2\nu}+f^{2\nu})
-\frac14  \left( b^{2\nu}+e^{2\nu} + c^{2\nu}+f^{2\nu}-a^{2\nu}-d^{2\nu}
\right) \right]^{-d/2}
\eea
This can still be factorized as is known from ancient Heron:
\begin{eqnarray}
& & \hspace{-0.3cm} \Bigg( \GJ \Bigg| \GB \Bigg)
	=	\nonumber \\
& & \Bigg[ \frac{1}{4} \left( \sqrt{a^{2 \nu}+d^{2 \nu}} +\sqrt{b^{2 \nu}+e^{2 \nu}}
	+\sqrt{c^{2 \nu}+f^{2 \nu}}\right) 
	\left( \sqrt{b^{2 \nu}+e^{2 \nu}} + \sqrt{c^{2 \nu}+f^{2 \nu}} 
          - \sqrt{a^{2 \nu}+d^{2 \nu}}\right) \nonumber \\
& &  \left( \sqrt{a^{2 \nu}+d^{2 \nu}} +\sqrt{c^{2 \nu}+f^{2 \nu}}
          -\sqrt{b^{2 \nu}+e^{2 \nu}}\right) 
   \left( \sqrt{a^{2 \nu}+d^{2 \nu}} +\sqrt{b^{2 \nu}+e^{2 \nu}}
          -\sqrt{c^{2 \nu}+f^{2 \nu}}
          \right) 
   \Bigg]^{-d/2} \nonumber \\
\label{e:ccr1:1'}
& & 
\end{eqnarray}

\section{Equation of Motion}
\label{s:Equation of Motion}
The equation of motion reflects the invariance of the functional integral
under an change of variables for the fields.
The expectation value of an observable 
$\cal O$ in the free theory is:
\be \label{e:EM1}
\left< {\cal O} \right>_0 = \frac{\int {\cal D} \left[r\right] 
{\cal O}\, \rme^{-\frac1{2-D} \int_x \GO}   }
{\int {\cal D} \left[r\right] \rme^{-\frac1{2-D} \int_x \GO}}
\ee
We now perform a global rescaling of $r(x)$:
\be
	r(x) \longrightarrow (1+\kappa) r(x) \ .
\ee
The expectation value  of $\cal O$, equation \eq{e:EM1} is unchanged.
The explicit form becomes up to first order in $\kappa$:
\be \label{e:EM2}
\left< {\cal O} \right>_0 = \frac{\int {\cal D} \left[r\right] 
{\cal O} \left(1+ \kappa \left[ {\cal O} \right]_r \right)
\left( 1-\frac{2 \kappa}{2-D} \int_x \GO   \right)
\rme^{-\frac1{2-D} \int_x \GO}   }
{\int {\cal D} \left[r\right] \left( 1-\frac{2 \kappa}{2-D} \int_x \GO   \right)\rme^{-\frac1{2-D} \int_x \GO}}	
\ee 
$\left[ {\cal O}\right]_r$ is the canonical dimension of the operator ${\cal O}$,
measured in units of $r$, i.e.\ that $\left[ r\right]_r=1$. 
Calculating the difference of \eq{e:EM1} and \eq{e:EM2} gives:
\begin{equation}
\bigg<{\cal O}\,\int \GO \bigg>_0^{\mbox{\scriptsize conn}} = \nu \left[ {\cal O}\right]_r \bigg<{\cal O} \bigg>_0^{\mbox{\scriptsize conn}}
\end{equation}
For several operators we have:
\begin{equation}
\bigg<{\cal O}_1{\cal O}_2 \,\int \GO \bigg>_0^{\mbox{\scriptsize conn}} = \nu (\left[ {\cal O}_1\right]_r+\left[ {\cal O}_2\right]_r )\bigg<{\cal O}_1{\cal O}_2 \bigg>_0^{\mbox{\scriptsize conn}}
\end{equation}
and in particular
\begin{eqnarray}
\bigg< \GB \,\int \GO \bigg>_0^{\mbox{\scriptsize conn}} &=& -\nu d \bigg<\GB \bigg>_0^{\mbox{\scriptsize conn}} \\
\bigg<{\cal O} \, \GO  \int \GO \bigg>_0^{\mbox{\scriptsize conn}} &=& \left(2 + \left[ {\cal O} \right]_r \right) \nu  \bigg<{\cal O} \GO \bigg>_0^{\mbox{\scriptsize conn}}
\end{eqnarray}
These relations are equivalently valid for non-connected expectation values.
(To prove this, remark that $\langle \GO \rangle_0=0$.) 

\noindent We can now apply these equations to calculate the pole-term of
$\bigg< \GX  \big| \GB \bigg>_L$. The equation of motion yields
\be \label{e:EM3}
\left<\int\!\!\!\int \int\!\!\!\int  \int \GB \ \GB \  \GO \right>_0 = -2 \nu d \left< \int\!\!\!\int \int\!\!\!\int \GB \ \GB \right>_0
\ee
On the l.h.s.\ the possible divergences proportional to $\GB$ come 
from the following integrals:
\be \label{e:EM4}
4 \left< \int\!\!\!\int \int\!\!\!\int \int \GX \right>_0 + 
2 \left<\int\!\!\!\int \int\!\!\!\int \int   \GM \GO \right>_0 
\ee
In the second term, $\GO$ has to have some distance  $L$ from the endpoints of
$\GM$. 

The pole-terms proportional to $\GB$ appearing on the r.h.s.\ of \eq{e:EM3} 
and in \eq{e:EM4} are:
\bea
\left< \int\!\!\!\int \int\!\!\!\int \GB \ \GB \right>_0 &=& 
\left( 2\bigg< \GM \bigg| \GB \bigg>_{\E^{-1}} +\cal{O} (\E^0) \right) 
\left< \int\!\!\!\int \GB \right>_0 
 \quad\\
\left< \int\!\!\!\int \int\!\!\!\int \int \GX \right>_0 
&=& \left( \bigg< \GX \bigg| \GB \bigg>_{\E^{-1}}
+\cal{O} (\E^0) \right) 
\left< \int\!\!\!\int \GB \right>_0
  \quad \weiter
\\
\left<\int\!\!\!\int \int\!\!\!\int \int   \GM \GO \right>_0
&=& -\nu d \left( \bigg< \GM \bigg| \GB \bigg>_{\E^{-1}} 
+\cal{O} (\E^0) \right)
\left< \int\!\!\!\int \GB \right>_0 
 \quad
\weiter
\label{e:EM7}
\eea
In the last equation, we used that 
\be
\left<\int\!\!\!\int \int\!\!\!\int \int   \GM \GO \right>_0
=\left< \int\!\!\!\int \int \FE \GO \right>_0
\times \left[ \bigg< \GM \bigg| \GB \bigg>_{\E^{-1}} +\cal{O} (\E^0) 
\right]
\ee
and that 
\be
\left< \int\!\!\!\int \int \FE \GO \right>_0
= \left< \int\!\!\!\int \int \GB \GO \right>_0
-2 \left< \int\!\!\!\int \int \FF \right>_0
\ee
The second term has no contribution proportional to $\GB$ and thus can 
be neglected. The first is evaluated using the equation of motion:
\be
\left< \int\!\!\!\int \int \GB \GO \right>_0 = -\nu d \left< \int\!\!\!\int
 \GB  \right>_0
\ee
Together this yields \eq{e:EM7}. 

Finally using equations \eq{e:EM3} to \eq{e:EM7} gives:
\be
\bigg< \GX \bigg| \GB \bigg>_{\E^{-1}} = -\frac{\nu d}2 \bigg< \GM \bigg| \GB \bigg>_{\E^{-1}} \cal{O} (\E^0) 
\ee
We further have to show that
\be \label{e:EM11}
\bigg<\GZ\bigg|\GO\bigg>_{\E^{-1}} = - \nu (d+2) 
\bigg< \GH \bigg| \GO \bigg>_{\E^{-1}}
+ \bigg< \GH \bigg| \GO \bigg>_{\E^{-1}}\bigg< \GW \bigg| \GO \bigg>_{\E^{0}}
\ .
\ee
The equation of motion yields
\be \label{e:EM5}
\left<\int\!\!\!\int \int \GB \  \GO \right>_0 
	= - \nu d \left<  \int\!\!\!\int \GB  \right>_0 \ .
\ee
The integral on the l.h.s.\ of \eq{e:EM5} can be decomposed as:
\be \label{e:EM8}
\left<\int\!\!\!\int \int \GZ \right>_0 
+ \left<\int\!\!\!\int \int \GH \  \GO \right>_0
\ee
The divergencies of the r.h.s.\ of \eq{e:EM5} and of 
the first term in \eq{e:EM8} are:
\bea
\left<  \int\!\!\!\int \GB  \right>_0 &=& 
	\bigg< \GH \bigg| \GO \bigg>_{\E^{-1}} 
\left< \int \GO \right>_0 
+\cal{O} (\E^0) \\ 
\left<\int\!\!\!\int \int \GZ \right>_0 &=& 
	\bigg< \GZ \bigg| \GO \bigg>_{\E^{-1}}\left< \int \GO \right>_0 
 +\cal{O} (\E^0) 
\eea
The divergence of the second term in \eq{e:EM8} is extracted
as 
\be \label{e:EM9}
 \left<\int\!\!\!\int \int \GH \  \GO \right>_0 =
 	\bigg< \GH \bigg| \GO \bigg>_{\E^{-1}} 
\left<\int \int \FG \  \GO \right>_0 
\ee
The second factor on the r.h.s. of \eq{e:EM9} is 
\bea
\left<\int \int \FG \  \GO \right>_0 
&=&
\left<\int \int \GO \  \GO \right>_0 
-	\left<\int \int \GW \right>_0 \nn \\
&=& 2 \nu \left< \int  \GO  \right>_0
-	\left<\int \int \GW \right>_0 \ ,
\label{e:EM10}
\eea
where again the equations of motion are used.
Using \eq{e:EM5} to \eq{e:EM10} yields \eq{e:EM11}.

\end{appendix}  

\addtocontents{toc}{\protect\contentsline{section}
{\protect\numberline{ }\protect\bf References}{\protect\pageref{references}}}
\newcommand{\titofart}[1]{}

\label{references}

\end{document}